\documentclass[11pt,urlcolor=blue, linkcolor=blue]{article} 
\usepackage{cite}
\usepackage{amsmath, amsthm, amssymb,slashed} 
\usepackage{ifpdf}
\ifpdf
  \usepackage[pdftex]{graphicx}
  \usepackage{epstopdf}
\else
  \usepackage[dvips]{graphicx}
\fi
\parskip=\baselineskip

\allowdisplaybreaks[1]

\usepackage{comment}

\newcommand{\cgrn}[1]{\textcolor{black}{#1}}
\newcommand{\cblue}[1]{\textcolor{black}{#1}}
\newcommand{\cred}[1]{\textcolor{black}{#1}}
\newcommand{\cblack}[1]{\textcolor{black}{#1}}
\usepackage{upgreek}

\newcommand{\GSD}{\text{GSD}}
\usepackage{tabularx}
\numberwithin{equation}{section}
\newcommand{\I}{\hspace{1pt}\mathrm{I}\hspace{1pt}}
\newcommand{\II}{\hspace{1pt}\mathrm{II}\hspace{1pt}}
\newcommand{\III}{\hspace{1pt}\mathrm{III}\hspace{1pt}}
\newcommand{\IV}{\hspace{1pt}\mathrm{IV}\hspace{1pt}}
\DeclareMathAlphabet{\mathpzc}{OT1}{pzc}{m}{it}

\newcommand{\bea}{\begin{eqnarray}}
\newcommand{\eea}{\end{eqnarray}}
\def\be{\begin{equation}}
\def\eeq{\end{equation}}

\usepackage{array,multirow}
\newcommand{\tD}{\mathrm{D}}
\usepackage{mathrsfs}

\newcommand{\ee}{\hspace{1pt}\mathrm{e}}
\def\bra#1{\left\langle#1\right|}
\def\ket#1{\left|#1\right\rangle}
\def\tx{{\mathrm{x}}}
\def\rh{{\mathrm{h}}}
\def\rf{{\mathrm{f}}}

\usepackage{color}
\usepackage[colorlinks,citecolor=blue]{hyperref}
\definecolor{red}{rgb}{1,0,0}
\definecolor{blue}{rgb}{0,0,1}
\definecolor{dblue}{rgb}{0,0,0.4}
\definecolor{green}{rgb}{0,1,0}
\definecolor{black}{rgb}{0,0,0}
\definecolor{white}{rgb}{1,1,1}

\definecolor{brn}{rgb}{.8,.4,.0}
\definecolor{redo}{rgb}{1,.5,.0}
\definecolor{ddgrn}{rgb}{0,0.4,0}
\definecolor{dgrn}{rgb}{0,0.6,0}
\definecolor{dbl}{rgb}{0,0,0.5}

\usepackage[bbgreekl]{mathbbol}
\usepackage{amscd}

\newcommand{\Z}{\mathbb{Z}}

\newcommand{\R}{\mathbb{R}}

\renewcommand{\v}[1]{\boldsymbol{#1}} 
\renewcommand{\t}[1]{\tilde{#1}} 
\newcommand{\ii}{\hspace{1pt}\mathrm{i}\hspace{1pt}}
\newcommand{\dd}{\hspace{1pt}\mathrm{d}}
\newcommand{\<}{\langle} 
\renewcommand{\>}{\rangle} 

\newcommand{\Ref}[1]{Ref.~\cite{#1}}

\newcommand{\eqn}[1]{eqn.~(\ref{#1})} 
\newcommand{\Eqn}[1]{eqn.~(\ref{#1})}

\renewcommand{\Im}{{\rm Im}}

\newcommand{\prt}{\partial}

\newcommand{\up}{\uparrow} 
\newcommand{\down}{\downarrow} 

\newcommand{\ie}{{i.e.~}}

\newcommand{\bpm}{\begin{pmatrix}}
\newcommand{\epm}{\end{pmatrix}}
\newcommand{\bmm}{\begin{matrix}}
\newcommand{\emm}{\end{matrix}}

\newcommand{\cH}{ {\cal H} } 
 
\newcommand{\cL}{ {\cal L} }

\newcommand{\cS}{ {\cal S} } 
\newcommand{\cT}{ {\cal T} } 
\newcommand{\cV}{ {\cal V} }

\newcommand{\del}{\delta}

\newcommand{\la}{\lambda} 
 
\newcommand{\om}{\omega} 
 
\renewcommand{\th}{\theta} 
 
\newcommand{\si}{\sigma}

\def\CM{{\cal M}}

\def\CU{{\cal U}}
\def\CV{{\cal V}}

\def\Z{{\mathbb{Z}}}

\def\R{{\mathbb{R}}}

\def\neg{\negthinspace}
\def\ta{{\mathrm{a}}}

\newcommand{\symfootnote}[1]{%
\let\oldthefootnote=\thefootnote%
\stepcounter{mpfootnote}%
\addtocounter{footnote}{-1}%
\renewcommand{\thefootnote}{\fnsymbol{mpfootnote}}%
\footnote{#1}%
\let\thefootnote=\oldthefootnote%
}

\newcommand{\sharpfootnote}[1]{%
\let\oldthefootnote=\thefootnote%
\stepcounter{mpfootnote}%
\addtocounter{footnote}{-1}%
\renewcommand{\thefootnote}{{$\sharp$}}
\footnote{#1}%
\let\thefootnote=\oldthefootnote%
}

\newcommand{\naturalfootnote}[1]{%
\let\oldthefootnote=\thefootnote%
\stepcounter{mpfootnote}%
\addtocounter{footnote}{-1}%
\renewcommand{\thefootnote}{{$\natural$}}
\footnote{#1}%
\let\thefootnote=\oldthefootnote%
}

\newcommand{\flatfootnote}[1]{%
\let\oldthefootnote=\thefootnote%
\stepcounter{mpfootnote}%
\addtocounter{footnote}{-1}%
\renewcommand{\thefootnote}{{$\flat$}}
\footnote{#1}%
\let\thefootnote=\oldthefootnote%
}

\begin{document}
\begin{titlepage}
\begin{flushright}
\end{flushright}
\vskip 1.05in
\begin{center}

{\bf\LARGE{Symmetric Gapped Interfaces of \\[2.5mm]
SPT and SET States: Systematic Constructions}}

\vskip0.5cm 
\Large{Juven Wang$^1$\sharpfootnote{e-mail: {\tt juven@ias.edu}}, 
Xiao-Gang Wen$^2$\naturalfootnote{e-mail: {\tt xgwen@mit.edu}}
 and Edward Witten$^1$\flatfootnote{e-mail: {\tt witten@ias.edu}\\[4mm] 
The arXiv:1705.06728 version is the original untrimmed version.\\
The Phys. Rev. X 8, 031048 (2018) published version is the trimmed version.
 } } 
\vskip.5cm
 {\small{\textit{$^1$School of Natural Sciences, Institute for Advanced Study, Einstein Drive, Princeton, NJ 08540, USA}}}
 \vskip.2cm
 {\small{\textit{$^2$Department of Physics, Massachusetts Institute of Technology, Cambridge, MA 02139, USA \\}}
}
\end{center}
\vskip.5cm
\baselineskip 16pt
\begin{abstract}

Symmetry protected topological (SPT) states have boundary 't Hooft anomalies that obstruct an effective boundary theory realized in its own dimension with UV completion and with an on-site $G$-symmetry. In this work, yet we show that a certain anomalous non-on-site $G$ symmetry along the boundary becomes on-site when viewed as an extended $H$ symmetry, via a suitable group extension $1\to K\to H\to G\to1$. Namely, a non-perturbative global (gauge/gravitational) anomaly in $G$ becomes anomaly-free in $H$. 
This guides us to construct exactly soluble lattice path integral and Hamiltonian of symmetric gapped boundaries,  
\emph{always existent} for \emph{any} SPT state in any spacetime dimension $d \geq 2$ of \emph{any} finite symmetry group, including on-site unitary and anti-unitary time-reversal symmetries. The resulting symmetric gapped boundary can be described either by an $H$-symmetry extended boundary of bulk $d \geq 2$, or more naturally by a topological {emergent} $K$-gauge theory with a global symmetry $G$ on a 3+1D bulk or above. The excitations on such a symmetric topologically ordered boundary can carry fractional quantum numbers of the symmetry $G$, described by representations of $H$. (Apply our approach to a 1+1D boundary of 2+1D bulk, we find that a deconfined gauge 
boundary indeed has \emph{spontaneous symmetry breaking} with long-range order. The deconfined symmetry-breaking phase crosses over smoothly to a confined phase without a phase transition.) In contrast to known gapped boundaries/interfaces obtained via \emph{symmetry breaking} (either global symmetry breaking or Anderson-Higgs mechanism for gauge theory), our approach is based on \emph{symmetry extension}. More generally, applying our approach to SPT states, topologically ordered gauge theories and symmetry enriched topologically ordered (SET) states, leads to generic boundaries/interfaces constructed with a mixture of \emph{symmetry breaking}, \emph{symmetry extension}, and \emph{dynamical gauging}.\\




\end{abstract}
\end{titlepage}



\tableofcontents   


\section{Introduction}

After the realization that a spin-1/2 antiferromagnetic {Heisenberg} chain in
1+1 dimensions (1+1D) admits a gapless state \cite{LSM6107,KMc9809163} that
``nearly'' breaks the spin rotation symmetry (\ie it has  ``symmetry-breaking''
spin correlation functions that decay algebraically), many physicists expected
that  spin chains with higher spin, having less quantum fluctuations, might
also be gapless with algebraic long-range spin order.  However, Haldane
\cite{H8364} first realized that antiferromagnetic Heisenberg spin  chains in 1+1D 
with integer spins have a gapped disordered phase with short-range
spin correlations.  At first, it was thought that those states are trivial
disordered states, like a product state of spin-0 objects.  Later, it was
discovered that they can have degenerate zero-energy  modes at the ends of the chain
\cite{AKL8877}, similar to the gapless edge states of quantum Hall systems.  This discovery 
led to a suspicion  that these gapped  phases of antiferromagnetic integer spin
chains might be topological phases.  

Are Haldane phases topological or not topological? What kind of ``topological'' is it? That was the question.  It
turns out that only odd-integer-spin Haldane phases (each site with an odd-integer spin) are topological, while the even-integer-spin Haldane phases (each
site with an even-integer spin) are really trivial (a trivial vacuum ground state
like the  product state formed by spin-0's).  The essence of nontrivial odd-integer-spin Haldane phases
was obtained in \Ref{GW0931}, based on a tensor network renormalization
calculation \cite{LN0701}, \cgrn{where simple fixed-point tensors
characterizing quantum phases can be formulated.  It was discovered that the spin-1
Haldane phase is characterized by a non-trivial fixed-point tensor --
a corner-double-line tensor. The corner-double-line structure implies that the spin-1
Haldane phase is actually equivalent to a product state, once we remove its global symmetry.  However \Ref{GW0931}
showed that the corner-double-line tensor is robust against any local
perturbations that preserve certain symmetries (namely, $SO(3)$ symmetry in the
case of the integer spin chain), but it flows to the trivial fixed point
tensor if we break the symmetry.}  This suggests that, in the presence of
symmetry, even a simple product state can be non-trivial (\ie, distinct from the
product state of spin-0's that has no corner-double-line structure), and such
non-trivial symmetric product states were named Symmetry Protected Topological
states (SPTs).  (Despite its name, an SPT state has no \emph{intrinsic
topological order} in the sense defined in \Ref{W9039,CGW1038}. By
this definition, an SPT state with no topological order cannot be deformed
into a trivial disordered gapped phase in a symmetry-preserving fashion.)

Since SPT states are equivalent to simple product states if we remove their global symmetry, one quickly obtained
their classification in 1+1D \cite{XieSPT1,FK1103,SPC1139}, in terms of
projective representations \cite{PBT1039} of the symmetry group $G$.  As
remarked above, one found that only the  {odd-integer-spin} Haldane phases are
non-trivial SPT states.  The  {even-integer-spin} Haldane phases are trivial
gapped states, just like the disordered product state of spin-0's \cite{PBT1225}.
Soon after their classification in 1+1D, bosonic SPT states in higher
dimensions were also classified based on group cohomology $\cH^{d+1}(G,U(1))$
and $\cH^{d+1}(G\times SO(\infty),U(1))$
\cite{XieSPT3,XieSPT4,XieSPT5,Wen1410.8477}, \footnote{ For $d+1$D SPT
states (possibly with a continuous symmetry), here we use the Borel group
cohomology $\cH^{d+1}(G,U(1))$ or $\cH^{d+1}(G\times SO(\infty),U(1))$ to
classify them \cite{XieSPT3,Wen1410.8477}. Note that $\cH^{d+1}(G,U(1)) =
H^{d+2}(BG,\Z)$, where $H^{d+2}(BG,\Z)$ is the topological cohomology of the
classifying space $BG$ of $G$.  When $G$ is a finite group, we have only the torsion part
$\cH^{d+1}(G,U(1))= H^{d+2}({BG,\Z})=H^{d+1}(BG,U(1))$.}  or based on
cobordism theory \cite{K1467, K1459, KTT1429}.  In fact, SPT states and
Dijkgraaf-Witten gauge theories \cite{DW9093} are closely related: Dynamically
gauging the global symmetry  \cite{LG1209,HW1267} in a bosonic SPT state  leads to a corresponding
Dijkgraaf-Witten bosonic topological gauge theory.

To summarize, SPT states are the \emph{simplest} 
of 
symmetric phases and, accordingly, have another name Symmetry Protected Trivial
states.  They are quantum-disordered product states that do not break the
symmetry of the Hamiltonian.  Naively, one would expect that such disordered product
states 
all have 
non-fractionalized bulk excitations.  What is nontrivial about an SPT state is more
apparent if one considers its possible boundaries.  For any bulk gapped theory
with $G$ symmetry, a $G$-preserving boundary is described by some effective
boundary theory with symmetry $G$.  However, the  boundary  theories of
different SPT states have different 't Hooft anomalies in the global symmetry $G$
\cite{RML1204,W1313,Kapustin:2014zva, Wang1405.7689}. %
A simple explanation follows:
While the bulk of SPT state of a symmetry group G has an onsite symmetry, 
the boundary theory of SPT state has an effective non-onsite G-symmetry. 
Non-onsite G-symmetry means that
the G-symmetry does not act in terms of a tensor product structure on each site,
namely the G-symmetry acts non-locally on several effective boundary sites. 
Non-onsite symmetry cannot be dynamically gauged ---
because conventionally the gauging process requires inserting gauge variables on the links between the local site variables of G-symmetry. 
Thus the boundary of SPT state of a symmetry G has an obstruction to gauging, as 't Hooft anomaly obstruction to gauging a global symmetry \cite{H8035}.
Such an anomalous boundary
is the essence of SPT states:  Different boundary anomalies characterize
different bulk SPT states. In fact, different SPT states classify gauge
anomalies and mixed gauge-gravity anomalies in one lower dimension \cite{W1313,
Kapustin:2014zva, Wang1405.7689}.\footnote{
Thus, more precisely, as explained above,
different SPT states have different 't Hooft anomalies on the boundary.
In this article, when we say gauge anomalies and mixed gauge-gravity anomalies on the boundaries of SPT states,
we mean the 't Hooft anomalies of global symmetries or spacetime diffeomorphisms, coupling to non-dynamical background probed field or background probed gravity. So the 
gauge anomalies and mixed gauge-gravity anomalies (on boundaries of SPTs) mean to be the 
\emph{background} gauge anomalies and mixed \emph{background} gauge-gravity anomalies:
Both the gauge fields and gravitational fields are \emph{background} non-dynamical probes.
}

From the above discussion, we realize that to understand the physical
properties of SPT states is to understand the physical consequence of anomalies
in the global symmetry $G$ on the boundary of SPT states, somewhat as in work
of 't Hooft on gauge theory dynamics in particle physics \cite{H8035}.  For a
1+1D boundary, it was shown that the anomalous global symmetry makes the
boundary gapless and/or symmetry breaking \cite{XieSPT3}.  However, in higher
dimensions, there is a third possibility: the boundary can be gapped,
symmetry-preserving, and topologically ordered. (This third option is absent
for a 1+1D boundary roughly because there is no bosonic topological
order in that dimension.\footnote{
Here we mean that there is no intrinsic 1+1D topological order in bosonic systems, neither in its own dimension nor on the boundary of any 2+1D bulk short-range entangled state.
(Namely, we may say that there is no 1+1D bosonic topological quantum field theory robust against any local perturbation.) 
However, the 1+1D boundary of a 2+1D bulk long-range entangled state may have an intrinsic topological order.
Moreover, in contrast, in a fermionic system, there is a 1+1D 
fermionic chain \cite{K0131} 
with an intrinsic fermionic topological order. 
})  
Concrete examples of  topologically ordered
symmetric boundaries have been constructed in particular cases \cite{VS1306,
BCF1372,CFV1350, WangPotterSenthil1306.3223, 1306.3286MMetlitskiKaneFisher,
1306.3230BondersonNayakQi, MV14063032, Mross1410.4201,WL151209111}.  In this
paper, we give a systematic construction that applies to any SPT state
with any finite\footnote{The symmetries may be ordinary unitary symmetries, or
may include anti-unitary time-reversal symmetries.} symmetry group $G$, for any
boundary of bulk dimension $2+1$ or more.  Namely, we show that symmetry-preserving
gapped boundary states \emph{always exist} for any $d+1$D bosonic SPT state
with a finite symmetry group $G$ when $d \geq 3$.  \cblue{We also study a few
examples, but less systematically, when SPT states have continuous compact Lie
groups $G$, and we study their symmetry-preserving gapped boundaries, which may or
may not exist.}

Symmetry-breaking gives a straightforward way to construct gapped boundary
states or interfaces, since SPT phases are completely trivial if one ignores
the symmetry.  For topological phases described by group cocycles
of a group $G$, the \emph{symmetry-breaking} mechanism can be described as
follows.  It is based on breaking the $G$ to a subgroup $G' \subseteq G$,
corresponding to an injective homomorphism $\iota$ as
\bea \label{inj}
G'  \overset{\iota}{\rightarrow} G.
\eea
Here $G'$ must be such that the cohomology class in $\cH^{d+1}(G,U(1))$ 
that characterizes the $d+1$D SPT or SET state becomes  trivial when pulled
back  (or equivalently restricted) to  $G'$.  The statement that the class is
``trivial'' does not mean that the relevant $G$-cocycle is 1 if we restrict its
argument from $G$ to $G'$, but that this cocycle becomes a coboundary when
restricted to $G'$.

Our approach to constructing \cgrn{exactly soluble gapped boundaries}
does not involve symmetry breaking but what one might call 
\emph{symmetry extension}:  
\bea\label{gext}
1 \to {K} \to H \overset{r}{\rightarrow} G \to 1.
\eea
Here we extend $G$ to a larger group $H$, such that $G$ is its quotient group,
$K$ is its normal subgroup, and $r$ is a surjective group homomorphism, more or
less opposite to the injective homomorphism $\iota$ related to symmetry
breaking (eqn. (\ref{inj})).  $H$ and $r$ must be such that the cohomology
class in $\cH^{d+1}(G,U(1))$ that characterizes the SPT or SET state becomes
trivial when pulled back  to $H$.  \cred{For any finite $G$ and any class in
$\cH^{d+1}(G,U(1))$, we show that suitable choices of $H$ and $r$ always exist, when the bulk space dimension $d \geq 1$.}  
Physically the gapped phases that we construct in this way have the property that boundary
degrees of freedom transform under an $H$ symmetry.  However, in condensed
matter applications, one should usually\footnote{See Sec.~\ref{bdry2w} for an
example in which it is natural in condensed matter physics to treat $K$ as a
global symmetry. See also a more recent work Ref.~\cite{2018arXiv180411236P} applying the idea to 1+1D bosonic/spin chains or fermionic chains.}  assume that the subgroup $K$ of $H$ is gauged, and then (in
the SPT case) the global symmetry acting on the boundary is $G$, just as in the
bulk.   So in that sense, when all is said and done the boundary states that we
construct simply have the same global symmetry as the bulk, and the boundaries
become topological since $K$ is gauged.  For 2+1D (or higher dimensional)
boundaries, such symmetry preserving topological boundaries may have excitations
with fractional $G$-symmetry quantum numbers.  The fact that the boundary
degrees of freedom are in representations of $H$ rather than $G$ actually
describes such a charge fractionalization.

The idea behind this work was described in a somewhat abstract way in Sec.~3.3 
of Ref.\cite{Witten:2016cio}, and a similar idea was used in
Ref.\cite{Seiberg:2016rsg} in examples.  In the present paper, we develop this
idea in detail and in a down-to-earth way, with both spatial lattice
Hamiltonians and spacetime lattice path integrals that are ultraviolet (UV)
complete at the lattice high energy scale.  We also construct a mixture
combining the \emph{symmetry-breaking} and \emph{symmetry-extension}
mechanisms.

We further expand our approach to construct anomalous gapped symmetry-preserving interfaces (i.e. domain walls) between 
bulk SPT states, topological orders (TO) and symmetry enriched topologically ordered states (SETs).
\footnote{We remark that our approach to constructing gapped boundaries may \emph{not} be applicable to some 
invertible topological orders (iTO, or the invertible topological quantum field theory [TQFT])  protected by \emph{no global symmetry}.
However, the gapped boundaries of certain iTO can still be constructed via our approach: For example,  the 4+1D 
 iTO with a topological invariant $(-1)^{\int w_2 w_3}$ has a boundary anomalous 3+1D $Z_2$ gauge theory.
 Here $w_i \equiv w_i(TM)$ is the $i$-th Stiefel-Whitney class of a tangent bundle $TM$ over spacetime $M$.
}
We will  recap the terminology for the benefit of some readers. SPTs are short-range entangled (SRE) states, which can be deformed to a trivial product state
under local unitary transformations at the cost of breaking some protected global symmetry. 
Examples of SPTs include topological insulators \cite{HasanKaneRMP,QiZhangRMP, bernevig2013topological}. 
Topological orders are long-range entangled (LRE) states, which cannot be deformed to a trivial product state under local unitarity transformations 
even if breaking all global symmetries.
SETs are topological orders  -- thus LRE states--  but additionally have some global symmetry.
Being long-range entangled, TOs and SETs have richer physics and mathematical structures than the short-range entangled SPTs. 
Examples of TOs and SETs include fractional quantum Hall states and quantum spin liquids \cite{SavaryBalents1601.03742}.
In this work, for TOs and SETs, we mainly focus on those that can be described by Dijkgraaf-Witten twisted gauge theories, possibly extended with global symmetries.
We comment on possible applications and generalizations to gapped interfaces of
bosonic/fermionic topological states obtained from beyond-group cohomology and cobordism theories in
  Sec.~\ref{sec:beyondGC}-\ref{sec:Cob}.
  
\subsection{Summary of physical results} 

In this article, we study a certain type of boundaries for $G$-SPT states with
a $G$-symmetry. This type of boundary is obtained by adding new degrees of
freedom along the boundary that transform as a representation of a properly
extended symmetry group $H$ via a group extension $1\to K\to H\to G\to 1$ with
a finite $K$.  Such an $H$-\emph{symmetry extended} (or \emph{symmetry
enhanced}) boundary can be fully gapped with an $H$-symmetry, but without any
topological order on the boundary. 
The last column in Tables \ref{table:example-app-1}-\ref{table:example-app-4}
describes such symmetry extension.

%

Moreover, there is another type of boundary, obtained by gauging the normal
subgroup $K$ of $H$. This type of boundary is described by a deconfined
$K$-gauge theory and has the same $G$-symmetry as the bulk.
We have constructed exactly soluble model to realize such type of boundaries
for any SPT state with a finite group $G$ symmetry, and for some SPT states
with a continuous group $G$ symmetry.  Tables \ref{table:example-app-1},
\ref{table:example-app-2},  \ref{table:example-app-3} and
\ref{table:example-app-4} summarize physical properties of this type of
boundaries obtained from exactly soluble models for various $G$-SPT states in
various dimensions.

\begin{table}[!h]
\makebox[\textwidth][c]{
\begin{tabular}{|c|c||c|c|c|}
\hline
Symm. group & \;  $\begin{matrix}\text{1+1D SPT Bulk inv.}\\ \text{($d$-cocycle $\omega_d$)} \end{matrix}$ & End-point states &\; $1\to K \to H \to G \to 1$   \\ \hline
$Z_2^2$: $\begin{matrix}\text{\ref{1+1DQ8Z22}}\\ \text{ \ref{1+1DD4Z22}}  \end{matrix}$
& $\omega_{2, \II}$, $\exp( \ii   \pi  \int a_1 a_2)$
& $\begin{matrix}\text{ 2-dim Rep($Q_8$)}\\\text{2-dim Rep($D_4$)}  \end{matrix}$ & 
$\begin{matrix}{1 \to Z_2 \to Q_8  \to (Z_2)^2 \to 1}  \\ {1 \to Z_2 \to D_4  \to (Z_2)^2 \to 1} \end{matrix}$  \\ \hline
$\begin{matrix}SO(3): \text{\ref{1+1DSU2SO3}}\\
\text{Haldane phase}\\\text{for odd-integer-spin}\end{matrix}$  & \parbox{1.5in}{Odd-integer AF\\ Heisenberg spin chain} & 2-dim Rep($SU(2)$) & $1 \to Z_2 \to SU(2) \to SO(3) \to 1$  \\ \hline
 \end{tabular}
}
\caption{The 1+1D $G$-SPT states and their 0+1D degenerate states at the open chain end.
The first column is the symmetry group $G$.  The second column is the
$G$-cocycle ({\it i.e.} the SPT invariant) that characterizes the SPT state.
The third column is the ground state degeneracy (GSD) of end-point states and
their $H$-representation.  The four column is the group extension that
trivialize the cocycle in the second column, which is used to construct the end
states (that give rise to group $H$).  Note that only Haldane phase for
odd-integer-spin chain corresponds to an SPT state.  }
  \label{table:example-app-1}
\end{table} 

\begin{table}[!h] 
\makebox[\textwidth][c]{
\begin{tabular}{|c|c||c|c|c|c|}
\hline
Symm. group & \;  $\begin{matrix}\text{2+1D SPT Bulk inv.}\\ \text{($d$-cocycle $\omega_d$)} \end{matrix}$ & $\begin{matrix}\text{Unbroken}\\ \text{edge symm.}\end{matrix}$ &  $\begin{matrix}\text{GSD}\\ \text{on $D^2$}\end{matrix}$ & $1\to K \to H \to G \to 1$   \\ \hline
$Z_2$: \ref{sec:examplesHZ4GZ2}/\ref{sec:examples2+1DHZ4GZ2} & $\omega_{3, \I}$, $\exp( \ii   \pi  \int (a_1)^3)$  & $G_\text{edge}=1$ & 2 & $1 \to {Z}_2 \to Z_4  \to Z_2 \to 1$ \\ 
 \ref{sec:examplesHZ4NGZ2} &  & $G_\text{edge}=1$ & $2N$ & $1 \to Z_{2N} {\to} Z_{4  N} {\to} Z_2 \to 1$  \\ 
  \ref{sec:examplesHQ8GZ2} &   & $G_\text{edge}=1$ & 4 & $1 \to Z_4 \to Q_8  \to Z_2 \to 1$  \\ \hline
$Z_2^2$: \ref{2+1DD4Z22}& $\omega_{3, \II}$, $\exp( \ii   \pi  \int a_1 \beta a_2  )$   & {$G_\text{edge}=Z_2$} & 2 & $1 \to Z_2 \to D_4  \to (Z_2)^2 \to 1$   \\ 
 \ref{2+1DD4Z2Z22} &   & {$G_\text{edge}=Z_2$} & 4 & $1 \to (Z_2)^2 \to D_4 \times Z_2  \to (Z_2)^2 \to 1$  \\ \hline
$Z_2^3$:  \ref{2+1DD4Z2Z23} & $\omega_{3, \III}$, $\exp( \ii   \pi  \int a_1 a_2 a_3)$  & {$G_\text{edge}=(Z_2)^2$} & 2 & $1 \to Z_2 \to D_4 \times Z_2  \to (Z_2)^3 \to 1$  \\ \hline
$\begin{matrix}U(1) \rtimes Z_2^T:\\
\text{BTI}\end{matrix}$ \ref{2+1DU1Z2T}  & $\exp( \ii   \pi  \int w_1 c_1)$   & $G_\text{edge}=U(1)$ & 2 & $\begin{matrix}1 \to Z_2 \to G  \to G \to  1,\\ G=U(1) \rtimes Z_2^T \end{matrix}$   \\ \hline
$\begin{matrix}Z_2 \rtimes Z_2^T:\\
\text{BTSC}\end{matrix}$ \ref{2+1DU1Z2T} & $\exp( \ii   \pi  \int w_1 (a_1)^2)$   & $G_\text{edge}=Z_2$ & 2 & $\begin{matrix}1 \to Z_2 \to G  \to G \to  1,\\ G=Z_2 \rtimes Z_2^T \end{matrix}$   \\ \hline
 \end{tabular}
}
\caption{The 2+1D $G$-SPT states, and their 1+1D gapped spontaneously symmetry
breaking edge states (or gapped symmetry extended edge states if we interpret
the $H$ as an extended symmetry in any (artificial) 
model similar to Sec.~\ref{bdry2w}
described later).  The first column is the symmetry group $G$.  The second
column is the $G$-cocycle ({\it i.e.} the SPT invariant) that characterizes the
SPT state.  The $\beta$ is the Bockstein homomorphism.  The third column is the
unbroken symmetry group on the edge.  The fourth column is GSD of the system on
$D^2$ (the edge is a single $S^1$).  The fifth column is the group extension
that we use to construct the exactly soluble edge (that give rise to above
results).  
Often we can construct many different exactly soluble edges (with different
physical properties) for the same bulk SPT state.  ``BTI'' stands for bosonic
topological insulator.  ``BTSC'' stands for bosonic topological superconductor.
(Here we intentionally omit gapless symmetry preserving edge states [e.g. edge
cannot be gapped enforced by symmetry and perturbative anomalies], see Table
\ref{table:example-app-4}, \ref{table:KfiniteHGcont} and
\ref{table:example-app} for such examples.)
 Here $w_i \equiv w_i(TM)$ is the $i$-th Stiefel-Whitney class of a tangent bundle $TM$ over spacetime $M$.
}
  \label{table:example-app-2}
\end{table} 

For 1+1D SPT states, their degenerate end states are
described by a representation of $H$, called Rep$(H)$. The Rep$(H)$ is also a
projective representation of $G$ when $K$ is Abelian (see Table
\ref{table:example-app-1}).  

For a 2+1D SPT state, 
the boundary $K$-gauge deconfined phase corresponds to a gapped spontaneously 
symmetry breaking boundary (breaking a part of $G$-symmetry), which is described by an unbroken edge symmetry
group $G_\text{edge}\subset G$ (see Table \ref{table:example-app-2}). 
(However, if we consider this boundary as an H-symmetry extended 1+1D gapped boundary of the 2+1D bulk G-SPT state, then the boundary has no spontaneous symmetry breaking. The full H symmetry is preserved.) 

For a 3+1D SPT state, the boundary $K$-gauge deconfined phase corresponds to a gapped
symmetry preserving topologically ordered boundary described by a $K$ gauge
theory (see Table \ref{table:example-app-3}).  Higher or arbitrary dimensional
results are gathered in Table \ref{table:example-app-4}.  
Note that the $d$-cocycle $\omega_d$ for a finite Abelian group $G$ with its
type Roman numeral index follows the notation defined in Ref.
\cite{Wang1405.7689}.

\begin{table}[!t] 
\makebox[\textwidth][c]{
\begin{tabular}{|c|c||c|c|c|c|}
\hline
Symm. group & \;  $\begin{matrix}\text{3+1D SPT Bulk inv.}\\ \text{($4$-cocycle $\omega_4$)} \end{matrix}$ & $\begin{matrix}\text{Gauge}\\ \text{$K$}\end{matrix}$  & $\begin{matrix}\text{GSD on}\\ D^2\times S^1\end{matrix}$ & $\begin{matrix}\text{Symm. frac. of}\\ \text{gauge charge}\\ \text{Rep($H$)}\end{matrix}$ & $ H/K= G $    \\ \hline
$\begin{matrix}Z_2^T: \text{BTSC} 
 \\ \text{\ref{sec:examplesHZ4TGZ2T}/\ref{sec:examples3+1DHZ4GZ2T}}  \end{matrix}$
& $ \exp( \ii   \pi  \int (w_1)^{4}) $ & $Z_2$ & 4 & $\begin{matrix}\text{Rep$(Z_4^T)$}\\ \text{ Kramer doublet}\end{matrix}$ & $Z_4^T/{Z}_2  = Z_2^T $ \\ 
\ref{sec:examples3+1Dw2sq}
& $\exp( \ii   \pi  \int (w_2)^2)$  & $Z_2$ & 4 & $\begin{matrix}{\text{Rep$(Z_2^T)$}}
\end{matrix}$ & 
$\frac{\text{Pin}^+(\infty)}{{Z}_2}  =O(\infty) $\\ 
\ref{sec:examples3+1Dw2sq} 
 & $ \exp( \ii   \pi  \int (w_1)^4+(w_2)^2) $  & $Z_2$ & 4 & Rep$(Z_4^T)$ & 
$\frac{\text{Pin}^-(\infty)}{{Z}_2}  =O(\infty) $\\ \hline
$\begin{matrix}U(1)\times Z_2^T: \\ \text{BTP}\end{matrix}$ 
& $ \exp( \ii   \pi  \int w_1^2 c_1)$ & $Z_2$ & 4 & $\begin{matrix}\text{Rep$(\t U(1)\times Z_2^T)$}\\ \text{$\frac12 U(1)$-charge}\end{matrix}$ & 
$  \frac{\t U(1)\times Z_2^T}{{Z}_2}  = U(1)\times Z_2^T $ \\[4mm] 
&  & $Z_2$ & 4 & $\begin{matrix}\text{Rep$(U(1)\times Z_4^T)$}\\ \text{Kramer doublet}\end{matrix}$ & 
$ \frac{U(1)\times Z_4^T}{{Z}_2}  = U(1)\times Z_2^T $ \\[4mm] 
&  & {$Z_2$}  & \cred{4} & $\begin{matrix}\text{Rep$(U(1)\times Z_4^T)$}\\ \text{$\frac12 U(1)$-charge is}\\ \text{Kramer doublet}\end{matrix}$ & 
$ \frac{ U(1)\times Z_4^T}{ {{Z}_2}}   = U(1)\times Z_2^T $ \\ \hline
$Z_2^2$: {\ref{3+1DD4Z22TypeII}} & $\begin{matrix} \omega_{4, \II},\\ \exp( \ii   \pi  \int a_1 a_2 \beta  a_2)\end{matrix}$  & $Z_2$ & 4 & Rep$( D_4 )$ & 
$D_4/Z_2 = (Z_2)^2 $    \\ \hline
$Z_2^3$: \ref{3+1DD4Z2Z23TypeIII} & $\begin{matrix} \omega_{4, \III},\\ \exp( \ii   \pi  \int a_1 a_2 \beta  a_3  )\end{matrix}$  & $Z_2$ & 4 & Rep$( D_4 \times Z_2)$ & 
$\frac{D_4 \times Z_2}{Z_2}=  (Z_2)^3 $  \\ \hline
$Z_2^4$: \ref{3+1DD4Z22Z24} & $\begin{matrix} \omega_{4, \IV},\\ \exp( \ii   \pi  \int a_1 a_2 a_3 a_4)\end{matrix}$  & $Z_2$ & 4 & Rep$(D_4 \times (Z_2)^2)$ & 
$ \frac{D_4 \times (Z_2)^2}{Z_2}  = (Z_2)^4 $  \\ \hline
 \end{tabular}
}
\caption{The 3+1D 
$G$-SPT states and  their 2+1D gapped symmetry
preserving topologically ordered
boundaries (or gapped symmetry extended boundary states if we interpret
the $H$ as an extended symmetry).  
The first column is the symmetry group $G$.  The second
column is the $G$-cocycle ({\it i.e.} the SPT inv.) that characterizes the SPT
state.  The third column is the gauge group $K$ for the boundary topologically
ordered gauge theory.  The fourth column is GSD of the system on the space $D^2\times S^1$. 
The fifth column is the symmetry fractionalization
of quasiparticle excitations on the boundary.  The sixth column is the group
extension that we use to construct the exactly soluble boundary (that give
rise to above results).  ``BTSC'' stands for bosonic topological
superconductor.  ``BTP'' stands for bosonic topological paramagnet.  The $\t
U(1)$ is a double-covering $U(1)$.
 Here $w_i \equiv w_i(TM)$ is the Stiefel-Whitney class of a spacetime tangent bundle $TM$.
}
  \label{table:example-app-3}
\end{table} 

\begin{table}[!h]
\makebox[\textwidth][c]{
\begin{tabular}{|c|c||c|c|c|c|}
\hline
Symm. group & \;  $\begin{matrix}\text{SPT Bulk inv.}\\ \text{(cocycle $\omega_{d+1}$)} \end{matrix}$ & $\begin{matrix}\text{Gauge}\\ \text{$K$}\end{matrix}$  & $\begin{matrix}\text{GSD on}\\ D^2\times (S^1)^{d-2}\end{matrix}$ & $\begin{matrix}\text{Symm. frac. of}\\ \text{gauge charge}\\ \text{Rep($H$)}\end{matrix}$ & $ H/K= G $    \\ \hline
\hline
$\begin{matrix}U(1): \text{\ref{6+1DU1U1}}\\6+1/5+1\tD  \end{matrix}$ & $\exp( \ii   \pi  \int w_2 w_3 c_1)$   & $Z_2$ & 32 & $\begin{matrix}\text{Rep$(\t U(1))$}\\ \text{$\frac12 U(1)$-charge}\end{matrix}$ & 
$ \frac{\t U(1)\times SO(\infty)}{Z_2}  = U(1)\times SO(\infty) $   \\ \hline
$\begin{matrix}Z_2: \text{\ref{sec:examplesHZ4GZ2-any-dim}}\\ d+1/d\tD\\ \text{(even $d$)} \end{matrix}$   & $\begin{matrix}  \omega_{d+1, \I},\\ \exp( \ii   \pi  \int (a_1)^{d+1})\end{matrix} $  & $Z_2$ & $2^{d-1}$ & $\begin{matrix}\text{Rep$(Z_4)$}\\ \text{ $\frac12$ $Z_2$-charge}\end{matrix}$ &  $ Z_4/{Z}_2  = Z_2 $ \\ \hline
$\begin{matrix}Z_2^T:  \text{\ref{sec:examplesHZ4TGZ2T-any-dim}}\\  
d+1/d\tD\\
\text{BTSC (odd $d$)}\end{matrix}$    & $\begin{matrix} \text{$Z_2^T$-cocycle,}\\ \exp( \ii   \pi  \int (w_1)^{d+1}) \end{matrix}$  & $Z_2$ & $2^{d-1}$ & $\begin{matrix}\text{Rep$(Z_4^T)$}\\ \text{ Kramer doublet}\end{matrix}$ &  $   Z_4^T/{Z}_2  = Z_2^T $ \\ \hline
$\begin{matrix} (Z_2)^{d+1}: \text{\ref{3+1DD4Z22Z24}}\\ d+1/d\tD \end{matrix}$  & $\begin{matrix}\omega_{d+1,\text{Top}},\\ {\exp( \ii   \pi  \int \cup_{i=1}^{d+1} a_i)}\end{matrix}$   & $Z_2$ & $2^{d-1}$ & Rep$(D_4 \times (Z_2)^{d-1})$ & $ \frac{D_4 \times (Z_2)^{d-1}}{Z_2} = (Z_2)^{d+1}$   \\ \hline
 \end{tabular}
}
\caption{Other/higher $d+1$D $G$-SPT states and their $G$-symmetry
preserving boundaries, or their $H$-symmetry extended boundaries if we interpret the $H$ as an extended symmetry.  
The caption follows the same set-up as in Table \ref{table:example-app-3}. 
The fourth column is GSD of the system on space $D^2\times (S^1)^{d-2}$.  
The fifth column is the symmetry fractionalization
of quasiparticle excitations on the boundary. 
All examples here have non-perturbative global anomalies allow symmetry-preserving gapped boundaries.
 ``BTSC'' stands for bosonic topological
superconductor. 
(However, there are other examples that enforce a gapless boundary with GSD=$\infty$,
such as $U(1)$-SPT states of $d+1$D of an even $d$ that has boundary perturbative Adler-Bell-Jackiw anomalies.  
See also Table \ref{table:example-app} and Appendix \ref{sec:examples} for further discussions.)
 Here $w_i \equiv w_i(TM)$ is the Stiefel-Whitney class of a spacetime tangent bundle $TM$.
}
  \label{table:example-app-4}
\end{table}

{The symmetry preserving gapped boundary has topological excitations that
carry fractional quantum numbers of the global symmetry $G$.  Such a symmetry
fractionalization is actually described by Rep$(H)$ in our theory (see Table
\ref{table:example-app-3}). In the following we will explain such a result.}


\begin{enumerate}
\item
First of all, we know that the $K$-gauge theory has gauge charges (point
particles) carrying the representation, Rep($K$), of its gauge group $K$.  Each
distinct representation Rep($K$) labels distinct gauge-charged particle
excitations.

\item Second, if  the $K$-gauge theory has a global $G$-symmetry, one may
naively think that a gauge charged excitation can be labeled by a pair
(Rep($K$), Rep($G$)), a representation Rep($K$) from the gauge group $K$ and a
representation Rep($G$) from the symmetry group $G$.  The label
(Rep($K$), Rep($G$)) is equivalent to Rep$(K\times G)$.  If gauge charged
excitations can be labeled by Rep$(K\times G)$, this will implies that the
gauge charged excitations do not carry any fractionalized quantum number of the
symmetry $G$ (\ie no fractionalization of the symmetry $G$). 

\item However, for the boundary $K$-gauge theory (e.g.  the 2+1D surface) of
$G$-SPT state, the gauge charge excitations are in general labeled by Rep$(H)$ with $H/K=G$,
instead of Rep$(K\times G)$.  $H$ is a ``twisted'' product of $K$ and $G$, which
is the so-called projective symmetry group (PSG) introduced in Ref.~\cite{PSG}. When a gauge charged excitation is described by Rep$(H)$ instead of
Rep$(K\times G)$, it implies that the particle carries a fractional quantum
number of global symmetry $G$.  We say there is a fractionalization of the
symmetry $G$.

\item Continued from the previous remark, if the gauge group $K$ is $Z_N$ or
$U(1)$, then Rep$(H)$ is also called the projective representation of $G$, named
Proj.Rep($G$). Projective representation of $G$ also corresponds to a
fractionalization of the symmetry $G$.

\item
The quantum dimension $d_\alpha$ (\ie the internal degrees of freedom) of a
gauge-charged excitation $\alpha$ labeled by Rep$(H)$ is given by the dimension
of the Rep$(H)$: $d_\alpha=\text{Dim}[\text{Rep}(H)]$.  For our gauge theoretic
construction, because the Rep$(H)$ always has an integer dimension, thus the
corresponding gauge charge always has an integer quantum dimension $d$.  More
general topological order may have an anyon excitation $\alpha'$ that has a
non-integer or irrational quantum dimension $d_{\alpha'}$.

\end{enumerate}

\cgrn{The first three rows in Table \ref{table:example-app-3} are for the three
3+1D time-reversal SPT states.  The first one is within group cohomology
\cite{XieSPT5} $\cH^4(Z_2^T,U(1))$, the second one is beyond
group cohomology \cite{VS1306}, and the third one the stacking of the previous
two.  The fourth rows in Table \ref{table:example-app-3} describes three
different boundaries of the same $U(1)\times Z_2^T$-SPT states (also known as
bosonic topological insulator).  Here we like to comment about the quantum
number on the symmetry preserving topological boundary of those SPT states.  }

\begin{enumerate}

\item 
When a particle carries the fundamental Rep($Z_4^T$), it means that the
particle is a Kramer doublet (since $T^2\neq +1$). The $Z_4^T$-symmetry is
related to $\text{Pin}^-(\infty)$ where the time reversal square to $-1$, say
$T^2=-1$.  \cgrn{This corresponds to the first and the third rows in Table
\ref{table:example-app-3}, where boundary excitations carry various
representations of $Z_4^T$, including Kramers doublets.}

\item 
When a particle carries the fundamental Rep($Z_2^T$), it means that the
particle is a Kramer singlet (since $T^2=+1$).  The $Z_2^T$-symmetry is related
to $\text{Pin}^+(\infty)$ where the time reversal (\ie the reflection) square to the
identity.  \cgrn{This corresponds to the second row in Table
\ref{table:example-app-3} which has no time-reversal symmetry
fractionalization. But the boundary topological particles in the boundary
$Z_2$-gauge theory are all fermions  \cite{VS1306}.}



\end{enumerate}



\subsection{Notations and conventions}

\noindent
Our notations and conventions are partially summarized here.
SPTs stands for Symmetry Protected Topological state, 
TOs stands for topologically ordered state, and
SETs stands for Symmetric-Enriched Topologically ordered state.
In addition, aSPT and aSET stand for the anomalous boundary version of an SPT or  SET state.
Also ``TI,'' ``TSC,'' and  ``TP'' stand for topological insulator, topological superconductor and topological paramagnet respectively.
 We may append ``B'' in front of ``TI,'' ``TSC'' and ``TP'' as ``BTI,'' ``BTSC'' and ``BTP'' for their bosonic versions, where underlying
 UV systems contain only bosonic degrees of freedom. 
 A proper theoretical framework for all these aforementioned states (SPTs, SETs, etc) is beyond the Ginzburg-Landau symmetry-breaking paradigm\cite{LandauBook5-1980, LandauBook9-1980}.
  
When we refer to ``symmetry,''  we normally mean the global symmetry. 
The gauge symmetry should be viewed as a gauge redundancy but not a symmetry.

The boundary theories of SPTs have anomalies  \cite{W1313, Kapustin:2014zva, Wang1405.7689}.
The possible boundary anomalies of SPTs include 
perturbative anomalies  \cite{AlvarezGaume:1983ig} and non-perturbative global anomalies \cite{Witten:1982fp, Witten:1985xe}.
The obstruction of gauging the global symmetries (on the SPT boundary) is known as the €™'t Hooft anomalies \cite{H8035}.
Although SPTs can have both perturbative and non-perturbative anomalies, our construction of symmetric gapped interfaces
is \emph{only} applicable to  SPTs with 
\emph{boundary non-perturbative anomalies}.\footnote{We note that there is a terminology clash between condensed matter and high energy/particle physics literature 
on ``Adler-Bell-Jackiw (ABJ) anomaly \cite{Adler, BellJackiw}.''
In condensed matter literature \cite{W1313}, the phrase ``ABJ anomaly \cite{Adler, BellJackiw}'' refers to ``perturbative'' anomalies
(with $\mathbb{Z}$ classes, captured by the free part of cohomology/cobordism groups),
regardless of further distinctions 
(e.g. anomalies in dynamical gauge theory, or anomalies in global symmetry currents, etc.).
In condensed matter terminology, the ABJ anomaly is captured by a 1-loop diagram that only involves a fermion Green's function (with or without dynamical gauge fields). 
Thus,  the 1-loop diagram can be viewed as a property of a free fermion system even without gauge field.
On the other hand, in high energy/particle physics literature,
the perturbative anomaly without dynamical gauge field captured by a 1-loop diagram 
is referred to as a perturbative 't Hooft anomaly, instead of the ABJ anomaly.
Here we attempt to use a neutral terminology to avoid any confusion.
}

We may use the long/short-range orders (LRO/SRO) to detect Ginzburg-Landau order parameters.
In particular, the LRO captures the two-point correlation function decaying to a constant value at a large distance or power-law decaying, 
that detects the spontaneous symmetry breaking or the gapless phases. 

On the other hand, we may use short/long-range entanglement (SRE/LRE) to
describe the gapped quantum topological phases.  
LUT stands for local unitary transformation.
A short-range entangled (SRE)
state is a gapped state that can be smoothly deformed into a trivial product
state by LUT without a phase transition (some global
symmetries may be broken during the deformation).  A long-range entangled (LRE)
state is a gapped state that is not any SRE state, namely that cannot be
smoothly deformed into a trivial product state  by LUT without a phase transition (even by breaking all global
symmetries during the deformation).  What are examples of SRE and LRE states?
SPTs are SRE states, which at low energy are closely related to invertible
topological quantum field theories, with an additional condition that there is
no perturbative or non-perturbative \emph{pure} gravitational anomalies\footnote{
We note that the definitions of gravitational anomalies in \cite{Wen1410.8477, KW1458} and \cite{K1467} are different.
This leads to different opinions,
between \cite{Wen1410.8477, KW1458} and \cite{K1467}, 
whether SPT states allow non-perturbative global gravitational anomalies or not along their boundaries, especially for SPT states with time reversal symmetries.} 
on the boundary
(e.g. for a 1+1D boundary, the chiral central charge $c_-=0$, or the thermal
Hall conductance vanishes $\sigma_H=0$).  TOs and SETs are LRE states.

\cblue{
The $d+1$D means the $d+1$ dimensional spacetime.
We may denote the $(d-1)$D boundary of a $d$D manifold $M^d$ as $\partial M^d \equiv \partial (M^d) \equiv (\partial M)^{d-1}$.
We denote Borel group cohomology of a group $G$ with $U(1)$ coefficients as $\cH^{d}(G,{U(1)})$ for the $d$-th cohomology group,
which is equivalent to a topological cohomology of classifying space $BG$ of $G$ as  $H^{d+1}(BG,\Z)$,
regardless whether the $G$ is a continuous or a discrete finite group.
The $d$-cocycles $\omega_d$ are the elements of a cohomology group $\cH^{d}(G,{U(1)})$ and satisfy a cocycle condition $\delta \omega_d =1$.
The above statements are true for both continuous and discrete finite $G$.
When $G$ is a continuous group,
we can either view the cocycle $\om_d(g)$ as a measurable function on $G^d$ which gives rise to Borel group cohomology, or
alternatively, view the cocycle $\om_d(g)$ as continuous around a trivial $g=1$ but more generally may contain branch cuts.
When $G$ is a finite group, we further have $\cH^{d+1}(G,U(1))= H^{d+2}({BG,\Z})=H^{d+1}(BG,U(1))$.
The $\beta$ denotes the Bockstein homomorphism.
 GSD stands for ground state degeneracy, which counts the number of the lowest energy ground states (so called the zero energy modes). 
}


In a $d$-dimensional spacetime, we write $\nu_d$ for homogenous cocycles, and $\mu_d$ for homogenous cochains.
We also write $\omega_d$ for inhomogeneous cocycles, 
and $\beta_d$ for inhomogeneous cochains. %
We write $\mathcal{V}_d$ for homogenous cocycles or cochains with both global symmetry variables and gauge variables, 
and  $\Omega_d$ for inhomogeneous cocycles or cochains with both global symmetry variables and gauge variables.
Homogeneous cochains/cocycles are suitable for SPTs and SETs that have global symmetries,
while the inhomogeneous cochains/cocycles are suitable for TOs that have only gauge symmetries with no global symmetries.
The $c_i$ is the $i$th Chern class and the $w_i$ is the $i$th Stiefel-Whitney (SW) class.
We generically denote the cyclic group of order $n$ as $Z_n$, but we write $\Z_n$ when we are referring to the distinct classes in a classification of topological phases or in a cohomology/bordism group.
When convenient, we use notation such as  $Z_n^K$, $Z_n^G$ and $Z_n^H$ to identify a particular copy of $Z_n$. 
We denote $\ii$ for the imaginary number where $\ii^2=-1$.

\subsection{The plan of the article} 

{
We aim to introduce a systematic construction of various gapped
boundaries/interfaces for bulk topological states based on the \emph{group
extension}. 
We had mentioned many examples with various bulk SPT states and different boundary states
in any dimension, in Table \ref{table:example-app-1} (1+1D bulk/0+1D boundary), Table \ref{table:example-app-2} (2+1D bulk/1+1D boundary), 
 Table \ref{table:example-app-3} and Table \ref{table:example-app-4} (3+1D bulk/2+1D boundary and higher dimensions).
Their properties are summarized in Table \ref{table:KHGfinite} (for a finite discrete symmetry group $G$) 
and Table \ref{table:KfiniteHGcont} (for a continuous symmetry group $G$),
and their constructions are summarized in Table \ref{table:summary1} .  
Furthermore, we can formulate more general gapped interfaces including
not only our proposal on \emph{symmetry-extension}, but also \emph{symmetry-breaking} and 
 \emph{dynamically gauging}  of topological states, in a framework, schematically shown in Table \ref{table:summary2}.
We will provide both
the spacetime lattice path integral (partition function) definition and the
wavefunction (as a solution to a spatial lattice Hamiltonian) definition, see
Table \ref{table:Tablelattice}, for our generic construction.
}

We begin in Sec.~\ref{bCZX}, by reviewing a model that realizes the 2+1D $Z_2$ SPT state, the CZX model.
Then in Sec.~\ref{sec:bdryCZX}, we construct various boundaries, both gapless and gapped, for the CZX model.  
In the process, we illustrate some of the main ideas of this paper.
In Appendix \ref{sec:lowEbdryCZX}, we examine various low-energy effective theories for the boundaries of CZX model.
Among the new phenomena, we find that in Appendix \ref{sec:deconfined-to-SSB}
the 1+1D boundary \emph{deconfined} and \emph{confined} $Z_2^K$-gauge states
belong to the same phase, {namely they are both \emph{spontaneous symmetry breaking} states
related by a crossover without phase transition.}
In Appendix \ref{sec:fCZX}, we study the fermionic version of CZX model, 
then we find anomalous boundary with emergent $Z_2^K$-gauge theory and anomalous global symmetry in
Appendix \ref{sec:fCZXbdryZ2-gauge}.
We expect that our approach can apply to other generic fundamentally fermionic many-body systems.

All these analyses have the advantage of illustrating our constructions in a completely explicit way, but they have a drawback.
The deconfined gapped boundary state that we construct for the CZX model (in the usual case that the global symmetry
is not extended along the boundary) is not really
a fundamentally new state with 1+1D topological order, but rather it can be interpreted in terms of \emph{broken global symmetry}.\footnote{
More generally, we find that,
 various 1+1D deconfined gauge theories (on the boundaries of 2+1D SPT states) 
are the \emph{spontaneous global symmetry breaking} states with either 
\emph{unitary}-symmetry or 
\emph{anti-unitary time reversal} $Z_2^T$-symmetry broken, 
 see Sec.~\ref{whichone} and Appendices \ref{sec:deconfined-to-SSB} and \ref{sec:SSB_field-theory}.}
This is consistent with the common lore  that ``there is no true topological order in 1+1D that is robust against any local perturbation.''  
However, we start with the CZX model because in that case everything can be stated
in a particularly simple and clear way.  Models in higher dimension ultimately realize the ideas of the present
paper in a more satisfying way -- as the phases we construct are essentially new -- but in higher dimension, it is hard to be equally simple.

Nevertheless, the extrapolation from our detailed treatment of the CZX model to a general discussion in higher dimensions is fairly
clear.
In Sec.~\ref{rev}, we distill the essence of {the non-on-site boundary symmetry of a generic SPT state in \emph{any} dimension}.
One key message of this section is the following: 
The symmetry-extended gapped boundary construction of a bulk $G$-SPTs relies on the fact that 
its boundary has a non-perturbative \emph{global} $G$-\emph{anomaly} 
(probed by gauge or gravitational fields, as a non-perturbative global gauge/gravitational anomalies)
which becomes an $H$-anomaly-free by pulling $G$ back to $H$.

In Sec.~\ref{sec:emergent-soft-gauge}, 
we introduce the concept of ``\emph{soft gauge theory}''
and introduce the cochains that encode the \emph{soft gauge} degree of freedom.
We consider an emergent soft gauge theory on the boundary, associated to a suitable group extension as eqn. (\ref{gext}),
as a way to construct  a symmetric gapped boundary state.
(The ``\emph{soft gauge theory}'' and the usual ``\emph{hard gauge theory}'' are contrasted later in Sec.~\ref{sec:general}.)
In Sec.~\ref{sec:extension-trivializeGcocycle}, we provide a method, in the context of an arbitrary SPT phase with symmetry $G$,
 to search for {an $H$-extension of $G$ that trivializes the $G$-cocycle}. 
We provide valid examples of symmetric gapped boundaries for SPTs with onsite unitary symmetry (Sec.~\ref{sec:examples2+1DHZ4GZ2}) and anti-unitary time reversal symmetry 
(Sec.~\ref{sec:examples3+1DHZ4GZ2T}) in 2+1D and 3+1D, more examples for \emph{any} dimensions are given in Appendix \ref{sec:examples}.

In Sec.~\ref{sec:beyondGC} and Sec.~\ref{sec:Cob}, we comment on the
application of our approach for gapped interfaces of topological states
obtained from beyond-symmetry-group cohomology and cobordism approach, for both
bosonic and fermionic systems.

In Sec.~\ref{sec:sym-enhanced:bdryDW}, 
we  consider generic gapped boundaries
and gapped interfaces (Sec.~\ref{interfaces})
with mixed \emph{symmetry-breaking} (Sec.~\ref{relsym} and Appendix \ref{sec:sym-breaking:bdry}), \emph{symmetry-extension} and \emph{dynamically-gauging} mechanisms (Sec.~\ref{mixed}).
Dynamically gauged gapped interfaces of topologically ordered gauge theories are explored in Sec.~\ref{sec:gauged-DW-interface} 
(more examples are relegated to Appendix \ref{sec:examples-gauge-sym-break}).
%

{
We also describe two different techniques to obtain symmetry-extended gapped boundaries / interfaces. 
One technique is in Sec.~\ref{sec:extension-trivializeGcocycle}: For a given symmetry (quotient) group $G$ and its $G$-cocycle,
we can determine the finite (normal subgroup) $K$ and then deduce the total group $H$, in order to obtain the trivialization of $G$-topological state via
the exact sequence $1 \to K \to H \to G \to 1$ in \eqn{gext}. 
Another technique in Appendix \ref{sec:trivializeGcocycle} is based on Lydon-Hochschild-Serre (LHS) spectral sequence method. 
Given the symmetry group $G$ and its $G$-cocycle, and suppose we assume the possible $H$ and $K$ within $1 \to K \to H \to G \to 1$,
the LHS method helps to construct an exact analytic function of the split $H$-cochain from the given $G$-cocycle and 
from the exact sequence \eqn{gext}. 
In short, both techniques have their own strengths:
The first technique in Sec.~\ref{sec:extension-trivializeGcocycle} has the advantage to search $K$ and $H$, for a given $G$ and a $G$-cocycle,
The second LHS's technique in Appendix \ref{sec:trivializeGcocycle} has the further advantage to construct the exact analytic split $H$-cochain.
}

In Sec.~\ref{sec:general}, we provide a systematic {general construction of lattice path integrals and Hamiltonians
for gapped boundaries/interfaces for topological phases 
in any dimension}.
More examples of \emph{symmetry-extended} gapped interfaces in various dimensions are provided in Appendix \ref{sec:examples}.
Many examples of gauge \emph{symmetry-breaking} gapped boundaries/interfaces via Anderson-Higgs mechanism are derived
based on our framework, and compared to the previous known results in the literature, in Appendix \ref{sec:examples-gauge-sym-break}.
We conclude in Sec.~\ref{sec:conclude}.

\subsection{Tables as the guide to the article}
\label{sec:tables}

\begin{figure}[!h]  
\centering
\includegraphics[scale=.67]{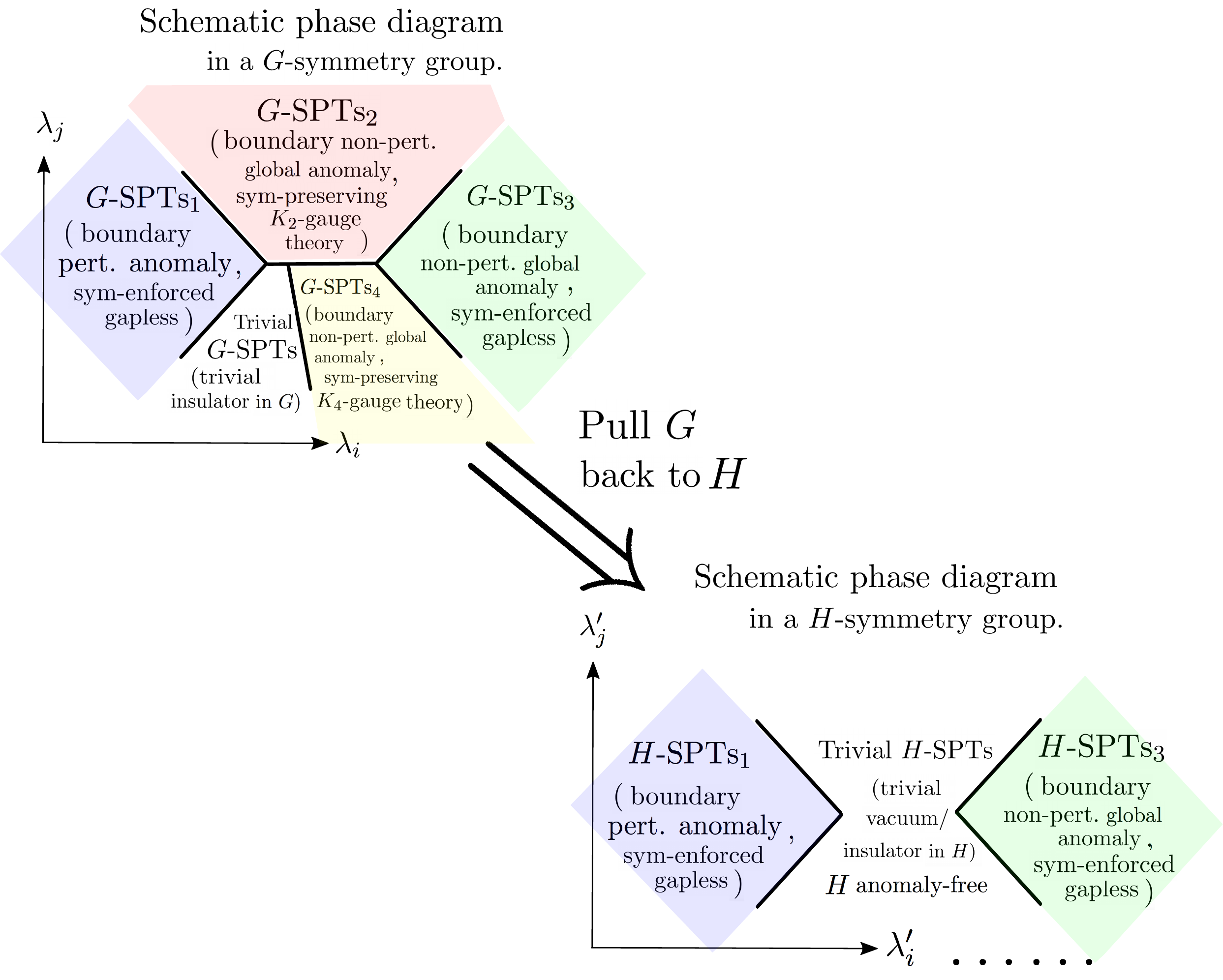} 
\caption{The schematic phase diagram of a symmetry $G$ (left) 
in the Hamiltonian $H(\lambda_i,\lambda_j,\dots)$'s coupling space $\{\lambda_i,\lambda_j,\dots\}$,
and the schematic phase diagram
of a symmetry $H$ (right)
in the Hamiltonian $H(\lambda_i',\lambda_j',\dots)$'s coupling space $\{\lambda_i',\lambda_j',\dots\}$.
Many $G$-SPT states may be trivialized to an $H$-trivial vacuum/insulator, by pulling $G$ back to $H$.
Some SPT states in $G$ (here G-SPTs$_1$ and G-SPTs$_2$) may be symmetry-enforced gapless on the physical boundary
(that may have either perturbative anomalies or non-perturbative global anomalies).
Other SPT states in $G$ (here G-SPTs$_2$ and G-SPTs$_4$) may have symmetry-preserving gapped boundaries 
(that must have non-perturbative global anomalies)
by pulling $G$  back to a larger symmetry group  (say $H_2$ and $H_4$, where $H_2 \neq H_4$ in general).  
We could consider the effective schematic phase diagram in a certain larger $H$. (We may choose $H \supseteq H_2$ and $H \supseteq H_4$.)
The effective Hilbert space for the whole systems in the $H$-symmetry may be different/larger than that in the $G$-symmetry.
The ``$\dots \dots$'' in the schematic $H$ phase diagram, implies other possible new phases that occur in a larger $H$-symmetry but do not occur in a $G$-symmetry.
The phase boundaries shown are schematically only, which could be the first order, second order or any continuous higher order phase transitions.
} 
\label{G-H-class-symm} 
\end{figure}

Readers can either find the following Tables \ref{table:KHGfinite},
   \ref{table:KfiniteHGcont},
   \ref{table:summary1}, \ref{table:summary2} and  \ref{table:Tablelattice} and Fig.~\ref{G-H-class-symm} as a quick tabular summary of partial results of the article, or 
 find them as a useful guide or menu to the later Sections. 
Readers may freely skip the entire Sec.~\ref{sec:tables} and Tables, 
then proceed to Sec.~\ref{bCZX} directly, and come back to these Tables later after visiting related materials in the later Sections.

\noindent
{\bf A (partial) summary of our constructions based on  Table 
   \ref{table:KHGfinite},
   \ref{table:KfiniteHGcont},
   \ref{table:summary1}, \ref{table:summary2},  \ref{table:Tablelattice},
and Fig.~\ref{G-H-class-symm}}:

{In Table \ref{table:KHGfinite} and \ref{table:KfiniteHGcont}, we discuss various boundary properties of a finite group symmetry $G$-SPTs (Table \ref{table:KHGfinite}) and
a continuous group symmetry $G$-SPTs (Table \ref{table:KfiniteHGcont}).
The discussions here parallel to the topological phase constructions  in Table \ref{table:summary1} and \ref{table:summary2},
and we enumerate the items in the similar orderings shown there.}

A schematic physical picture is shown in Fig.~\ref{G-H-class-symm}. 
{Conceptually, we could ask how a phase diagram of the Hamiltonian's coupling space in a symmetry $G$ (the left figure of Fig.~\ref{G-H-class-symm})
evolves if we consider the phase diagram of the Hamiltonian's coupling space in a larger symmetry $H$ (the left figure of Fig.~\ref{G-H-class-symm}).
The effective Hilbert space for the whole system in the $H$-symmetry may be larger than that in the $G$-symmetry. Thus
one may need to modify Hamiltonians as well as Hilbert spaces to consider such a phase-diagram evolution, which is difficult in practice.
But as a thought experiment, we could expect that several distinct SPT states in $G$ may become the same trivial insulator/vacuum in $H$.
Those $G$-SPT states contain certain non-perturbative global $G$-anomalies along physical boundaries, that become anomaly-free in $H$.
We note that the \emph{phase boundaries} in the phase diagrams shown in Fig.~\ref{G-H-class-symm} are schematic only and
are \emph{not} equivalent to the \emph{physical boundaries} to a trivial vacuum in the spacetime.
}

In Table \ref{table:summary1} and \ref{table:summary2}, we show various gapped boundaries (bdry) and interfaces of topological states in $d$-dimensional spacetime
with their interfaces in $(d-1)$-dimensions: 

\noindent
In Table \ref{table:summary1} (i), $G$-SPTs has an anomalous boundary with an anomalous non-onsite symmetry in $G$ (Sec.~\ref{sec:non-on-site-G}). 
However, the non-onsite $G$-symmetry can be made to be onsite in $H$, thus the $G$-anomaly becomes anomaly free 
(denoted as anom.~free) in $H$ (Sec.~\ref{sec:on-site-H}).  
This also gives us a way to obtain a $H$-symmetry-extended gapped boundary of $G$-SPTs.

\noindent
In Table \ref{table:summary1} (ii), $G$-SPT state's above boundary in (i) can be dynamically gauged on its normal subgroup $K$ on the boundary.
We denote such a boundary state as $H/K$-aSETs, 
which  means that it has a full group $H$, a dynamical gauge group $K$, and with a $G$-anomaly.

\begin{table}[!h]
\begin{center}
\makebox[\textwidth][c] 
{
\begin{tabular}{c| l | l}
\hline
\multicolumn{3}{c}{
\begin{minipage}{8in}
Symmetry extension construction: $1 \to {K} \to H \overset{r}{\rightarrow} G \to 1$, with  $G,H$ and $K$ \emph{finite} groups.\\[1mm]
A $G$-topological state (e.g. $G$-cocycle $\nu_d^{G}$ or $G-$bundle) is trivialized in $H$.
For a bulk $G$-SPTs, its boundary has a non-perturbative global $G$-anomaly from $\cH^d(G,U(1))=H^{d+1}(BG,\Z)=H^d(BG,U(1))$ 
(for a finite $G$, containing only the \emph{torsion}), 
which becomes an $H$-anomaly free by pulling $G$ back to $H$.
Formally, we can \emph{prove} that, given a $G$-cocycle $\nu_d^{G} \in  \cH^d(G,U(1))$, certain finite $K$ and $H$ \emph{exist}, 
so ${r}^* \nu_d^{G}= \nu^H_d =\delta  \mu^H_{d-1} \in \cH^d(H,U(1))$,  split to $H$-cochains $\mu^H_{d-1}$.\\[-2mm] 
\end{minipage}}\\
\hline
\hline
$\begin{matrix}{\text{The first gapless boundary}:}\\{\text{Non-universal complicated dynamics.}}\\
{\text{$G$ is a global symmetry for}}\\
{\text{the whole bulk and boundary.}}\\
{\text{Sec.~\ref{bdry1}, \ref{sec:1st-complx-bdry}, \ref{sec:non-on-site-G}}.}  \end{matrix}$ 
 & 
 \multicolumn{2}{c}{
 \begin{minipage}[c]{4.8in}
 {$\bullet$ The bulk+boundary theory has an on-site $G$-symmetry.}\\
  {{$\bullet$ The effective boundary theory has a local Hilbert space,}}\\
  {{but has a \emph{non-on-site} $G$-symmetry (\ref{sec:non-on-site-G-Hamiltonian}).  
  }}
  \end{minipage}
  } \\
\hline\\[-2mm]
$\begin{matrix}{\text{The second gapped boundary}:}\\
{\text{$G$ global symmetry}}\\
{\text{is extended to}}\\
{\text{$H$ symmetry}}\\ 
{\text{on the boundary.}}\\
{\text{Sec.~\ref{bdry2w}, \ref{sec:extSymm}}.}\\ 
{\text{Table  \ref{table:summary1} (i)}.}
 \end{matrix}$
 &  \multicolumn{2}{l}{  
 \begin{minipage}[c]{5.in}
 {$\bullet$ The bulk+boundary theory has an on-site symmetry.}\\
  {{$\bullet$ The effective boundary theory has a local Hilbert space,}}
  {{and has an \emph{on-site} $H$-symmetry (\ref{mainpoint}).}} \\ 
  {{$\bullet$ The boundary $G$-anomaly becomes $H$-anomaly free (\ref{mainpoint}).}}\\ 
  {{Interpretation:}}\\
  {{$\bullet$ (i) Extending $G$ to $H$-symmetry \emph{only} on the boundary,}}
  {{but the model is artificial in condensed matter (\ref{bdry2w}, \ref{1+1DSU2SO3}).}}\\ 
  {{$\bullet$ (ii) A nontrivial bulk $G$-SPTs becomes a trivial bulk}}
   {{$H$-SPTs (trivial vacuum), when pulling $G$ back to $H$ (\ref{sec:non-on-site-G-Hamiltonian}).}}\\[-2mm]  
  \end{minipage}
  }\\
  \hline
$\begin{matrix}
{\;}\\ 
{\;}\\ 
{\text{The third gapped boundary}:}\\
{\text{$G$ is a global symmetry, but}}\\
{\text{$K$ is hard gauged}}\\ 
{\text{on the boundary.}}\\
{\text{Sec.~\ref{bdry3w}, \ref{sec:hard-gauge}}.}\\ 
{\text{Table  \ref{table:summary1} (ii)}.}\\
{\;}\\ 
{\;}\\ 
  \end{matrix}$
&
\begin{minipage}{1.5in}
  $\bullet$ The effective boundary theory 
  has a non-local Hilbert space,
  therefore its on-site or
  non-on-site symmetry is
  ill-defined ({\ref{sec:cocycle-violate-local}}). 
  {}
\end{minipage}
&  \multirow{-5}{*}{
\begin{minipage}{3.3in}
$\bullet$ For 2+1D bulk/1+1D boundary,
a deconfined boundary $K$-gauge
 theory has a spontaneous symmetry 
    breaking (SSB) long-range order in $G$, 
    either breaking unitary (e.g. $Z_2$) or 
    anti-unitary $Z_2^T$ time reversal (Sec.~\ref{bdry3w}, \ref{sec:deconfined-to-SSB}, \ref{sec:SSB_field-theory})
    subgroup in $G$. The SSB states smoothly cross over to confined states. 
    We find no robust intrinsic
        topological order even on a
        1+1D boundary of SPTs. \\[2mm]
        $\bullet$ For 3+1D bulk/2+1D boundary or higher dimensions,
        there \emph{always exists} a symmetry-preserving deconfined boundary $K$-gauge
            theory with a robust intrinsic
            topological order. The $K$-gauge charge
   carries a representation Rep($H$).
   The $G$-anomaly is a non-perturbative \emph{global} (gauge/gravitational) \emph{anomaly} that becomes absence in $H$.
      Given a finite $G$, we can find a \emph{finite Abelian} $K$ to achieve this (Sec.~\ref{sec:extension-trivializeGcocycle}).  
\\[-3mm]
\end{minipage}
}\\[-2mm]
\cline{1-2}
$\begin{matrix}
{\;}\\ 
{\;}\\ 
{\text{The fourth gapped boundary}:}\\
{\text{$G$ is a global symmetry, but}}\\
{\text{$K$ is soft gauged}}\\ 
{\text{on the boundary.}}\\
{\text{Sec.~\ref{bdry4w}, \ref{sec:cocycle-emergent}}.}\\ 
{\;}\\ 
{\;}\\ 
{\;}\\ 
 \end{matrix}$
&
\begin{minipage}{1.5in}
{{$\bullet$ The bulk+boundary theory
has an on-site symmetry.}}\\[2mm]
  {{$\bullet$ The effective boundary theory
  has a local Hilbert space,
  but has a \emph{non-on-site}
  $G$-symmetry (\ref{sec:cocycle-preserve-local}).}} 
 \end{minipage} \\[-3mm]
   \hline\\[-3mm]
$\begin{matrix}{\text{The fifth gapped boundary}:}\\
{\text{$G$ is partly/fully gauged in the bulk,}}\\
{\text{and $H$ is partly/fully gauged on the}}\\
{\text{boundary. Sec.~\ref{ito}, \ref{sec:ZSPTSETG}}}\\ 
{\text{Table  \ref{table:summary1} (iii) and (iv)}.}  \end{matrix}$
&
\multicolumn{2}{l}{
\begin{minipage}{4.8in}
  $\bullet$ The effective boundary theory has a non-local Hilbert space,
    which cannot be local even by soft-gauging,
  therefore its on-site or non-on-site symmetry
   is ill-defined (for SETs). This relates to the fact that
      an intrinsic bulk topological order has long-range entanglements
            and gravitational anomaly.
\end{minipage} 
  }
 \\
  \hline
 \end{tabular}
 }
 \hspace*{0mm}
\end{center}
\caption{
Symmetry extended boundary construction of topological states  in any dimension for $G,H$ and $K$ all finite groups. 
}
  \label{table:KHGfinite}
\end{table} 

\begin{table}[!h]
\begin{center}
\makebox[\textwidth][c] 
{
\begin{tabular}{c| l }
\hline
\multicolumn{2}{c}{
\begin{minipage}{7.6in}
Symmetry extension construction: $1 \to {K} \to H \overset{r}{\rightarrow} G \to 1$,
with  $G$ and $H$ as \emph{continuous} groups (in particular 
compact Lie groups) 
but $K$ as a \emph{finite} group.
The finite group extension from a continuous $G$ to a continuous $H$ by a finite $K$ is the finite \emph{covering} of $G$.
For a bulk $G$-SPTs (classified by $\cH^d(G,U(1)) \equiv H^{d+1}(BG,\Z)$), its boundary may either have a perturbative $G$-anomaly 
from the \emph{free} part of $\cH^d(G,U(1))$ (e.g. a perturbative anomaly with a $\Z$ class),
or a non-perturbative global $G$-anomaly from the \emph{torsion} part of $\cH^d(G,U(1))$ 
(e.g. global anomaly with a product of $\Z_n$ classes).
Only a non-perturbative global $G$-anomaly from the torsion part may become $H$-anomaly free. 
Similarly, only the corresponding $G$-SPTs may be trivialized in $H$.
Our approach suggests a method to find a continuous $H$ to construct symmetry-preserving gapped boundaries:  
either an $H$-symmetry extended gapped boundary, or a deconfined finite $K$-gauge theory, for such as a $G$-SPTs.
\\[-2mm]
\end{minipage}
}\\
\hline
\hline\\[-2mm]
$\begin{matrix}
{\;}\\ 
{\;}\\
{\text{$G$ is a global}}\\
{\text{symmetry}}\\ 
{\text{(G-SPTs),}}\\ 
{\text{but there is}}\\ 
{\text{a $K$ gauge theory}}\\ 
{\text{on the boundary,}}\\ 
{\text{for a total}}\\ 
{\text{group $H$.}}\\
{\text{The third/fourth}}\\
{\text{gapped boundary.}}\\
{\text{Sec. 3.3, 4.6}.}\\
{\text{Table \ref{table:summary1} (ii)}.}\\
{\;}\\ 
{\;}\\ 
{\;}\\ 
  \end{matrix}$ & 
  {
\begin{minipage}{6in}
$\bullet$ For 2+1D bulk/1+1D boundary,
if a deconfined boundary $K$-gauge
 theory exists, it has a spontaneous symmetry 
    breaking long-range order in $G$, 
    either breaking unitary (e.g. $Z_2$ in Sec.~\ref{bdry3w}, \ref{sec:deconfined-to-SSB}, \ref{sec:SSB_field-theory}) or 
    anti-unitary $Z_2^T$ time reversal
    discrete finite subgroup in the full $G$. 
    We find no spontaneous global symmetry breaking for the continuous subgroup sector in $G$ on the 1+1D boundary, 
    consistent with Coleman-Mermin-Wagner theorem. 
    We also find no robust intrinsic
        topological order on a
        1+1D boundary of SPTs. (e.g. $U(1) \rtimes Z_2^T$ and $Z_2 \rtimes Z_2^T$ in \ref{sec:SSB_field-theory})\\[-2mm]
        \hrule
        \medskip
        $\bullet$ For 3+1D bulk/2+1D boundary or higher dimensions,
        there \emph{may or may not exist} a symmetry-preserving deconfined boundary $K$-gauge
            theory. Our construction depends on the properties of continuous $G$ and $H$: \\[-1mm]
            - - - - - - - - - - - - - - - - - - - - - - - - - - - - - - - - - - - - - - - - - - - - - - - - - - - - - - - - \\[-1mm]
            $\diamond$ When a continuous Lie group $G$ is connected but not simply-connected, there exists a finite extension of $G$ as a finite covering of $G$
            from $\pi_1(G)\neq 0$ (e.g. $G=U(1), SO(n), etc.$). 
            When $G$ is disconnected, there may still exist a finite covering (e.g. $G=O(n)$).
             Two scenarios:\\
             \;\; $(i).$ If $G$-anomaly is a perturbative anomaly, 
             then  the group extension of $G$ to $H$ \emph{cannot} make this $H$-anomaly free.
             There exists \emph{no} boundary deconfined gauge theory with topological orders for such a $G$-SPTs. (e.g. {\ref{2+1DU1U1}}'s $U(1)$ chiral anomaly.) 
             \\
             \;\; $(ii).$ If $G$-anomaly is a non-perturbative global anomaly, we can check whether the group extension of $G$ to $H$, by a finite $K$, makes it $H$-anomaly free.
            If yes, then deconfined-$K$ gauge theories exist with robust intrinsic topological orders. For examples, 
            \ref{6+1DU1U1}'s $U(1)$-global anomaly, 
            \ref{sec:examples3+1Dw2sq}'s $Z_2^T$-BTSC with $G=O(\infty)$, or 3+1D $U(1)\times Z_2^T$-BTP (shown in Table \ref{table:example-app-3}),
            which finite coverings are allowed from $\pi_1(U(1))=\Z$ or $\pi_1(SO(\infty))=\pi_1(O(\infty))=\Z_2$.
            \\[-1mm]
                        - - - - - - - - - - - - - - - - - - - - - - - - - - - - - - - - - - - - - - - - - - - - - - - - - - - - - - - - \\[-1mm]
$\diamond$ When a continuous Lie group $G$ is simply-connected, then there is no finite extension of $G$ because there is no finite covering of $G$. 
 (e.g. $\pi_1(SU(n))=0$ for $n \geq 2$.)
Thus our construction fails to construct symmetry-preserving gapped boundary of such a $G$-SPTs. This implies either that
 boundary states must be symmetry-enforced gapless, or one needs to seek other construction (different from ours) for 
 symmetry-preserving gapped boundary.\\[-2.5mm]
\hrule
\medskip
$\bullet$ The properties of global symmetry (on-site or non-on-site) for continuous $G$ and $H$ still follow the similar discussions as in Table \ref{table:KHGfinite}.
\\[-3mm]
\end{minipage}
}\\
  \hline
 \end{tabular}
 }
 \hspace*{0mm}
\end{center}
\caption{
Symmetry extended boundary construction of topological states in any dimension for $G$ and $H$ as continuous groups. We require a finite gauge group $K$ for a deconfined $K$-gauge theory.
Through our construction,
the $G$-SPTs of a connected but not simply-connected compact Lie group $G$ may have a 
symmetry-preserving surface deconfined boundary gauge theory,
but that of a simply-connected compact Lie group $G$ cannot have such a boundary.
}
  \label{table:KfiniteHGcont}
\end{table} 

\begin{center}
\begin{table}[!h]
\noindent
\makebox[\textwidth][c] 
{
\begin{tabular}{ccc} 
\hline
$\begin{matrix}
\text{(I). Systems (schematic)} \\
\text{Bulk$\mid$Boundary}\\
\text{Bulk$\mid$Interface$\mid$Bulk}\\
\end{matrix}$
& 
$\begin{matrix}
\text{(II). Description,} \\
\text{Cocycle/cochain expressions} 
\end{matrix}$
& 
$\begin{matrix}
\text{(III). Realization}\\
\text{Criteria and Comments} 
\end{matrix}$ \\
\hline\\[-4mm]
(i) 
 $\begin{matrix}
\includegraphics[scale=0.45]{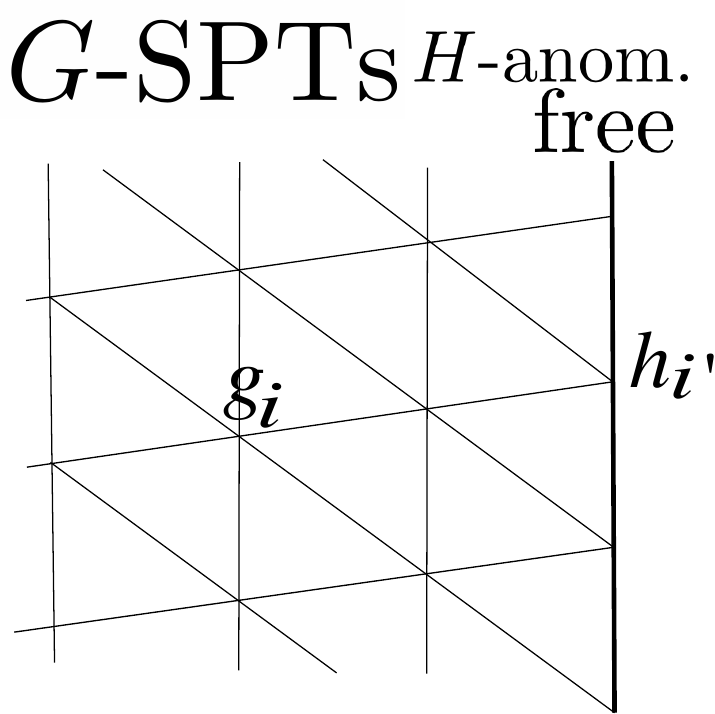}  
\end{matrix}$ 
&  $\begin{matrix}
\text{$G$-SPTs $\nu_d^G$ in the bulk.} \\
\text{$\mu_{d-1}^H$ on the boundary.} \\
\text{Boundary $G$-anomaly}\\ 
\text{becomes $H$-anomaly-free.}
\end{matrix}$ 
&
 $\begin{matrix}
 \text{Sec.~\ref{bdry2w}, \ref{sec:extSymm}'s second boundary.}\\
\text{Sec.~\ref{mixed}, \ref{sec:ZSPTSETG}, \ref{sec:generalGBDWHamiltonian}, and \ref{sec:sym-enhanced:bdry}.}\\
1 \to {K} \to H \overset{r}{\rightarrow}  G \to 1.\\
\nu_d^G(r(h))=\nu_d^H(h)= \delta \mu_{d-1}^H (h).
\end{matrix}$
\\
\hline
(ii) 
$\begin{matrix}
\includegraphics[scale=0.45]{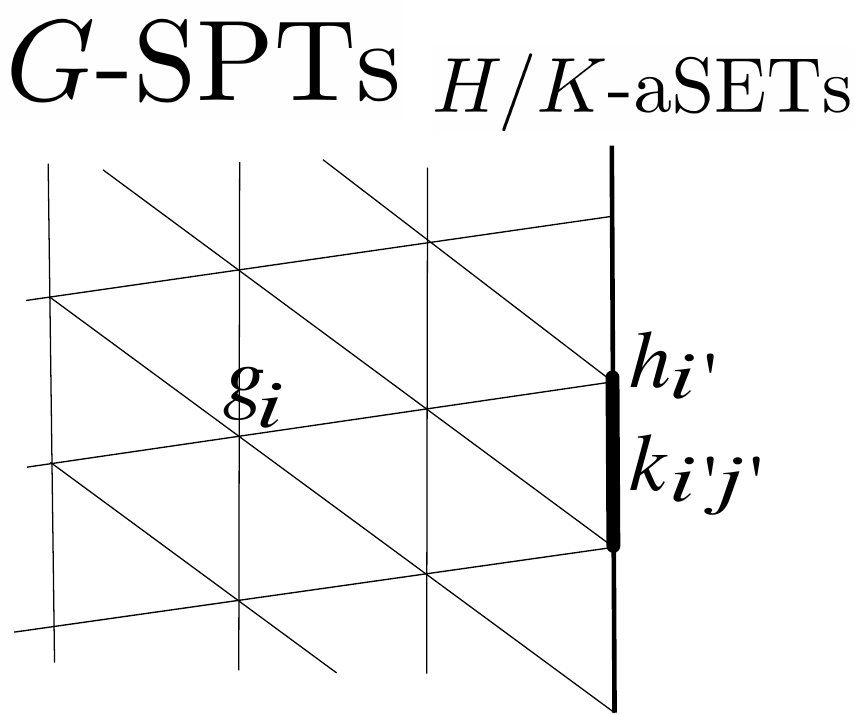}  
\end{matrix}$ & 
$\begin{matrix}
\text{$G$-SPTs $\nu_d^G$ in the bulk.} \\
\text{$\mathcal{V}_{d-1}^{H,K}$ on the boundary,} \\
\text{with a total $H$, a gauge $K$,}\\ 
\text{and a $G$-anomaly.}
\end{matrix}$ 
& 
$\begin{matrix}
 \text{Sec.~\ref{bdry3w}, \ref{sec:hard-gauge}'s third boundary,}\\ 
 \text{Sec.~\ref{bdry4w}, \ref{sec:cocycle-emergent}'s fourth boundary.}\\
\text{Sec.~\ref{mixed}, \ref{sec:ZSPTSETG}, \ref{sec:generalGBDWHamiltonian}, and \ref{sec:sym-enhanced:bdry}.}\\
1 \to {K} \to H \overset{r}{\rightarrow}  G \to 1.\\
\text{Gauge $\mu_{d-1}^H (h)$ to $\mathcal{V}_{d-1}^H (h;k)$.}
\end{matrix}$
\\
\hline
(iii)$\begin{matrix} \includegraphics[scale=0.45]{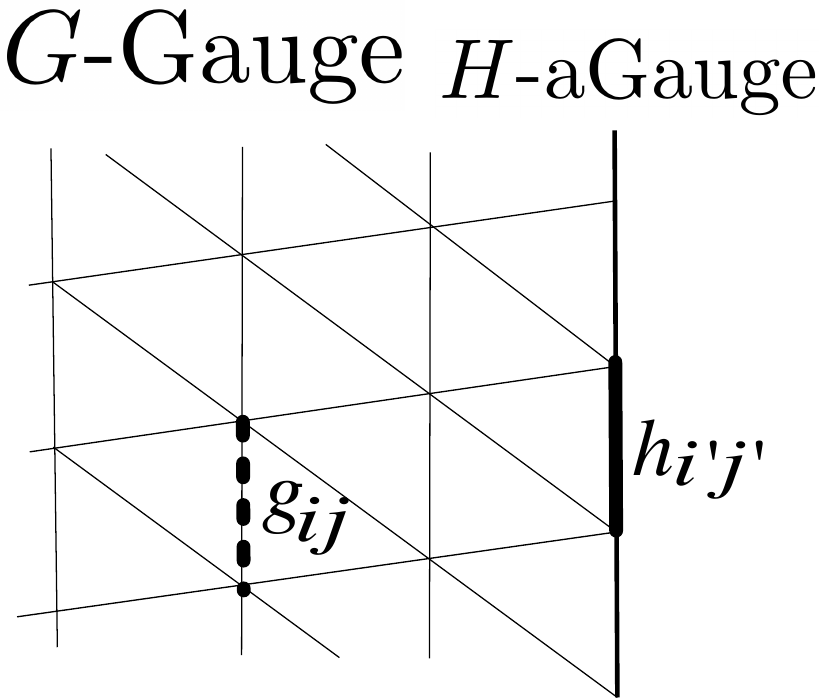} \end{matrix}$   
&
 $\begin{matrix}
\text{$G$-TO $\omega_d^G$ in the bulk.} \\
\text{$\Omega_{d-1}^{H}$ on the boundary,} \\
\text{with a total gauge group $H$.}
\end{matrix}$ 
&
 $\begin{matrix}
  \text{The fifth boundary.}\\
\text{Sec.~\ref{ito}, \ref{sec:ZSPTSETG}, \ref{sec:generalGBDWHamiltonian}, and \ref{sec:sym-enhanced:bdry}.}\\
1 \to {K} \to H \overset{r}{\rightarrow} G \to 1.\\
\omega_d^G(r(h))=\omega_d^H(h)= \delta \Omega_{d-1}^{H}(h).
\end{matrix}$
\\
\hline
(iv)$\begin{matrix} \includegraphics[scale=0.45]{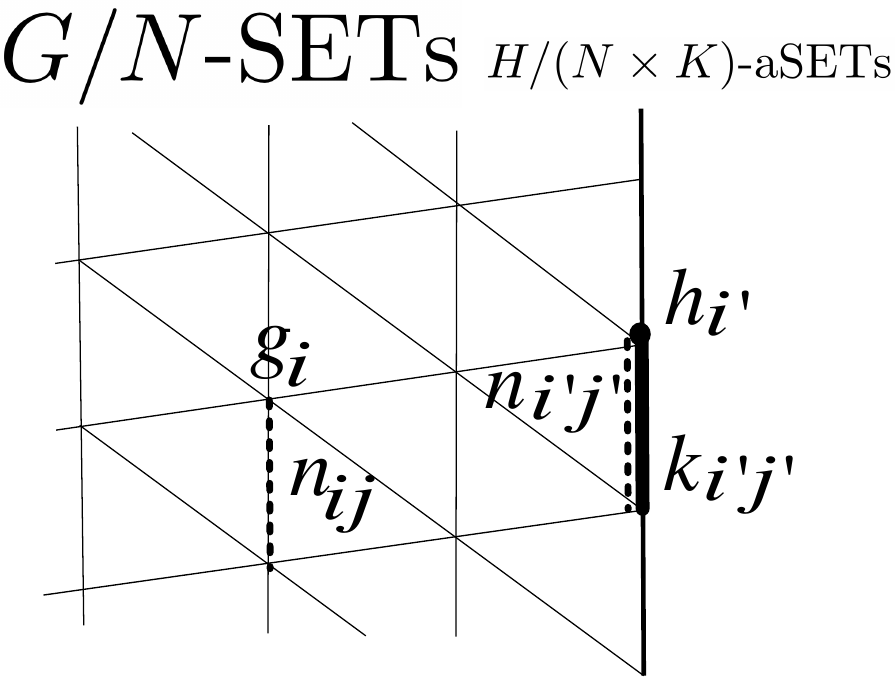} \end{matrix}$  
& 
 $\begin{matrix}
\text{$G/N$-SETs $\cV_d^{G,N}$ in the bulk.} \\
\text{$\cV_{d-1}^{H, N, K}$ on the boundary}\\
\text{with a total $H$, a gauge $(N \times K)$,}\\
\text{and an $H/(N \times K)=G/N$-anomaly.} 
\end{matrix}$ 
&
 $\begin{matrix}
\text{Sec.~\ref{mixed}, \ref{ito},}\\
\text{ \ref{sec:ZSPTSETG}, \ref{sec:generalGBDWHamiltonian} and \ref{sec:sym-enhanced:bdry}.}\\
1 \to N \to G  \to Q \to 1.\\
1 \to {K \times N}\to H \to Q \to 1.
\end{matrix}$ \\
\hline
(v)$\begin{matrix} \includegraphics[scale=0.45]{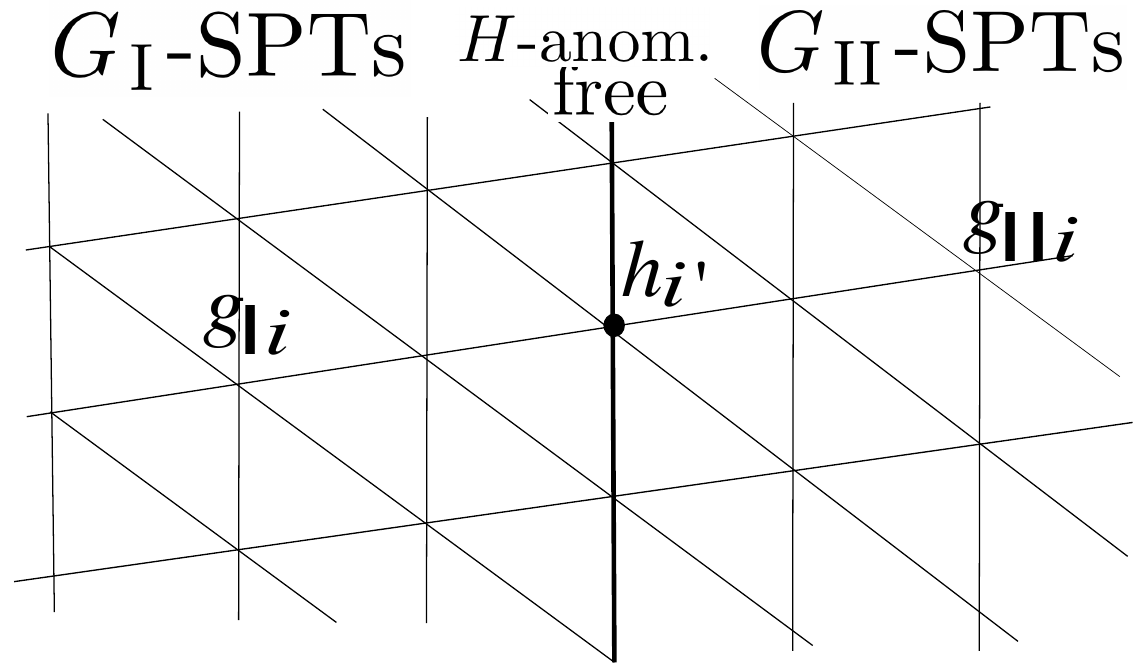} \end{matrix}$  
&
$\begin{matrix}
\text{$G_{\I}$- and $G_{\II}$-SPTs} \\
\text{$\nu_d^{G_{\I}}$ and $\nu_d^{G_{\II}}$ in the bulk.}\\
\text{$\mu_{d-1}^H$ on the interface}\\
\text{with $H$-anomaly free.} 
\end{matrix}$
&
 $\begin{matrix}
\text{Sec.~\ref{interfaces}, \ref{sec:ZSPTSETG1G2}, \ref{sec:generalGBDWHamiltonian} and \ref{sec:sym-enhanced:DW}.} \\
1 \to K \to H \to  {G_{\I} \times G_{\II}}  \to 1.
\end{matrix}$ 
\\
\hline
(vi)$\begin{matrix} \includegraphics[scale=0.45]{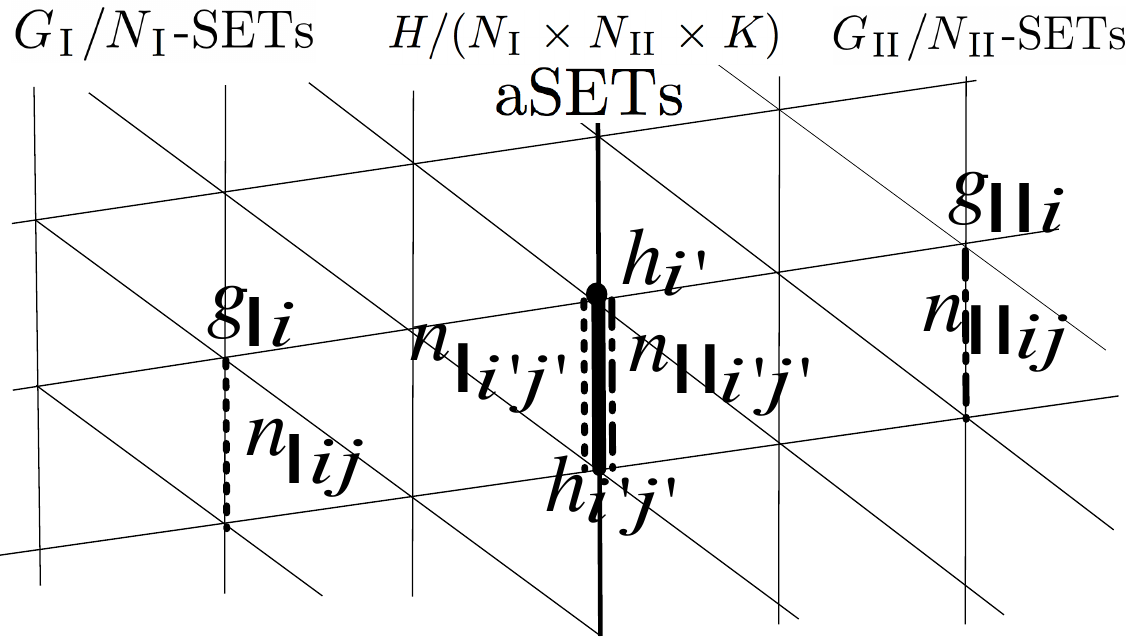} \end{matrix}$  
& 
$\begin{matrix}
\text{$G_{\I}/N_{\I}$- and $G_{\II}/N_{\II} $-SETs} \\
\text{ $\cV_d^{{G_{\I}},{N_{\I}}}$ and $\cV_d^{{G_{\II}},{N_{\II}}}$ in the bulk.}\\
\text{$\cV_{d-1}^{H, N_{\I}  \times N_{\II}, K}$ on the interface,}\\ 
\text{a total $H$, a gauge $(N_{\I}  \times N_{\II}  \times K)$,}\\
\text{and an $H/(N_{\I}  \times N_{\II}  \times K)$-anomaly.} 
\end{matrix}$
&
 $\begin{matrix}
\text{Sec.~\ref{interfaces}, \ref{ito},}\\  
\text{\ref{sec:ZSPTSETG1G2}, \ref{sec:generalGBDWHamiltonian}, and \ref{sec:sym-enhanced:DW}.}\\
1 \to N_{\I} \times N_{\II}  \to G_{\I} \times G_{\II}  \to Q \to 1.\\
1 \to {K \times N_{\I} \times N_{\II}}\to H \to Q \to 1.
\end{matrix}$ 
\\
\hline
\end{tabular}
} \hspace*{35mm}
\caption{A quick guide for
general constructions of \emph{symmetry-extended} gapped interfaces for 
symmetric-protected topological states (SPTs), topological orders (TOs), 
and symmetric-enriched topologically ordered states (SETs) in $d$D spacetime,
see the menu links to Sec.~\ref{sec:sym-enhanced:bdryDW} and \ref{sec:general}.
}
\label{table:summary1}
\end{table}
\end{center}

\begin{center}
\begin{table}[!h]
\noindent
\makebox[\textwidth][c] 
{
\begin{tabular}{ccc} 
\hline
(vii)$\begin{matrix} \includegraphics[scale=0.45]{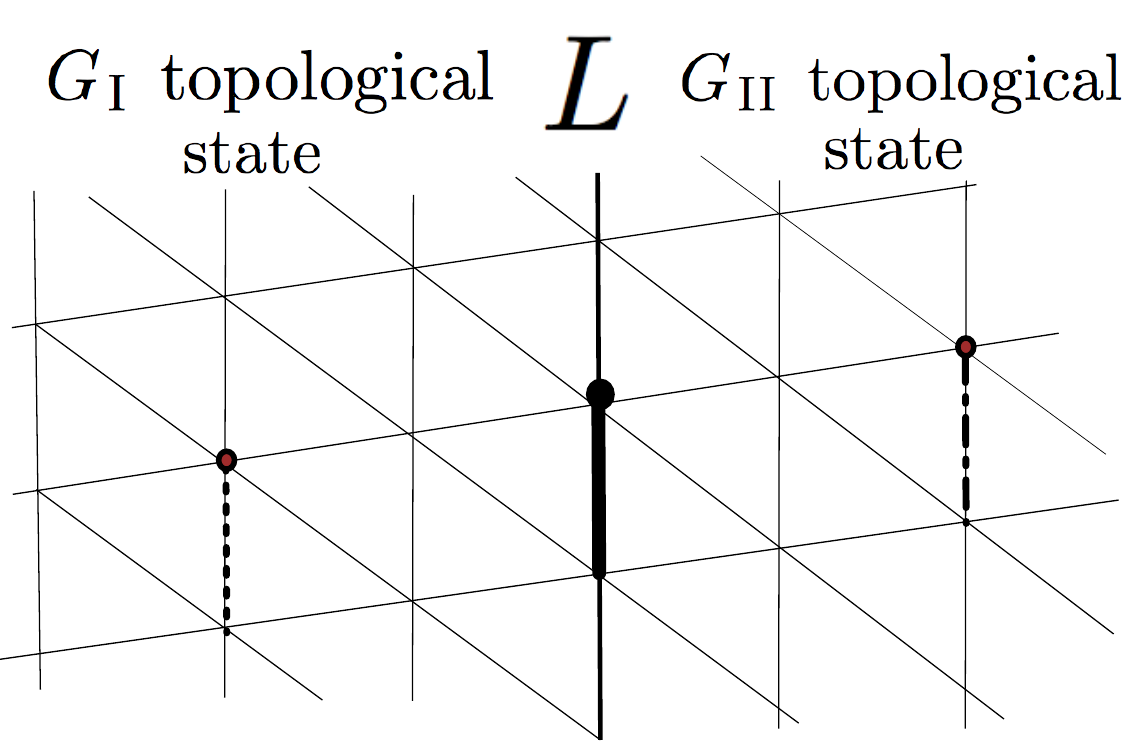} \end{matrix}$  
& 
$\begin{matrix}
\text{$G_{\I}$- and $G_{\II}$-topological states:} \\
\text{ $\cV_d^{{G_{\I}}}$ and $\cV_d^{{G_{\II}}}$ in the bulk,}\\
\text{$L$-interfaces:} \\
\text{$\cV_{d-1}^{L}$ on the interface.}
\end{matrix}$
&
 $\begin{matrix}
\text{Sec.~\ref{mixed},
\ref{ito},}\\
\text{and \ref{sec:gauged-DW-interface}.} \\
L  \to G_{\I} \times G_{\II}.
\end{matrix}$ 
\\
\hline
\end{tabular}
} \hspace*{35mm}
\caption{The schematic figure shows a generic combined framework including not only
\emph{symmetry-extension}, but also \emph{symmetry-breaking} and 
 \cred{\emph{dynamically gauging}}  of topological states (e.g. SPTs, TOs or SETs).
 The $G_{\I}$- and $G_{\II}$-topological states are described by some cocycle $\cV_d^{{G_{\I}}} (\cV_d^{{G_{\II}}})^{-1}$ (or a nontrivial $G_{\I} \times G_{\II}$-bundle, 
 which can encode both global symmetry group and gauge group), 
 when pulling this cocycle back to $L$, it becomes a trivial coboundary in $L$.
 Here we only require the map $L  \to G_{\I} \times G_{\II}$ to be group homomorphism, but not necessarily surjective nor injective.
Sec.~\ref{mixed} discusses the mixed mechanisms. 
Dynamically gauged gapped interfaces of topologically ordered gauge theories are explored in Appendix \ref{sec:gauged-DW-interface}.}
\label{table:summary2}
\end{table}
\end{center}

\begin{table}[h!]
\footnotesize
\begin{center}
    \begin{tabular}{| c | c|c |  }
    \hline
   Dim & Spacetime lattice (for path integral $Z$)  & Spatial lattice (for Hamiltonian $\widehat{H}$)  \\ \hline\hline
1+1D & 1+1D \includegraphics[scale=.3]{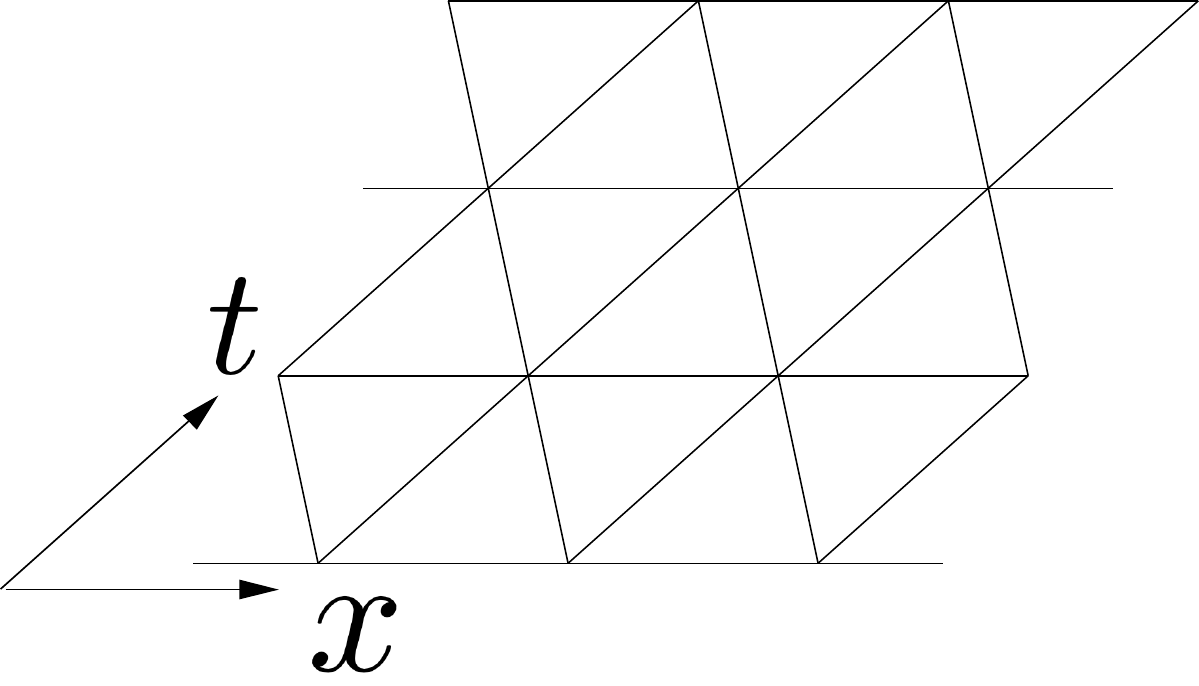} & \dots 
\\ \hline 
2+1D & 2+1D \includegraphics[scale=.3]{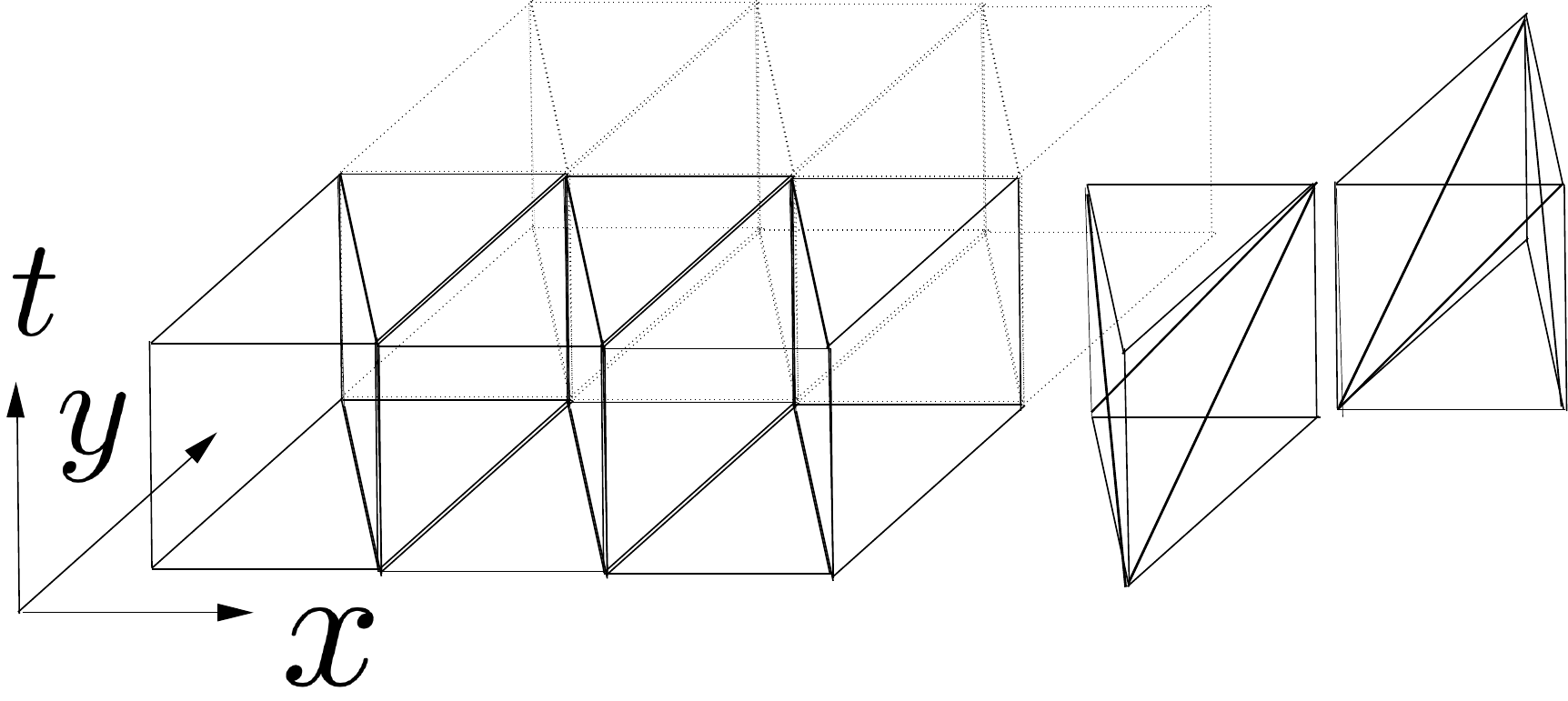} & 2D \includegraphics[scale=.3]{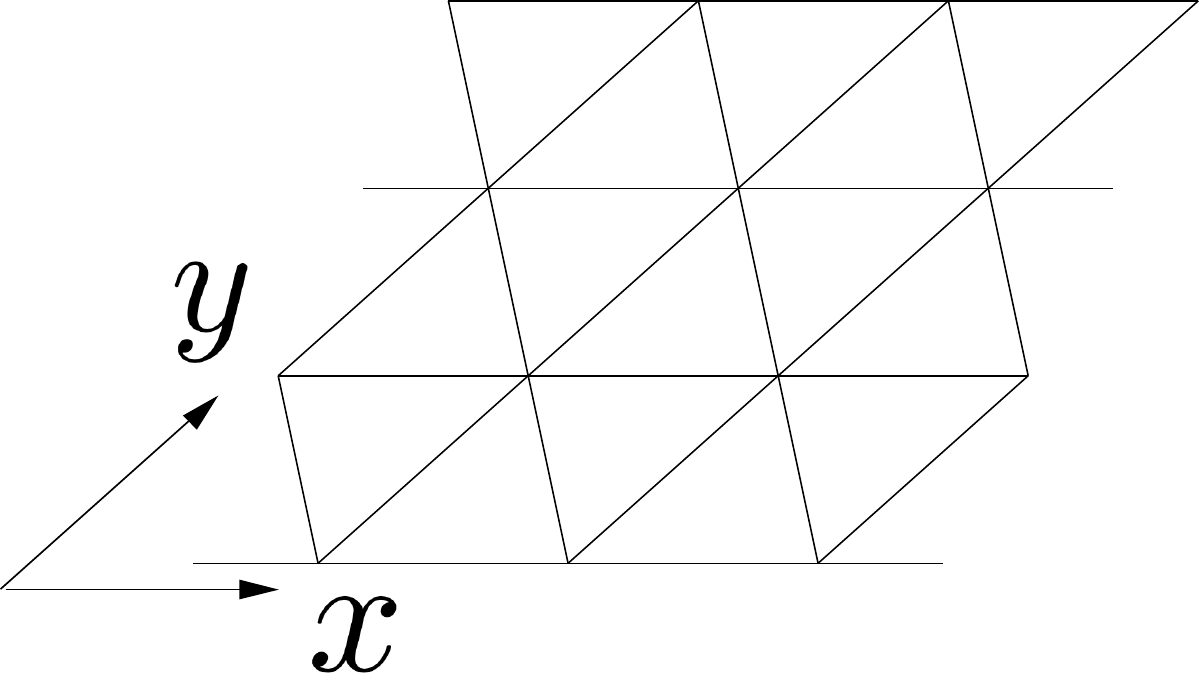}  
 \\ \hline    
3+1D  & \dots & 3D \includegraphics[scale=.3]{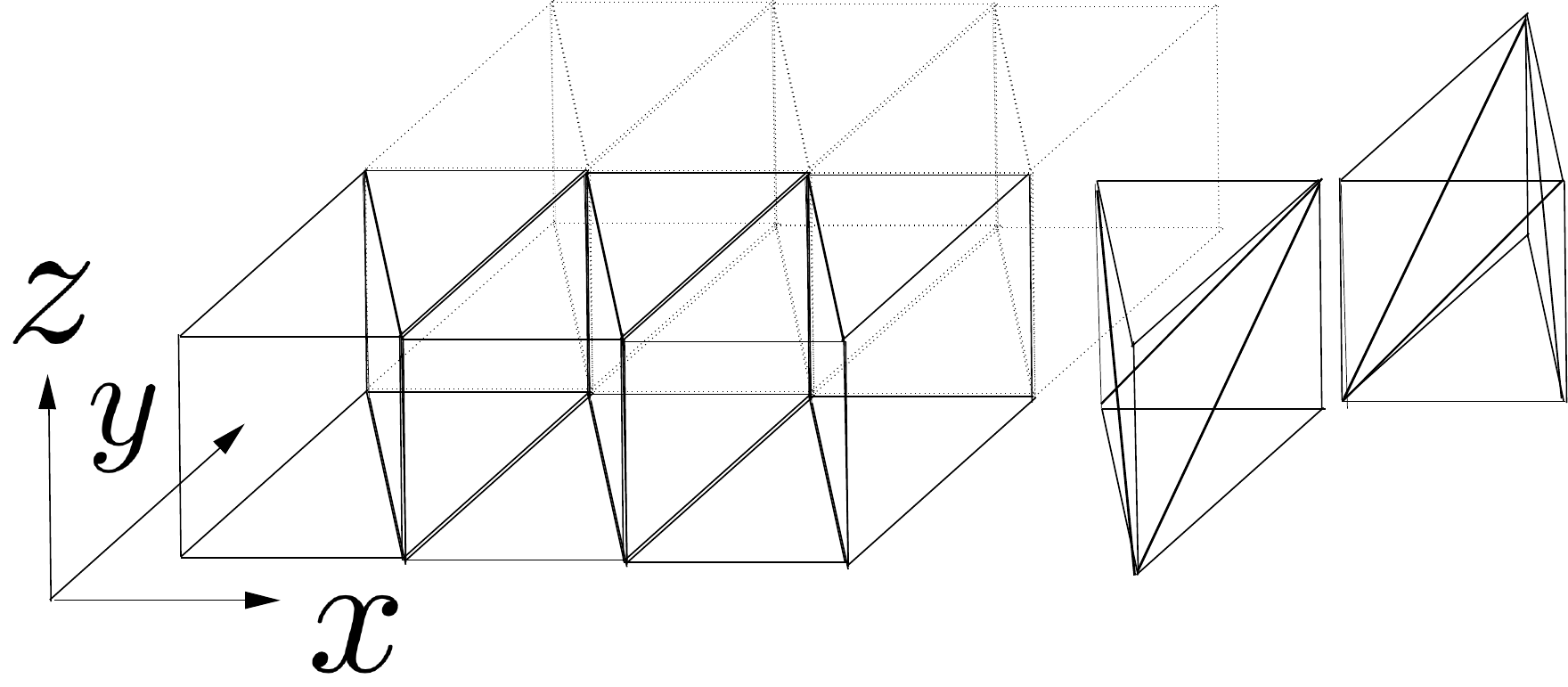}
\\ \hline    
\dots & \dots& \dots
\\ 
\hline
    \end{tabular}
    \end{center}
\caption{An introduction of spatial lattices and spacetime lattices based on packing of $d$-simplices, 
suitable for the general construction in Sec.~\ref{sec:general}. 
The second column shows that the packing of simplices can be used for defining the spacetime triangulation, in order to define the spacetime path integral $Z$.
The third column shows that, the spatial lattice for Hamiltonian systems on a 2D $x$-$y$ plane is filled with 2-simplices, 
each vertex has 6 nearest neighbor vertices. (For instance, the Hamiltonian $\widehat{A}_v$ term in \Eqn{eq:Avgn} thus contains a product of 6 of 3-cocycles.) 
The spatial lattice on a 3D $x$-$y$-$z$ space is filled with 3-simplices. 
(For instance, the Hamiltonian $\widehat{A}_v$ term in \Eqn{eq:Avgn} thus contains a product of 24 of 4-cocycles)}
\label{table:Tablelattice}
\end{table}

\noindent
In Table \ref{table:summary1} (iii), $G$-gauge means a generic twisted $G$-gauge theory (possibly with a Dijkgraaf-Witten cocycle).
The $H$-aGauge on the boundary means that 
it has a full dynamical gauge group $H$ on the boundary. 
But the boundary theory is not the usual gauge theory in its own dimensions described by $(d-1)$-cocycles, but by special $(d-1)$-cochains 
with additional gauge holonomy conservation constraints.

\noindent
In Table \ref{table:summary1} (iv), $G/N$-SETs means a SET state with a full group $G$, a dynamical gauge group $N$, and a global symmetry $G/N=Q$.
The $H/(N\times K)$-aSETs means a symmetry-enriched boundary state with
a full group $H$, a dynamical gauge group $N \times K$  and it has a boundary $G$-anomaly (from anomalous non-onsite $G$-global symmetry transformation on the boundary).

\noindent
In Table \ref{table:summary1} (v), $G_{\I}$- and $G_{\II}$-SPTs with non-onsite  $G_{\I}$- and $G_{\II}$- symmetries can have an onsite $H$-symmetry on the shared gapped interface.
Thus the two topological states become anomaly free by pulling them back to a certain larger $H$.\\

\noindent
In Table \ref{table:summary1} (vi), 
the $G_{\I}/N_{\I}$-SETs means a SET state with a full group $G_{\I}$, a dynamical gauge group $N_{\I}$, and individually has a global symmetry $G_{\I}/N_{\I}$.  
The $G_{\II}/N_{\II}$-SETs means a SET state with a full group $G_{\II}$, a dynamical gauge group $N_{\II}$, and individually has a global symmetry $G_{\II}/N_{\II}$.
A global symmetry on the whole system including the left and right sectors become $(G_{\I} \times G_{\II})/ (N_{\I} \times N_{\II})=Q$.
The $H/(N_{\I} \times N_{\II}\times K)$ aSETs means a symmetry-enriched boundary state, with a full group $H$, a dynamical gauge group $(N_{\I} \times N_{\II}\times K)$. 
The boundary has a $Q$-anomaly, where $Q=H/(N_{\I} \times N_{\II}\times K)$ (from an anomalous non-onsite $Q$-global symmetry transformation on the boundary).

\noindent
In Table \ref{table:summary2} (vii), we consider generic $G_{\I}$ topological state and $G_{\II}$ topological state (of SPTs, TOs or SETs), 
and construct generic gapped interfaces based on mixed mechanisms of \emph{symmetry-extension},  \emph{symmetry-breaking} and \cred{\emph{dynamically gauging}}.
The interface of $L$ is found by trivializing the nontrivial cocycle or bundle associated to  $G_{\I} \times G_{\II}$
via pulling back to $L$ from a generic group homomorphism $L  \to G_{\I} \times G_{\II}$.

In Table \ref{table:Tablelattice}, we show a schematic systematic lattice construction of the above systems, suitable for spacetime path integrals and
Hamiltonian/wavefunctions. Their details are in Sec.~\ref{sec:general}.


\section{A model that realizes the 2+1D $Z_2$ SPT state: CZX model} \label{bCZX}

\begin{figure}[!h] 
\begin{center}
\includegraphics[scale=.8]{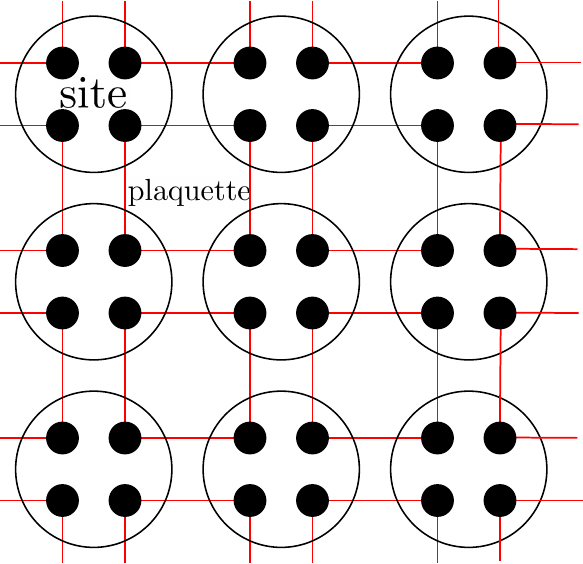} 
\end{center}
\caption{
The CZX model.  Each site (a large disc) contains four qubits or objects of spin 1/2 (shown
as small black dots).  The squares, formed by red links, are plaquettes, introduced later.
} 
\label{bulk} 
\end{figure}

The first lattice model that realizes a 2+1D SPT state (the $Z_2$-SPT state)
was introduced by Chen-Liu-Wen \cite{XieSPT3}, and was named the CZX model.  The CZX model is
a model on a square lattice (Fig. \ref{bulk}), where each lattice site contains
four qubits, or objects of spin-1/2.  For each spin, we use a basis  $|\up\>$
and  $|\down\>$ of $\si^z$ eigenstates.  Thus a single site has a Hilbert space
of dimension $2^4$.  

Now let us introduce a $Z_2$ symmetry transformation. An obvious choice is the operator that acts
on each site $s$ as
\begin{align}
\label{UX}
U_{X,s}= 
\prod_{j=1}^4 \sigma^x_j ,\ \ \ \ U_{X,s}^2=1,
\end{align}
 which simply flips the four spins in site $s$.  However, to construct the CZX model, a more subtle choice is made.
In this model, in the basis $| \uparrow \>$, $| \downarrow \>$, the flip operator $U_{X,s}$ is modified with $\pm $ signs.
For a pair of spins $i,j$, we define an operator\footnote{The name $CZ$ is read ``controlled $Z$'' and is suggested by quantum computer
science.  The operator $U_{CZ,ij}$ measures $\sigma_z$ of spin $j$ if spin $i$ is in state $|\down\rangle$ and otherwise does
nothing.} $U_{CZ,ij}$ that acts as $-1$ if spins $i,j$ are both in state $|\down\>$, and otherwise
acts as $+1$.   There are various ways to describe $U_{CZ,ij}$ by a formula:
\begin{align}
U_{CZ,ij} 
&= \frac{1+\si^z_{i}+\si^z_{j}  -\si^z_{i}\si^z_{j}}{2} \nonumber\\
&=\ii^{(\si^z_i+\si^z_j - \si^z_i\si^z_j-1)/2}. 
\end{align}
Now for a site $s$ that contains four spins $j=1,\dots,4$ in  cyclic order, we define
\be\label{spinsym}
U_{CZ,s}=\prod_{j=1}^4 U_{CZ,j\,j+1}.
\eeq
The $Z'_2$ symmetry of the spins at site $s$ is defined as
\be\label{defsym}
U_{CZX,s}=U_{X,s}U_{CZ,s}.\eeq
 By a short exercise, one can verify that  $U_{X,s}$ and $U_{CZ,s}$ commute and accordingly
 that $U_{CZX,s}^2=1$.   The $Z_2$ symmetry generator of the CZX model is defined as a product over all sites
 of $U_{CZX,s}$:
 \be\label{wels}U_{CZX}=\prod_s U_{CZX,s}. \eeq
 
 Clearly this is an on-site symmetry, that is, it acts separately on the Hilbert space associated to each site.  Being onsite, the symmetry
 is gaugeable and anomaly-free.  We have not yet picked a Hamiltonian for the CZX model, but whatever $U_{CZX}$-invariant
 Hamiltonian we pick, the $Z_2$ symmetry can be gauged by coupling to a $Z_2$ lattice gauge field that will live on links that connect
 neighboring sites.
 
 What we have done so far is trivial in the sense that, by a change of basis on each site, we could have put $U_{CZX,s}$ in a more
 standard form.  However this would complicate the description of the Hamiltonian and ground state wavefunction of the CZX model,
 which we come to next. 
 
\begin{figure}[!h] 
\begin{center}
\includegraphics[scale=.8]{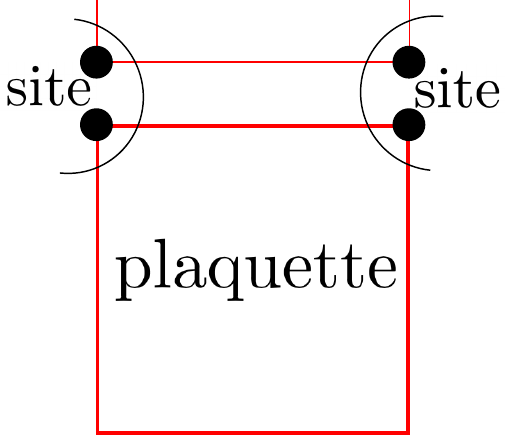} 
\end{center}
\caption{ 
{A pair of adjacent spins:}
To preserve the symmetry $U_{CZX}$, we choose a Hamiltonian that only flips the spins
in a plaquette if pairs of adjacent spins in neighboring plaquettes are equal.  Thus the spins shown here at the top of this plaquette are only flipped
if the two spins just above them are equal.  Both the spins in the plaquette and the ones just above them are in different sites, as shown.} 
\label{neighbors} 
\end{figure}

It is easier to first describe the desired ground state wavefunction of the model and then describe a Hamiltonian that has that
ground state. In Fig. \ref{bulk}, we have drawn squares that contain four spins,
one from each of four neighboring sites.  We call these squares ``plaquettes.''   For each plaquette $p$, we define the wavefunction
$ |\Psi_p\> = \frac{1}{\sqrt{2}} ( |\up\up\up\up\> +
|\down\down\down\down\> )$.  The  ground state of the CZX model  in the bulk is given by a product over all plaquettes of this
wavefunction for each plaquette:
\begin{align}\label{gstate}
 |\Psi_{\mathrm{gs}}\> &= \prod_p |\Psi_p\>
= \prod_p\frac{1}{\sqrt{2}} ( |\up\up\up\up\> +   |\down\down\down\down\> )
.\end{align}
This state is  $U_{CZX}$-invariant, 
\begin{align}
U_{CZX} |\Psi_{\mathrm{gs}}\>=|\Psi_{\mathrm{gs}}\>,
\end{align}
if we define the whole system on a torus without boundary (i.e., with periodic boundary conditions).
But that fact is not completely trivial: It depends on cancellations among $CZ_{ij}$ factors
for adjacent pairs of spins, see Fig. \ref{neighbors}.

 Clearly, the entanglement in this wavefunction is short-range, and this wavefunction describes a gapped state.  Moreover, if
 we would regard the plaquettes (rather than the large discs in Fig. \ref{bulk}) as ``sites,'' then this wavefunction would be a trivial
 product state.  But in that case the $Z_2$ symmetry of the model would not be on-site.  The subtlety of the model comes from the
 fact that we cannot simultaneously view it as a model with on-site symmetry and a model with a trivial product ground state.

 \begin{figure}[!h]
\begin{center}
(1) \includegraphics[scale=1.2]{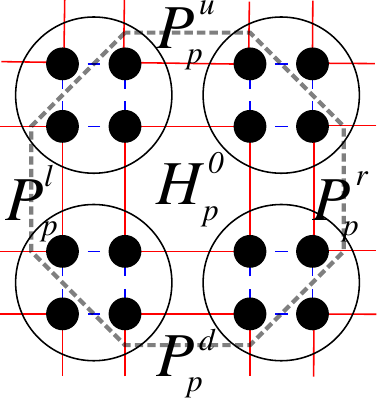} \;\;\;\;\;\;\;\;\;\;\;\;\;\;\;\;\;\; (2) \includegraphics[scale=0.8]{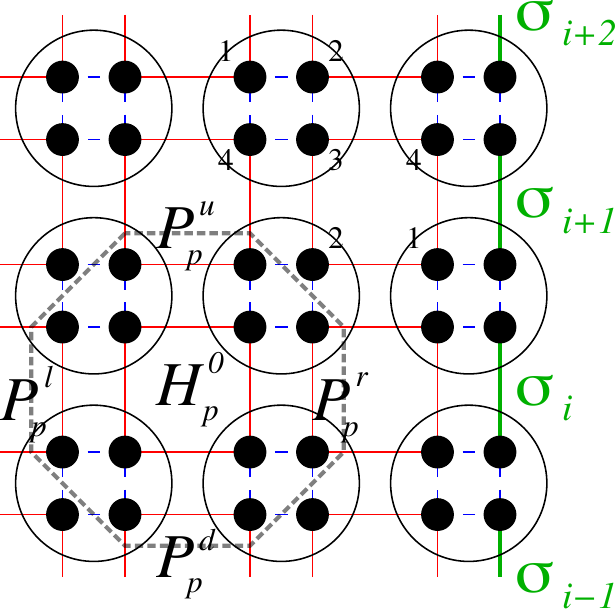} 
\end{center}
\caption{
Each plaquette Hamiltonian $H_p$ acts on the spins contained in an octagon, as depicted in dashed gray line in the left subfigure (1) and
also in the lower left of the right subfigure (2).
In the subfigure (2),
the octagon in the lower left contains the four spins in plaquette $p$ and four adjacent pairs of spins.
In the case of a finite sample made of complete sites, as depicted here, most of the spins can be grouped in plaquettes,
but there is a row of spins on the boundary -- shown here on the right of the figure -- that are not contained in any
plaquette.  However, the Hamiltonian  acts on these boundary spins through the projection operators $P_p^\alpha$ from
a neighboring plaquette.
} 
\label{bdryA1} 
\end{figure}

 The most obvious Hamiltonian with $|\Psi_{\mathrm{gs}}\>$ as its ground state would be a sum over all plaquettes $p$ of an operator $H_p^0$
 that flips all spins in plaquette $p$:
 \be\label{obvious}H^0=\sum_p H_p^0,~~~H_p^0=-\bigl(|\neg \up\up\up\up\rangle\langle\down\down\down\down\neg|+|\neg \down\down\down\down\rangle
 \langle  \up\up\up\up\neg|\bigr).\eeq
 This Hamiltonian commutes with the obvious $Z_2$ symmetry that flips all the spins, but does not commute with the more
 subtle  symmetry  $U_{CZX}$.  To commute with $U_{CZX}$, we modify $H^0$ to only flip the spins
 in a plaquette if adjacent pairs of spins in the neighboring plaquettes are equal (Fig. \ref{neighbors}). 
For a plaquette $p$, we define operators 
\cblue{$P_p^\alpha \equiv |\neg \up\up\rangle\langle\up\up\neg|+|\neg \down\down\rangle \langle\down\down\neg|$} 
that project onto states
in which the two spins adjacent to $p$ in the $\alpha$ direction (where $\alpha$ equals up, down, left, or right,
\cblue{denoted as $u$, $d$, $l$, or $r$}) are equal.  
Then the CZX Hamiltonian is defined to be
\begin{align} \label{hdec}
H &=\sum_p H_{p}
\nonumber\\
H_{p}&=-\bigl(|\neg \neg\up\up\up\up\rangle\langle\down\down\down\down\neg\neg|+|\neg \neg\down\down\down\down\rangle
 \langle  \up\up\up\up\neg\neg|\bigr)\otimes_\alpha P_p^\alpha .
\end{align}
Thus each $H_p$ acts on the spins contained in an octagon (Fig. \ref{bdryA1}(1)), flipping the spins in a plaquette if all adjacent pairs
of spins are equal. 
\cblue{This Hamiltonian is  $U_{CZX}$-invariant, 
\begin{align}
[ U_{CZX}, H]=0,
\end{align}
in the case of a system without boundary (an infinite system or a finite system with periodic
boundary conditions).
}
The state $|\Psi_{\mathrm{gs}}\rangle$ is a symmetry-preserving ground state with short-range entanglement.
However, it is a nontrivial symmetry-protected topological or SPT state.  This becomes clear if we examine
possible boundaries of the CZX model.

\section{Boundaries of the CZX model}

\label{sec:bdryCZX}

\subsection{The first boundary of the CZX model -- \\
{1+1D symmetry-preserving gapless boundary
with a non-on-site global $Z_2$-symmetry}
}

\label{bdry1}
The boundary of the CZX model that was studied in the original paper is a very natural one in which one simply considers
a finite system with an integer number of sites (Fig. \ref{bdryA1}(2)).  One groups the spins into plaquettes, as before, but as shown
in the figure, there is a row of spins on the boundary that are not contained in any complete plaquette.  We call these the boundary spins.

We define the Hamiltonian as in eqn. (\ref{hdec}), where now the sum runs over complete plaquettes only.   Because the boundary
spins are not contained in any complete plaquette, the system is no longer gapped.  However, the boundary spins are not completely
free to fluctuate at no cost in energy.  The reason is that, to minimize the energy, a pair of boundary spins that are adjacent to
a plaquette $p$ are constrained to be equal.  This is because of the projection operators $P_p^\alpha$ in the definition of $H_p$.

Hence, in a state of minimum energy, the boundary spins are locked together in pairs.  These pairs are denoted as $\upsigma_i$,
$\upsigma_{i+1}$, etc., in Fig. \ref{bdryA1}(2), and one can think of them as composite spins.

How does the $Z_2$ symmetry generated by $U_{CZX}$ act on the composite spins?  Evidently, $U_{CZX}$ will flip each composite
spin.  However, $U_{CZX}$ also acts by a $CZ$ operation on each adjacent pair of composite spins $\upsigma_i$, $\upsigma_{i+1}$.
That is because, for example,  in Fig. \ref{bdryA1}(2), the ``upper'' spin making up the composite spin $\upsigma_i$ and the ``lower'' spin making 
up $\upsigma_{i+1}$ are adjacent spins contained in the same site $s$ in the underlying square lattice.   Accordingly, in the $Z_2$ generator
$U_{CZX,s}$ for site $s$, there is a $CZ$ factor linking these two spins.  

Therefore, the effective $Z_2$ generator for the composite spins on the boundary is
\begin{align}
 \widehat U_{Z_2} = \prod_i \si^x_i U_{CZ,i\, i+1} .
\end{align}
The product runs over all composite spins $\upsigma_i$; $\widehat U_{Z_2}$ is the product of operators $\si^x_i$ that flip $\upsigma_i$
and operators $U_{CZ,i\, i+1}$ that give the usual $CZ$ sign factors for each successive pair of composite spins.
Clearly, this effective $Z_2$ symmetry is not on-site.  No matter how we group a finite set of composite spins into boundary sites,
the operator $U_{Z_2}$ will always contain $CZ$ factors linking one site to the next.\footnote{In the
case of a compact ring boundary, $(\widehat U_{Z_2})^2=+1$ for an even-site boundary, while $(\widehat U_{Z_2})^2=-1$ for an odd-site boundary. 
To avoid the even or odd lattice site effect, from now on we assume the even-site boundary system throughout our work for simplicity.
If there are no corners or spatial defects or curvature -- which would lead to corrections in these statements -- then the
number of odd-site boundary components is always even, so overall $\widehat U_{Z_2}^2=1$.}

With the Hamiltonian as we have described it so far, all states labeled by any values of the composite spins $\upsigma_i$,
but with complete bulk plaquettes placed in their ground state $|\Psi_p\rangle$, are degenerate.  
Of course, it is possible to add perturbations that partly lift the degeneracy.   However, it has been shown in Ref.\cite{XieSPT3}
that the non-onsite nature of the effective $Z_2$ symmetry gives
an obstruction to making the boundary gapped and symmetry-preserving.

\subsection{The second boundary of the CZX model -- \\
{1+1D gapped boundary by
extending the $Z_2$-symmetry to a $Z_4$-symmetry}}

\label{bdry2w}

\begin{figure}[!h]
\begin{center}
\includegraphics[scale=0.8]{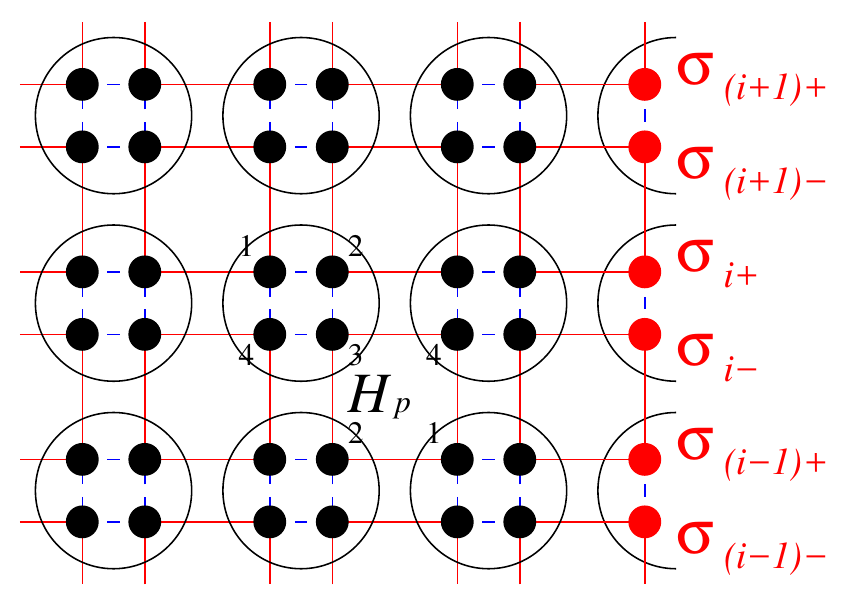}
\end{center}
\caption{
By omitting  the right row of spins from the boundary of Fig. \ref{bdryA1}(2), we get an alternative boundary of the CZX model.
Now all spins are contained in plaquettes, but on the boundary there are ``incomplete sites,'' shown as semicircles on the right
of the figure, that contain only two spins instead of four.  The ``upper'' and ``lower'' spins of the $i^{th}$ boundary site
have been labeled $\upsigma_{i+}$ and $\upsigma_{i-}$. 
}
\label{bdryB1}
\end{figure}

The main idea of the present paper can be illustrated by a simple alternative boundary of
the CZX model.  To construct this boundary, we simply omit the boundary spins from the previous discussion.
This means that now, the system is made of complete plaquettes, even along the boundary (Fig. \ref{bdryB1}),
but there is a row of boundary spins that are not in complete sites.  As indicated in the figure, we combine
the boundary spins in pairs into boundary sites.  Thus a boundary site has only two spins while a bulk site has four.
In the figure, we have denoted the ``upper'' and ``lower'' spins in the $i^{th}$ boundary site as $\upsigma_{i+}$ and $\upsigma_{i-}$.

To specify the model, we should specify what the Hamiltonian looks like near the boundary and how the global symmetry is defined
for the boundary spins.  First of all, now that all spins are in complete plaquettes, we can look for a gapped system with the
same ground state wave function as in eqn. \ref{gstate}:
\begin{align}\label{gdstate}
 |\Psi_{\mathrm{gs}}\> &= \prod_p |\Psi_p\>
= \prod_p\frac{1}{\sqrt{2}} (|\neg\up\up\up\up\> +   |\neg\down\down\down\down\> )
.\end{align}
To get this ground state, we define the Hamiltonian by the same formula as in eqn. (\ref{hdec}).  Only one very small
change is required: A boundary plaquette is adjacent to only three pairs of spins instead of four, so in the definition of $H_p$
in eqn. (\ref{hdec}), if $p$ is a boundary plaquette, the product of projection operators $\otimes_\alpha P_p^\alpha$ contains
only three factors and not four.

The last step is to define the action of the global ``$Z_2$'' symmetry for boundary sites.  
We have put ``$Z_2$'' in quotes for a reason that will be clear
in a moment.  Once we have chosen the Hamiltonian as above, the choice of the global symmetry generator is forced on us.  The symmetry
generator at the $i^{th}$ boundary site will have to flip the two spins $\upsigma_{i+}$ and $\upsigma_{i-}$, of course, but it also
needs to have a $CZ$ factor linking these two spins.   So the symmetry generator of the $i^{th}$ boundary site will have to be
\be\label{symg} U_{CZX,i}=\sigma^x_{i+}\sigma^x_{i-}U_{CZ,i+\,i-}. \eeq
The full symmetry generator is 
\be\label{wymb}U_{CZX} =\prod_s U_{CZX,s},\eeq
where the product runs over all bulk or boundary sites $s$, and $U_{CZX,s}$ is defined in the usual way for bulk sites, and as in eqn.
(\ref{symg}) for boundary states.

We have found a gapped, symmetry-preserving boundary state for the CZX model.  There is a catch, however.
The global symmetry is no longer $Z_2$.  Although the operator $U_{CZX,s}$ squares to 1 if $s$ is a bulk site,
this is not so for boundary sites.  Rather, from (\ref{symg}), we find that for a boundary site,
\be\label{nysg} U_{CZX,i}^2=-\sigma^z_{i+}\sigma^z_{i-}. \eeq
This operator is $-1$ if the two spins $\upsigma_{i+}$ and $\upsigma_{i-}$ in the $i^{th}$ boundary site are both up or both
down, and otherwise $+1$.   Clearly $U_{CZX,i}^2\not=1$,  
so the full global symmetry generator $U_{CZX}$ does not obey
$U_{CZX}^2=1$ but rather
\be\label{ysg} U_{CZX}^4=1. \eeq
Thus, rather than the symmetry being broken by our choice of boundary state, it has been {enhanced from $Z_2$ to $Z_4$.}
But a $Z_2$ subgroup of $Z_4$ generated by $U_{CZX}^2$ acts only on the boundary, since $U_{CZX}^2=1$ for bulk sites.

What we have here is a group extension 
\begin{align}\label{gextt}
 1 \to K \to H \to G \to 1.
\end{align}
$G=Z_2 \equiv Z_2^G$ is the global symmetry group of the bulk theory, $H=Z_4  \equiv Z_4^H$ is the global symmetry of the complete system including
its boundary, and $K=Z_2  \equiv Z_2^K$ (or a different $Z_2'$) is the subgroup of $H$ that acts only along the boundary.  In this case, we denote the exact sequence eqn.(\ref{gextt})
also as
$$
{
 0 \to Z_2^K \to Z_4^H \to Z_2^G \to 0.}
$$
As was explained from an abstract point of view in Sec.~3.3 of Ref.\cite{Witten:2016cio}
and as we will explain more concretely later
in this paper, when \emph{certain conditions} are satisfied, such a group extension along the boundary gives a way to construct
gapped boundary states of a bulk SPT phase.  (As we explain in detail later, the
 relevant condition is that the cohomology class of $G$ that characterizes
the SPT state in question should become trivial if it is ``lifted'' or ``pulled back'' from $G$ to $H$, or more concretely
if certain fields are regarded as elements of $H$ rather than as elements of $G$.)

From a mathematical point of view, this gives another choice in the usual
paradigm that says that the boundary of an SPT phase either is gapless, has
topological order on the boundary, or breaks the symmetry.  {Another
possibility is that the global symmetry of the bulk SPT phase might be extended
(or enhanced) to a larger group along the boundary, satisfying certain
conditions.}  In $1+1$ dimensions, this is a standard result: The usual
symmetry-preserving boundaries of $1+1$-dimensional bulk SPT phases have a
group extension along the boundary. The novelty is that a gapped boundary can
be achieved above $1+1$ dimensions via such a group extension.

Let us pause to explain more fully the assertion that what we have just
described extends a standard $1+1$-dimensional phenomenon to higher dimensions.
In the usual formulation of the $1+1$-dimensional Haldane or Affleck-Lieb-Kennedy-Tasaki (AKLT) spin chain,
one considers a chain of spin 1 particles with $SO(3)$ symmetry.  The boundary
is not gapped and carries spin 1/2.  Alternatively, one could attach a spin 1/2
particle to each end of such a chain.  Then the system can be gapped, with a
unique ground state, but the global symmetry is extended from $SO(3)$ to
$SU(2)$ at the ends of the chain. What we have described is an analog of such
symmetry extension in $2+1$ dimensions.

\cgrn{ In general, a bulk SPT state protected by a symmetry $G$, can also be
viewed as a many-body state with a symmetry $H$, where the subgroup $K$ acts
trivially in the bulk (\ie the bulk degrees of freedom are singlets of $K$).
For example, we may view the CZX model to have a $Z_4^H$ symmetry in the bulk.
By definition, two states in two different $G$-SPT phases cannot smoothly
deform into each other via deformation paths that preserve the $G$-symmetry.
However, two such $G$-SPT states may be able to smoothly deform into each other
if we view them as systems with the extended $H$-symmetry and deform them along
the paths that preserve the $H$-symmetry.  For example, the non-trivial
$Z_2^G$-SPT state of the CZX model can smoothly deform into the trivial
$Z_2^G$-SPT state along a deformation path that preserves the extended
$Z_4^H$-symmetry.  In other words, when viewed as a $Z_4^H$ symmetric state,
the ground state of the CZX model has a trivial $Z_4^H$-SPT order.  Since it
has a trivial $Z_4^H$-SPT order, it is not surprising that the CZX model can
have a gapped boundary that preserves the extended $Z_4^H$ symmetry, as
explicitly constructed above.  In general, if two $G$-SPT states are connected
by an $H$-symmetric deformation path, then we can always construct a
$H$-symmetric domain wall between them by simply using the  $H$-symmetric
deformation path.  This is the physical meaning behind a $G$-SPT state having a
gapped boundary with an extended symmetry $H$.  
}

From the point of view of condensed matter physics, however, the sort of gapped boundary that we have described so far
will generally not be 
physically sensible.
Microscopically, condensed matter systems generally do not have extra symmetries that act only along their boundary.  (There
can be exceptions like the case just mentioned, which is conceivable in any dimension: a
 system that, in bulk, is made from particles of integer spin but has half-integer spin particles attached on the surface.  Then a $2\pi$
 rotation of the spins is nontrivial only along the boundary.) 
 
In a system microscopically without 
an extended symmetry along the boundary, one might be tempted to interpret
$K$ as a group of emergent global symmetries, not present microscopically.  But
there is a problem with this.  In condensed matter physics, one may often run
into emergent global symmetries in a low-energy description.  But these are
always {\it approximate} symmetries, explicitly broken by operators that are
irrelevant at low energies in the renormalization group sense.

That is not viable in the present context.  Since the global symmetry that is
generated by $U_{CZX}$ is supposed to be an exact symmetry, we cannot
explicitly violate the boundary symmetry group generated by $U_{CZX}^2$.
Obviously, any interaction that is not invariant under $U_{CZX}^2$ is also not
invariant under $U_{CZX}$.

What we can do instead is to {\it gauge} the boundary symmetry group $K$.
Then, the global symmetry group that acts on gauge-invariant operators and on
physical states is just the original group $H/K=G$.  This way, we do not break
nor extend the symmetry on the boundary.  Since $K$ is an on-site symmetry
group, there is no difficulty in gauging it; we explain two approaches in
Sec.~\ref{bdry3w} and \ref{bdry4w}.

In $3+1$ (or more) dimensions, a procedure along these lines starting with a bulk SPT phase with symmetry group $G$ and a group
extension as in eqn. (\ref{gextt})  that satisfies the appropriate cohomological condition will lead
to a gapped boundary state with topological order along the boundary.  The topological order is a version of gauge theory with gauge
group $K$ (possibly twisted by a cocycle).   We will give a general description of such gapped boundary states  in Sec.~\ref{sec:general}.    
In $2+1$ dimensions, the boundary has dimension $1+1$ and one runs into the fact
that topological order is not possible in $1+1$ dimensions.  As a result, what we will actually get in the CZX model
by gauging the boundary symmetry $K$ is not really a fundamentally new boundary state.

\subsection{The third boundary of the CZX model -- 
Lattice $Z_2^K$-gauge theory on the boundary}\label{bdry3w}

\begin{figure}[h!]
\begin{center}
\includegraphics[scale=0.8]{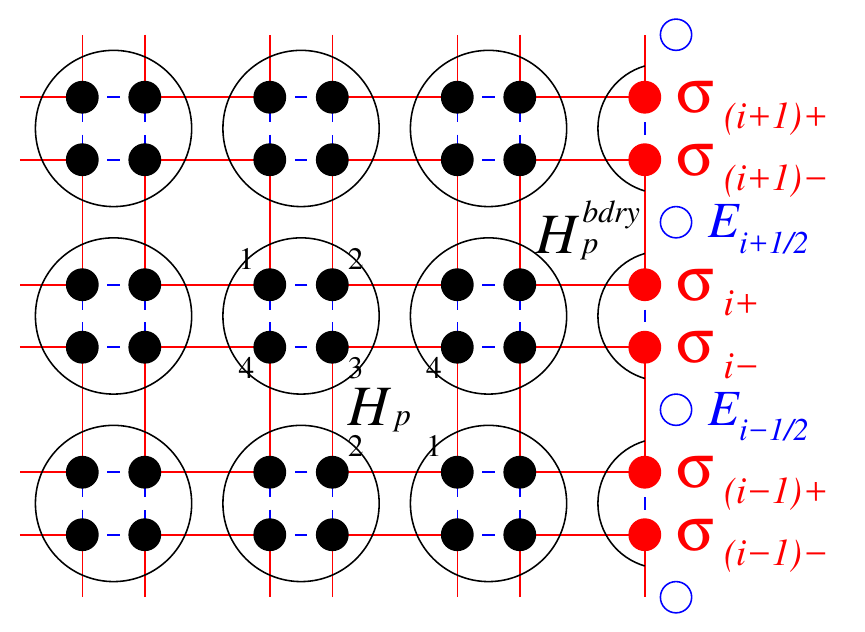}
\end{center}
\caption{
Gauging the boundary symmetry $K=Z_2  \equiv Z_2^K$ of the boundary state of Fig. \ref{bdryB1}  is accomplished
by placing on each boundary link a $Z_2$-valued gauge field.  We label the link between boundary sites $i$ and $i+1$
by the half-integer $i+\frac12$.  We associate to this link a new qubit with a discrete holonomy (as discussed in the text)
and a discrete electric field $E_{i+\frac12}$.
}
\label{bdryB1g}
\end{figure}

We will describe two ways to gauge the boundary symmetry $K=Z_2  \equiv Z_2^K$.  The most straightforward way, although as we will
discuss ultimately less satisfactory for condensed matter physics, is to simply incorporate a boundary gauge field. 

As indicated in Fig. \ref{bdryB1g}, we label the link between boundary sites $i$ and $i+1$ by the half-integer $i+\frac12$.
Placing a $Z_2$-valued gauge field on this link means introducing a qubit associated
to this link with operators $V_{i+\frac12}$, $E_{i+\frac12}$ that obey
\be\label{relns} V_{i+\frac12}^2=E_{i+\frac12}^2=1,~~~E_{i+\frac12}V_{i+\frac12}=-V_{i+\frac12}E_{i+\frac12}.\eeq
Here $V_{i+\frac12}$ describes parallel transport between sites $i$ and $i+1$ and $E_{i+\frac12}$ is a discrete electric field that
flips the sign of $V_{i+\frac12}$.

Now let us discuss the gauge constraint at site $i$.  A gauge transformation that acts at site $i$ by the nontrivial element
 in $Z_2^K$ is supposed to flip the signs of $V_{i\pm \frac12}$, the holonomies on the two links connecting to site $i$.
To do this, it will have a factor $E_{i+\frac12}E_{i-\frac12}$.   It should also act on the spins as $U_{CZX,i}^2=-\sigma^z_{i+}\sigma^z_{i-}$.
Thus the gauge generator on site $i$ is
\be\label{ggen}\Omega_i  = E_{i+\frac12}E_{i-\frac12}U_{CZX,i}^2. \eeq
A physical state $|\Psi\rangle$ in the gauge theory must be gauge-invariant, that is, it must obey
\be\label{ngen}\Omega_i|\Psi\rangle =|\Psi\rangle. \eeq
However, as $E_{i+\frac12}^2=1$ for all $i$, if we take the product of $\Omega_i$ over all boundary sites, the factors of $E_{i+\frac12}$
cancel out, and we get
\be\label{genprod}\prod_i\Omega_i=\prod_i U_{CZX,i}^2.\eeq
Hence eqn. (\ref{ngen}) implies that a physical state $|\Psi\rangle$ satisfies
\be\label{enprod}\prod_i U_{CZX,i}^2|\Psi\rangle =|\Psi\rangle.\eeq
But this precisely means that a physical state is invariant under the global action of $K$, so that the global symmetry group
that acts on the system reduces to the original global symmetry $G$.

The Hamiltonian  $H =\sum H_{p}$ must be slightly modified to be gauge-invariant, that is, to commute with $\Omega_i$.
To see the necessary modification, let us look at the plaquette Hamiltonian $H_p$ for the
boundary plaquette shown in the figure, which contains the boundary link labeled $i+\frac12$.  $H_p$ as defined in eqn. (\ref{hdec})
anticommutes with $\Omega_i$ and $\Omega_{i+1}$ because the operator $|\neg\up\up\up\up\> \langle\down\down\down\down\neg|+   |\neg\down\down\down\down\>\langle\up\up\up\up\neg|$ has
that property.  (It flips one of the spins at boundary site $i$ and one at boundary site $i+1$, so it anticommutes with
$U_{CZX,i}^2=-\sigma^z_{i+}\sigma^z_{i-}$ and similarly with $U_{CZX,i+1}^2$.) To restore gauge-invariance is surprisingly simple:
We just have to multiply $H_p$ by $V_{i+\frac12}$, which also anticommutes with $\Omega_i$ and $\Omega_{i+1}$.  So  we can
take the Hamiltonian
for a boundary plaquette containing the boundary link $i+\frac12$ to be
\be\label{bph} H^{\text{bdry}}_{p,i+\frac12} =-\bigl(|\neg\up\up\up\up\rangle\langle\down\down\down\down\neg|+|\neg\down\down\down\down\rangle
 \langle  \up\up\up\up\neg|\bigr)\otimes V_{i+\frac12} \otimes_\alpha P_p^\alpha .
\eeq
For a gauge-invariant and $G$-invariant Hamiltonian, we can take the sum of all bulk and boundary plaquette Hamiltonians.

This Hamiltonian $H$  commutes with all the discrete gauge fields $V_{i+\frac12}$, so in looking for an eigenstate of $H$ (ignoring
for a moment the gauge constraint),
we can specify arbitrarily the eigenvalues of the $V$'s.  Let $|v_{i+\frac12}\rangle$ be a state of the gauge fields with eigenvalue 
$v_{i+\frac12}$ for $V_{i+\frac12}$.  (Of course these eigenvalues are $\pm 1$ since $V_{i+\frac12}^2=1$.)
 The ground state of $H$ with these eigenvalues of the $V_{i+\frac12}$ is simply
\begin{align}
\bigotimes_\text{bulk}  \frac{
|\neg\up\up\up\up\>+|\neg\down\down\down\down\>}{\sqrt 2} 
\bigotimes_\text{bdry} \frac{
|\neg\up\up\up\up\>+ V_{i+\frac12} |\neg\down\down\down\down\>}{\sqrt 2}
\otimes|v_{i+\frac 12} \>\label{grdstate}
\end{align}
Let us denote this state as $||v_{i+\frac12}\rangle\rangle$.    If the boundary has $L$ links, there are $2^L$ of these states.

The states $||v_{i+\frac12}\rangle\rangle$ are degenerate, and these are the ground states of $H$.  However, to make
states that satisfy the gauge constraint, we must take linear combinations of the $||v_{i+\frac12}\rangle\rangle$.  Since a gauge
transformation at site $i$ flips the signs of $v_{i\pm \frac12}$, the only gauge-invariant function of the $v_{i+\frac12}$ is
their product.  Assuming that the boundary is compact and thus is a circle, this product is the holonomy of the
$Z_2^K$ gauge field around the circle.  (With periodic boundary conditions along the boundary, there are no corners along the boundary
circle; otherwise, our discussion can be slightly modified to incorporate corners.)  Thus there are two gauge-invariant ground states,
depending on the sign of the holonomy $\prod_i v_{i+\frac12}$.  They are 
\begin{align} \label{eq:gs+}
|\Psi_{\mathrm{gs}}(+)\> = \sum_{\{v_{i+\frac 12}\}, \prod_iv_{i+\frac 12}=1} c_{\{v_{i+\frac 12}\}}
 ||v_{i+\frac12}\rangle\rangle
\end{align}
and
\begin{align} \label{eq:gs-}
|\Psi_{\mathrm{gs}}(-)\> = \sum_{\{v_{i+\frac 12}\}, \prod_iv_{i+\frac 12}=-1} c_{\{v_{i+\frac 12}\}}
||v_{i+\frac12}\rangle\rangle.
\end{align}
(Here the signs $c_{\{v_{i+\frac 12}\}}=\pm 1$ are determined by the gauge constraints.  With our choice of sign in the gauge constraints $\Omega_i$,
flipping two of the $v_i$ that are separated by $n$ lattice states multiplies the amplitude by $(-1)^n$. This could be avoided by changing
the sign of $\Omega_i$, but that creates complications elsewhere.)

Now let us study the transformation of these states under the global symmetry
group $G=Z_2 \equiv Z_2^G$.  When we apply $ U_{CZX}$ to the states
$|\Psi_{\mathrm{gs}}(\pm)\>$, we find that all the sign factors $CZ_{ij}$
cancel each other.   This occurs by the same cancellation as in the original
bulk version of the CZX model.  However, the wavefunction is no longer
trivially invariant under flipping the spins; rather, the wavefunction
$|\up\up\up\up\>+ V_{i+\frac12} |\down\down\down\down\>$ for a boundary
plaquette is multiplied by $V_{i+\frac12}$ when the spins in this plaquette are
flipped.  So taking into account all the boundary plaquettes,
\begin{align}\label{preft}
 U_{CZX} |\Psi_{\mathrm{gs}}(\pm)\>
=
\pm |\Psi_{\mathrm{gs}}(\pm)\>.
\end{align}  
Thus, the transformation of a state under the global symmetry $Z_2^G$ is locked to its
holonomy under the gauge symmetry $Z_2^K$.

The formula (\ref{preft}) has been written as if the boundary of the system consists of a single circle; for example,
the spatial topology may be a disc.  More generally, we can consider a system whose boundary consists of several
circles.  Each boundary component has its own $Z_2^K$-valued holonomy, and the action of $U_{CZX}$ on a ground
state is the product of all of these holonomies.

\def\Psig{\Psi_{\mathrm{gs}}}
\def\gs{{\mathrm{gs}}}

Now let us look for a local operator with a nonzero matrix element between the two ground states $|\Psi_{\mathrm{gs}}(\pm) \rangle$.
For this, we need first of all an operator that changes the sign of the holonomy around the boundary.  The simplest
operator with this property is simply $E_{i+1/2}$ (for some $i$).  Because it flips the sign of $V_{i+1/2}$, it reverses
the sign of the holonomy.   However, the operator $E_{i+1/2}$ is invariant under the global symmetry
group $Z_2^G$, and therefore, it cannot possibly have a nonzero matrix element between the two
ground states, which transform oppositely under the global symmetry.

Concretely, $E_{i+1/2}$ does not map $|\Psig(\pm) \rangle$ to $|\Psig(\mp) \rangle$ because it anticommutes with 
$V_{i+1/2}$, which appears
in one factor in the definition of the state $||v_{i+1/2}\rangle\rangle$ in eqn. (\ref{grdstate}), namely 
\be\label{onefactor}|\up\up\up\up\>+ V_{i+1/2} |\down\down\down\down\> .\eeq
(Instead, $E_{i+1/2} |\Psig(+) \rangle$
is a new state that has the same holonomy as $|\Psig(-) \rangle$,
but  differs from it by the presence of an additional 
quasiparticle carrying a nontrivial global $Z_2^G$-charge localized near the  link at $i+1/2$.)
However, we can get a local operator that reverses the holonomy and commutes with this $V_{i+1/2}$ if we just
replace $E_{i+1/2}$ by
\be\label{newop}X_{i+1/2}=E_{i+1/2}\sigma^z_{i+}.\eeq
(We could equally well use $\sigma^z_{i+1-}$ instead of $\sigma^z_{i+}$.)   This operator leaves invariant the expression
in eqn. (\ref{onefactor}), and, accordingly, it simply exchanges the states \cblue{$|\Psig(\pm) \rangle$}:
\be\label{excha}X_{i+1/2} \cblue{|\Psig(\pm)\rangle=|\Psig(\mp)\rangle}. \eeq

The operator $X_{i+1/2}$ is odd under the global $Z_2^G$ symmetry, because of the factor of $\sigma^z_{i+}$.
This of course is consistent with the fact that this operator exchanges the states \cblue{$|\Psig(\pm) \rangle$}.  However, the existence
of a $Z_2^G$-odd local operator that exchanges the two ground states means that we must interpret the boundary
state that we have constructed as one in which the global $Z_2^G$ symmetry is spontaneously 
broken along the
boundary.  Indeed, although $\langle \Psig(+)|X_{i+1/2}|\Psig(+)\rangle=0$, the two-point function of
the operator $X_{i+1/2}$ in the state \cblue{$|\Psig(+) \rangle$}  exhibits {the} \emph{long-range order} that signals the $Z_2^G$-spontaneous symmetry breaking.  In fact, 
\be\label{zerof}
\langle \Psig(+)|X_{i+1/2}X_{j+1/2}|\Psig(+)\rangle = 1
\eeq
for any $i,j$. Similarly, $\langle \Psig(-)|X_{i+1/2}X_{j+1/2}|\Psig(-)\rangle = 1$.

This result is somewhat disappointing, since it 
 is certainly already known that any SPT phase in any dimension can have a gapped boundary state
in which the symmetry is explicitly or spontaneously broken.  However, as we will see starting in \cblue{Sec.~\ref{rev},} 
similar gapped boundary states can be constructed for SPT phases in any dimension, and in $3+1$ (or more)
dimensions, the gapped boundary states constructed this way are genuinely novel: They have topological
order along the boundary, rather than symmetry breaking.  What we have run into here is that the $1+1$-dimensional
boundary of a $2+1$-dimensional system does not really support topological order.  Discrete gauge symmetry
(such as the $Z_2^K$ considered here) can describe topological order in dimensions $\geq 2+1$, but not in $1+1$
dimensions.  

By contrast, the gapped boundary state described in Sec.~\ref{bdry2w}, in which the symmetry is extended
along the boundary rather than being spontaneously broken, is genuinely new even in $2+1$ dimensions.
But as we have noted, such a symmetry extension along the boundary is physically sensible in condensed
matter physics only in  particular circumstances.

Going back to the case that the boundary symmetry is gauged, where
 does the state that we have described fit into the usual classification of gapped phases of discrete gauge theories?
Since the states \cblue{$|\Psig(\pm)\rangle$} with opposite holonomies are degenerate, this would usually be called \cblue{a deconfined} 
phase.  But it differs from a standard \cblue{deconfined} 
phase in the following way.  Typically, in $1+1$-dimensional  gauge theory
with discrete gauge group, the degeneracy between states with different holonomy can be lifted by a suitable
perturbation such as
\be\label{perturb} -u\sum_i E_{i+1/2}, \eeq
with a constant $u$
(or more generally $-\sum_i u_i E_{i+1/2}$ with any small parameters $u_i$; a small local
perturbation is enough).  In an ordinary $Z_2^K$ gauge theory, such a term would induce an effective Hamiltonian
density   $-u\begin{pmatrix} 0 & 1\cr 1& 0\end{pmatrix}$ acting on the two states $\begin{pmatrix}\Psig(+)\cr \Psig(-)\end{pmatrix}$.  
The ground state would then be (for $u>0$)  a
superposition of $| \Psig(+) \rangle$ and $|\Psig(-) \rangle$.  A  discrete gauge theory with a non-degenerate ground state that
involves such a sum over holonomies is said to be confining.  

In the present context, 
the global $Z_2^G$ symmetry under which the states \cblue{$|\Psig(\pm) \rangle$} transform oppositely prevents such an effect.  On the
contrary, it ensures that the degeneracy among these two states cannot be lifted by any local perturbation that
preserves the $Z_2^G$ symmetry.  The above remarks demonstrating the spontaneous breaking of the global $Z_2^G$ symmetry
makes \cblack{the issue} 
clear.   The spontaneously broken symmetry leads to a two-fold degeneracy of the ground state that is 
exact in the limit of a large system.

The remarks that we have just made have obvious analogs in the construction described in the emergent gauge theory construction
of Sec.~\ref{bdry4w}, and they will not be repeated there.

\subsection{The fourth boundary of the CZX model -- 
Emergent lattice $Z_2^K$-gauge theory on the boundary}

\label{bdry4w}

\begin{figure}[h!]
\begin{center}
\includegraphics[scale=0.8]{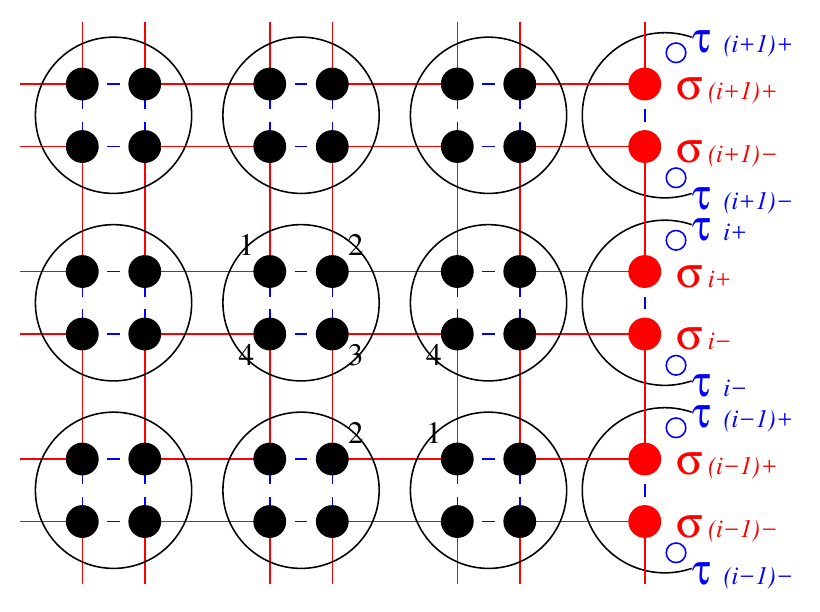}
\end{center}
\caption{
The filled dots are qubits (or spin-1/2's).  A (half-)circle (with dots inside)
represents a site.  The dashed blue line connecting dots $i,j$ represents the
phase factor $CZ_{ij}$ in the $Z_2^G$ or $Z_4^H$ global symmetry transformation.  
 The open dots on
the boundary are the $Z'_2 \equiv Z_2^K$-gauge degrees of freedom $E_{i+\frac 12}$.
}
\label{bdryB1ge}
\end{figure}

The model constructed in Sec.~\ref{bdry3w} using lattice $Z_2^K$ gauge fields reduces the global symmetry to the original
$Z_2^G$.  However, it has one flaw from the point of view of condensed matter physics.  In condensed matter physics, not only
are the symmetries on-site, but more fundamentally the Hilbert space can be assumed to be on-site: that is, the full 
Hilbert space is a tensor product of
local factors, one for each site. (In fact, the Hilbert space has to be on-site before it makes sense to say that the symmetries
are on-site.)  

The purpose of the present section is to explain how to construct a model with on-site Hilbert space and symmetries that
has the same macroscopic behavior as found in Sec.~\ref{bdry3w}.

The reason that the model in Sec.~\ref{bdry3w} does not have this property is that the variables $V_{i+\frac12}$ and $E_{i+\frac12}$ are
associated to boundary links, not to boundary sites.  One could try to cure this problem by associating these link variables
to the site just above (or just below) the link in question.  The trouble with this is that then although the full Hilbert space is on-site,
the gauge-symmetry generators $\Omega_i$ are not on-site (they involve operators acting at two adjacent sites).  Accordingly
the space of physical states, invariant under the $\Omega_i$, is not an on-site Hilbert space.

By analogy with various constructions in condensed matter physics, one might be tempted to avoid this problem by relaxing
the physical state constraint $\Omega_i|\Psi\rangle=|\Psi\rangle$ and instead adding to the Hamiltonian a term
\be\label{relaxed}\Delta H=-c \sum_i\Omega_i, \eeq
with a positive constant $c$.  Then minimum energy states satisfy the constraint \cblue{$\Omega_i|\Psi \rangle=|\Psi \rangle$} as assumed in
Sec.~\ref{bdry3w}, and on the other hand the full Hilbert space and the global $Z_2^G$ symmetry are  on-site.

In the present context, this approach is not satisfactory.  Once we relax the constraint that physical states are invariant under
$\Omega_i$, the global symmetry of the model is extended along the boundary from $G=Z_2^G$ to $H=Z_4^H$, and we have really
not gained anything by adding the gauge fields.

Instead what we have to do is to replace the ``elementary'' $Z_2'=Z_2^K$ gauge fields of Sec.~\ref{bdry3w} by ``emergent'' 
gauge fields, by which we mean simply gauge fields that emerge in an effective low-energy description from a microscopic
theory with an on-site Hilbert space.   There are many ways to do this, and it does not matter exactly which approach we pick.  In this section,
we will describe one simple approach.

 We start with the boundary
obtained in Sec.~\ref{bdry2w}, and add to each boundary site a pair of qubits described by Pauli matrices $\v \tau_{i_\pm}$
(see Fig. \ref{bdryB1ge}).  Since each boundary site already contained the two qubits $\upsigma_{i\pm}$, this gives a total
of four qubits in each boundary site, and a local Hilbert space $\mathscr{H}_i^0$ of dimension $2^4$.  However, we define the
Hilbert space $\mathscr{H}_i$ of the $i^{th}$ boundary site to be the subspace of $\mathscr{H}_i^0$ of states that satisfy 
the local gauge constraint 
\be\label{constr} \cblue{\widehat{U}^\text{gauge}_i}|\Psi\rangle =|\Psi\rangle ,\eeq
where 
\begin{align}\label{gc}
\widehat U^\text{gauge}_i =-
\si^z_{i_+}\si^z_{i_-}  
\tau^z_{i_+}\tau^z_{i_-}  .
\end{align}
The constraint is on-site so $\mathscr{H}_i$ is on-site.

Now we add to the Hamiltonian a gauge-invariant boundary perturbation 
\begin{align}\label{pertu}
 - U \sum_i \tau^z_{i_+}\tau^z_{(i+1)_-}
\end{align}
with a large positive coefficient $U$.
At low-energies, this  will lock $\tau^z_{i_+}=\tau^z_{(i+1)_-}$.  In this low-energy subspace, $\tau^z_{i_+}=\tau^z_{(i+1)_-}$ will play the role of $E_{i+\frac
12}$ in the last subsection.  What will now play the role of the conjugate gauge field is
\be\label{conjg}V_{i+\frac12}= \tau_{i_+}^x \tau_{(i+1)_-}^x 
\eeq which anticommutes with  $\tau^z_{i_+}=\tau^z_{(i+1)_-}$. The Hamiltonian for a boundary
plaquette is defined as in eqn. (\ref{bph}), but with this ``composite'' definition of $V_{i+\frac12}$, 
and commutes with the gauge constraint operator (\ref{gc}).

The global $Z_2$-symmetry generator on the $i^{th}$ boundary site is now given by 
\begin{align}
\label{UZ2g}
\widehat U_{Z_2,i} = \si^x_{i_-} \si^x_{i_+} \cblue{U_{CZ,{i_-,i_+}}}
\ee^{\ii \frac{\pi}{4} \tau^z_{i_-}}
\ee^{-\ii \frac{\pi}{4} \tau^z_{i_+}}
.
\end{align}
We find that
\begin{align}
 \widehat U^2_{Z_2,i} = -\si^z_{i_-} \si^z_{i_+}\tau^z_{i_+}\tau^z_{i_-} =\widehat U_i^{\mathrm{gauge}}.
\end{align}
So $\widehat U_{Z_2,i}^2=1$ on states that satisfy the gauge constraint.  This is true for every bulk or boundary state,
so the full global symmetry generator, obtained by taking the product of the symmetry generators over all bulk or boundary sites,
generates the desired symmetry group $Z_2^G$.

The low-energy dynamics can be analyzed precisely as in Sec.~\ref{bdry3w}, and with the same results.  The first
step is to observe that, even in the presence of the perturbation of eqn. (\ref{pertu}), the Hamiltonian commutes with the operators
$V_{i+\frac12}$.  Just as in Sec.~\ref{bdry3w}, one diagonalizes these operators with eigenvalues $v_{i+\frac12}$, finds the ground state 
for given $v_{i+\frac12}$, and then takes linear combinations of these states to satisfy the gauge constraint.

We remind the readers that
Appendix \ref{sec:lowEbdryCZX} of this paper contains more details on boundaries of the CZX model and their 1+1D boundary effective theories. For a fermionic
version of the CZX model, see Appendix \ref{sec:fCZX}.
The boundary of the fermionic CZX model with emergent $Z_2^K$-gauge theory with anomalous global symmetry
is detailed in Appendix \ref{sec:fCZXbdryZ2-gauge}.

For the generalization of what we have done to arbitrary SPT phases in any dimension,
we can now proceed to Sec.~\ref{rev}.

\section{Boundaries of generic SPT states in any dimension}

\label{rev}

What we have done for the CZX model in $2+1$ dimensions has an analog
for a general SPT state in any dimension.  To explain this will require a more abstract approach.
We work in the framework of the group cohomology approach to SPT states, 
with a Lagrangian on a spacetime lattice.
So we
first introduce our notation for that subject. We generically write $\nu_d$ for a
homogeneous $d$-cocycle, and $\mu_d$ for a  homogeneous $d$-cochain.  We similarly write
$\omega_d$ for an inhomogeneous $d$-cocycle, and $\beta_d$ for an inhomogeneous $d$-cochain.
Finally, we write $\mathcal{V}_d$ for homogeneous $d$-cocycles or $d$-cochains with both
global symmetry variables and gauge variables, 
and denote $\Omega_d$ as inhomogeneous $d$-cocycles or $d$-cochains with both global
symmetry variables and gauge variables. 

\subsection{An exactly soluble path integral model that realizes a generic SPT
state} 

\begin{figure}[h!]
\begin{center}
\includegraphics[scale=0.6]{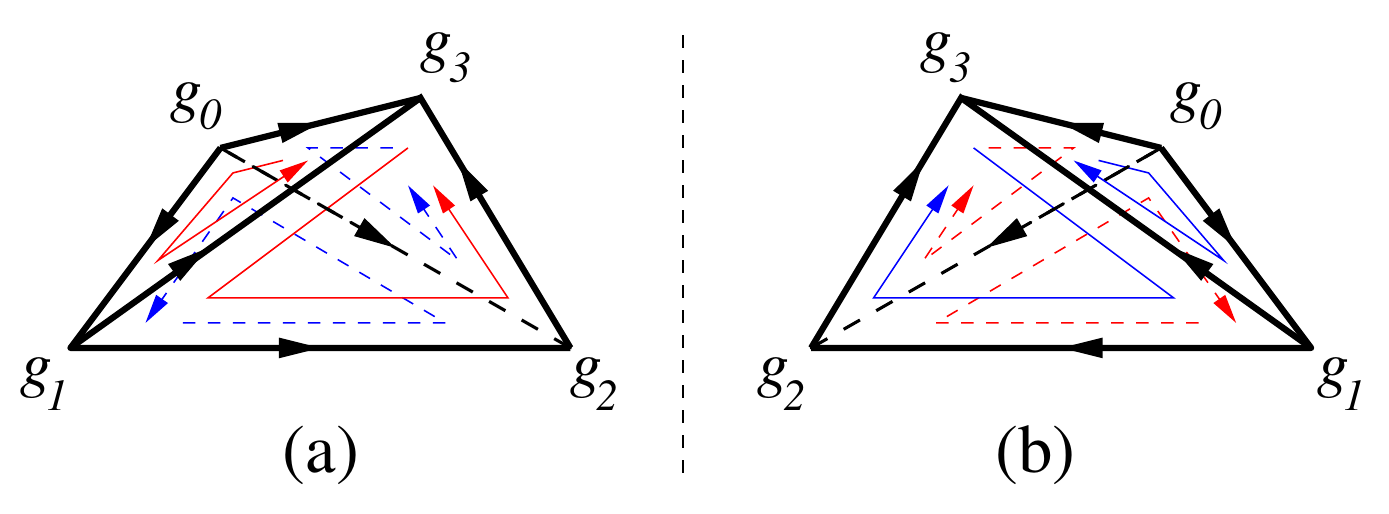}
\end{center}
\caption{
The triangles with red (blue) loops have positive orientation $s_{ijk}=1$
(negative orientation $s_{ijk}=-1$), with an outward (inward) area vector
through the right-hand rule.  The orientation of a tetrahedron (i.e. the
3-simplex) is determined by the orientation of the triangle not containing the
{first} vertex.  So (a) has a {positive} orientation
$s_{01234}={+1}$ and (b) has a {negative} orientation
$s_{01234}={-1}$.
} \label{mir} \end{figure}

A generic SPT state with a finite symmetry group $G$ can be described by a path
integral on a space-time lattice, or more precisely, a space-time complex with
a \emph{branching structure}.  A branching structure can be viewed as an
ordering of all vertices.  It gives each link an orientation -- which we can
think of as  an arrow that runs from the  smaller vertex on that link to the
larger one, as in Fig. \ref{mir}.  More generally, a branching structure
determines an orientation of each $k$-dimensional simplex, for every $k$,
including the top-dimensional ones that are glued together to make the full
spacetime.

To each vertex $i$, we attach a $G$-valued variable $g_i$.  (Later we may also
assign group elements $g_{ij}$ to each edges  $\overline{ij}$.) An assignment
of group elements to vertices or edges will be called a \emph{coloring}.  For a
discrete version of the usual  \emph{path integral} of quantum mechanics,  we
will to \emph{sum over all the colorings}.  (See
Sec.~\ref{sec:generalPathintegral}.) 
On a closed oriented space-time, 
{in the Euclidean signature,}
the ``integrand'' of the path integral  is given by 
\begin{align}
\label{actampSPT}
e^{-\int_{M^3} \cL_{\text{Bulk}} {d^3x}} 
& =\prod_{M^3} \nu^{s_{ijkl}}_{3}( g_i,  g_j, g_k,g_l).
\end{align}
{The argument of the path integral is a complex number with a nontrivial phase and thus
it can produce complex Berry phases.}
We have written this formula for the case of $2+1$ dimensions, but it readily
generalizes to any dimension.  Here, $s_{ijkl}=\pm 1$ for a given simplex with
vertices $ijkl$ depending on whether the orientation of that simplex that comes
from the branching structure agrees or disagrees with the orientation of $M$.
The symbol $\prod_{M^3}$ represents a product over all $d$-simplices.

Finally, and most importantly, the $U(1)$-valued $ \nu_{d}( g_0, \cdots  ,g_d) $ is a homogeneous
cocycle representing an element of  $\cH^d(G,U(1))$.  
This means $\nu_d(g_0,\cdots,g_d)$ satisfy the
cocycle condition $\delta  \nu_{d}=1$, where
\begin{align}
(\delta  \nu_{d})(g_0,\cdots,g_{d+1})\equiv
 \frac{
\prod_{i=\text{even}} \nu_{d}(g_0,\cdots,\widehat g_i, \cdots,g_{d+1}) 
}{
\prod_{i=\text{odd}} \nu_{d}(g_0,\cdots,\widehat g_i, \cdots,g_{d+1}) 
}.
\end{align} 
(The symbol $\widehat g_i$ is an instruction to omit $g_i$ from the sequence.)

We regard the complex phase
$\nu_d^{s}$  as a  quantum amplitude assigned to a  $d$-simplex in a
$d$-dimensional spacetime.  

First, the path-integral model defined by the action amplitude eqn.
(\ref{actampSPT}) has a $G$-symmetry 
\begin{align}
 \prod_{M^3} \nu^{s_{ijkl}}_{3}( g_i,  g_j, g_k,g_l)
=\prod_{M^3} \nu^{s_{ijkl}}_{3}( gg_i,  gg_j, gg_k,gg_l), \ \  \ g\in G,
\end{align}
since the homogeneous cocycle satisfies
\begin{align}
  \nu_{3}( g_i,  g_j, g_k,g_l)= \nu_{3}( gg_i,  gg_j, gg_k,gg_l).
\end{align}
Second, because of the cocycle condition, one can show that
\begin{align}
e^{-\int_{M^3} \cL_{\text{Bulk}} {d^3x}} 
 =\prod_{M^3} \nu^{s_{ijkl}}_{3}( g_i,  g_j, g_k,g_l) = 1,
\end{align}
for any set of $g$'s, when the spacetime $M^3$ is an orientable closed manifold.  This implies
that the  model is trivially soluble on a closed spacetime, and describes a  state in which all  local operators have short-range correlations.
This state is symmetric and gapped.  It realizes an SPT state with symmetry
$G$.  The state is determined up to equivalence by the cohomology class of $\nu_3$.

\subsection{The first boundary of a generic SPT state -- A simple model but
with complicated boundary dynamics} \label{sec:1st-complx-bdry}

So far, we have described a discrete system with $G$ symmetry on a closed
$3$-manifold $M^3$.  What happens if $M^3$ is an open manifold that has a boundary $\partial M^3 = M^2$?  The simplest
path-integral model that we can construct is simply to use all of the above
formulas, but now on a manifold with boundary.  Thus, the argument of the path integral is still given by  eqn. (\ref{actampSPT}), but now, this
is no longer trivial:
\begin{align}
e^{-\int_{M^3} \cL_{\text{Bulk}} {d^3x}} 
 =\prod_{M^3} \nu^{s_{ijkl}}_{3}( g_i,  g_j, g_k,g_l) \neq  1 .
\end{align}
Because of the properties of the cocycle, this amplitude only depends
on the $g_i$ on the boundary, so it can be viewed as the integrand of the path integral  of a
boundary theory.

\begin{figure}[t]
\begin{center}
\includegraphics[height=1.2in]{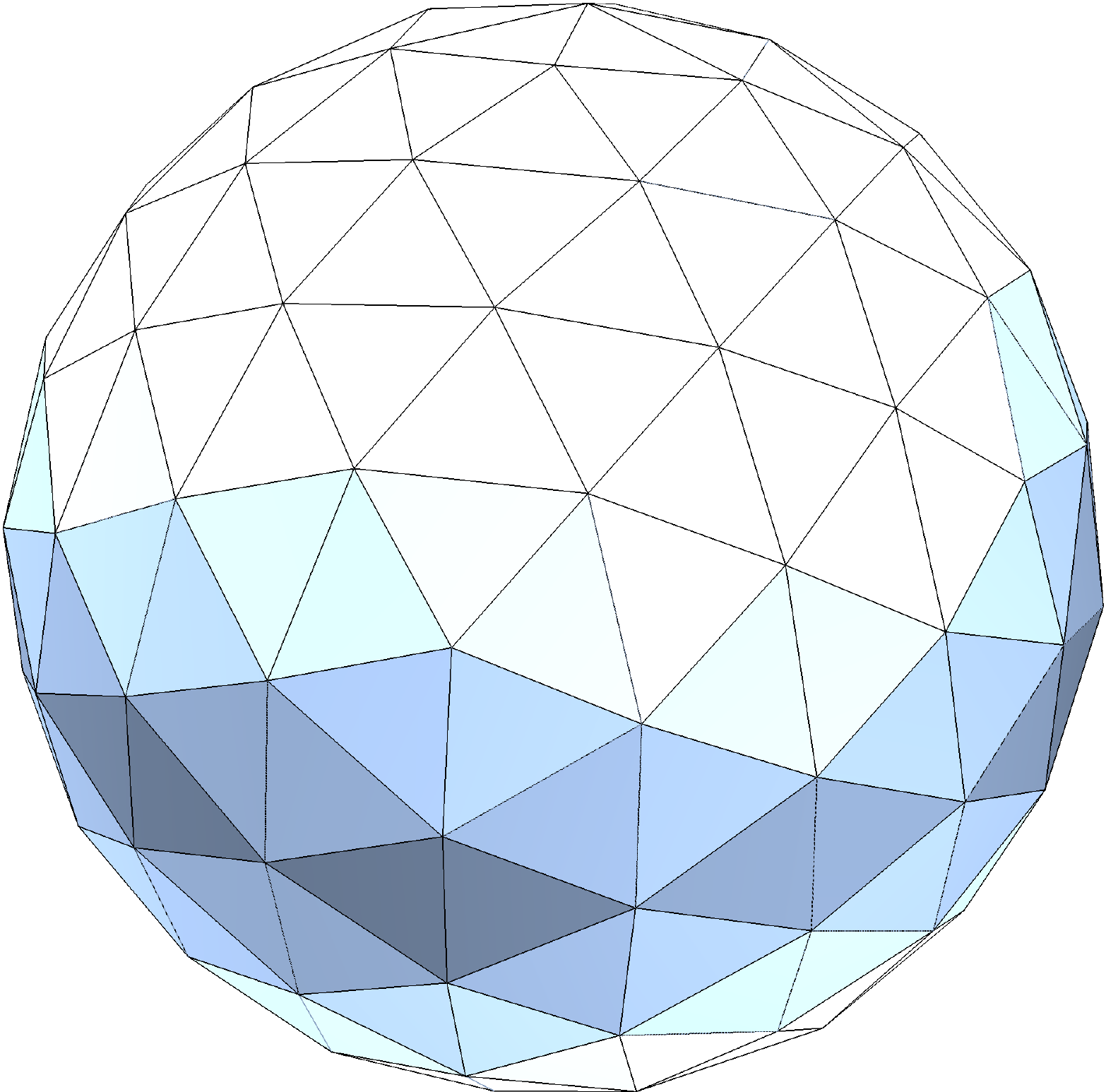}
~~~~~~~~~~~~~~~~~~~~~~~~~~~~~~~~~
\\[-1.1in]
\hfill 
\includegraphics[height=1.2in]{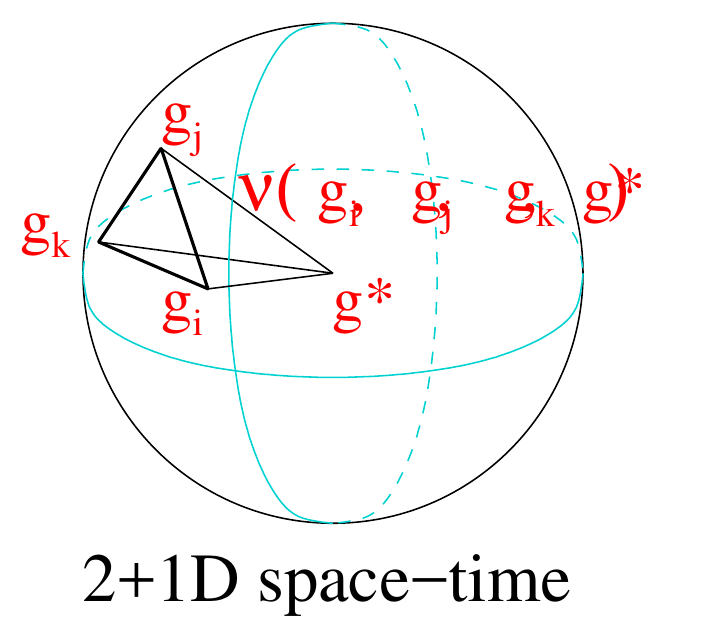} ~~~~~~~~~~~~~~~~~~~~~~~~~~~~~~
\end{center}
\caption{
The space-time $D^3$, with a triangulation of the boundary and a construction
of 3-simplices (or 4-cells) in the bulk.  Such a triangulation is used to construction a low-energy
effective path integral for the boundary.
}
\label{trisphere}
\end{figure}

To calculate the path integral amplitude of the boundary theory, we can {simplify} the
bulk so that it contains only one vertex $g^*$ (see Fig. \ref{trisphere}). In
this case, the effective boundary theory is described by a path integral based on the
following amplitude: \begin{align}
\label{bdry1path}
e^{-\int_{\prt M^3} \cL_{\text{Bdry},{\prt M^3}}  {d^2x}}
& =\prod_{\prt M^3} \nu^{s_{ijk}}_{3}( g_i,  g_j, g_k,g^*).
\end{align}
This depends only on the boundary spins $g_i,g_j,g_k,\cdots$, and not 
on $g^*$ in the bulk. (This follows from the cocycle condition for $\nu_3$.  Readers who are not familiar with
this statement can find the proof in Sec.~\ref{sec:general}.) 
Here, $s_{ijk}=\pm 1$ depending on whether the
orientation of a given triangle that comes from the branching structure agrees
with the orientation that comes from the triangle as part of the
boundary of the oriented manifold $M^3$.  (Symbols like $d^3x$ and similar
notation below are shorthands for products over simplices, as written
explicitly in the right hand side of eqn.~(\ref{bdry1path}).)

Since the path integral amplitude of the boundary theory is path dependent and not equal to 1, the dynamics
of the simple model is hard to solve, and we do not know if the boundary is
gapped, symmetry breaking, or topological.  In fact, for cocycles $\nu_3$ that
are in the same equivalence class but differ by coboundaries, the boundary
 amplitudes are different, which may lead to different boundary dynamics.
In Sec.~\ref{bdry1}, for the case of the CZX model, we have chosen a particular cocycle in an equivalence
class. This choice of cocycle leads to a gapless boundary.

\cgrn{
In general, given only a generic cocycle, the dynamics of this model is unclear and possibly non-universal.}
We will describe more fully the anomalous symmetry realization in this boundary state in Sec.~\ref{sec:non-on-site-G}, 
and then we will introduce alternative boundary states in Sec.~\ref{sec:extSymm}.

\subsection{ 
Non-on-site (anomalous) $G$-symmetry transformation on the boundary effective theory}

\label{sec:non-on-site-G}

\subsubsection{Symmetry transformation on a spacetime boundary in Lagrangian formalism}

\label{sec:non-on-site-G-Lagrangian}

We continue to assume that the spacetime manifold $M^3$ has a boundary ${\prt M^{3}} = M^2$, which can be regarded 
as a fixed-time slice on the closed space region ${\prt M^{3}}$.
The effective theory eqn. (\ref{bdry1path}) possesses the $G$
symmetry:
\begin{align}
\label{zolt}
e^{-\int_{\prt M^3} \cL_{\text{Bdry},{\prt M^3}}  {d^2x}} 
& =\prod_{\prt M^3} \nu^{s_{ijk}}_{3}( g_i,  g_j, g_k,g^*)
\nonumber\\
&=\prod_{\prt M^3} \nu^{s_{ijk}}_{3}( gg_i,  gg_j, gg_k,g^*) .
\end{align}
But this $G$ symmetry in the presence of a  boundary is in fact anomalous (\ie
non-on-site).  The anomalous nature of the symmetry along the boundary is the most important property
of SPT states.

To understand such an anomalous (or non-on-site) symmetry, we note that locally
(that is, for a particular simplex) the action amplitude is not invariant under
the $G$-symmetry transformation:
\begin{align}
\nu_{3}( gg_i,  gg_j, gg_k, g^*)
&\neq \nu_{3}( g_i,  g_j, g_k,g^*).
\end{align}
Only the total action amplitude on the whole boundary (here the boundary ${\prt M^3}=M^2$ of an open manifold is a closed
manifold) is invariant under the $G$-symmetry transformation. 
(Readers who are not familiar with
this statement can read the proof in Sec.~\ref{sec:general}.)  
Such a symmetry is an  anomalous (or non-on-site) symmetry.

\begin{figure}[!t]
\begin{center}
\includegraphics[scale=0.6]{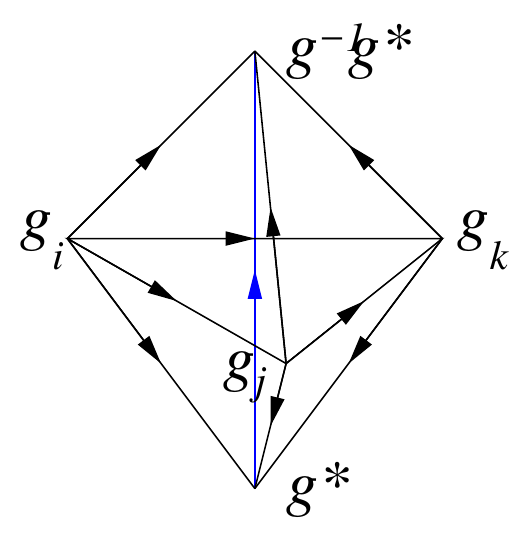}
\end{center}
\caption{
Graphic representations of $f_2(g_i,g_j,g_k)= \frac{\nu_{3}( g_i,  g_j,
g_k,g^{-1}g^*)} {\nu_{3}( g_i,  g_j, g_k,g^*)} $, which is actually a
coboundary. See eqn.~(\ref{eq:f2f1coboundary}).
}
\label{f2}
\end{figure}

Since the action amplitude is not invariant locally, but
invariant on the whole boundary ${\prt M^3}=M^2$,
thus under the symmetry transformation, the
Lagrangian may change by a total derivative term:
\begin{align}
\cL_{\text{Bdry},{\prt M^3}}[g g(x)]= \cL_{\text{Bdry},{\prt M^3}}[ g(x)]+ \dd \cL'[ g(x)]. 
\end{align}
The presence of $\dd \cL'[ g(x)]$ is another sign of the anomalous symmetry.
To understand the symmetry transformation on the boundary in more detail, we
note that, in our case, $\dd \cL'[ g(x)]$ is given by
\begin{align}
 \ee^{-\int_{M^2} \dd \cL'[ g(x)] {d^2x}} = \prod_{M^2} 
\frac{\nu^{s_{ijk}}_{3}( g_i,  g_j, g_k,g^{-1}g^*)}
{\nu^{s_{ijk}}_{3}( g_i,  g_j, g_k,g^*)}.
\end{align}
If we view
\begin{align}
 f_2(g_i,g_j,g_k) \equiv \frac{\nu_{3}( g_i,  g_j, g_k,g^{-1}g^*)}
{\nu_{3}( g_i,  g_j, g_k,g^*)}
\end{align}
as a 2-cochain, it is actually a 2-coboundary (see Fig. \ref{f2}):
\begin{align} \label{eq:f2f1coboundary}
 f_2(g_i,g_j,g_k) &= \frac{\nu_{3}( g_i,  g_j, g_k,g^{-1}g^*)}
{\nu_{3}( g_i,  g_j, g_k,g^*)}
\\
&=\frac{
\nu_{3}( g_i,  g_j, g^*,g^{-1}g^*)
\nu_{3}( g_j,  g_k, g^*,g^{-1}g^*)
}
{\nu_{3}( g_i,  g_k, g^*,g^{-1}g^*)}
= \dd f_1 \nonumber
\end{align}
with a 1-cochain $f_1$ as
\begin{align} \label{eq:f2f1coboundary-2}
 f_1(g_i,g_j)=\nu_{3}( g_i,  g_j, g^*,g^{-1}g^*).
\end{align}
Thus
\begin{align}\label{cotto}
 \ee^{-\int_{M^2} \dd \cL'[ g(x)] {d^2x} } 
= \prod_{M^2} f_2^{s_{ijk}}( g_i,  g_j, g_k)
= \prod_{\prt M^2} f_1^{s_{ij}}( g_i,  g_j) {.}
\end{align} 
In some sense $\cL'$ is given by $f_1$.  When the spacetime boundary $M^2=\partial M^3$,
we have $\partial M^2=\partial^2 M^3= \varnothing$, and therefore eqn.~(\ref{cotto}) simplifies to 
\begin{equation}\label{otto} \ee^{-\int_{M^2} \dd \cL'[ g(x)] {d^2x}}      =1.\end{equation}
Thus, globally there is a global symmetry, as was claimed in eqn.~(\ref{zolt}), though it holds only up to a lattice version
of a total derivative.  

\subsubsection{Symmetry transformation on a spatial boundary in Hamiltonian formalism}

\label{sec:non-on-site-G-Hamiltonian}

In the above, we have discussed the effective symmetry transformation on the
\emph{spacetime boundary} in Lagrangian formalism.  Now we will proceed with a Hamiltonian formalism.

\cgrn{
What we mean by a Hamiltonian formalism is to choose a fixed space $M^2$, and
use the path integral on $M^2\times I$ to construct the imaginary-time
evolution {unitary} operator $\ee^{-\widehat H_{M^2}}$, where $I=[0,1]$ represents the
time direction (see Fig. \ref{bdrybulk}).  The matrix elements of the imaginary-time evolution operator
is $(\ee^{-\widehat H_{M^2}})_{\{g_i'', ...\}, \{g_i', ...\}}$, where $\{g_i',
...\}$ are the degrees of freedom on $M^2\times \{0\}$, and $\{g_i'', ...\}$
on $M^2\times \{1\}$.  We may choose $M^2\times I$ to represent just one time
step of evolution, so that there are no interior degrees of freedom to sum over.
In this case, the {unitary} operator are
\begin{align}
(\ee^{-\widehat H_{M^2}})_{\{g_i'', ...\}, \{g_i', ...\}}
 =\prod_{M^2\times I} \nu^{s_{ijkl}}_{3}( g_i,  g_j, g_k,g_l).
\end{align}
}

%

\cgrn{
When the space $M^2$ has a boundary, then some degrees of freedom live on the
boundary $\prt M^2$ and others live in the interior of $M^2$.}
We can ask about the properties of global symmetry transformations in two scenarios:
The first is the symmetry of \emph{the whole bulk and the boundary} 
included together, which is an \emph{on-site symmetry}.
The second is the symmetry of \emph{the effective boundary theory} only,
which turns out to be a \emph{non-on-site symmetry}. 

\begin{enumerate}

\item
For the first scenario, the symmetry of the whole bulk and the boundary together, we have
$$(\ee^{-\widehat H_{M^2}})_{\{gg_i'', ...\}, \{gg_i', ...\}} = (\ee^{-\widehat H_{M^2}})_{\{g_i'', ...\}, \{g_i', ...\}},$$
because every homogeneous cochain satisfies
$ \nu_{3}( g g_i,  g g_j,  gg_k, gg_l)= \nu_{3}( g_i,  g_j, g_k,g_l)$.
If we write the evolution operator $\ee^{-\widehat H_{M^2}}$ explicitly, including the matrix elements and basis projectors,
we see that
$$
|\{g  g_i'', ...\} \rangle(\ee^{-\widehat H_{ M_2}})_{\{g g_i'', ...\}, \{gg_i', ...\}}
\langle \{g g_i', ...\} |
=
\widehat U_0( g)
|\{  g_i'', ...\} \rangle 
(\ee^{-\widehat H_{M_2}})_{\{ g_i'', ...\}, \{g_i', ...\}}
\langle \{g_i', ...\} |
\widehat U_0^\dag ( g),
$$
where
$\widehat U_0(g)$ generates the usual {on-site} $G$-symmetry transformation 
$|\{ g_i, ...\} \rangle \to |\{g g_i, ...\} \rangle$.
Thus, the $G$-symmetry transformation on the whole system (with bulk and boundary included) is an \emph{on-site} symmetry,
as it reasonably should be as in condensed matter.

\item
For the second scenario, to obtain the symmetry of the effective boundary theory,
\cgrn{
 we can
simplify all the interior degrees of freedom into a single one $g^*$, then
the degrees of freedom on $M^2$ are given by $\{g_1,g_2,\cdots,g^*\}$ where $g_i$
live on the boundary $\prt M^2$ and $g^*$ lives in the interior of $M^2$ (see Fig. \ref{bdrybulk}).
Now the imaginary-time evolution operator
is given by
\begin{align}
(\ee^{-\widehat H_{\prt M^2}})_{\{g_i'', ... \}, \{g_i', ...\}}
 =\prod_{M^2\times I} \nu^{s_{ijk*}}_{3}( g_i,  g_j, g_k,g^*) .
\end{align}
which defines an effective Hamiltonian for the boundary.
Now, we are ready to ask: What is the symmetry of the
effective boundary Hamiltonian, or effectively the symmetry of time evolution operator $\ee^{-\widehat H_{\prt M^2}}$?
}

\begin{figure}[h!] 
\begin{center}
\includegraphics[scale=0.8]{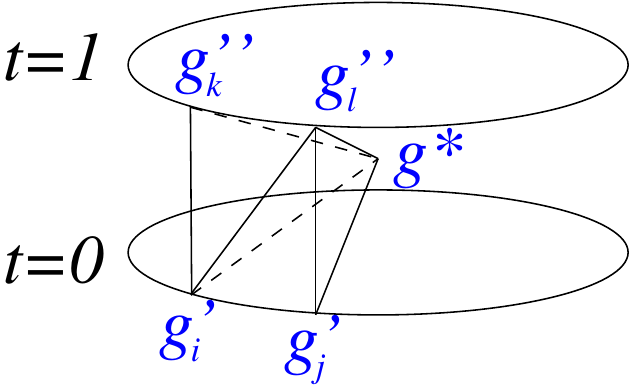}
\end{center}
\caption{
$M^2\times I$ representing one step of imaginary time evolution, for
the effective boundary theory.
The space $M^2$ is given by the disk.
}
\label{bdrybulk}
\end{figure}

\cgrn{
The analysis of global symmetry in Sec.~\ref{sec:non-on-site-G-Lagrangian} no longer applies.
The discrete time evolution operator
does not have the usual global symmetry:
\begin{align} \label{eq:eHamiltonianHM2-1}
&\ \ \ \
(\ee^{-\widehat H_{\prt M^2}})_{\{gg_i'', ...\}, \{gg_i', ...\}}
\neq 
(\ee^{-\widehat H_{\prt M^2}})_{\{g_i'', ...\}, \{g_i', ...\}},
\end{align}
since
\begin{align}
 \prod_{M^2\times I} \nu^{s_{ijk*}}_{3}( gg_i,  gg_j, gg_k,g^*)
=\prod_{M^2\times I} \nu^{s_{ijk*}}_{3}( g_i,  g_j, g_k, g^{-1}g^*)
\neq \prod_{M^2\times I} \nu^{s_{ijk*}}_{3}( g_i,  g_j, g_k, g^*).
\end{align}
The difference between two matrix elements
$(\ee^{-\widehat H_{\prt M^2}})_{\{gg_i'', ...\}, \{gg_i', ...\}}$
and 
$(\ee^{-\widehat H_{\prt M^2}})_{\{g_i'', ...\}, \{g_i', ...\}}$
is just a $U(1)$ phase factor
\begin{align} \label{eq:eHamiltonianHM2-diff}
&\prod_{M^2\times I} \frac{\nu^{s_{ijk}}_{3}(  g_i,   g_j,  g_k,g^{-1}g^*)}{\nu^{s_{ijk}}_{3}( g_i,  g_j, g_k,g^*)}
= \prod_{\prt M^2\times I} f_2^{s_{ijk}}( g_i,  g_j, g_k)
= \prod_{\prt M^2\times \prt I} f_1^{s_{ij}}( g_i,  g_j) \nonumber \\
&
=\prod_{\prt M^2 \times \prt I} \nu^{s_{ij}}_{3}( g_i,  g_j, g^*,g^{-1}g^*)
=
\frac{\prod_{(ij)} \nu^{s_{ij}}_{3}( g_i'',  g_j'', g^*,g^{-1}g^*)}
{ \prod_{(ij)} \nu^{s_{ij}}_{3}( g_i',  g_j', g^*,g^{-1}g^*)}
{,}
\end{align}
where $\prod_{M^2\times I}$ multiplies over all the 3-simplices in Fig.~\ref{bdrybulk}, $\prod_{\prt M^2\times I}$ over all the 2-simplices on $\prt
M^2\times I$, $\prod_{\prt M^{2}\times \prt I}$ over all the 1-simplices on the
top and the bottom boundaries of $\prt M^2 \times  I$.  
Note that many oppositely oriented $\nu_{3}$ terms are cancelled out in order to derive the last form of the above eqn.~(\ref{eq:eHamiltonianHM2-diff}).
This means that the
boundary time evolution operator is invariant 
$$
|\{g  g_i'', ...\} \rangle(\ee^{-\widehat H_{ \prt M^2}})_{\{g g_i'', ...\}, \{gg_i', ...\}}
\langle \{g g_i', ...\} |
=
\widehat U( g)
|\{  g_i'', ...\} \rangle 
(\ee^{-\widehat H_{\prt M^2}})_{\{ g_i'', ...\}, \{g_i', ...\}}
\langle \{g_i', ...\} |
\widehat U^\dag ( g),
$$
under a modified $G$-symmetry
transformation
%
\begin{align} 
\widehat U(g) & \equiv   \widehat U_0(g)\; U_{\{g_i, ...\}},
\end{align}
where
\begin{align}
 U_{\{g_i, ...\}} =\prod_{(ij)} \nu^{s_{ij}}_{3}( g_i,  g_j, g^*,g^{-1}g^*)
\end{align}
and $\widehat U_0(g)$ generates the usual {on-site} $G$-symmetry transformation 
$|\{ g_i, ...\} \rangle \to |\{g g_i, ...\} \rangle$.  
The phase factor $U_{\{g_i, ...\}}$ makes the $G$-symmetry to be \emph{non-on-site}
at the boundary.
}

We have written these formulas in 2+1 dimensions, but they all can be
generalized.  In $d$ dimensions, we have \cgrn{an effect boundary symmetry
operator $\widehat U(g)$ acting on $\prt M^{d-1}$  for the effective boundary
Hamiltonian $\ee^{-\widehat H_{\prt M^{d-1}}}$}:
\begin{align} \label{eq:U(g)}
\widehat U(g) &\equiv   \widehat U_0(g)\; U_{\{g_i, ...\}}
= \widehat U_{0}(g) \prod_{(ij\cdots \ell)  \in \prt M^{d-1} } 
\nu^{s_{ij\cdots \ell}}_{d}( g_i,  g_j,\cdots, g_\ell, g^*,g^{-1}g^*),
\end{align}

\end{enumerate}

\subsection{The second  boundary of a generic SPT state -- Gapped boundary by extending the $G$-symmetry to an $H$-symmetry}
\label{sec:extSymm}


In Sec.~\ref{bdry1} and also in Sec.~\ref{sec:1st-complx-bdry}, we considered the path integral of a $G$-SPT state
described by a homogeneous cocycle $\nu_d \in \cH^d(G,U(1))$.   The path integral
that we studied in that section remained $G$-symmetry invariant even on a manifold with
boundary, where the $G$-symmetry is an on-site symmetry in the bulk.  However,
if we integrate out the bulk degrees of freedom, the effective boundary theory
will have an effective $G$-symmetry, which must be non-on-site (\ie anomalous)
on the boundary. This anomalous $G$-symmetry on the boundary forces the
boundary to have some non-trivial dynamical properties.

However, the simple model introduced in Sec.~\ref{sec:1st-complx-bdry} has a complicated
boundary dynamics, which is hard to solve.  There are several standard ways to
modify the construction in  Sec.~\ref{sec:1st-complx-bdry} to get a boundary that can be
solved exactly.  One way to do so is to constrain the group variables $g_i$ on
boundary sites to all equal 1, or at least to take values in a subgroup
$G'\subseteq G$ such that the cohomology class of $\nu_d$ becomes trivial when
restricted to $G'$.  Given this, after possibly modifying $\nu_d$ by a coboundary, we can
assume that  $\nu_d=1$ when the group variables $g_i$
all belong to $G'$.  In this case, the action amplitudes for the boundary
effective theory eqn.~(\ref{bdry1path}) are always equal to 1 (after choosing
$g^* \in G'$).  So the boundary constructed in this way is exactly soluble, and is gapped.  This construction
amounts to spontaneous or explicit breaking of the
symmetry from $G$ to $G'$.  

In this section, we will explain another procedure to construct a model with
the same bulk physics and an exactly soluble gapped boundary.  This will be accomplished
by {\it extending} (rather than breaking) the global symmetry along the
boundary.   Then, as in our explicit example of the CZX model in 
Sec.~\ref{bdry2w}, we get a boundary state that is gapped and symmetric, but the
symmetry along the boundary is enhanced relative to the bulk.


\subsubsection{A purely mathematical setup on that $G$-cocycle is trivialized in $H$}\label{pure}

To describe the symmetry extended boundary, let us introduce a purely
mathematical result.  We consider an extension of $G$,
\begin{align}
1 \rightarrow K \overset{}{\rightarrow}  H  \overset{{r}}{\rightarrow} G \rightarrow 1
\end{align}
where $K$ is a normal subgroup of $H$, and $H/K=G$. Here $r$ is a surjective
group homomorphism from $H$ to $G$.  A ``$G$-variable'' $G$-cocycle
$\nu_d(g_0,\cdots,g_d)$ can be ``pulled back'' to an ``$H$-variable'' $H$-cocycle
$\nu_d^H(h_0,\cdots,h_d)$, defined by
\begin{align}
\label{vphih}
 \nu^H_d(h_0,\cdots,h_d) = \nu_d({r}(h_0),\cdots,{r}(h_d)) \equiv \nu_d^G({r}(h_0),\cdots,{r}(h_d)).
\end{align}
The case of interest to us is that $\nu_d^H$ 
is trivial in $\cH^d(H,U(1))$.  This means $\nu^H_d(h_0,\cdots,h_d)$ can be 
rewritten as a coboundary, namely
\begin{align}
\label{Omom}
 \nu^H_d(h_0,\cdots,h_d) = \delta  \mu^H_{d-1}(h_0,\cdots,h_d)\equiv
 \frac{
\prod_{i=\text{even}} \mu^H_{d-1}(h_0,\cdots,\widehat h_i, \cdots,h_d) 
}{
\prod_{i=\text{odd}} \mu^H_{d-1}(h_0,\cdots,\widehat h_i, \cdots,h_d) 
}.
\end{align} 
(The symbol $\widehat h_i$ is an instruction to omit $h_i$ from the sequence.)

\cgrn{
For the convenience and the preciseness of the notation, we can also shorten the above eqn.~(\ref{Omom}) to
\bea
\nu^G_d(r(h))=\nu^H_d(h)=\delta  \mu^H_{d-1}(h),
\eea
where the variable $h$ in the bracket is a shorthand of many copies of group elements in a direct product group of $H$.
By pulling back a $G$-cocycle $ \nu^G_d$ back to $H$, it becomes an $H$-coboundary $\delta  \mu^H_{d-1}$.
Formally, we mean that a nontrivial $G$-cocycle
\bea
\nu_d^G \in \cH^d(G,U(1))
\eea
becomes a trivial element when it is pulled back (denoted as ${}^*$) to $H$
\bea
{r}^* \nu_d^{G}= \nu^H_d =\delta  \mu^H_{d-1} \in \cH^d(H,U(1)).
\eea
{Saying} that this element is trivial means that the corresponding cocycle is a coboundary.}

Here $\mu^H_{d-1}(h_0, \cdots,h_{d-1})$ is a homogeneous $(d-1)$-cochain:
\begin{align}
 \mu^H_{d-1}(hh_0, \cdots,hh_{d-1})=\mu^H_{d-1}(h_0, \cdots,h_{d-1}).
\end{align}
The definition of $\nu_d^H$ also ensures that
\begin{align}
\label{Kgauge}
 \nu^H_d(v_0h_0,\cdots,v_dh_d)=\nu^H_d(h_0,\cdots,h_d),\ \ \ \
v_i\in K,
\end{align}
since $r(v_i)=1$ is trivial in $G$ for any $v_i\in K$.
In particular, 
$\nu^H_d(v_0,\cdots,v_d) =1,\ v_i\in K$, and therefore
\begin{align}
 \frac{
\prod_{i=\text{even}} \mu^H_{d-1}(v_0,\cdots,\widehat v_i, \cdots,v_d) 
}{
\prod_{i=\text{odd}} \mu^H_{d-1}(v_0,\cdots,\widehat v_i, \cdots,v_d) 
}
=1. 
\end{align}
Thus when we restrict to $K$, the cochain $\mu^H_{d-1}(v_0, \cdots,v_{d-1})$ becomes a cocycle {$\mu^K_{d-1}$} in
$\cH^{d-1}(K,{U(1)})$.  
An important detail is that in general
the cohomology class of {$\mu^K_{d-1}$} is {\it not} uniquely determined by the original cocycle $\nu_d$.  
In general, it can depend
on the choice of cochain $\mu^H_{d-1}$ that was used to trivialize $\nu_d^H$.  

In fact,
let $\mu^H_{d-1}$ and $\widetilde\mu^H_{d-1}$ be two cochains, either of which could be used to trivialize $\nu_d^H$:
\begin{equation}\label{bogo}\nu_d^H=\delta \mu^H_{d-1}=\delta \widetilde \mu^H_{d-1}.\end{equation}
Then $\nu_{d-1}^H=\mu^H_{d-1}(\widetilde \mu^H_{d-1})^{-1}$ is a cocycle, $\delta \nu_{d-1}^H=1$.   So $\nu^H_{d-1}$ has
a class in $\cH^{d-1}(H,U(1))$. If this class is nontrivial,
the gapped boundary states that we will construct using $\mu^H_{d-1}$ and $\widetilde \mu^H_{d-1}$ are inequivalent.
Thus the number of inequivalent gapped boundary states that we can make by the construction
described below (keeping fixed $H$ and $K$) is the order of  the finite group
$\cH^{d-1}(H,U(1))$.\footnote{It is not true that these states can be classified canonically by $\cH^{d-1}(H,U(1))$, because there is no
natural starting point, that is, there is no natural choice of $\mu^H_{d-1}$ to begin with.   Once one makes such a choice,
the boundary states that we will construct can be classified by $\cH^{d-1}(H, U(1))$.}  

A nontrivial class in $\cH^{d-1}(H,U(1))$ may or may not remain nontrivial after restriction from $H$ to $K$, so 
in general as stated above the cohomology class of $\nu^K_{d-1}$ can depend on the choice of $\mu^H_{d-1}$.

\subsubsection{
$H$-symmetry extended boundary --- By extending $G$-symmetry to  $H$-symmetry}\label{mainpoint}

To construct the second  boundary of generic SPT state, we allow the degrees of
freedom on the vertices at the boundary 
to be labeled by
$h_i \in H$.  This amounts to adding new degrees of freedom along the boundary.
The degrees of freedom on the vertices in the bulk are still labeled by $g_i
\in G$.
With this enhancement of the boundary variables, we can
write down the action amplitude for the second construction as
\begin{align}
\label{bndry2ndL}
e^{-\int_{M^d} \cL_{\text{Bulk}} {d^dx}} 
 =
\frac{\prod_{M^d} \nu_{d}^{s_{01\cdots d}}( g_0,  g_1, \cdots,g_d) }{
\prod_{\prt M^d} (\mu^H_{d-1})^{s_{01\cdots (d-1)}}( h_0,  h_1, \cdots,h_{d-1}) }
\end{align}
where $\nu_d$ and $\mu^H_{d-1}$ are the cochains introduced in the last section
and $M^d$ may have a boundary.  Here, if a vertex in $\nu_{d}( g_0,  g_1,
\cdots,g_d)$ is on the boundary, the corresponding $g_i$ is given by
$g_i=r(h_i)$.

We note that, since $r$: $H\to G$ is a group homomorphism, the action $h$:
$H\to H$, $h_i \to h h_i$, induces an action $r(h)$: $G\to G$, $g_i \to r(h)
g_i$.  Therefore, the total action amplitude \eqn{bndry2ndL} has $H$
symmetry:
\begin{align}
\frac{\prod_{M^d} \nu_{d}^{s_{01\cdots d}}( g_0,  g_1, \cdots,g_d) }{
\prod_{\prt M^d} (\mu^H_{d-1})^{s_{01\cdots (d-1)}}( h_0,  h_1, \cdots,h_{d-1}) }
=
\frac{\prod_{M^d} \nu_{d}^{s_{01\cdots d}}( r(h)g_0,  r(h)g_1, \cdots,r(h)g_d) }{
\prod_{\prt M^d} (\mu^H_{d-1})^{s_{01\cdots (d-1)}}( hh_0,  hh_1, \cdots,hh_{d-1}) }
\end{align}
where $h\in H$.   In the bulk, the symmetry is $G$, but along the boundary it is extended to $H$.  Such a total action amplitude defines our second construction of the
boundary of a $G$-SPT state, which has a symmetry extension $G$ lifted to $H$ on the
boundary.
{We return to more details on this model in Sec.~\ref{sec:general}.}

The bulk of the constructed model is described by the same group cocycle
$\nu_{d}$, which give rise to the $G$-SPT state.  But the boundary has an
extended symmetry $H$. In this case, we should view the whole
system (bulk and boundary) as having an extended $H$-symmetry,  with the
$K$ subgroup acting trivially in the bulk.  So the effective symmetry in the
bulk is $G=H/K$.

The dynamics of our second boundary is very simple, since the total action
amplitude \eqn{bndry2ndL} is always equal to 1 by construction:  
\begin{align}\label{usefulone}
\prod_{M^d} \nu_{d}^{s_{01\cdots d}}( g_0,  g_1, \cdots,g_d) 
&=
\prod_{M^d} (\nu^H_{d})^{s_{01\cdots d}}( h_0,  h_1, \cdots,h_d) 
\nonumber\\
&=
\prod_{\prt M^d} (\mu^H_{d-1})^{s_{01\cdots (d-1)}}( h_0,  h_1, \cdots,h_{d-1}),
\end{align}
where $g_i = r(h_i)$.  Thus, the ground state is always gapped and there is no
ground state degeneracy regardless of whether the system has a boundary or not.
In other words, the second boundary of the $G$-SPT state is gapped with $H$
symmetry and no topological order.  The gapped boundary with $H$ symmetry and
no topological order is possible, since we have chosen $H$ so that when we view
the $G$-SPT state as an $H$-SPT state, the non-trivial $G$-SPT state becomes a
trivial $H$-SPT state.

\subsection{
On-site (anomaly-free) $H$-symmetry transformation on the boundary effective theory}

\label{sec:on-site-H}

{
Now we show that symmetry extension, as described in Sec.~\ref{mainpoint},
gives a boundary state with on-site (anomaly-free) $H$-symmetry, based on the
Hamiltonian formalism on the boundary.  This section directly parallels  the
previous discussion in Sec.~\ref{sec:non-on-site-G}, where a non-trivial
$G$-cocycle gives rise to a non-on-site effective $G$-symmetry on the boundary.
After extending the symmetry to $H$, the non-trivial $G$-cocycle $\nu_d$
becomes a trivial $H$-cocycle $\nu_d^H$, which in turn gives rise to an \emph{on-site}
effective $H$-symmetry for the boundary effective theory.  }

%

\begin{figure}[!h]
\begin{center}
\includegraphics[scale=0.6]{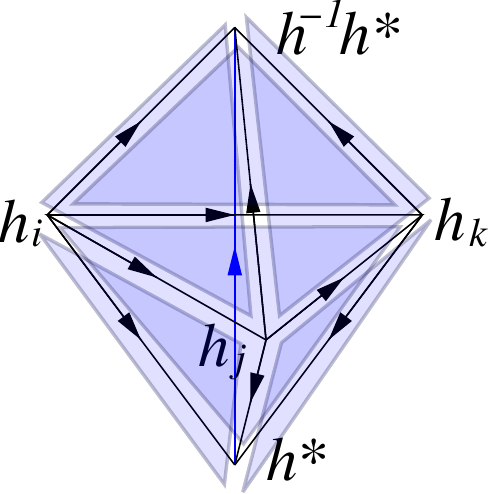}
\end{center}
\caption{
Graphic representation of 
$f_2(h_i,h_j,h_k)$ $=$ $\frac{\mu_{2}^H( h_i,  h_j, h^{-1}h^*)}{\mu_{2}^H( h_i,  h_j, h^*)} $
$\frac{\mu_{2}^H( h_j,  h_k, h^{-1}h^*)}{\mu_{2}^H( h_j,  h_k, h^*)}$ 
$\frac{\mu_{2}^H( h_i,  h_k, h^*)}{\mu_{2}^H( h_i,  h_k, h^{-1}h^*)}$ $= \dd f_1$, again as a
coboundary. Each shaded blue triangle is assigned with a split cochain $\mu_{2}^H$. See eqns.~(\ref{eq:f2f1Hcoboundaryi}) and (\ref{eq:f2f1coboundary-2-nu}).
}
\label{f2H_split}
\end{figure}

Taking $d=3$ as an example, Eqns. (\ref{eq:f2f1coboundary}), (\ref{eq:f2f1coboundary-2}) and (\ref{cotto}) of Sec.~\ref{sec:non-on-site-G-Lagrangian}  still hold. 
Furthermore, when $h_i$, $h_j$ and $h_k$ are boundary degrees of freedom in $H$, eqn.(\ref{eq:f2f1coboundary}) becomes
\begin{align} 
 f_2(h_i,h_j,h_k) &=
\frac{{\mu_{2}^H( h_i,  h_j, h^{-1}h^*)} {\mu_{2}^H( h_j,  h_k, h^{-1}h^*)}{\mu_{2}^H( h_i,  h_k, h^{-1}h^*)}^{-1}}{
{\mu_{2}^H( h_i,  h_j, h^*)} {\mu_{2}^H( h_j,  h_k, h^*)}{\mu_{2}^H( h_i,  h_k, h^*)}^{-1}}= \dd f_1.
\label{eq:f2f1Hcoboundaryi}
\end{align}
See Fig.~\ref{f2H_split} for an illustration.
Here $\mu_{2}^H$ is a homogeneous 2-cochain that splits $\nu^H_3$ (or $\nu_3^G(\{r(h)\})$)
and satisfies ${\mu_{2}^H( h_i,  h_j, h^{-1}h^*)}={\mu_{2}^H( h h_i,  h h_j, h^*)}$.
Now the split 2-cochain $f_1$ in eqn.(\ref{eq:f2f1coboundary-2}) has a new form:
\begin{align} \label{eq:f2f1coboundary-2-nu}
 f_1(h_i,  h_j)=\frac{{\mu_{2}^H( h_i,  h_j, h^{-1}h^*)}}{{\mu_{2}^H( h_i,  h_j, h^*)}}.
 \end{align}

%

\begin{figure}[h!] 
\begin{center}
\includegraphics[scale=0.8]{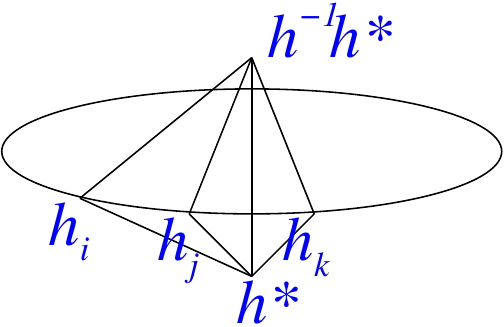}
\end{center}
\caption{
Geometric picture to explain the calculation from \eqn{HsymmB} to \eqn{HsymmB1}
(for the $d=3$ case).
$\prod_{(ij\cdots \ell)  \in \prt M^{d-1} }(\nu_d^H)^{s_{ij\cdots \ell}}( h_i,  h_j,\cdots, h_\ell, h^*,h^{-1}h^*) $ in \eqn{HsymmB} 
is a product over all the 3-simplices in the figure.
$\prod_{(ij\cdots \ell)  \in \prt M^{d-1} } 
(\mu_{d-1}^H)^{s_{ij\cdots \ell}}( hh_i,  hh_j,\cdots, hh_\ell, h^*)=
\prod_{(ij\cdots \ell)  \in \prt M^{d-1} } 
(\mu_{d-1}^H)^{s_{ij\cdots \ell}}( h_i,  h_j,\cdots, h_\ell, h^{-1}h^*)
 $ is a product over all the 2-simplices on the top surface, and
$\prod_{(ij\cdots \ell)  \in \prt M^{d-1} } 
(\mu_{d-1}^H)^{s_{ij\cdots \ell}}( h_i,  h_j,\cdots, h_\ell, h^*)$
is a product over all the 2-simplices on the bottom surface.
}
\label{Hphase}
\end{figure}

{
To show more clearly that $H$-symmetry can be made on-site and anomaly-free in any dimension $d$, we note that the action amplitude \eqn{bndry2ndL}
can be rewritten as
\begin{align}
e^{-\int_{M^d} \cL_{\text{Bulk}} {d^dx}} 
 =
\frac{\prod_{M^d} (\nu_{d}^H)^{s_{01\cdots d}}( h_0,  h_1, \cdots,h_d) }{
\prod_{\prt M^d} (\mu^H_{d-1})^{s_{01\cdots (d-1)}}( h_0,  h_1, \cdots,h_{d-1}) }. 
\end{align}
Each local term $(\mu^H_{d-1})^{s_{01\cdots (d-1)}}( h_0,  h_1, \cdots,h_{d-1})$ is
already invariant under $H$-symmetry transformation on the boundary. So we will
drop it.  The term $(\nu_{d}^H)^{s_{01\cdots d}}( h_0,  h_1, \cdots,h_d)$ may
not be invariant under $H$-symmetry transformation on the boundary, although
their product $\prod_{M^d} (\nu_{d}^H)^{s_{01\cdots d}}( h_0,  h_1,
\cdots,h_d)$ is. This may lead to a non-on-site $H$-symmetry.  Repeating the
calculation in Sec.~\ref{sec:non-on-site-G}, we found that the discrete time evolution 
operator $\ee^{-\widehat H_{\prt M^{d-1}}}$ does not have the usual global symmetry,
where their matrix elements follow:
\begin{align} \label{eq:eHamiltonianHM2-h}
&\ \ \ \
(\ee^{-\widehat H_{\prt M^{d-1}}})_{\{h h_i'', ...\}, \{h h_i', ...\}}
\neq 
(\ee^{-\widehat H_{\prt M^{d-1}}})_{\{h_i'', ...\}, \{h_i', ...\}}.
\end{align}
But it is invariant 
$$
|\{h  h_i'', ...\} \rangle(\ee^{-\widehat H_{ \prt M^{d-1} }})_{\{h h_i'', ...\}, \{h h_i', ...\}}
\langle \{h h_i', ...\} |
=
\widehat U( h)
|\{  h_i'', ...\} \rangle 
(\ee^{-\widehat H_{ \prt M^{d-1} }})_{\{ h_i'', ...\}, \{h_i', ...\}}
\langle \{h_i', ...\} |
\widehat U^\dag (h),
$$
under a modified symmetry transformation operator 
\begin{align} 
\label{HsymmB}
\widehat U(h) &\equiv   \widehat U_0(h)
\prod_{(ij\cdots \ell)  \in \prt M^{d-1} } 
(\nu_d^H)^{s_{ij\cdots \ell}}( h_i,  h_j,\cdots, h_\ell, h^*,h^{-1}h^*),
\end{align}
which appears to be non-on-site.
However, since $\nu_d^H = \del \mu_{d-1}^H$ is a coboundary,
the above can be rewritten as (see Fig.~\ref{bdrybulk} and ~\ref{Hphase})
\begin{align}
\label{HsymmB1}
 \widehat U(h) &= \widehat U_0(h)
\frac{
\prod_{(ij\cdots \ell)  \in \prt M^{d-1} } 
(\mu_{d-1}^H)^{s_{ij\cdots \ell}}( hh_i,  hh_j,\cdots, hh_\ell, h^*)
}{
\prod_{(ij\cdots \ell)  \in \prt M^{d-1} } 
(\mu_{d-1}^H)^{s_{ij\cdots \ell}}( h_i,  h_j,\cdots, h_\ell, h^*)
}.
\end{align}
After a local unitary transformation 
$|\{h_i\}\> \to W(\{h_i\})|\{h_i\}\> \equiv |\{h_i\}'\>$ with
$$W(\{h_i\})\equiv \prod_{(ij...) \in \prt M^{d-1} } {{\mu_{d-1}^H( h_i,  h_j, ..., h^*)}},$$
we can change the above $H$-symmetry transformation 
to
\begin{align} 
\widehat U(h) &\to W^\dag \widehat U(h) W =  \widehat U_0(h)
\end{align}
which indeed becomes \emph{on-site}. The \emph{on-site} symmetry $\widehat U_0(h)$ makes the time evolution operator invariant under
$$
|\{h  h_i'', ...\}' \rangle(\ee^{-\widehat H_{{ \prt M^{d-1} }}})_{\{h h_i'', ...\}, \{h h_i', ...\}}
\langle \{h h_i', ...\}' |
=
\widehat U_0( h)
|\{  h_i'', ...\}' \rangle 
(\ee^{-\widehat H_{{ \prt M^{d-1} }}})_{\{ h_i'', ...\}, \{h_i', ...\}}
\langle \{h_i', ...\}' |
\widehat U_0^\dag (h).
$$
}

The subtle difference between Sec.~\ref{sec:non-on-site-G}  and Sec.~\ref{sec:on-site-H} is that
the $\nu_{d}( g_i,  g_j,\cdots, g_\ell, g^*,g^{-1}g^*)$ cannot be absorbed through \emph{local unitary transformations}, but
its split form ${{\mu_{d-1}^H( h_i,  h_j, ..., h^*)}}$ can be absorbed. 
Namely, one can think of $\mu_{d-1}^H$ as an output of a local unitary matrix acting on local nearby sites with input data $h_i,  h_j, \dots$ in a quantum circuit.

To summarize what we did in Sec.~\ref{sec:non-on-site-G} and \ref{sec:on-site-H},
 the $G$-symmetry transformation on the boundary was \emph{non-on-site} thus \emph{anomalous}.
The $H$-symmetry transformation on the boundary is now made to be \emph{on-site}, by pulling back $G$ to $H$, thus, it is \emph{anomaly-free} in $H$.


\subsection{The third boundary of a generic SPT state: A gapped symmetric boundary that violates locality with (hard) gauge fields }

\label{sec:hard-gauge}

In the last section, we constructed a gapped symmetric boundary of an SPT state
such that the global symmetry is extended from $G$ to $H$ along the boundary.  Such boundary enhancement of the symmetry
is usually\footnote{In Sec.~\ref{bdry2w}, we described a situation in which it is natural.} not natural in condensed matter physics.   Just as in our discussion of the CZX model
in Sec.~\ref{bdry3w}, \ref{bdry4w}, the way to avoid symmetry extension is to gauge the boundary symmetry $K$, giving a construction in which the full global symmetry
group is $G$ (or $G'$ in the more general mixed breaking and extension construction described in Sec.~\ref{mixed}).

As in the CZX model, there are broadly two approaches to gauging the $K$ symmetry.  One may use ``hard gauging'' in which one introduces
(on the boundary) elementary fields that gauge the $K$ symmetry, or ``soft gauging'' in which the boundary gauge fields are emergent.
Hard gauging is generally a little quicker to describe, so we begin with it, but soft gauging, which will be the topic of Sec.~\ref{sec:cocycle-emergent}, is more natural
in condensed matter physics because it can be strictly local or ``on-site.''   Our discussion here and in the next section is roughly parallel to 
Sec.~\ref{bdry3w} and \ref{bdry4w} on the CZX model.

\begin{figure}[!h]
\begin{center}
\includegraphics[scale=0.7]{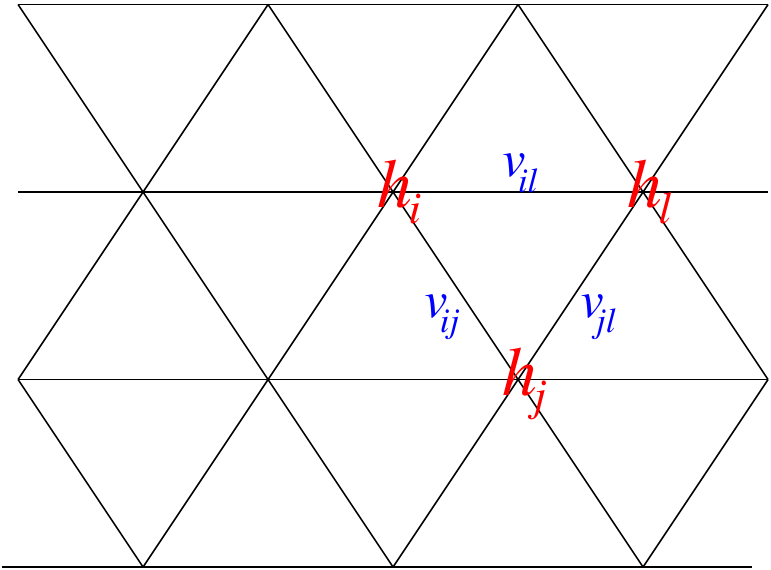}
\end{center}
\caption{
A boundary of a $G$-SPT state. A vertex $i$ on the boundary carries $h_i \in H$,
and a link $(ij)$ carries $v_{ij}\in K$.
}
\label{bdryhv}
\end{figure}

To construct a new boundary, let us consider a system on a $d$-dimensional
space-time manifold $M^d$, with a triangulation that has a branching structure.  A
vertex $i$ inside $M^d$ carries a degree of freedom $g_i \in G$.  A vertex $i$
on the boundary $\prt M^d$ carries a degree of freedom $h_i \in H$.  A link
$(ij)$ on the boundary $\prt M^d$ carries a degree of freedom $v_{ij} \in K$.
See Fig.~\ref{bdryhv}

We choose the action amplitude of our new model to be
\begin{align}
\label{actamp}
&\ee^{-\int_{M^d} \cL {d^dx}} = \prod_{(i_0\cdots i_d) \in M^d}
\nu_d^{s_{i_0\cdots i_d}}(g_{i_0},\cdots,g_{i_d}) \times
\\
&
\prod_{(i_0\cdots i_{d-1}) \in \prt M^d}
(\cV^{H,K}_{d-1})^{-s_{i_0\cdots i_{d-1}}}(h_{i_0},\cdots,h_{i_{d-1}};
v_{i_0i_1},v_{i_0i_2},\cdots)
\nonumber 
\end{align}
where $\prod_{(i_0\cdots i_d)}$ is a product over $d$-dimensional simplices
$(i_0\cdots i_d)$ in the bulk, and $\prod_{(i_0\cdots i_{d-1})}$ is a product
over $(d-1)$-dimensional simplices $(i_0\cdots i_{d-1})$ on the boundary.
$s_{i_0\cdots i_d}=\pm 1$ is the orientation of the $d$-simplex $(i_0\cdots i_d)$,
and $s_{i_0\cdots i_{d-1}}=\pm 1$ is the orientation of the $(d-1)$-simplex
$(i_0\cdots i_{d-1})$.   Finally, $\cV^{H,K}_{d-1}$ will be defined in Sec.~\ref{defmo}, using
$\mu^H_{d-1}$ introduced in Sec.~\ref{pure}, as well as 
 ``\emph{hard gauge fields}''  $v_{ij}$ along boundary links.

In the action amplitude \eqn{actamp}, $\nu_d\in \cH^d(G,U(1))$ is the cocycle
describing the $G$-SPT state.  We have assumed that if a vertex $i$ in
$\nu_d(g_{0},\cdots,g_{d})$ is on the boundary, then the corresponding $g_i$ is
given by $g_i = {r}(h_i)$.  

\subsubsection{A cochain that encodes ``hard gauge fields''}\label{defmo}

The generalized cochain $ \cV^{H,K}_{{d-1}}(h_{i_0},\cdots,h_{i_{d-1}};
v_{i_0i_1},v_{i_0i_2},\cdots)$ will be defined for boundary simplices.  It will depend on $H$-valued boundary
spins $h_i$ as well as $K$-valued boundary link variables $v_{ij}$.  As usual in lattice gauge theory, we can regard $v_{ij}$ as
a $K$ gauge connection on the link $ij$.

First, we assume  that $\cV^{H,K}_{{d-1}}(h_{i_0},\cdots,h_{i_{d-1}};
v_{i_0i_1},v_{i_0i_2},\cdots)=0$ for any configurations $v_{ij}$ that do not
satisfy $ v_{i_1i_2} v_{i_2i_3}= v_{i_1i_3}$, for some $i_1$, $i_2$, $i_3$.  So
only the $v_{ij}$ configurations that satisfy
\begin{align}
\label{flatC}
 v_{i_1i_2} v_{i_2i_3} = v_{i_1i_3}, 
\end{align}
on every triangle can contribute to the path integral. This means that only flat
$K$ gauge fields are allowed.

For a flat connection on a simplex with vertices $i_0,\dots,i_{d-1}$, all of the $v_{i_ji_k}$ can be expressed in terms of
$v_{01},v_{12},v_{23},\cdots,v_{d-2,d-1}$.  So likewise
$\cV^{H,K}_{d-1}(h_0,\cdots,h_{d-1}; v_{01},v_{02},v_{12},\cdots)$ can be expressed as
$\cV^{H,K}_{d-1}(h_0,\cdots,h_{d-1}; v_{01}, v_{12}, \cdots,v_{d-2,d-1})$. We define $\cV^{H,K}_{d-1}$ in terms of 
the homogeneous cochain $\mu^H_{d-1}$ of Sec.~\ref{pure} by
\begin{align}
\label{defV}
&\ \ \
\cV^{H,K}_{d-1}(h_{0},\cdots,h_{{d-1}};
v_{01},v_{02},v_{12},\cdots)
\nonumber\\
&=
 \cV^{H,K}_{d-1}(h_0,\cdots,h_{d-1}; v_{01}, v_{12}, \cdots,v_{d-2,d-1})
\nonumber\\
&=
\mu^H_{d-1}(h_0, v_{01}h_1, v_{01}v_{12}h_2, \cdots).
\end{align}
In other words
\begin{align}
 \cV^{H,K}_{d-1}(h_0,\cdots,h_{d-1}; v_{01}, v_{12}, \cdots,v_{d-2,d-1})
=
\mu^H_{d-1}(\t h_0, \t h_1, \t h_2, \cdots).
\end{align}
where $\t h_i$ is given by $h_i$ parallel transported from site-$i$ to site-$0$
using the connection $v_{ij}$:
\begin{align}
 \t h_i = v_{01}v_{12}\cdots v_{i-1,i}h_i .
\end{align}

We note that $\cV^{H,K}_{d-1}$
has a local $K$-symmetry generated by $v_0, v_1, \cdots \in K$:
\begin{align}
\label{lsym}
&\ \ \ \
 \cV^{H,K}_{d-1}(v_0h_0,\cdots,v_{d-1}h_{d-1}; v_{01}, v_{12}, \cdots,v_{d-2,d-1})
\nonumber\\
& =
\cV^{H,K}_{d-1}(h_0,\cdots,h_{d-1}; v_0^{-1}v_{01}v_1, v_1^{-1}v_{12}v_2, \cdots)\cred{.}
\end{align}
Next we will view such a boundary local symmetry as a $K$-gauge redundancy
by viewing two boundary configurations $(h_i,v_{ij})$ and $(h'_i,v'_{ij})$ as
the same configuration if they are related by a gauge transformation
\begin{align}
h'_i = v_i h_i,\ \ \ \ 
v'_{ij} = v_i v_{ij} v_j^{-1},\ \ \ \ v_i \in K. 
\end{align} 
Eqn. \ref{lsym} ensures the gauge-invariance of the boundary action.

Now that we have gauged the $K$ symmetry, the global symmetry of the full
system, including its boundary, is $G$.  \cgrn{However, viewing two boundary
configurations $(h_i,v_{ij})$ and $(h'_i,v'_{ij})$ as the same configuration
makes the gauged theory no longer  a local bosonic system.
This is because the number of different
(\ie gauge inequivalent) configurations on the space-time boundary
$\prt M^d$ is given by\footnote{The formula works when all groups are Abelian. 
For non-Abelian groups, there could be additional constraints on this formula, for example, in terms of conjugacy classes.}
\begin{align}
\frac{|H|^{N_v} |K|^{N_l}}{|K|^{N_v}} |K|^{|\pi_0(\prt M^d)|},
\end{align}
where $N_v$ is the number of vertices, $N_l$ is the number of links on the
boundary $\prt M^d$, and $|\text{Set}|$ is the number of elements in the
$\text{Set}$.  
Here we count all the \emph{distinct} configurations of vertex variables of $H$ and link variables of $K$, 
 identifying them up to $K$-gauge transformations on the vertices.
 We consider all higher energetic configurations, which include both flat and locally non-flat configurations, much more than just ground state sectors.
Constant gauge transformations yield an additional factor $|K|^{|\pi_0(\prt M^d)|}$.
The appearance of the factor $|K|^{|\pi_0(\prt M^d)|}$ whose
exponent is not linear in $N_v$ and $N_l$ implies a non-local system.  } So the
third boundary is no longer local in that strict sense.  In  subsection
\ref{sec:cocycle-violate-local}, we show that this non-local boundary is gapped
and symmetric.  In  Sec.~\ref{sec:cocycle-emergent}, we will replace hard gauging
with soft gauging and thereby get a boundary that is fully local and on-site,
while still gapped and symmetric.

\subsubsection{A model that violates the locality {for the boundary theory}}

\label{sec:cocycle-violate-local}

In the path integral, we  only sum over gauge distinct configurations:  
\begin{align}
\label{Zgauge}
& Z =\sum_{\{g_i,[h_i,h_{ij}]\}} \prod_{(i_0\cdots i_d) \in M^d}
\nu_d^{s_{i_0\cdots i_d}}(g_{i_0},\cdots,g_{i_d}) \times
\\
&
\prod_{(i_0\cdots i_{d-1}) \in \prt M^d}
(\cV^{H,K}_{d-1})^{s_{i_0\cdots i_{d-1}}}(h_{i_0},\cdots,h_{i_{d-1}};
v_{i_0i_1},v_{i_1i_2},\cdots)
\nonumber 
\end{align}
where $[h_i,v_{ij}]$ represents the gauge equivalence classes.   (Equivalently, we can sum over all configurations and divide by \cgrn{the number of 
equivalent configurations in each gauge equivalence class}.) 

  We emphasize:
 
\emph{``Since the boundary theory is non-local respect to the boundary sites, it is no longer meaningful to distinguish on-site from
non-on-site symmetry, or anomaly-free from anomalous symmetry}.''

However, this system does have a global $G$ symmetry.
To see this,
let us consider a transformation generated by $h\in H$ 
given by
\begin{align}
(h_i,v_{ij})  \to (hh_i,hv_{ij}h^{-1})  
\end{align}
if $i$ is on the boundary, and
\begin{align}
g_i  \to  \cred{r}(h)g_i 
\end{align}
if $i$ is in the bulk.  Clearly, such a transformation is actually a $G$
transformation in the bulk.  On the boundary, since $(h_i,v_{ij})$ and
$(vh_i,vv_{ij}v^{-1})$ are gauge equivalent for $v\in K$, $h$ and $hv$ generate the
same transformation.  So the transformation on the boundary is given by the
equivalence class $[h]$ under the equivalence relation $h\sim hv$, $v\in K$.
Since $K$ is a normal subgroup of $H$, the  equivalence classes form a group
$H/K=G$. Thus, the transformation is also a $G$ transformation on the boundary.
Such a transformation is a symmetry of the model since
\begin{align}
\cV^{H,K}_{d-1}(hh_{i_0},\cdots,hh_{i_{d-1}}; hv_{i_0i_1}h^{-1},hv_{i_1i_2}h^{-1},\cdots) 
=
\cV^{H,K}_{d-1}(h_{i_0},\cdots,h_{i_{d-1}}; v_{i_0i_1},v_{i_1i_2},\cdots),
\end{align}
where we have used the definition \eqn{defV}.
We note that $hv_{ij}h^{-1} \in K$ since $K$ is a normal subgroup of $H$.
So the partition function \eqn{Zgauge} give{s} us a boundary effective
theory that still has the $G$ global symmetry.

Now we can ask
whether the ground state at the boundary breaks the $G$-symmetry or not. More
generally, what is the dynamical property of such a boundary?  Is it gapped? To
answer such a question, we note that on a triangulated $M^d$, in general
\begin{align}
\prod_{(i_0\cdots i_d) \in M^d}
\nu_d^{s_{i_0\cdots i_d}}(g_{i_0},\cdots,g_{i_d}) \neq 1
\end{align}
since $M^d$ has a boundary. But we can show that if the boundary is simply-connected, then
\begin{align}
&\ee^{-\int_{M^d} \cL {d^dx}} = \prod_{(i_0\cdots i_d) \in M^d}
\nu_d^{s_{i_0\cdots i_d}}(g_{i_0},\cdots,g_{i_d}) \times
\\
&
\prod_{(i_0\cdots i_{d-1}) \in \prt M^d}
(\cV^{H,K}_{d-1})^{-s_{i_0\cdots i_{d-1}}}(h_{i_0},\cdots,h_{i_{d-1}};
v_{i_0i_1},v_{i_1i_2},\cdots) =1.
\nonumber 
\end{align}
To show this, we first recall that only flat connections on the boundary contribute to the path integral.
If the boundary is simply-connected, this means that we can assume that $v_{ij}$ is pure gauge.
So  by a gauge transformation \eqn{lsym}, we can set 
all $v_{ij}$ to $1$ on the boundary:
\begin{align}
&\prod_{(i_0\cdots i_d)}
\nu_d^{s_{i_0\cdots i_d}}(g_{i_0},\cdots,g_{i_d}) 
\prod_{(i_0\cdots i_{d-1})}
( \cV^{H,K}_{d-1})^{-s_{i_0\cdots i_{d-1}}}(h_{i_0},\cdots,h_{i_{d-1}};
v_{i_0i_1},v_{i_1i_2},\cdots)
\nonumber\\
&=\prod_{(i_0\cdots i_d)}
\nu_d^{s_{i_0\cdots i_d}}(g_{i_0},\cdots,g_{i_d}) 
\prod_{(i_0\cdots i_{d-1})}
(\cV^{H,K}_{d-1})^{-s_{i_0\cdots i_{d-1}}}(\t h_{i_0},\cdots,\t h_{i_{d-1}};
1,1,\cdots)
\nonumber\\
&=\prod_{(i_0\cdots i_d)}
\nu_d^{s_{i_0\cdots i_d}}(g_{i_0},\cdots,g_{i_d}) 
\prod_{(i_0\cdots i_{d-1})}
(\mu^{H}_{d-1})^{-s_{i_0\cdots i_{d-1}}}(\t h_{i_0},\cdots,\t h_{i_{d-1}})
\end{align}
where $\t h_i $ is obtained from $h_i$ by the gauge transformation that sets the $v_{ij}$ to 1.
But this is 1 by virtue of eqn. (\ref{usefulone}).


The fact that the action amplitude of our theory on $M^d$ is always one if the
boundary of $M^d$ is simply-connected is enough to show that  the system on
$M^d$ is in a gapped phase both in the bulk and on the boundary.  Such a gap
state is the $K$-gauge {deconfined} state, described by the flat $K$-connection
$v_{ij} \in K$ on each link.  Also $h_i$ and $g_i$ are strongly fluctuating and
are quantum-disordered as well.  This is because the action amplitude is always
equal to 1 regardless the values of $h_i$ and $g_i$ (say, in the $v_{ij}=1$
gauge discussed above).  So the partition function \eqn{Zgauge} {gives} us a
boundary of the SPT state that is in the deconfined phase of $K$-gauge theory,
and does not break the $G$ symmetry.  



\subsection{The fourth boundary of a generic SPT state: A gapped symmetric boundary that preserves locality with emergent (soft) gauge fields}

\label{sec:cocycle-emergent}

\begin{figure}[h!]
\begin{center}
\includegraphics[scale=0.7]{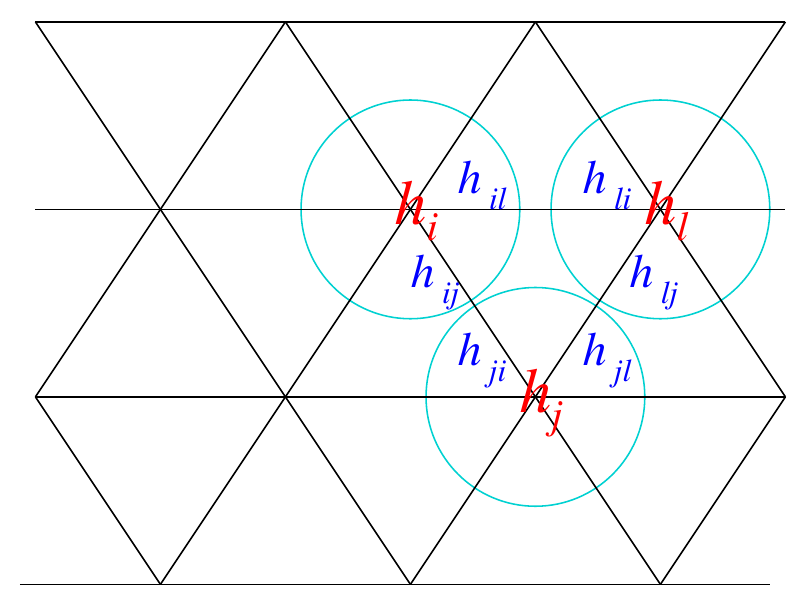}
\end{center}
\caption{
A boundary of $G$-SPT state. A vertex $i$ on the boundary carries $h_i \in H$,
and a link $(ij)$ carries $h_{ij}$ and $h_{ji}$.
The degrees of freedom in a circle, $h_i$, $h_{ij}$, $h_{il},\cdots$,
 belong to the same site 
 {labeled by $i$}.
}
\label{bdryhk}
\end{figure}

In the last section, we constructed a gapped symmetric boundary of an SPT state
by making its boundary non-local.  In this section, we are going to fix this
problem, by constructing the fourth gapped symmetric boundary of an SPT state
without changing the symmetry and without destroying the locality.  The new
gapped symmetric boundary has emergent gauge fields and topological order on
the boundary.  By this explicit construction, we show that:  

``\emph{In 3+1D and any higher dimensions, an SPT state with a finite group symmetry, regardless unitary or anti-unitary symmetry, always\footnote{To complete the argument,
we need to know that for every SPT phase with $G$ symmetry, a suitable extension $1\to K\to H\overset{r}{\to}G\to 1$ exists.
This is shown in Sec.~\ref{sec:extension-trivializeGcocycle}.}
 has a gapped local boundary with the same symmetry.}''  

The  construction in this section is a
generalization of the construction in Sec.~\ref{bdry4w}.

To construct a local boundary, we replace $v_{ij}$ on a link by two degrees of
freedom $h_{ij} \in H$ and $h_{ji} \in H$.   In other words, a link $(ij)$ on
the boundary $\prt D^d$ now carries two degrees of freedom $h_{ij} \in H$ and
$h_{ji} \in H$ (see Fig. \ref{bdryhk}).  We regard $h_i, h_{ij},h_{il},\cdots$
as the degrees of freedom on site-$i$ of the boundary (see Fig. \ref{bdryhk}).
In the bulk, a site-$i$ only carries a degree of freedom described by $g_i$.


We choose the action amplitude for our fourth boundary to be
\begin{align}
\label{actampS}
&\ee^{-\int_{D^d} \cL {d^dx}} = \prod_{(i_0\cdots i_d) \in D^d}
\nu_d^{s_{i_0\cdots i_d}}(g_{i_0},\cdots,g_{i_d}) \times
\\
&
\prod_{(i_0\cdots i_{d-1}) \in \prt D^d}
(\cV^{H,K}_{d-1})^{-s_{i_0\cdots i_{d-1}}}(h_{i_0},\cdots,h_{i_{d-1}};
h_{i_0i_1},h_{i_1i_0},\cdots) .
\nonumber 
\end{align}
In the following, we will define $\cV^{H,K}_{d-1}$.
{We introduce
a new form of cochain $\cV^{H,K}_{d-1}$ encoding ``\emph{soft gauge fields}'' emergent from 
the local boundary sites that we prescribe below.}

\subsubsection{A new cochain that encodes ``emergent soft gauge fields''}
\label{sec:emergent-soft-gauge}

First, we assume that 
$\cV^{H,K}_{{d-1}}(h_{i_0},\cdots,h_{i_{d-1}};
h_{i_0i_1},h_{i_1i_0},\cdots)=0$ 
for any configurations $h_{ij}$ that  
do not satisfy
\begin{align}
\label{vhh}
 v_{ij}\equiv h_{ij}h^{-1}_{ji} \in K
\end{align}  
for every link
or do not satisfy
\begin{align}
 v_{i_1i_2} v_{i_2i_3}= v_{i_1i_3},
\end{align}
for every triangle.
So only the $h_{ij}$ configurations 
that satisfy
\begin{align}
\label{flatCS}
 v_{i_1i_2} v_{i_2i_3} = v_{i_1i_3}, \ \ \ \
 v_{ij}=h_{ij}h^{-1}_{ji} \in K
\end{align}
on every triangle contribute to the path integral.  
Here $v_{ij}$ corresponds to the $K$-gauge
connection introduced in the last section.

\cgrn{
The $K$-gauge symmetry will impose the equivalence relation 
\begin{align}
\label{kg} (h_i, h_{ij}) \sim (k_i h_i, k_i h_{ij}), 
\end{align} 
for any $k_i\in K$. The total number of inequivalent configurations on space-time
boundary $\prt M^d$ is given by 
\begin{align} 
\frac{|H|^{N_v+2N_l}}{|K|^{N_v}}.
\end{align} 
The exponent in the number of configurations is linear in $N_v$ and
$N_l$, implying that the system is local.  }

Let us further assume that $\cV^{H,K}_{d-1}(h_{0},\cdots,h_{{d-1}};
h_{01},h_{10},\cdots)$ depends on $h_{ij}$ only via $
v_{ij}=h_{ij}h^{-1}_{ji}$.  So we can express
$\cV^{H,K}_{d-1}(h_{0},\cdots,h_{{d-1}}; h_{01},h_{10},\cdots)$ as
$\cV^{H,K}_{d-1}(h_{0},\cdots,h_{{d-1}}; v_{01},v_{02},v_{12},\cdots)$.  We can
{simplify} this further: The non-zero
$\cV^{H,K}_{d-1}(h_{0},\cdots,h_{{d-1}}; v_{01},v_{02},v_{12},\cdots)$ can be
expressed via $\cV^{H,K}_{d-1}(h_0,\cdots,h_{d-1}; v_{01}, v_{12},
\cdots,v_{d-2,d-1})$.  In other words, $v_{ij}$ on all the links of a
$(d-1)$-simplex can be determined from a subset $v_{01}, v_{12},
\cdots,v_{d-2,d-1}$.  

At this stage, we simply define ${\mathcal V}^{H,K}_{d-1}$ via eqn.~(\ref{defV}), but using the effective gauge fields $v_{ij}$ defined in 
eqn.~(\ref{flatCS}) to replace the hard gauge fields that were assumed previously.
The resulting model is manifestly gauge invariant, just as it was before.
However, hard gauging has now been replaced with soft gauging, making the model
\cgrn{completely local, both in the bulk and on the boundary.  In this case,
the global symmetry $G$ is on-site for the whole system (including bulk and boundary).
But if we integrate out the
gapped bulk, and consider only the effective boundary theory, we would like to ask if
the effective global symmetry $G$ on the boundary is on-site or not?  Since
this point is important, we elaborate on it in the next section.}

\subsubsection{The locality  and effective non-on-site symmetry for the
boundary theory }

\label{sec:cocycle-preserve-local}

\cgrn{We have shown that the model obtained by soft gauging is local both in
the bulk and on the boundary.
If we
integrate out the bulk degrees of freedom, we get an effective boundary theory,
whose action amplitude is given by a product of terms defined for each
boundary simplex.  The total boundary action amplitude  is invariant under the
$G$-symmetry transformation on the boundary, but each local term on a single
boundary simplex may not be.  This leads to a possibility that the effective
boundary $G$-symmetry is not on-site.  We have constructed two boundaries that
are local in Sec.~\ref{sec:1st-complx-bdry} and \ref{sec:extSymm}. The first
boundary in Sec.~\ref{sec:1st-complx-bdry} has a non-on-site effective
$G$-symmetry on the boundary, while the second boundary in Sec.~\ref{sec:extSymm} has an on-site effective $H$-symmetry on the boundary.}  

In the path integral, we only sum over gauge distinct configurations:  
\begin{align}
\label{Zgauge-1}
& Z =\sum_{\{g_i,[h_i,h_{ij}]\}} \prod_{(i_0\cdots i_d) \in D^d}
\nu_d^{s_{i_0\cdots i_d}}(g_{i_0},\cdots,g_{i_d}) \times
\\
&
\prod_{(i_0\cdots i_{d-1}) \in \prt D^d}
(\cV^{H,K}_{d-1})^{s_{i_0\cdots i_{d-1}}}(h_{i_0},\cdots,h_{i_{d-1}};
h_{i_0i_1},h_{i_1i_0},\cdots)
\nonumber 
\end{align}
where $[h_i,h_{ij}]$ represents the gauge equivalence classes.  

Such a lattice gauge theory with soft gauging will have an \emph{on-site} global symmetry $G$.
To see this, let us consider a transformation generated by $h\in H$ on site
$i$.  It is given by, if $i$ is on the boundary,
\begin{align}
(h_i,h_{ij})  \to (hh_i,hh_{ij})  
\end{align}
and, if $i$ is in the bulk,
\begin{align}
g_i  \to  {r}(h)g_i .
\end{align}
Such a transformation is a $G$ transformation in the bulk.  On the boundary,
since $(h_i,h_{ij})$ and $(vh_i,vh_{ij})$ are gauge equivalent for $v\in K$,
$h$ and $hv$ generate the same transformation.  So the transformation on the
boundary is given by the equivalence class $[h]$ under the equivalence relation
$h\sim hv$, $v\in K$.  Since $K$ is the normal subgroup of $H$, the  equivalence
classes form a group $H/K=G$. Thus, the transformation is also a $G$
transformation on the boundary.  Such a transformation is on-site, and is a
symmetry of the model since
\cgrn{each term in the action amplitude, such as
$\nu_d^{s_{i_0\cdots i_d}}(g_{i_0},\cdots,g_{i_d})$
and $(\cV^{H,K}_{d-1})^{s_{i_0\cdots i_{d-1}}}(h_{i_0},\cdots,h_{i_{d-1}};
h_{i_0i_1},h_{i_1i_0},\cdots)$, is invariant under the $G$-symmetry transformation: $\nu_d^{s_{i_0\cdots i_d}}(gg_{i_0},\cdots,gg_{i_d})=\nu_d^{s_{i_0\cdots i_d}}(g_{i_0},\cdots,g_{i_d})$ and
\begin{align}
\label{BsymmG}
&\ \ \
\cV^{H,K}_{d-1}(hh_{i_0},\cdots,hh_{i_{d-1}}; hh_{i_0i_1},hh_{i_1i_0},\cdots)
\nonumber \\
&=
\cV^{H,K}_{d-1}(hh_{i_0},\cdots,hh_{i_{d-1}}; hv_{i_0i_1}h^{-1},hv_{i_1i_2}h^{-1},\cdots) 
&&
\text{[used the definition \eqn{vhh}]} 
\nonumber \\
&=
\cV^{H,K}_{d-1}(h_{i_0},\cdots,h_{i_{d-1}}; v_{i_0i_1},v_{i_1i_2},\cdots)
&& 
\text{[used the definition \eqn{defV}]} 
\nonumber \\
&=
\cV^{H,K}_{d-1}(h_{i_0},\cdots,h_{i_{d-1}}; h_{i_0i_1},h_{i_1i_0},\cdots) \cred{.}
\end{align}
}

\cgrn{
To see if the effective boundary $G$-symmetry is on-site or not, we first note
that the term in the total action amplitude, $\prod_{(i_0\cdots i_{d-1}) \in
\prt D^d} (\cV^{H,K}_{d-1})^{s_{i_0\cdots i_{d-1}}}(h_{i_0},\cdots,h_{i_{d-1}};
h_{i_0i_1},h_{i_1i_0},\cdots)$, is purely a boundary term.  Each contribution
from a single boundary simplex is already invariant under the $G$-symmetry
transformation (see \eqn{BsymmG}).  So, such a term will not affect the
on-site-ness of the effective boundary symmetry, and we can ignore it in our
discussion. 
}

\cgrn{
The other term $\prod_{(i_0\cdots i_d) \in D^d}
\nu_d^{s_{i_0\cdots i_d}}(g_{i_0},\cdots,g_{i_d})$ may lead to
non-on-site effective boundary symmetry.
But the calculation is identical to that in Sec.~\ref{sec:non-on-site-G}.
We find that the resulting effective boundary $G$-symmetry
is indeed non-on-site if the $G$-cocycle
$\nu_d(g_{i_0},\cdots,g_{i_d})$ is not a coboundary.
}

So the partition function \eqn{Zgauge-1} gives us a boundary effective theory
that still has the $G$ symmetry, as well as a local Hilbert space.  (The
boundary does not break or extend the $G$-symmetry.) \cgrn{But the effective
$G$ symmetry on the boundary is non-on-site (\ie anomalous)}.

The dynamical properties of the soft gauging model in
Sec.~\ref{sec:cocycle-emergent} are the same as in the hard gauging case in
Sec.~\ref{sec:hard-gauge}, since the two path integrals are manifestly the
same. In particular, this is a physically satisfactory construction of a
symmetry-preserving gapped boundary of a bulk SPT phase with global symmetry
$G$.  The boundary is topologically ordered with emergent $K$-gauge symmetry.
The $K$-gauge theory is in a deconfined phase, which we discuss further in
Sec.~\ref{whichone}.  The boundaries of the CZX model discussed in Sec.~\ref{bdry4w} and Appendix \ref{bdry4l} are examples of this general construction.

\subsection{Gapped boundary  gauge theories:
$G$-symmetry preserving (2+1D boundary or above) or $G$-spontaneous symmetry breaking (1+1D boundary)
}\label{whichone}

To identify the boundary $K$-gauge theory, we look more closely at the boundary
factors in the path integral (\ref{actamp}).  To understand the boundary theory
in isolation, it is convenient to consider the case that all $g_i$ are equal to
1, which ensures that the boundary spins are $K$-valued.  The boundary theory
is now just a theory of $K$-valued variables with an action amplitude that is
given by the product over all boundary simplices of the generalized cochain
${\mathcal V}^{H,K}_{d-1}$ that was defined in eqn. (\ref{defV}).

\cgrn{
If we choose the spacetime to be a $d$-ball $D^d$, then the action amplitude in
\eqn{Zgauge-1} is always equal to one regardless the values of $\{g_i\}$ in the
bulk and $\{h_i,h_{ij}\}$'s on the boundary (that satisfy \eqn{flatCS}).  Thus
the system on a spacetime $D^d$ is in a gapped phase both in the bulk and on
the boundary.  Such a gapped state is the $K$-gauge deconfined state, since the
$K$-connections $v_{ij}=h_{ij}h^{-1}_{ji} \in K$ are always flat
thus $v_{ij}v_{jk}v_{ki}=1$.  
}

\cgrn{
Does such a $K$-gauge deconfined state spontaneously break the $G$-symmetry?
We note that, except the combinations $v_{ij}v_{jk}v_{ki}$ that are not
fluctuating, other combinations of $h_{ij}$'s are strongly fluctuating and
quantumly disordered.  Also $h_i$ and $g_i$ are strongly fluctuating and
quantumly disordered.  In fact, the model described by  \eqn{Zgauge-1} has a
\emph{local $G$ symmetry}\footnote{Here 
the \emph{local $G$
symmetry} does \emph{not} mean the gauge symmetry.
On one hand, the \emph{local $G$ symmetry} is that physically distinct
configurations [note that in the main text discussion, two distinct
configurations are $(g_i,h_i,h_{ij})$ and $(g'_i,h'_i,h_{ij}')$] have the same
action amplitude.  On the other hand, the gauge symmetry is 
not a (global) symmetry but indeed a gauge redundancy.  The gauge symmetry is a gauge
redundancy that two (redundant) configurations are indeed the same equivalent
physical configuration, and are related to each other through gauge
transformations.}: The  action amplitude for configuration $(g_i,h_i,h_{ij})$
is the same as the  action amplitude for configuration $(g'_i,h'_i,h_{ij}') =
(r(\t h_i)g_i,\t h_i h_i,\t h_i h_{ij} )$ where $\t h_i \in H$ generate the
local $G$-symmetry on gauge-invariant states.  This is because the action
amplitude is always equal to 1 regardless of the values of $h_i$, $g_i$ and
$h_{ij}$ on a spacetime $D^d$ (as long as $v_{ij}v_{jk}v_{ki}=1$ is satisfied).
This local $G$-symmetry allows us to show that any $G$-symmetry breaking order
parameter that can be expressed as a local function of $(g_i,h_i,h_{ij})$ will
have a short-range correlation.  }

\cgrn{
However, such a result is not enough for us to show all $G$-symmetry breaking
order parameters that are local operators to have short-range correlations.
This is because some local operators are not local functions of
$(g_i,h_i,h_{ij})$, such as the operator that corresponds to a breakdown of
the flat-connection condition $v_{ij}v_{jk}v_{ki}=1$.  On a 1+1D boundary, such
kinds of local operators can change the holonomy of the $K$-gauge field around
the space $S^1$ of the boundary.  As discussed in Sec.~\ref{bdry4w}, it is
the order parameter that changes the holonomy that acquires a long-range
correlation.  }

\cgrn{
Therefore, we need to find a more rigorous way to test the spontaneous
breaking of the $G$-symmetry.  One way to do so is to calculate the partition
function \eqn{Zgauge} on a spacetime $M^d$, which is given by the
number of configurations that satisfy that the flat-connection condition
$v_{ij}v_{jk}v_{ki}=1$ and the condition $v_{ij}\in K$.
When $K$ is Abelian, we find the partition function to be\footnote{
Here let us focus on the case that $K$ is Abelian (while $H$ and $G$ may be non-Abelian), for the simplicity of the formulas. 
One may generalize the situation to non-Abelian groups as well.} 
\begin{align}
\label{ZMd}
 Z(M^d)= 
\frac{|G|^{N^\text{Bulk}_v} |H|^{N^\text{Bdry}_v}}{|K|^{N^\text{Bdry}_v}} 
 |H|^{N^\text{Bdry}_l} 
\frac{|K|^{N^\text{Bdry}_v}}{|K|^{|\pi_0(\prt M^d)|}}
 |\text{Hom}[\pi_1(\prt M^d), K]|.
\end{align}
Let us explain the above result.  The $g_i$'s on the vertices in the bulk
contribute the factor $|G|^{N^\text{Bulk}_v}$ to the total configurations,
where $N^\text{Bulk}_v$ is the number of vertices in the bulk (not including
the boundary).  The $h_i$'s on the vertices on the boundary contribute the
factor $|H|^{N^\text{Bdry}_v}$ to the total configurations, where
$N^\text{Bdry}_v$ is the number of vertices on the boundary.  The
$(h_{ij},h_{ji})$ of the link on the boundary can be labeled by
$(h_{ij},v_{ij})$, where $h_{ij} \in H$ and $v_{ij}\in K$.  The $h_{ij}$'s
contribute the  factor $|H|^{N^\text{Bdry}_l}$, where $N^\text{Bdry}_l$ is the
number of links on the boundary. 
The $v_{ij}\in K$ needs to satisfy flat-connection condition $v_{ij}v_{jk}v_{ki}=1$, and the counting is
complicated.
When $K$ is Abelian, $v_{ij}$'s contributes to a factor
$\frac{|K|^{N^\text{Bdry}_v}}{|K|^{|\pi_0(\prt M^d)|}}$ which comes from
$v_{ij}$ of the form $v_{ij}=v_i v_j^{-1}$, $v_i,v_j \in K$.  But those are
only contributions from the ``pure gauge'' configurations. There is another
factor $ |\text{Hom}[\pi_1(\prt M^d), K]|$ which is the number of
inequivalent $K$-gauge flat connections on $\prt M^d$.  Last, we need to divide
out a factor $|K|^{N^\text{Bdry}_v}$ due to the $K$-gauge redundancy \eqn{kg}.
}

\cgrn{
The volume-independent partition function is given by
\begin{align}
  Z^\text{top}(M^d)= 
\frac{|\text{Hom}[\pi_1(\prt M^d), K]|}{|K|^{|\pi_0(\prt M^d)|}},
\end{align}
which is a topological invariant on spacetime with a vanishing Euler number \cite{KW1458}.  If
we choose $M^d=S^1\times D^{d-1}$, then $Z^\text{top}(S^1\times D^{d-1})$ will be
equal to the ground state degeneracy on $D^{d-1}$ space:
\begin{align}
 \text{GSD}(D^{d-1}) =
Z^\text{top}(S^1\times D^{d-1}) 
=\begin{cases}
 |K|, &\ \ \ \text{ if } d=3 \; \text{  (2+1D)};\\
 1, &\ \ \ \text{ if } d>3.
\end{cases}
\end{align}}

Our strategy here is to test the ground state degeneracy caused by spontaneous symmetry breaking, 
based on the degeneracy of a spatial sphere $S^{d-2}$ on the boundary of a spatial bulk $D^{d-1}$. 
Namely, we compute $ \text{GSD}(D^{d-1})=Z^\text{top}(S^1\times D^{d-1})$.
Our argument relies on

``\emph{No ground state degeneracy on a spatial boundary sphere $S^{d-2}$ means no spontaneous symmetry breaking}.''

Here we show that on a 1+1D spatial boundary $S^1$ of a 2+1D bulk, the GSD is $|K|$, and we cannot exclude the
possibility of spontaneous $G$-symmetry breaking.  
On a 2+1D spatial boundary $S^2$ of a 3+1D bulk, or any higher dimensions, the GSD is $1$, and there is no spontaneous
$G$-symmetry breaking.

We note our result here on the spontaneous symmetry breaking of 1+1D deconfined $K$-gauge theory
 is consistent with other independent checks from a Hamiltonian approach of Sec.~\ref{bdry3w} and Appendix \ref{sec:deconfined-to-SSB}, 
 and a field theory approach of Appendix ~\ref{sec:SSB_field-theory}.

As explained in Sec.~\ref{pure}, once all the variables are $K$-valued,
$\mu^H_{d-1}$ reduces to a cocycle $\mu^K_{d-1}$ appropriate for a $K$ gauge
theory.  As a result, the boundary factor in the path integral in \eqn{Zgauge} or \eqn{Zgauge-1}, 
when the $g_i$ are 1, is just the action amplitude of a
$K$-gauge theory deformed with the cocycle $\mu^K_{d-1}$, as in
Dijkgraaf-Witten theory.   This is the boundary state that has been
coupled to the bulk $G$-SPT phase to give a gapped symmetric boundary.

In general, not all variants of $K$ gauge theory can occur in this way, because
there may be some $\mu^K_{d-1}$ that do not come from any $\mu^H_{d-1}$.
Restriction from $H$ to $K$ gives a map $s: \cH^{d-1}(H,U(1))\to
\cH^{d-1}(K,U(1))$.   The versions of $K$-gauge theory that arise in our
construction are the ones associated to classes that are in the image of $s$.
In general, if a given version of $K$-gauge theory can arise by our
construction as the gapped boundary of a given $G$-SPT state, it can arise in
more than one way.  The number of ways that this can happen is the kernel of
$s$, which equals the number of classes in $\cH^{d-1}(H,U(1))$ that map to a
given class in $\cH^{d-1}(K,U(1))$.

\section{Find a group extension of $G$ that trivializes a $G$-cocycle} 
\label{sec:extension-trivializeGcocycle}

\begin{figure}[!h] 
\begin{center}
\includegraphics[scale=0.7]{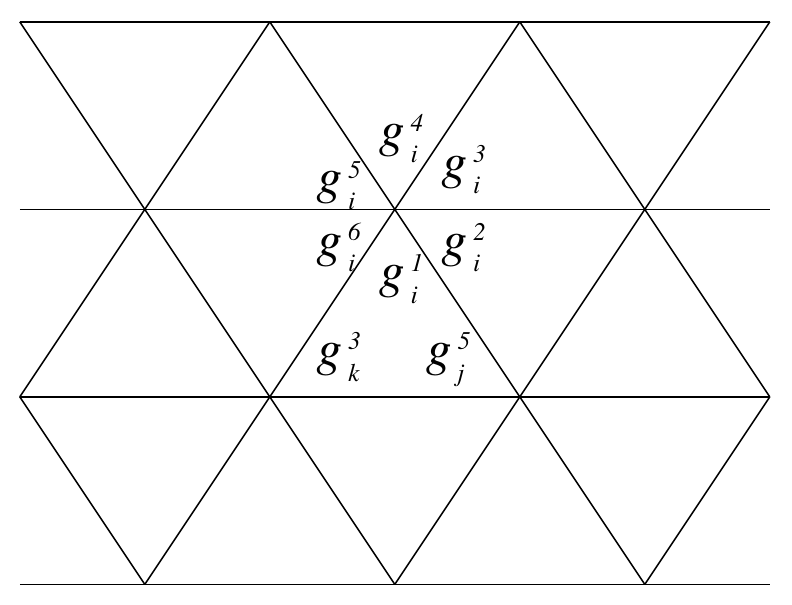}
\end{center}
\caption{
On the boundary, we can split $g_i$ on each vertex into several $g_i^1$,
$g_i^2$, $\cdots$, etc., one for each attached simplex.
}
\label{split}
\end{figure}

\subsection{Proof: Existence of a finite $K$-extension trivializing any finite $G$'s $d$-cocycle in $H$ for $d \geq 2$}
\label{sec:proof}

The construction in the last section gives a symmetric gapped boundary for the $G$-SPT
state associated to a
 $G$-cocycle $\nu_d \in
\cH^d(G,{U(1)})$, provided that we can find an extension of $G$,
\begin{align}
1 \rightarrow K \overset{}{\rightarrow}  H  \overset{{r}}{\rightarrow} G \rightarrow 1,
\end{align}
such that the $G$-cocycle $\nu_d$ becomes trivial when pulled back to an
$H$-cocycle by ${{r}}$.
In this section, we will give an explicit construction of such an extension
for any finite group $G$, and 
for any $G$-cocycle $\nu_d$ when $d \geq 2$. 
This approach works for $d$-cocycles with $d \geq 2$, thus the bulk dimension of $G$-SPT state has to be greater than or equal to $1+1$D.
Based on this method, 
below we show that a suitable group extension \emph{always exists}, 
thus we prove that within group cohomology construction,

{\bf Statement 1}: ``\emph{Any bosonic 
SPT state with a finite onsite symmetry group $G$, including both unitary and anti-unitary symmetry, can have an 
$H$-symmetry-extended (or $G$-symmetry-preserving) gapped boundary 
via a nontrivial group extension by a finite $K$, given the bulk spacetime dimension $d \geq 2$}.''

To motivate the construction, we start with the  non-on-site symmetry discussed in Sec.~\ref{rev}.
We can make the non-on-site symmetry to be on-site by splitting $g_i$ on each
vertex on the boundary into several variables $g_i^1$, $g_i^2$, $\cdots$, etc., one for
each attached simplex (see Fig.~\ref{split}).
{In the Euclidean signature,} we take the new evolution operator
\begin{align}
(\ee^{-\widehat H_\text{Bdry}})_{\{\t g_i^m, ...\}, \{g_i^m, ...\}}
\end{align}
to be non-zero only when  $g_i^1 = g_i^2 = g_i^3 = \cdots$ on each vertex.  In
other words, if the condition $g_i^1 = g_i^2 = g_i^3 = \cdots$ is not satisfied
on some vertices, then the configuration will correspond to high energy
boundary excitations on those vertices.

In the new boundary Hilbert space spanned by $\otimes_{i,m} |g_i^m\>$,
the symmetry transformation
\begin{align}
\widehat U(g)=\prod_{(ij\cdots k)} \widehat U_0(g) 
\nu^{s_{ij\cdots k}}_{d}( g_i^{m_i},  g_j^{m_j},\cdots, g_k^{m_k}, g^*,g^{-1}g^*)
\end{align}
becomes on-site (or on-cell, or on-simplex).
{
On each simplex, the  symmetry transformation $\widehat U(g)$ is given by
\begin{align}
&\ \ \ \
\widehat U(g)|g_i,g_j,\cdots ,g_k\>
\\
&=
\widehat U_0(g) 
\nu^{s_{ij\cdots k}}_{d}(g_i,  g_j,\cdots, g_k, g^*,g^{-1}g^*)
|g_i,g_j,\cdots ,g_k\>
\nonumber\\
&=
\nu^{s_{ij\cdots k}}_{d}(g_i,  g_j,\cdots, g_k, g^*,g^{-1}g^*)
|gg_i,gg_j,\cdots ,gg_k\>.
\nonumber 
\end{align}
}
Thus we can make any non-on-site symmetry on the boundary into an
on-site symmetry, by redefining the boundary sites.  This seems to contradict
our picture that the non-on-site symmetry on the boundary captures the bulk SPT
state, which should not be convertible into on-site boundary symmetry by any
boundary operations (that have the local site structure).

In fact, there is no contradiction since $\widehat U(g),\ g\in G$ may not generate
the group $G$.  They may generate a bigger group $H$ -- an extension of $G$ by
an \emph{Abelian} group $K$.  So after we split $g_i$ into $g_i^1$, $g_i^2$, etc. on
the boundary, the symmetry of our model is no longer $G$. It is changed into
$H$.  Since the symmetry transformation generated by $H$ is on-site, such a
symmetry transformation is not anomalous.  The bulk $G$-SPT state can also be
viewed as an $H$-SPT state. But as an $H$-SPT state, it is the trivial one,
since the the $H$-symmetry is on-site on the boundary.

So, we have found an extension of $G$, under
$
1 \rightarrow K \rightarrow  H  \overset{r}{\rightarrow} G \rightarrow 1
$,
where $K$ is an Abelian normal subgroup of $H$, such that
\begin{align}
 \nu^H_d(h_0,\cdots,h_d) \in \cH^d(H,U(1))
\end{align}
defined as
\begin{align}
 \nu^H_d(h_0,\cdots,h_d) = \nu_d(r(h_0),\cdots,r(h_d))
\end{align}
is trivial in $\cH^d(H,U(1))$.  We also note that $K$ is a local symmetry (on
each simplex) of the effective boundary Hamiltonian.

\begin{figure}[tb]
\begin{center}
\includegraphics[scale=0.6]{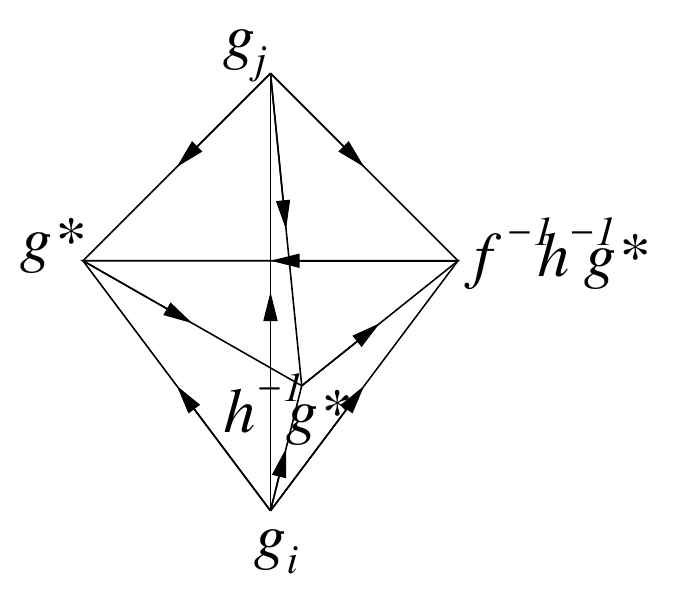}
\end{center}
\caption{
Visualization for guiding the calculation in eqn.~(\ref{eq:trivialGtoH-hf}), shown here as 
three symmetry transformations (say, $h$, $f$, and $(fh)^{-1}$) on a 1+1D boundary of a 2+1D bulk.
}
\label{grpH}
\end{figure}

To calculate $K$ from $\nu_{d}( g_i,  g_j,\cdots, g_k, g^*,g^{-1}g^*)$,
we consider three symmetry transformations $h$, $f$, and $(fh)^{-1}$.
We find that (see Fig.~\ref{grpH})
\begin{align} \label{eq:trivialGtoH-hf}
&\ \ \ \ \widehat U((fh)^{-1}) \widehat U(f) \widehat U(h)
\nonumber\\
& =
\nu_{d}( fhg_i,  fhg_j,\cdots, fhg_k, g^*,fh g^*)\times
\nu_{d}( hg_i,  hg_j,\cdots, hg_k, g^*,f^{-1}g^*)\times
\nu_{d}( g_i,  g_j,\cdots, g_k, g^*,h^{-1}g^*)
\nonumber\\
& =
\nu_{d}( g_i,  g_j,\cdots, g_k, h^{-1}f^{-1}g^*, g^*)\times
\nu_{d}( g_i,  g_j,\cdots, g_k, h^{-1}g^*,h^{-1}f^{-1}g^*)\times
\nu_{d}( g_i,  g_j,\cdots, g_k, g^*,h^{-1}g^*)
\nonumber\\
&\equiv \Phi_{h,f}(g_i,  g_j,\cdots, g_k).
\end{align}
The above phase factor $\Phi_{h,f}(g_i,  g_j,\cdots, g_k)$, as a function of
$g_i,  g_j,\cdots, g_k$,  is a generator of
the group $K$.  We can obtain all the generators by choosing different $h$ and
$f$, and in turn obtain the full group $K$.  
We note that, due to the geometry nature of Fig.~\ref{grpH} and its generalization in dimensions $d$, the above construction is true only for $d \geq 2$.

Thus, this concludes our proof of {\bf Statement 1}. We can rephrase it to the equivalent proved statements: 

{\bf Statement 2}: ``\emph{Any $G$-cocycle $\nu_d^G \in \cH^d(G,U(1))$ of a finite group $G$
(a bosonic SPT state with a finite onsite, unitary or anti-unitary symmetry, symmetry group $G$), can be
pulled back to a finite group $H$ via 
a certain group extension $1 \rightarrow K \rightarrow  H  \overset{r}{\rightarrow} G \rightarrow 1$ by a finite $K$, 
such that ${r}^* \nu_d^{G}= \nu^H_d =\delta  \mu^H_{d-1} \in \cH^d(H,U(1))$. Namely,
a $G$-cocycle becomes a $H$-coboundary, split to $H$-cochains $\mu^H_{d-1}$, 
given the dimension $d \geq 2$} (q.e.d).''

{\bf Statement 3}: ``\emph{Any $G$-anomaly in $(d-1)$D given by $\nu_d^G \in \cH^d(G,U(1))$ of a finite group $G$, 
can be pulled back to a finite group $H$ via 
a certain group extension $1 \rightarrow K \rightarrow  H  \overset{r}{\rightarrow} G \rightarrow 1$ by a finite $K$, 
such that $G$-anomaly becomes $H$-anomaly free, given the dimension $d \geq 2$} (q.e.d).''

Unfortunately, we do not have a systematic understanding of what $K$ will be generated by this construction.
In particular,   $K$ may
be different for cocycles $\nu_{d}$ that differ only by coboundaries.
Another drawback of this method is that we cannot obtain the exact analytic function of the split $H$-cochain easily.

However, we provide a different method that helps to derive the analytic $H$-cochain, based on the Lydon-Hochschild-Serre spectral sequence in Appendix \ref{sec:LHS}.
Readers can find more systematic examples in Appendix \ref{sec:examples}.  Finally, we remark that very recently Ref.~\cite{Tachikawa1712.09542} has proven statements
related to ours in a more mathematical setup.\footnote{After the appearance of our preprint on arXiv, one of the authors (J.W.) thanks 
Yuji Tachikawa for informing the recent
Ref.~\cite{Tachikawa1712.09542}'s Sec. 2.7 of  as a mathematical proof,  reproducing and  obtaining similar results as our Sec.~\ref{sec:proof}.}

\subsection{2+1/1+1D and $d+1$/$d$D Bosonic SPTs for an even $d$: 
The $d$D $Z_2^K$-gauge theory boundary of $d+1$D bulk invariant $(-1)^{\int (a_1)^{d+1}}$ via $0 \to {Z}_2 \to Z_4  \to Z_2 \to 0$} \label{sec:examples2+1DHZ4GZ2}

We would like to apply the above method to some cocycles that describes SPT states.
For example, we can consider a non trivial cocycle in $\nu_3 \in \cH^3(Z_2,U(1))$.
\begin{align}
 \nu_3(-,+,-,+)= \nu_3(+,-,+,-)=-1,\ \ \ \ \text{others }=1
\end{align}
where $Z_2=\{+,-\}$.
Choose $g^*=+$, $h=-$ and $f=-$, we find
\begin{align}
\Phi_{--}(g_i,  g_j)
& =
\nu_{3}( g_i,  g_j, -,+)
\nu_{3}( g_i,  g_j, +,-).
\end{align}
In fact, $ \Phi_{--}(g_i,  g_j) = \Phi_{-+}(g_i,  g_j) = \Phi_{+-}(g_i,  g_j)
$, and 
\begin{align}
 \Phi_{h,f}(-,+)= \Phi_{h,f}(+,-)=-1,\  \text{others }=1.
\end{align}
So $K=Z_2$ and $H=Z_4$.  The short exact sequence 
$0 \rightarrow Z_2 \rightarrow  Z_4  \rightarrow Z_2 \rightarrow 0$ trivializes the cocycle $\nu_3
\in \cH^3(Z_2,U(1))$.

See Appendix \ref{sec:examplesHZ4GZ2} for further illumination of this example.
In general, we find that  in any odd spacetime dimension, there is a $Z_2$-SPT phase
and that a gapped symmetric boundary for this phase can be obtained from the extension
$0 \to {Z}_2^K \to Z_4^H  \to Z_2^G \to 0$. See Appendix \ref{sec:examplesHZ4GZ2-any-dim}.
The bulk SPT phase is associated to the invariant
$\exp({\ii \pi \int a_1 \cup a_1 \cup \dots \cup a_1})\equiv \exp({\ii \pi \int (a_1)^{d+1}})$ with a cup product form of $a_1 \cup a_1 \cup \dots \cup a_1$, 
a nontrivial element in $\cH^{d+1}( Z_2,U(1))$ for an even $d$.
The $a_1$ here is a $\Z_2$-valued 1-{cocycle} in $\cH^1(M^{d+1}, \Z_2)$ on the spacetime complex $M^{d+1}$.

\subsection{3+1/2+1D and $d+1$/$d$D Bosonic topological superconductor with $Z_2^T$ time-reversal symmetry for an odd $d$:
The $d$D $Z_2^K$-gauge theory boundary of $d+1$D bulk invariant $(-1)^{\int (w_1)^{d+1}}$ via 
$0 \to {Z}_2 \to Z_4^T  \to Z_2^T \to 0$} \label{sec:examples3+1DHZ4GZ2T}

{
Next, we consider a non trivial cocycle $\nu_4 \in \cH^4(Z_2^T,U_T(1))=Z_2$ \cite{XieSPT4}. 
The $\nu_4$ represents a nontrivial class of bosonic SPTs with an anti-unitary $G=Z_2^T$ time-reversal symmetry. 
This SPTs is also named as bosonic topological superconductor or bosonic topological paramagnet with $G=Z_2^T$.
Here $Z_2$ and $Z_2^T$ are the same group mathematically. 
However, the generator in $Z_2^T$ provides a non-trivial action on the $G$-module $U(1)$, denoted as $U_T(1)$.
The subscript $T$ in the module $U_T(1)$ indicates that the group $Z_2^T$ has a non-trivial action on the module.

More generally when a group $G$ contains an anti-unitary operation such as time-reversal $Z_2^T$,
we define a nontrivial $G$-module $U(1)$ as $U_T(1)$.
We stress that $U(1)$ and $U_T(1)$ are the same Abelian group.
The group action is only non-trivial when
$g \cdot \nu = \nu^{s(g)}$, for $g \in G$, $\nu \in U_T(1)$, 
such that $s(g)=-1$ if $g$ contains an anti-unitary element, and
$s(g) = 1$ if $g$ contains no anti-unitary element.
The formalism developed in this paper up to this point is applicable to this case, for
models that fit in the group cohomology framework.

The group cocycle of this SPT phase is given by
\begin{align}
 \nu_4(-,+,-,+,-)= \nu_4(+,-,+,-,+)=-1,\  \text{others }=1,
\end{align}
where $Z_2^T=\{+,-\}$.  Choose $g^*=+$, $h=-$ and $f=-$, we find
\begin{align}
\Phi_{--}(g_i,g_j,g_k)
& =
\nu_{4}( g_i,g_j,g_k, -,+)
\nu_{4}( g_i,g_j,g_k, +,-).
\end{align}
and $ \Phi_{--}(g_i,g_j,g_k) = \Phi_{-+}(g_i,g_j,g_k) =
\Phi_{+-}(g_i,g_j,g_k)$. 
In fact, we obtain 
\begin{align}
 \Phi_{h,f}(-,+,-)= \Phi_{h,f}(+,-,+)=-1,\  \text{others }=1.
\end{align}
So $K=Z_2$ and $H=Z_4^T$.  The short exact sequence $0 \rightarrow Z_2
\rightarrow  Z_4^T  \rightarrow Z_2^T \rightarrow 0$ trivializes the cocycle $\nu_4
\in \cH^4(Z_2^T,U_T(1))$.
This means that $\nu_4$ becomes a coboundary in $\cH^4(Z_4^T,U_T(1))$ for a larger group $H=Z_4^T$.
Thus, we find that the 3+1D bosonic SPTs with $Z_2^T$ symmetry (the bosonic topological superconductor of $G=Z_2^T$)
has a 2+1D symmetry-preserving surface $Z_2$ topological order.} 

For the boundary $K$-gauge theory of a 
$G$-SPT state, the gauge charge excitations are labeled by Rep$(H)$=Rep($Z_4^T$) with $H/K=G=Z_4^T/Z_2=Z_2^T$,
instead of Rep$(K\times G)$=Rep$(Z_2\times Z_2^T)$.  $H$ is a ``twisted'' product of $K$ and $G$, the so-called projective symmetry group (PSG) introduced in Ref.~\cite{PSG}. When a gauge charged excitation is described by Rep$(H)$ instead of
Rep$(K\times G)$, it implies that the particle carries a fractional quantum
number of global symmetry $G$.  We say there is a fractionalization of the
symmetry $G$.

{We note that the $e_T m_T$ surface topological order first proposed in \cite{VS1306} on the surface of 3+1D $Z_2^T$-bosonic topological superconductor
is also a 2+1D deconfined $Z_2$ gauge theory.
}

See Appendix \ref{sec:examplesHZ4TGZ2T} for further illumination of this example.
In general, we find that the $0 \to {Z}_2^K \to Z_4^T  \to Z_2^T  \to 0$ construction can provide a boundary $d$D $Z_2^K$ gauge theory
on $d+1$D bosonic $Z_2^T$-SPTs, when $d$ is odd, see Appendix \ref{sec:examplesHZ4TGZ2T-any-dim}.
The bulk SPT invariant is equivalent to the partition function $\exp({\ii 2 \pi \int \frac{1}{2} w_1^{d+1}})$  for an odd $d$, a nontrivial element in $\cH^{d+1}( Z_2^T,U_T(1))=\Z_2$.
The $w_1$ here is a $\Z_2$-valued, the first Stiefel-Whitney (SW) class in $\cH^1(M^{d+1}, \Z_2)$ on the spacetime complex $M^{d+1}$.
Here $w_1= w_1(TM^{d+1})$ is the $w_1$ of a spacetime tangent bundle over $M^{d+1}$.
The $w_1\neq 0$ holds on a non-orientable manifold.

More examples of symmetry-extended gapped boundaries are provided in Appendix \ref{sec:examples}.

\section{Boundaries of SPT states with finite/continuous symmetry groups and beyond group cohomology}
\label{sec:beyondGC}

\cgrn{
In the above Sec.~\ref{sec:extension-trivializeGcocycle}, we  described a method that constructs exactly soluble boundary for any
within-group-cohomology SPT states with a finite symmetry group $G$, via a nontrivial group extension by a finite group $K$.  Those
boundaries preserve the $G$-symmetry and have topological orders if the
boundary dimension is 2+1D and higher.  Such a result can be generalized to SPT
states with a continuous compact symmetry group $G$, provided that the group
cocycle that describes the $G$-SPT state can be trivialized by a \emph{finite
extension} $1\to K\to H\to G\to 1$, namely, with a finite group $K$.  This is
because even for a continuous compact symmetry group $G$, the action amplitude
in \eqn{Zgauge} is still always equal to one regardless of the values of $\{g_i\}$
in the bulk and $\{h_i,h_{ij}\}$'s on the boundary. Thus  \eqn{ZMd} is still
valid if we treat $|H|$ and $|G|$ as the volumes of the continuous group $H$ and
$G$.  When $K$ is finite, the flat condition $v_{ij}v_{jk}v_{ki}=1$ makes the
$K$-gauge theory in a gapped deconfined phase.  Therefore, \emph{for both a finite group $G$ and
a continuous compact group $G$, a $d+1$D $G$-SPT state within group cohomology
can have a symmetry preserving gapped boundary if the $G$-group cocycle can be
trivialized by a finite extension of $G$ and when $d \geq 3$.} }

The SPT states within group cohomology have pure gauge $G$-anomalies  on the
boundary corresponding to the global symmetry group $G$.  More general SPT
states exist that have mixed gauge-gravitational anomalies on the boundary
\cite{Wen1410.8477}. Those SPT states are referred to as beyond-group-cohomology
SPT states \cite{VS1306}.  Those beyond-group-cohomology SPT states can be
constructed using group cohomology of $G\times SO(\infty)$.  More precisely,
using the action amplitude constructed from the group cocycle $\nu_{d+1} \in
\cH^{d+1}(G\times SO(\infty), U(1))$, we can construct models that realize the
beyond-group-cohomology SPT states (as well as within group-cohomology SPT
states) in $d+1$D \cite{Wen1410.8477}.  However, the correspondence between
$G\times SO(\infty)$-cocycle $\nu_{d+1}$ and a $d+1$D $G$-SPT state is not
one-to-one: Several different cocycles can correspond to the same SPT state.

\cgrn{
We note that \cite{Wen1410.8477}
\begin{align}
 \cH^{d+1}(G\times SO(\infty), U(1))=
\cH^{d+1}(SO(\infty), U(1)) \oplus \bigoplus_{k=1}^{d+1} 
\cH^{k}\big(G, \cH^{d+1-k}(SO(\infty), U(1)) \big).
\end{align}
The cocycles in the first term $\cH^{d+1}(SO(\infty), U(1))$ describe
invertible topological orders which do not need the symmetry group $G$.  The
cocycles in the second term $\bigoplus_{k=1}^{d+1} \cH^{k}\big(G,
\cH^{d+1-k}(SO(\infty), U(1)) \big)$ will describe $G$-SPT states in a
many-to-one fashion.
}

\cgrn{
When $G$ is finite, a cocycle in $\bigoplus_{k=1}^{d+1} \cH^{k}\big(G,
\cH^{d+1-k}(SO(\infty), U(1)) \big)$ can always be trivialized by an Abelian
extension $K$: $1\to K\to H\to G\to 1$.  This is because when
$\cH^{d+1-k}(SO(\infty), U(1))=\Z_N$, 
then the $\cH^{k}(G, \Z_N)$ can be viewed as a part of $\cH^{k}(G, U(1))$, 
and we can use the approach in Sec.~\ref{sec:extension-trivializeGcocycle} to show that the cocycles in $\cH^{k}(G,
\Z_N)$ can always be trivialized by a finite extension of $G$.  When
$\cH^{d+1-k}(SO(\infty),$ $U(1))=\Z$, we note that $\cH^{k}(G, \Z) \cong
\cH^{k-1}(G, U(1))$.  Using the  approach in Sec.~\ref{sec:extension-trivializeGcocycle}, we can show that the cocycles in
$\cH^{k-1}(G, U(1))$ can always be trivialized, which in turn allows us to show
that the cocycles in $\cH^{k}(G, \Z)$  can always be trivialized.
}

This allows us to conclude that the bosonic $d+1$D beyond-group-cohomology
$G$-SPT states described by $\bigoplus_{k=1}^{d+1} \cH^{k}\big(G,
\cH^{d+1-k}(SO(\infty), U(1)) \big)$ always have a symmetry preserving gapped
boundary when $G$ is finite and when the bulk space dimension $d\geq 3$.
Here $G$ can contain anti-unitary symmetries including time-reversal symmetry.




\section{Boundaries of bosonic/fermionic SPT states: Cobordism approach}
\label{sec:Cob}

In principle, the philosophy of our approach should also work for the cobordism group description of topological states.
For example, based on Ref.\cite{K1467},
one can consider bosonic SPTs in a $d+1$-dimensional spacetime with a finite internal onsite symmetry group $G$ via a cobordism theory.
Such an SPT state is proposed to be classified by 
\bea \label{eq:CobordismSO}
\Omega_{\text{tors}}^{d+1,SO}(BG,U(1)) \equiv \Omega^{d+1,{SO}}(BG,U(1))/ \text{im}(e_G)=\text{Hom}(\Omega_{d+1,{\text{tors}}}^{SO}(BG),U(1)),
\eea 
which is called the Pontryagin-dual of the torsion subgroup of the oriented bordism group $\Omega_{d+1}^{SO}(BG)$.
In the first equality of \eqn{eq:CobordismSO}, the $\Omega^{d+1,SO}(BG,U(1))$ is called the oriented cobordism group of 
$BG$ with $U(1)$ coefficient, it is defined as
$\Omega^{d+1,{SO}}(BG,U(1)) \equiv \text{Hom}(\Omega_{d+1}^{SO}(BG),U(1))$,
the space (here an Abelian group) of homomorphisms from $\Omega_{d+1}^{SO}(BG)$ to $U(1)$.
The $e_G$ is a map defined as $e_G: \text{Hom}(\Omega_{d+1}^{SO}(BG),\R) \to \text{Hom}(\Omega_{d+1}^{SO}(BG),U(1))$.
The image of the $e_G$ map is composed by elements of $\Omega^{d+1,{SO}}(BG,U(1))$ that vanish on 
the torsion subgroup of bordism group, $\Omega_{d+1,{\text{tors}}}^{SO}(BG)$.
Effectively, this yields the second equality, the $\Omega_{\text{tors}}^{d+1,SO}(BG,U(1))$
is equivalent to $\text{Hom}(\Omega_{d+1,{\text{tors}}}^{SO}(BG),U(1))$, namely
the space (here again an Abelian group) of homomorphisms from the torsion subgroup of bordism group $\Omega_{d+1,{\text{tors}}}^{SO}(BG)$ to $U(1)$.

To determine the \emph{symmetry-extended} gapped interface of a $G$-SPT state,
we need to find a larger total group $H$ that forms a group extension $1 \to {K} \to H \overset{r}{\rightarrow} G \to 1$ by a finite group $K$.
By pulling $G$ back to $H$,
we require that the nontrivial element in $\Omega_{\text{tors}}^{d+1,SO}(BG,U(1))$ specifying a $G$-SPT state,
becomes a trivial identity element in the Cobordism group
$\Omega_{\text{tors}}^{d+1,SO}(BH,U(1)) \equiv \Omega^{d+1,{SO}}(BH,U(1))/ \text{im}(e_H)=\text{Hom}(\Omega_{d+1,{\text{tors}}}^{SO}(BH),U(1))$,
where $e_H:$  $\text{Hom}(\Omega_{d+1}^{SO}(BH),\R)$ $\to$ $\text{Hom}$ $(\Omega_{d+1}^{SO}(BH),U(1))$.
In short,  the $G$-SPT state within Cobordism group $\Omega_{\text{tors}}^{d+1,SO}(BG,U(1))$ 
becomes a trivial $H$-SPT state (a trivial vacuum in $H$)
within Cobordism group $\Omega_{\text{tors}}^{d+1,SO}(BH,U(1))$.
The boundary of such a $G$-SPT state should allow a $G$-symmetry-preserving gapped interfaces with a deconfined 
topologically ordered $K$-gauge theory (where $K$ is a finite discrete group),
if the spacetime dimensions of bulk dimension $d+1 \geq 4$, above or equal to $3+1$D.

The above procedure is for bosonic SPT states including only fundamental bosons.
For fermionic SPT states including fundamental fermions, in principle, we can replace the oriented $SO$ in Cobordism groups
$\Omega^{d+1,SO}(BG,U(1))$ and $\Omega^{d+1,SO}(BH,U(1))$,
to the Spin version of Cobordism groups for the fermionic SPT states (namely $\Omega^{d+1,\text{Spin}}(BG,U(1))$ and $\Omega^{d+1,\text{Spin}}(BH,U(1))$),
and to the Pin$^{\pm}$ version of Cobordism groups for the  fermionic SPT states with time reversal symmetries 
(namely $\Omega^{d+1,\text{Pin}^{\pm}}(BG,U(1))$ and $\Omega^{d+1,\text{Pin}^{\pm}}(BH,U(1))$), where $T^2=(-1)^F$ for $\text{Pin}^+$ 
or $T^2=+1$  for $\text{Pin}^-$, respectively \cite{KTT1429}. The $F$ is the fermion-number parity. 
In this setup, our approach for symmetric gapped interfaces should be applicable to both bosonic and fermionic SPT states.
The underlying idea again is related to the fact that a certain global anomaly associated to $G$ on the boundary of $G$-SPT states
becomes anomaly-free in a larger group $H$.

It will be interesting to find more concrete examples and figure out the explicit analytic (exactly soluble or not) lattice Hamiltonian construction for such 
symmetry-preserving gapped boundaries within the cobordism setup
in the future.


\section{Generic gapped boundaries/interfaces: 
Mixed symmetry breaking, symmetry extension and dynamically gauging} \label{sec:sym-enhanced:bdryDW}

In this section, we will give an overview of how the symmetry extension construction we have described is related to what may be more familiar gapped boundary states.
We will also describe the generalizations of the ideas to interfaces between SPT states, and to the case that the bulk phase has intrinsic topological order.
We will further develop their path integrals, lattice Hamiltonians and wave functions suitable for many-body quantum systems in Sec.~ \ref{sec:general}.

\subsection{Relation to Symmetry Breaking}\label{relsym}

The most familiar type of gapped boundary state for a $G$-SPT phase is obtained by explicitly or spontaneously breaking the $G$ symmetry on the boundary
to a subgroup $H$ of $G$.  Here $H$ must have the property that the cocycle defining the $G$-SPT phase becomes a coboundary when the variables are restricted
from $G$ to $H$. \cgrn{For the notational distinction, we call this unbroken subgroup $H$ of $G$ as $H=G'$.}

From the point of view of this paper, the statement that \cgrn{$G'$} is a subgroup of $G$ means that there is an injective homomorphism 
\cgrn{$\iota: G' \to G$}.  
A gapped boundary
state can be constructed if the given cohomology class in $\nu_d^G\in \cH^d(G,U(1))$ is trivial when pulled back to \cgrn{$G'$}.
See Appendix \ref{sec:examples-gauge-sym-break} for explicit examples.

\subsection{{Symmetry Extension and Mixed Symmetry Breaking/Extension}}\label{mixed}

Our construction on the symmetry extension
in this paper is instead based on a surjective, rather than injective, homomorphism $r:H\to G$.  Because $r$ is surjective, the symmetry is extended
(from $G$ to $H$) along the boundary, rather than being broken.  By gauging $K=H/G$, one can arrange so that the global symmetry of the full system is $G$.
\cgrn{Many examples of symmetry-extended gapped boundaries are shown in Appendix \ref{sec:examples}.}

It is straightforward to combine the two cases.  We can construct a gapped  boundary state associated to any homomorphism $\cgrn{\varphi}:H\to G$, such that the cohomology
class in $\cH^d(G,U(1))$ becomes trivial when pulled back to $H$.   The construction proceeds exactly as we have explained in earlier sections of this paper,
without any substantial modification.  In this boundary state, $G$ is spontaneously or explicitly broken to the subgroup $G'=\cgrn{\varphi}(H)$, and then $G'$ is extended to $H$.

\cgrn{
More explicitly, one could also imagine arranging the above procedure in a two-stage process.   
Assume that in a layer within a distance $\ell$ from the boundary, 
$G$ is spontaneously broken down to $G'$. 
Then near the boundary the global/gauge symmetry is only $G'$ and the boundary
condition is defined by the choice of a group $H$, with a surjective map $r$ to $r(H)=G'$, such that the cocycle of $G'$ becomes trivial by lifting to $H$: 
via $1 \rightarrow K' \overset{}{\rightarrow}  H  \overset{{r}}{\rightarrow} G' \rightarrow 1$.
In other words, to construct a boundary condition in a mixed symmetry breaking/extension case, what we need is that the cocycle of $G$ that defines the bulk topological state, 
when restricted to $G'$ and then pulled back to $H$, becomes trivial.}


In all of these cases, one has to actually pick a trivialization of the  pullback of $\nu_d^G$ to $H$.  
The possible choices differed by a class in $\cH^{d-1}(H,U(1))$ correspond
to an $H$-topological state on the boundary. This corresponds roughly to appending an $H$-topological state on the boundary.

\subsection{Gapped Interfaces}\label{interfaces}

One can similarly consider the case of an interface (i.e. domain wall) between two SPT phases.  In general, we may have one symmetry group $G_{\I}$ on one side
of the interface, with a cohomology class $\nu_{\I}$, and a second symmetry group $G_{\II}$ on the other side, with its own cohomology class $\nu_{\II}$.
(The gapped boundary of $G$-topological state can be regarded as a gapped interface 
between a $G$-topological state and a trivial vacuum.) 
We shall describe gapped interfaces between these two states. 

Interfaces can be reduced to boundary states by a well-known folding trick.  Instead of saying that there is $G_{\I}$ on one side and $G_{\II}$ on the other
side, one ``folds'' along the interface and considers a system with a combined symmetry group $G=G_{\I}\times G_{\II}$, and a cohomology class $\nu_{\I}\times \nu_{\II}^{-1}$.
(Folding inverts one of the two cohomology classes.)   Then we can construct gapped interfaces associated as above to any homomorphism $\cgrn{\varphi}:H\to G_{\I}\times G_{\II}$.

An interesting special case is that the same group $G$ is supposed to be unbroken on both sides and also along the interface.  This means that $G_{\I}=G_{\II}=G$,
and that the unbroken subgroup $\cgrn{\varphi}(H)$ is a diagonal subgroup $G'$ of $G_{\I}\times G_{\II}$.    The cohomology class $\nu_{\I}\times \nu_{\II}^{-1}$ of $G_{\I}\times G_{\II}=G\times G$
restricts to a class of $G'$ that we can denote by the same name.  $H$ can be any finite extension of $G'\cong G$ that trivializes this class.

\subsection{Intrinsic Topological Order}\label{ito}

Though our emphasis in this paper has been on gapped boundary states for SPT phases, a similar construction applies to bulk phases with intrinsic
topological order.

We can construct such a phase simply by gauging the $G$ symmetry of a given $G$-SPT state.  Then since $G$ is extended to $H$ along the boundary,
for consistency we have to gauge the full $H$ symmetry along the boundary.  All our formulas make sense in that context.  

SET phases can be treated in a similar way.  For this, we gauge a subgroup $G_0$ of $G$.  The most significant case is that $G_0=N$ is a normal subgroup
of $G$.   Then gauging $N$ gives a state with intrinsic topological order of an $N$-gauge theory, 
in which $Q=G/N$ is a quotient group of global symmetries.  Along the boundary, we have to gauge
the inverse image of $N$ in $H$.  If the map $\cgrn{\varphi}:H\to G$ is surjective, then the $Q$ symmetry remains as a symmetry of the boundary state and is extended
along the boundary to the inverse image of $Q$ in $H$.  For details see again Sec.~\ref{sec:general}.  It is again possible to consider more general cases in which the $Q$
symmetry may be partly broken along the boundary and partly extended.

There is no essential loss of generality in assuming here that $G_0$ is a normal subgroup $N$ of $G$, for the following reason.  If $G_0$ is not normal,
then gauging $G_0$ will explicitly break $G$ to a subgroup $G^*$, the normalizer of $G_0$ in $G$.  Then $G_0$ is normal in $G^*$.  After replacing
$G$ by $G^*$, everything proceeds as before.

We provide other details of path integral/Hamiltonian models in Sec.~\ref{sec:general}.
Many examples of dynamically gauging gapped boundaries/interfaces are provided in Appendix \ref{sec:gauged-DW-interface}.

%
%

\section{General construction of exactly soluble lattice path integral and Hamiltonian of gapped boundaries/interfaces for topological phases 
in any dimension} \label{sec:general}

We consider the spacetime-lattice path integral formulation in Sec.~\ref{sec:generalPathintegral} and the spatial lattice Hamiltonian formulation in Sec.~\ref{sec:generalHamiltonian}
for a systematic construction of gapped boundaries/interfaces for topological phases in any dimension.
See Table \ref{table:Tablelattice} for an example of the lattice formed by simplices on a space or spacetime complex.

\subsection{Path integral} \label{sec:generalPathintegral}

In the following subsections, we systematically construct the path integral $Z$ defined for various topological phases 
(including SPT, gauge theory, SET, gapped boundary/interfaces, etc) and contrast their properties.
We shall clarify the gauge equivalent configuration briefly mentioned in \Eqn{Zgauge}, and the precise mod-out factor to remove the symmetry/gauge redundancy. 
In Sec.~\ref{sec:cocycle-emergent}, we showed the construction of cocycle 
$(\cV^{H,K}_{d-1})^{s_{i_0\cdots i_{d-1}}}(h_{i_0},\cdots,h_{i_{d-1}};
h_{i_0i_1},h_{i_1i_0},\cdots)$
 that contains the emergent gauge fields.
We call this type of gauge field is \emph{soft gauged}, which means that the Hilbert space of the gauge theory is still a tensor product form defined on each local site.
$\mathscr{H}_\text{tot} = \otimes_i \mathscr{H}_i$, because the $h_i,h_{ij},h_{il}$ are variables assigned 
to the site $i$ (see Fig. \ref{bdryhk}).
Below we discuss the \emph{hard gauged} theory,
where the total Hilbert space $\mathscr{H}_\text{tot} \neq \otimes_i \mathscr{H}_i$
is not a tensor product form of Hilbert spaces $ \mathscr{H}_i$ on each local site $i$ since we require additional link variables.

We should note that we can easily formulate a soft gauge theory from a hard gauge theory, based on Sec.~\ref{sec:cocycle-emergent}.
One reason to consider the hard gauge theory in the following Secs.~\ref{sec:Zgaugeclosed} and \ref{sec:ZSETclosed}
is for the simplicity of notation and calculation, and for its smaller Hilbert space.

\begin{figure}[h!] 
\begin{center}
\includegraphics[scale=0.335]{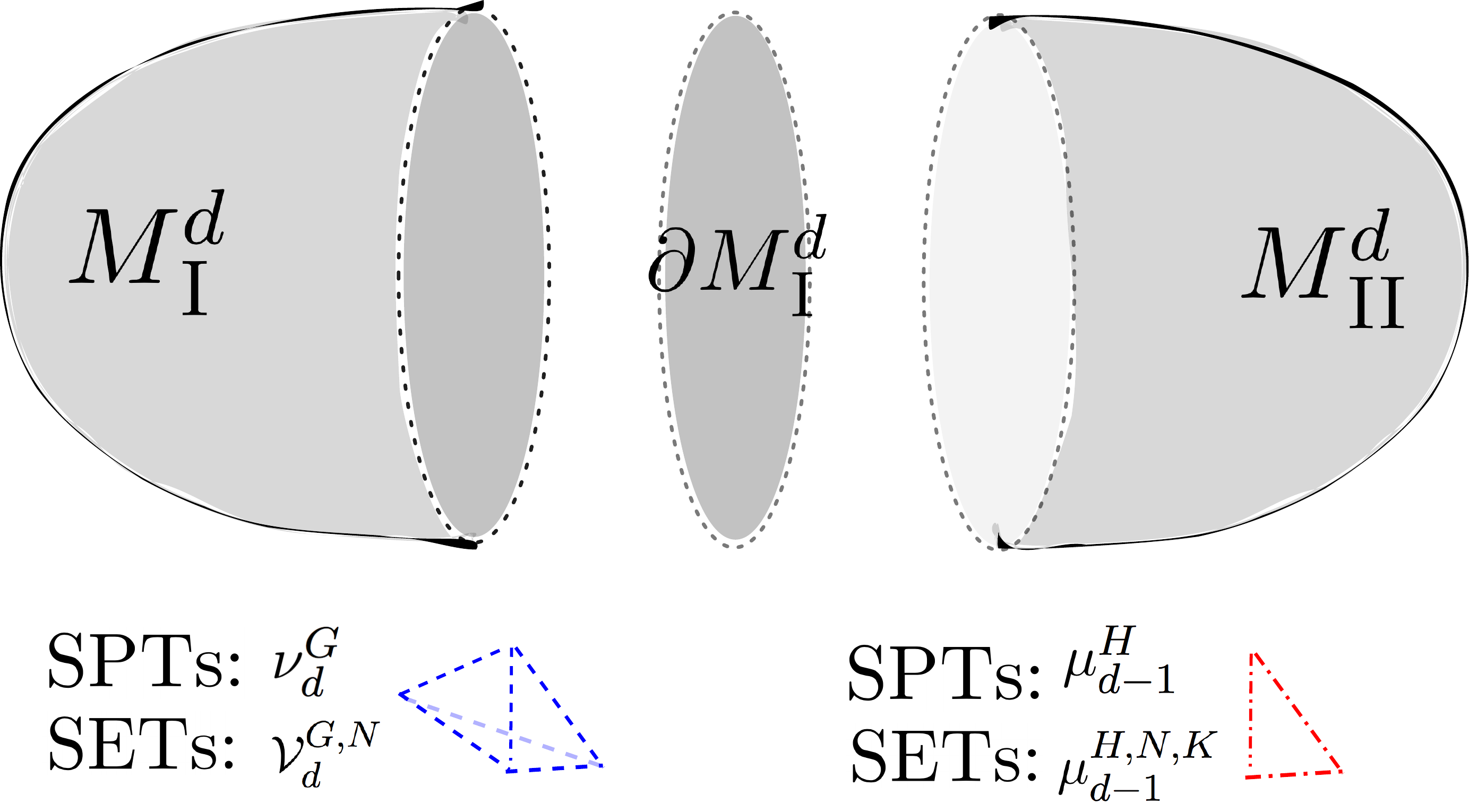} \\
\end{center}
\caption{In Sec.~\ref{sec:generalPathintegral}, we define
a lattice path integral on a $d$-dimensional spacetime manifold by triangulating the manifold to $d$-simplices. 
If the spacetime is closed, as in Sec.~\ref{sec:ZSPTclosed}, \ref{sec:Zgaugeclosed} and \ref{sec:ZSETclosed},
we assign $d$-simplices with cocycles $\nu^{G}_{d}$ for SPTs or with $\cV^{G, N}_{d}$ for SETs.
In this figure, 
the spacetime $M^d$ is obtained as the gluing of two manifolds $M_{\I}^d \cup M_{\II}^d$ with a common boundary $\partial M_{\I}^d$.
For simplicity, we draw the $d=3$ case. One example of the $M^3=S^3$ is a 3-sphere,
then we can choose $M_{\I}^3=D^3$ and $M_{\II}^3=D^3$, where the gapped spacetime boundary is on a 2-sphere $\partial M_{\I}^3=S^2$.
We would like to define the path integral on an open manifold $M_{\I}^d$ with a gapped boundary $\partial M_{\I}^d$,
where details are discussed in Sec.~\ref{sec:ZSPTSETG}.
In our construction, we  assign lower dimensional split cochains $\mu^{H}_{d-1}$ (or $\cV^{H,K}_{d-1}$) for SPTs and $\mu^{H, N, K}_{d-1}$ for SETs
to $(d-1)$-simplices paved onto a gapped boundary $\partial M_{\I}^d$.
}
\label{fig:lattice-path-integral-1}
\end{figure}

\begin{figure}[h!] 
\begin{center}
(1)\includegraphics[scale=0.3]{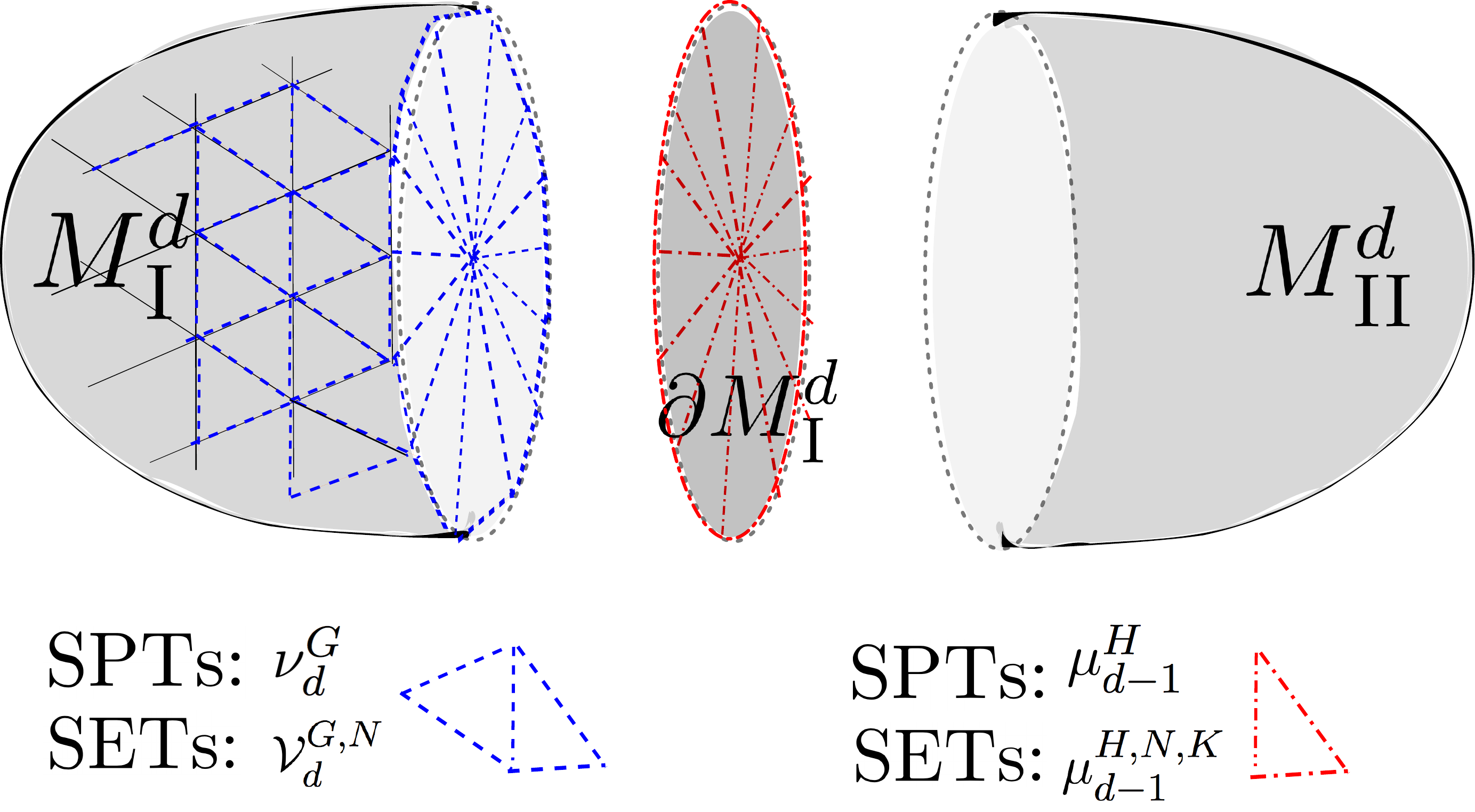}\;\;\;\;\;\;
(2)\includegraphics[scale=0.3]{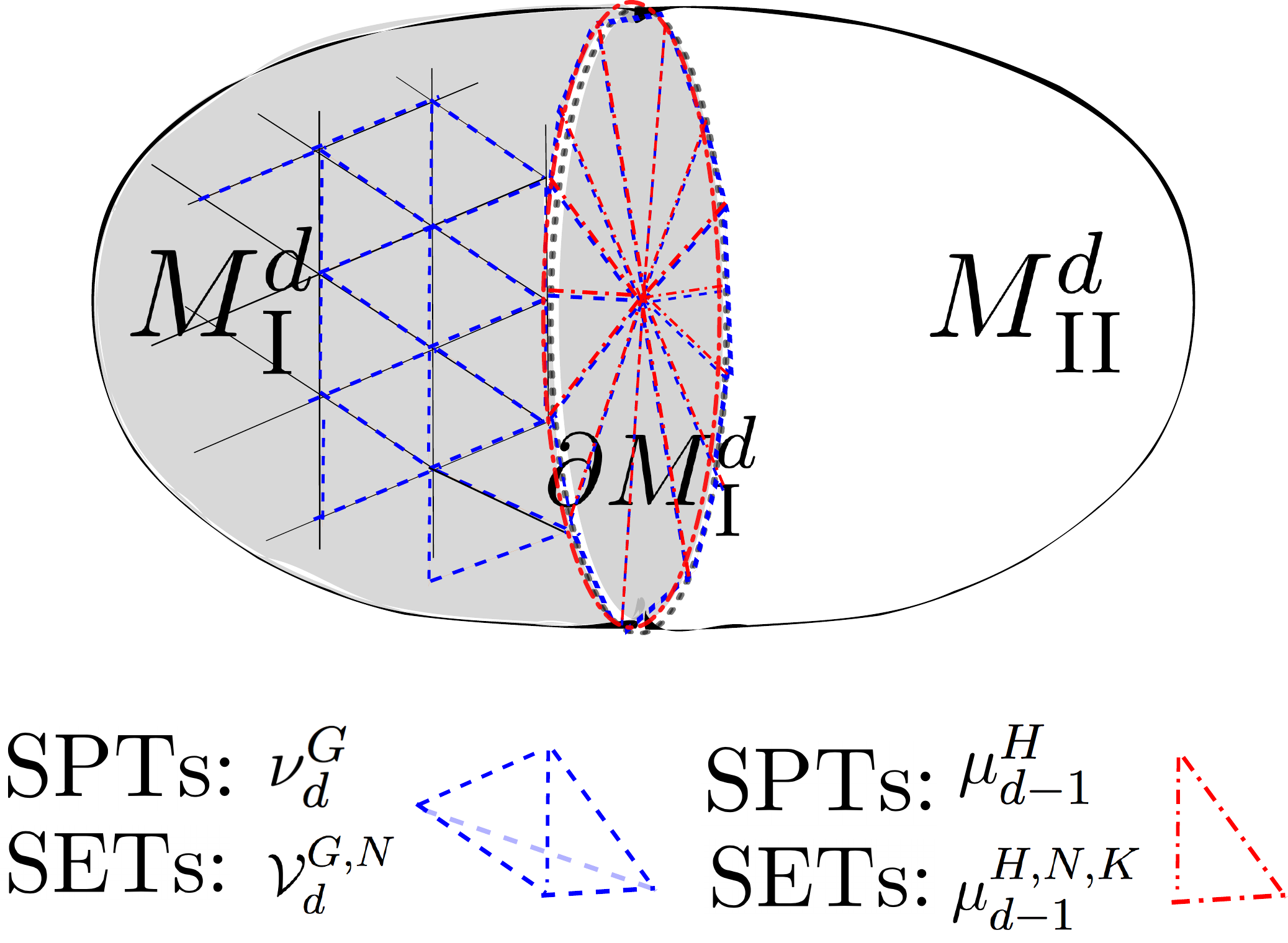}
\end{center}
\caption{Follow Fig.\ref{fig:lattice-path-integral-1}, 
the fig.(1) shows the filling of $d$-cocycles into the gapped bulk in $M_{\I}^d$, 
and the filling of $(d-1)$-cochains onto a gapped boundary $\partial M_{\I}^d$.
The combined result contributes to the topological amplitude shown in fig.(2). Then we need to sum over all the allowed group element configurations onto
each vertex/link (the so-called ``sum over all the colorings'') to obtain the path integral $Z$.
The explicit formula is derived in Sec.~\ref{sec:ZSPTSETG}.}
\label{fig:lattice-path-integral-2}
\end{figure}

Schematically
Figs.\ref{fig:lattice-path-integral-1} and Fig.\ref{fig:lattice-path-integral-2} summarize how to define an exactly soluble partition function or path integral on a triangulated spacetime complex.
Normally, a path integral of gapped topological phase is well-defined on a closed spacetime manifold.
However, here in particular, some path integral of fully gapped topological phase
is also well-defined in the gapped bulk on $M_{\I}^d$ with a gapped interface $\partial M_{\I}^d$.

\subsubsection{SPTs on a closed manifold} \label{sec:ZSPTclosed}

We start from reviewing and strengthening the understanding of SPT path integral defined by homogeneous $d$-cocycles $\nu_d(g_{i_0},\cdots,g_{i_d})$ 
of a cohomology group $\cH^d(G,U(1))$
for a global symmetry group $G$ \cite{XieSPT4} on a closed manifold, 
\begin{align}
\label{ZSPT}
& Z ={\frac{1}{|G|^{N_{v,\text{Bulk}}}} } \sum_{\{g_i\}} \prod_{(i_0\cdots i_d) \in M^d}
\nu_d^{s_{i_0\cdots i_d}}(g_{i_0},\cdots,g_{i_d}).  
\end{align}
We first assign the ordering of vertices as the \emph{branching structure} ,
then we assign a group element for each vertex as \emph{coloring}.
The sum over all possible \emph{colorings}, by summing over all assignments of group elements, is done by $\sum_{\{g_i\}}$.
On any closed manifold $M^d$, say with a number of vertices ${N_{v,\text{Bulk}}}$, we can prove that the amplitude $\prod_{(i_0\cdots i_d) \in M^d} \nu_d^{s_{i_0\cdots i_d}}(g_{i_0},\cdots,g_{i_d}) =1$ for any choice of ${\{g_i\}}$. Here is the proof: 
First, recall that the cocycle condition imposes that the cocycle 
$\prod \nu_d^{s_{i_0\cdots i_d}}(g_{i_0},\cdots,g_{i_d}) =1$ on any closed sphere $S^d$.
Second, we can simply connect every vertex $g_{j}$ on $M^d$ to an additional new point assigned with $g_0$ through a new edge $\overline{0 {j}}$, and we can view the 
amplitude as  
\begin{align}
& \prod_{(i_0\cdots i_d) \in M^d} \nu_d^{s_{i_0\cdots i_d}}(g_{i_0},\cdots,g_{i_d}) =
\prod_{(i_0\cdots i_d) \in M^d} \nu_d^{s_{i_0\cdots i_d}}(g_{i_0},\cdots,g_{i_d})  
\prod_{(j_0\cdots j_{d-1}) \in M^d} \nu_d^{s_{j_0\cdots j_{d-1}, 0}}(g_{j_0},\cdots,g_{j_{d-1}}, g_0) \nonumber \\
&\;\;\;\;\;\; =
\prod_{(i_0\cdots i_d) \in M^d} 
\nu_d^{s_{i_0\cdots \widehat{i} \cdots i_d 0}}(g_{i_0},\cdots, \widehat{g_i}, \cdots  ,g_{i_d} , g_0)  \nonumber \\
&\;\;\;\;\;\; =
\prod_{(i_0\cdots i_d) \in M^d} 
(\delta \nu_d^{s_{i_0\cdots {i} \cdots i_d 0}}(g_{i_0},\cdots, {g_i}, \cdots  ,g_{i_d} , g_0)  ) 
=
\prod_{(i_0\cdots i_d) \in M^d}  1=1.
\end{align}
The first equality computes the amplitude from all vertices on $M^d$ and $g_0$,
We use the fact that there are two terms under the same form 
$\nu_d(g_{j_0},\cdots,g_{j_{d-1}}, g_0)$ overlapping the same $d$-simplex
with opposite orientations that cancel out.
The second equality takes the product of each $d$-simplex where $\widehat{g_i}$ is a removed entry, 
where $i$ ranges from $\{i_0,\cdots, i_d, 0\}$. 
Moreover, the vertices $\{i_0,\cdots, i_d, 0\}$ and their connected edges also form a 
$d+1$-simplex. There are $d+1$ number of $d$-cocycles $\nu_d$ assigned to $d$-simplices paving on the surface of
the $d+1$-simplex. Effectively, the surface $d+1$-simplex is a closed $S^{d+1}$ sphere,
and the amplitude on $S^{d+1}$ yields a $d$-cocycle condition 
$(\delta \nu_d^{s_{i_0\cdots {i} \cdots i_d 0}}(g_{i_0},\cdots, {g_i}, \cdots  ,g_{i_d} , g_0)  )=1$ in the third line.
In \Eqn{ZSPT}, the product of amplitudes is 1, and the summation $\sum_{\{g_i\}}$ yields a factor ${|G|^{N_{v,\text{Bulk}}}}$ exactly canceling with the 
mod-out factor.
We thus show that $Z=1$ on any closed manifold for SPT defined by homogeneous cocycles.

\emph{Global symmetry}: We note that the global symmetry of SPT also manifests in the path integral.
We first define the global symmetry transformation $\mathbf{g} \in G$ 
of SPT as sending each group element $g_{i} \to \mathbf{g} g_{i}$ on every vertex $i$.
Through  the homogeneous cocycle condition
$\mathbf{g} \cdot \nu_d( g_{i_0},\cdots,  g_{i_d})=\nu_d(\mathbf{g} \cdot g_{i_0},\cdots, \mathbf{g} \cdot g_{i_d})=\nu_d(  g_{i_0},\cdots,g_{i_d})$ \cite{XieSPT4};
thus, $Z$ is invariant under the global symmetry transformation.

\subsubsection{Gauge theory with topological order on a closed manifold}  \label{sec:Zgaugeclosed} 
The gauge theory of a gauge group $K$ in this subsection is topological gauge theory \cite{DW9093}, suitable for certain topological orders.
The path integral defined by inhomogeneous $d-$cocycles
$\omega_d({k_{i_0 i_{1}}},\cdots,{k_{i_{d-1} i_{d}}}) \in \cH^d(K,U(1))$ is 
\begin{align}
\label{ZTgauge}
& Z ={\frac{1}{|K|^{N_{v,\text{Bulk}}}} } \sum_{\{k_{i_j i_{j+1}}\}} \prod_{(i_0\cdots i_d) \in M^d}
\omega_d^{s_{i_0\cdots i_d}}({k_{i_0 i_{1}}},\cdots,{k_{i_{d-1} i_{d}}}).  
\end{align}
on any closed manifold $M$. Each triangle (more generally any contractible 2-face or 2-plaquette) must satisfy 
${k_{12}} {k_{23}} {k_{31}}=1$ as a trivial element in $K$, which means a zero flux through a 2-surface.

We note that the gauge theory $Z$ is \emph{not} equal to 1 in general. The reason is that
on a manifold with non-contractible cycles such as $S^1$ circles, the inhomogeneous cocycles allow distinct gauge group elements winding through each cycle (that
does not occur in homogeneous cocycles).
This fact also reflects in nontrivial holonomies along non-contractible cycles for gauge field theory.
However, we can show that $Z=1$ on $S^{d-1} \times S^1$.
By considering the minimum triangulation that $S^{d-1}$ is the surface of a $d$-simplex, and another
$S^1$ connects each point back to itself. Each cocycle amplitude turns out to be $1$, but the $\sum_{\{k\}}$
sums over group elements. 
The minimum triangulation of $S^{d-1} \times S^1$ has $N_v=d+1$ vertices and $N_e=d+1$ independent edge variables,
thus $Z={|K|^{N_e}}/{|K|^{N_v}}=1$ on $S^{d-1} \times S^1$.

\emph{Gauge symmetry}: We note that the gauge symmetry also manifests in the path integral.
We first define the local gauge-symmetry transformation  
$
\mathbf{k}  \in K
$
on a particular site $j$
sending each group element on all the neighbor links through
$$
k_{i_j i_{j+1}}  \to (\mathbf k)^{-1} k_{i_j i_{j+1}}, \;\;\;\;\; k_{i_{j-1} i_{j}}  \to  k_{i_{j-1} i_{j}} (\mathbf k).
$$
Effectively what we do is equivalent to a Pachner move shifting the vertex ${i_j}$ to a new vertex ${i_{j'}}$ with a new triangulation near this vertex,
and we assign the link $\overline{{i_j}{i_{j'}}}$ with a gauge transformation variable
$\mathbf k = k_{i_j {i_{j'}}}  \in K$.
We can focus on a local gauge transformation on a single site ${i_j}$, one can easily generalize to apply
gauge transformations on every site.
To prove that the $Z$ is gauge invariant, 
we show that
$\prod_{(i_0\cdots i_d) \in M^d}
\omega_d^{s_{i_0\cdots i_d}}({k_{i_0 i_{1}}},\cdots,{k_{i_{d-1} i_{d}}})$ is gauge invariant.
The ratio of amplitudes before and after gauge transformations is:
\begin{align} \label{eq:trick-gauge-inv}
& \frac{
\prod_{(i_0\cdots i_d) \in M^d}
\omega_d^{s_{i_0\cdots i_d}}({k_{i_0 i_{1}}},\cdots, k_{i_{j-1} i_{j}}, k_{i_j i_{j+1}},  \cdots,{k_{i_{d-1} i_{d}}})
}{
\prod_{(i_0\cdots i_d) \in M^d}
\omega_d^{s_{i_0\cdots i_d}}({k_{i_0 i_{1}}},\cdots,  k_{i_{j-1} i_{j}} (\mathbf k), (\mathbf k)^{-1} k_{i_j i_{j+1}},  \cdots,{k_{i_{d-1} i_{d}}})
}\nonumber \\
&\;\;\;\;\;\; =
\prod_{(\cdots i_{j} i_{j'} \cdots) \in S^{d}}
\omega_d^{s_{i_{\cdots}}} (\cdots)
=(\delta \omega)_{d+1}=1.
\end{align}
In the first equality, we 
find that amplitudes around the vertex $i_{j}$ and $i_{j'}$ are left over that cannot be directly canceled.
There are two local patches centered around $i_{j}$ and $i_{j'}$ as two $d$-dimensional disks $D^d$ and $D^d$. 
The two disks share the same boundary and can be glued to a sphere $S^d$.
Thus we can apply the $d$-cocycle condition $\delta \omega_d=1$ that the amplitude on $S^d$ is 1, to prove that each amplitude in $Z$ is invariant.  
 Local gauge transformation can be applied on every site, and the $Z$ is still gauge invariant by the same proof above. 

\subsubsection{SETs on a closed manifold via $1 \to N \to G \to Q \to 1$,\\ 
and a relation between SPTs and topologically ordered gauge theory}  \label{sec:ZSETclosed}

Consider an anomaly-free SET path integral on a closed manifold under $1 \to N \to G \to Q \to 1$ \cite{W0213, HW1267}.
Here $G$ is a total symmetry group named a projective symmetry group (PSG), 
$N$ is a normal subgroup that can be dynamically gauged, and $Q$ is a quotient group of the remaining global symmetry \cite{W0213}.
We can regard the anomaly-free SET (well-defined in its own dimensions) as gauging the $N$ normal subgroup in $G$-SPT in Sec.~\ref{sec:ZSPTclosed}.
\begin{align}
\label{ZgaugeSETb}
& Z ={\frac{1}{|G|^{N_{v,\text{Bulk}}}} \frac{1}{|N|^{N_{v,\text{Bulk}}} } } \sum_{\{g_i, n_{ij}\}} 
\prod_{(i_0\cdots i_{d}) \in \prt M^d}
(\cV^{G,N}_{d})^{s_{i_0\cdots i_{d}}}(g_{i_0},\cdots,g_{i_{d}};
n_{i_0i_1}, n_{i_1i_2},\cdots, n_{i_{d-1} i_d}),
\end{align}
with hard-gauge variables $n_{i_j i_{j+1}} \in N$ defined on the link/edge.
%
%
The cocycle $\cV^{G,N}$ can be rewritten in terms of homogeneous $G$ cocycle $\nu$ and inhomogeneous $G$ cocycle $\omega$: 
\begin{align} \label{eq:inhomonuSET}
&\cV^{G,N}_{d}(g_{i_0},\cdots,g_{i_{d}};
n_{i_0i_1}, n_{i_1i_2},\cdots, n_{i_{d-1} i_d}) \nonumber\\
&= \nu^{G}_{d}(g_{i_0}, n_{i_0i_1} g_{i_1}, n_{i_0i_1} n_{i_1i_2} g_{i_2}, \cdots, n_{i_0i_1} \cdots n_{i_{d-1}i_d} g_{i_{d}}) \nonumber\\
&= \omega^{G}_{d}(g_{i_0}^{-1} n_{i_0i_1} g_{i_1}, g_{i_1}^{-1}  n_{i_1i_2} g_{i_2}, \cdots, g_{i_{d-1}}^{-1} n_{i_{d-1}i_d} g_{i_{d}}).
\end{align}

\emph{Gauge symmetry}:
The cocycle $\cV^{G,N}_{d}$ is invariant under the local gauge symmetry transformation $\mathbf n_j \in N$ on each site for a gauge group $N$:
\bea \label{eq:gauge_symmetry}
g_{i_j} \to (\mathbf n_{i_j}) \cdot g_{i_j}, \;\;\;\; n_{i_j i_{j+1}} \to (\mathbf n_{i_j})  n_{i_j i_{j+1}} (\mathbf n_{i_{j+1}})^{-1}, 
\;\;\;\; n_{ i_{j-1} i_j} \to (\mathbf n_{i_{j-1}})  n_{ i_{j-1} i_j} (\mathbf n_{i_{j}})^{-1}.
\eea
So the $Z$ is invariant under the local gauge symmetry transformation.

\emph{Global symmetry}: 
The cocycle $\cV^{G,N}_{d}$ is invariant under a \emph{total} symmetry transformation $\mathbf g$ of the symmetry group $G$:
\bea \label{eq:global_symmetry}
g_{i_j} \to \mathbf g \cdot g_{i_j}, \;\;\;\; n_{i_j i_{j+1}} \to (\mathbf g)  n_{i_j i_{j+1}} (\mathbf g)^{-1}.
\eea
So the $Z$ is invariant under the global symmetry transformation.
The true \emph{global symmetry} that does not include the \emph{gauge symmetry} is the quotient group $G/N \equiv Q$.

The normalization in \Eqn{ZgaugeSETb} has the $({|G|^{N_{v,\text{Bulk}}}})^{-1}$ modding out the site variables to make the path integral independent to the number of sites.
The additional $({|N|^{N_{v,\text{Bulk}}}})^{-1}$ mods out the gauge transformation on each site through $\forall \; (\mathbf n_j) \in N$ 
to remove the gauge redundancy. It is easy to check that $Z[S^{d-1} \times S^1]$ as a path integral on $S^{d-1} \times S^1$ is always 1, but in general $Z \neq 1$ for generic closed manifolds. If we choose that $N=1$ is trivial, then we reduce to a $G$-symmetric SPTs in Sec.~\ref{sec:ZSPTclosed}.
If we choose that all $g_j=1$ are trivial, then we reduce to the gauge theory in Sec.~\ref{sec:Zgaugeclosed} of a gauge group $N$.

We can find a mapping between a $G$-symmetric SPTs and a topologically ordered $G$-gauge theory, by the above $1 \to N \to G \to Q \to 1$ construction.
For a $G$-symmetric SPTs, we choose $N=1$ and $Q=G$.
For a $G$-gauge theory, we choose $N=G$ and $Q=1$.
This is a more general version of the relation between SPTs and topological order studied by Levin and Gu \cite{LG1209}.

\subsubsection{Symmetry-extended boundary of a $G/N$-SET state 
via $1 \to N \to G \to Q \to 1$
and
$1 \to {K \times N}\to H \to Q \to 1$} \label{sec:ZSPTSETG}

Consider the $1 \rightarrow K \overset{}{\rightarrow}  H  \overset{{r}}{\rightarrow} G \rightarrow 1$ formulation with $H/K=G$ in Appendix \ref{sec:sym-enhanced:bdry}.

\begin{enumerate}

\item \emph{Bulk $G$-SPTs on an open manifold with gapped boundary with extended $H$-symmetry action} (schematically shown in Table \ref{table:summary1} (i)): 

We consider a closed manifold $\CM^d$ glued from two open manifolds: $M^d$ and its complement space $\CM^d \smallsetminus  M^d$. Namely $M^d \cup (\CM^d \smallsetminus  M^d )=\CM^d$, with a common $(d-1)$D boundary $\partial M^d$. We denote ${N_{v, \text{Bulk}}}$ as the number of vertices in $M^d$ but not on the boundary $\partial M^d$ nor on the complement $(\CM^d \smallsetminus  M^d )$,
each of these vertices has a dimension of Hilbert space $|G|$ on each site. 
We denote ${N_{v,\text{Bdry}}}$ as the number of vertices only on the boundary $\partial M^d $,
each of these vertices has a dimension of Hilbert space $|H|$ on each site.
We denote ${N_{v,\text{Complt}}}$ as the number of vertices on the complement $(\CM^d \smallsetminus  M^d )$ but  excluding the boundary $\partial M^d $,
each of these vertices has again a dimension of Hilbert space $|H|$ on each site.
The path integral is:
\begin{align}
\label{ZSPTGSPTHmod}
& Z =
\frac{1}{|G|^{N_{v, \text{Bulk}}}} {\frac{1}{|H|^{N_{v,\text{Bdry}}+ {N_{v,\text{Complt}}} }   }}
\sum_{\{g_i, h_i\}} \prod_{\underset{(j_0\cdots j_d) \in \partial M^d\text{or}  \CM^d \smallsetminus  M^d }{(i_0\cdots i_d) \in M^d}}
\nu_d^{s_{i_0\cdots i_d}}(g_{i_0},\cdots,g_{i_d})  (\nu_d^H)^{s_{j_0\cdots j_d}}(h_{j_0},\cdots,h_{j_d}) 
\nonumber\\
&
=
\frac{1}{|G|^{N_{v, \text{Bulk}}}} {\frac{1}{|H|^{N_{v,\text{Bdry}}+ {N_{v,\text{Complt}}} }   }}
\sum_{\{g_i, h_i\}} \prod_{\underset{(j_0\cdots j_d) \in \partial M^d\text{or}  \CM^d \smallsetminus  M^d }{(i_0\cdots i_d) \in M^d}}
\nu_d^{s_{i_0\cdots i_d}}(g_{i_0},\cdots,g_{i_d})  (\mu_{d-1}^H)^{s_{j_0\cdots j_{d-1}}}(h_{j_0},\cdots,h_{j_{d-1}}) \nonumber\\
&
=
\frac{1}{|G|^{N_{v, \text{Bulk}}}} {\frac{1}{|H|^{N_{v,\text{Bdry}} }   }}
\sum_{\{g_i, h_i\}} \prod_{\underset{(j_0\cdots j_{d-1}) \in  \partial M^d}{(i_0\cdots i_d) \in M^d}}
\nu_d^{s_{i_0\cdots i_d}}(g_{i_0},\cdots,g_{i_d})  (\mu_{d-1}^H)^{s_{j_0\cdots j_{d-1}}}(h_{j_0},\cdots,h_{j_{d-1}}). 
\end{align}
Above we had applied \Eqn{Omom}, and the fact that the homogeneous cocycle
$ \nu^H_d(h_0,\cdots,h_d)= \nu_d(r(h_0),\cdots,r(h_d))$ which then split it to lower-dimensional homogeneous cochains $\mu_{d-1}^H$.
Here ${(i_0\cdots i_d) \in M^d}$ means the vertices in the bulk $M^d$ (with a total number ${N_{v, \text{Bulk}}}$) as well as on the boundary (with a total number $N_{v,\text{Bdry}}$).
Here ${(j_0\cdots j_d) \in \partial M^d}$ {or} $\CM^d \smallsetminus  M^d$ means the vertices on the boundary $\partial M^d$ or in the complement $\CM^d \smallsetminus  M^d$
with a total number
${N_{v,\text{Bdry}}+ {N_{v,\text{Complt}}} }$.
The cochains inside the volume of the complement $\CM^d \smallsetminus  M^d$ cancel out to 1 due to overlapping terms with opposite orientations.
An overall sum ${(j_0\cdots j_d) \in \CM^d \smallsetminus  M^d}$ contributes a factor $|H|^{N_{v,\text{Complt}}}$ canceling with a normalizing factor
to obtain \Eqn{ZSPTSETmod}.

\item \emph{Bulk $G$-SPT on an open manifold with gapped boundary anomalous SET (with a $G$-anomaly) of gauge group $K$} 
(schematically shown in Table \ref{table:summary1} (ii)):  

Consider an SPT path integral on an open manifold $M^d$ with gapped boundary anomalous SET on the $\prt M^d$.
We can directly start from \Eqn{ZSPTGSPTHmod}, and introduce gauge variables $k_{j j'}\in K$ 
on the links between boundary sites on $\prt M^d$. 
After properly modding out the gauge redundancy, both obtain:
\begin{align}
\label{ZSPTSETmod}
& Z =
\frac{1}{|G|^{N_{v,\text{Bulk}}}} {\frac{1}{|H|^{N_{v,\text{Bdry}}}} \frac{1}{|K|^{N_{v,\text{Bdry}}} } }
\sum_{\{g_i, h_i,h_{ij}\}} \prod_{(i_0\cdots i_d) \in M^d}
\nu_d^{s_{i_0\cdots i_d}}(g_{i_0},\cdots,g_{i_d}) 
\\
&
\prod_{(j_0\cdots j_{d-1}) \in \prt M^d}
(\cV^{H,K}_{d-1})^{s_{j_0\cdots j_{d-1}}}(h_{j_0},\cdots,h_{j_{d-1}};
k_{j_0 j_1}, k_{j_1 j_2},\cdots, k_{j_{d-2} j_{d-1}}).
\nonumber 
\end{align}
The $\cV^{H,K}_{d-1}(h_{j_0},\cdots,h_{j_{d-1}};
k_{j_0 j_1}, k_{j_1 j_2},\cdots, k_{j_{d-2} j_{d-1}})
=
\nu^{H}_{d-1}(h_{j_0}, k_{j_0 j_1} h_{j_1}, \cdots, k_{j_0 j_1}k_{j_1 j_2}  \cdots h_{j_{d-1} })
$ 
can be evaluated as homogeneous cochains by absorbing link variables to site variables.

\item {\emph{Bulk $G$-gauge theory on an open manifold with gapped boundary anomalous $H$-gauge theory}
(schematically shown in Table \ref{table:summary1} (iii)):   

We can gauge the global symmetry $G$ of \Eqn{ZSPTSETmod} in the bulk to obtain the bulk $G$-gauge theory, while the boundary has 
an $H$-gauge theory as an anomalous gapped boundary. 
\begin{align}
\label{ZgaugeHKmod}
& Z =
\frac{1}{|G|^{N_{v,\text{Bulk}}}} {\frac{1}{|H|^{N_{v,\text{Bdry}}}}  }
\sum_{\{g_{ij},h_{ij}\}} \prod_{(i_0\cdots i_d) \in M^d}
\omega_d^{s_{i_0\cdots i_d}}(g_{i_0i_1},\cdots,g_{{i_{d-1}}{i_d}}) 
\\
&
\prod_{(j_0\cdots j_{d-1}) \in \prt M^d}
(\Omega^{H}_{d-1})^{s_{j_0\cdots j_{d-1}}}(
h_{j_0 j_1}, h_{j_1 j_2},\cdots, h_{j_{d-2} j_{d-1}}).
\nonumber 
\end{align}
The $\omega_d$ and $\Omega_{d-1}$ are an inhomogeneous cocycle and cochain suitable for gauge theories.}

\item \emph{Bulk SET on an open manifold with gapped boundary anomalous SET} 
(schematically shown in Table \ref{table:summary1} (iv)):   

Alternatively, we can partially gauge a normal subgroup $N \subseteq G$ in the bulk $G$-SPTs and also on the boundary.
Let us name the quotient group
$$\frac{H}{K \times N}=\frac{G}{N}\equiv Q.$$
This gives us a bulk SET with global symmetry $Q$ and gauge symmetry ${N}$ via:
\bea \label{eq:1NGQ1}
1 \to N \to G \to Q \to 1.
\eea
 The boundary anomalous SET  with global symmetry $Q$ and gauge symmetry ${K \times N}$ is
\bea \label{eq:1KNHQ1}
1 \to {K \times N}\to &H& \to Q \to 1.\\
\text{Note that: } 1 \to {K}\to &H& \to G \to 1.
\eea

\end{enumerate}

\subsubsection{Symmetry-extended interface between two topological phases $G_{\I}$ and $G_{\II}$} \label{sec:ZSPTSETG1G2}

We construct a path integral of topological phases $G_{\I}$ and $G_{\II}$ following Appendix \ref{sec:sym-enhanced:DWG1G2}
under
$1 \rightarrow K \overset{}{\rightarrow}  H  \overset{{r}}{\rightarrow} G_{\I} \times  G_{\II}\rightarrow 1$.
First, consider a closed manifold $\CM^d$ glued from two open manifolds: $M^d$ and its complement space $\CM^d \smallsetminus  M^d$, 
with a common $(d-1)$D boundary $\partial M^d$. 
The $M^d$ is assigned with a Hilbert-space dimension $G_{\I} \times G_{\II}$ on each degree of freedom (on site or edge).
The $\CM^d \smallsetminus  M^d$ is originally assigned with $G_{\I} \times G_{\II}$-cocycles, but lifted to $H$ to become trivial coboundaries.
Using the folding trick, given $ \omega^{G_{\I} \times G_{\II}}_d(g) =\omega_{\I}^{G_{\I}}(g_{\I} ) \cdot \omega_{\II}^{G_{\II}}(g_{\II})^{-1}$,
we can fold $\omega_{\II}^{G_{\II}}(g_{\II})$ to $-M^d$ with an opposite orientation, while we keep
$\omega_{\I}^{G_{\I}}(g_{\I})$ to $M^d$. The $M^d \cup (-M^d)$ can be glued to a closed manifold because they share the same boundary.
We can define the path integral on a closed $M^d \cup (-M^d)$.
More generally, we can call $M^d$ as $M_{\I}^d$, while we can modify the amplitude on $-M^d$ to a new amplitude 
on any open manifold $M_{\II}^d$ 
provided that 
$\partial M_{\I}^d =\partial M_{\II}^d=\partial M^d$ is the same common boundary.
We denote the number of vertices ${N_{v,\I}}$  on $M_{\I}^d$  but not on $\partial M_{\I}^d$, and the similar definition for ${N_{v,\II}}$  with $\I \to \II$.
We denote the number of vertices ${N_{v,\partial}}$  on $\partial M_{\I}^d =\partial M_{\II}^d$.
We define this path integral on a closed spacetime $M_{\I}^d \cup M_{\II}^d$ below.
\begin{enumerate}

\item \emph{Bulk $G_{\I}$ and $G_{\II}$-SPTs with gapped $H$-interface} (schematically shown in Table \ref{table:summary1} (v)): 
{
\bea
Z 
=&& \frac{1}{{|G_{\I}|^{N_{v,\I}}} {|H|^{N_{v,\partial}}} {|G_{\II}|^{N_{v,\II}}}}
\sum_{\{g_{{\I},i}\},\{h_i\}, \{g_{{\II},i}  \}} \prod_{(i_0\cdots i_d) \in M^d_{\I}}
{\nu_d^{G_{\I}}}^{s_{i_0\cdots i_d}}(g_{{\I},i_0},\cdots,g_{{\I},i_d})\;\;\;\;\;\;\;\;\;  \\
&& \prod_{(i_0\cdots i_{d-1}) \in \partial M^{d}}
{ \mu^{H {s_{i_0\cdots i_{d-1}}}}_{d-1} (h_{i_0},\cdots,h_{i_{d-1}})}
\prod_{(i_0\cdots i_d) \in M^d_{{\II}}}
{\nu_d^{G_{\II}}}^{s_{i_0\cdots i_d}}(g_{{\II},i_0},\cdots,g_{{\II},i_d}). \nonumber
\eea
}
\item \emph{Bulk $G_{\I}$ and $G_{\II}$-SPTs with gapped boundary anomalous SET of gauge group $K$} 
(schematically shown in Table \ref{table:summary1} (vi)):  
{
\bea
Z 
=&& \frac{1}{{|G_{\I}|^{N_{v,\I}}} {|H|^{N_{v,\partial}}} {|G_{\II}|^{N_{v,\II}}}}
\sum_{\{g_{{\I},i}\},\{h_i\}, \{g_{{\II},i}  \}} \prod_{(i_0\cdots i_d) \in M^d_{\I}}
{\nu_d^{G_{\I}}}^{s_{i_0\cdots i_d}}(g_{{\I},i_0},\cdots,g_{{\I},i_d}) \;\;\;\;\;\;\;\;\;  \\
&& \prod_{(j_0\cdots j_{d-1}) \in \prt M^d}
{ 
(\cV^{H,K}_{d-1})^{s_{j_0\cdots j_{d-1}}}(h_{j_0},\cdots,h_{j_{d-1}};
k_{j_0 j_1}, k_{j_1 j_2},\cdots, k_{j_{d-2} j_{d-1}})
} \nonumber \;\;\;\;\;\;\;\;\;  \\
&& \prod_{(i_0\cdots i_d) \in M^d_{\II} }
{\nu_d^{G_{\II}}}^{s_{i_0\cdots i_d}}(g_{{\II},i_0},\cdots,g_{{\II},i_d}). \nonumber
\eea}
Here we dynamically gauged the normal subgroup $K=H/(G_{\I} \times  G_{\II})$ on $\prt M^d$ by introducing the link variables along $\prt M^d$, thus
we rewrote $\mu^{H}_{d-1}$ into $(\cV^{H,K}_{d-1})$.
\item \emph{Bulk SETs  with gapped interface anomalous SET of enhanced gauge symmetry} (schematically shown in Table \ref{table:summary1} (vi)): 

Developed from the above case 2. ``{Bulk $G_{\I}$ and $G_{\II}$-SPTs with gapped boundary anomalous SET of gauge group $K$},''
we can partially gauge normal subgroups of $G_{\I}$ and $G_{\II}$-SPTs, so that the bulk has SETs while the interface has an anomalous SETs.
\end{enumerate}
%

\subsection{Wavefunction and Lattice Hamiltonian} \label{sec:generalHamiltonian}

We would like to formulate a lattice Hamiltonian on the space lattice, whose time-dependent Schr\"odinger equation gives rise to the same low energy physics governed by 
the path integral definition in the previous Sec.~\ref{sec:generalPathintegral}.
We motivate the Hamiltonian construction by thinking of ground-state wavefunctions. 
The lattice Hamiltonian below will be a SET generalization from the SPTs of Ref.\cite{XieSPT4} and the topological orders/gauge theories of Ref.\cite{Wan1211.3695,Wan:2014woa}.
Our Hamiltonian in Sec.~\ref{sec:generalSETHamiltonian} is also a generalization of SETs of Ref.\cite{Ran1212.0835} to include a projective symmetry group under $G/N=Q$.
We further implement anomalous SET gapped boundaries/interfaces in Sec.~\ref{sec:generalGBDWHamiltonian}.

\begin{figure}[h!] 
\begin{center}
\includegraphics[scale=0.335]{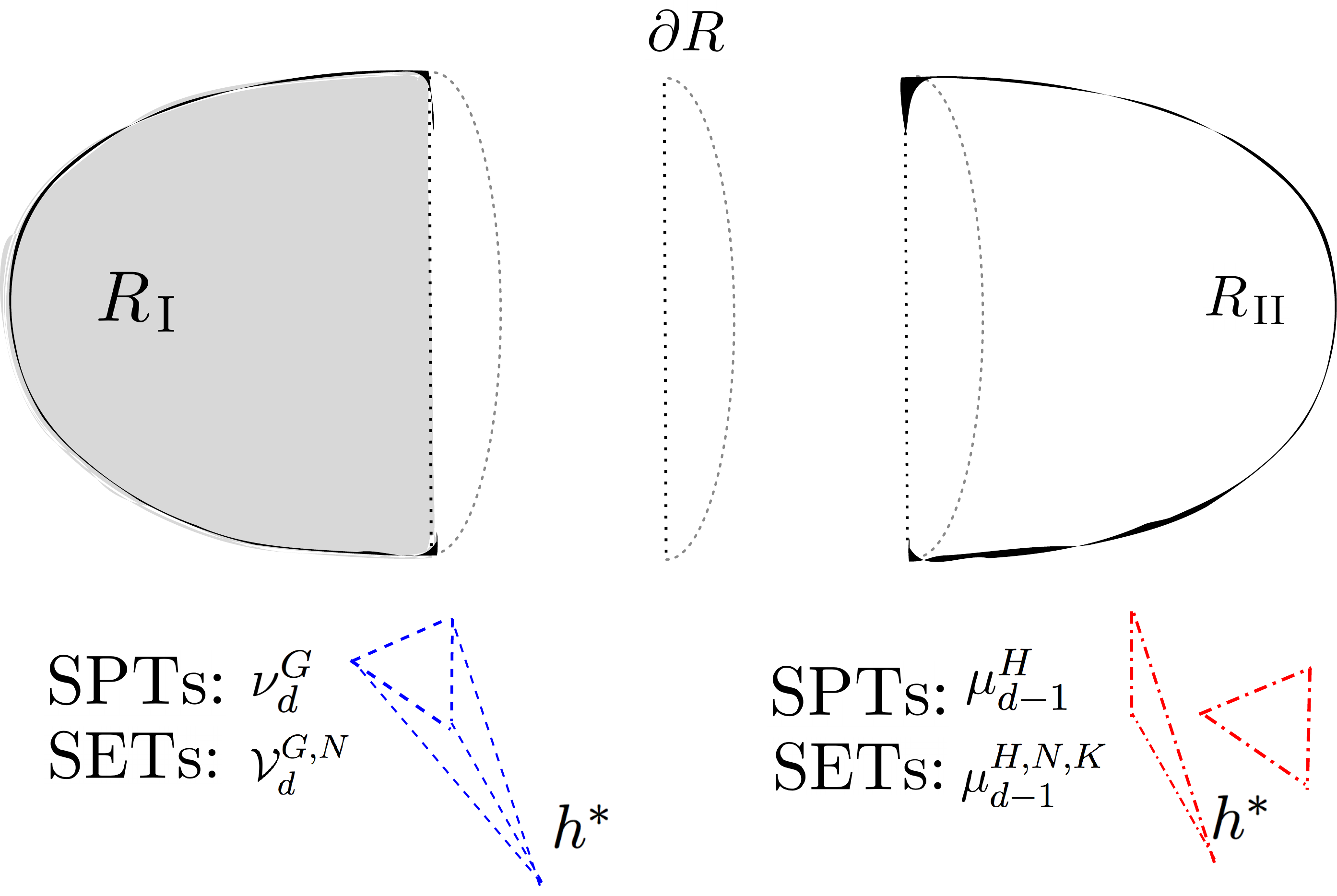} \\
\end{center}
\caption{In Sec.~\ref{sec:generalHamiltonian}, we define
wavefunctions and lattice models on a $(d-1)$-dimensional space manifold by triangulating the manifold to $(d-1)$-simplices. 
If the space is closed, as in Sec.~\ref{sec:generalSETHamiltonian},
we assign ``$(d-1)$-simplices together with an extended vertex $h^*$,'' with cocycles $\nu^{G}_{d}$ for SPTs or with $\cV^{G, N}_{d}$ for SETs.
In this figure, 
the space is obtained as the gluing of two spatial manifolds $R_{\I} \cup R_{\II}$ with a common boundary $\partial R$.
For simplicity, we draw the $d=3$ case. One example of the $R_{\I} \cup R_{\II}=S^2$ is a 2-sphere,
then we can choose $R_{\I}=D^2$ and $R_{\II}=D^2$, where the gapped spacetime boundary is on a 1-circle $\partial R=S^1$.
We would like to define the wavefunction on an open manifold $R_{\I}$ (shown in gray) with a gapped boundary $\partial R$ (shown as a dotted curve),
 details of which are discussed in Sec.~\ref{sec:generalGBDWHamiltonian}.
In our construction, we  assign lower-dimensional split cochains $\mu^{H}_{d-1}$ (or $\cV^{H,K}_{d-1}$) for SPTs and $\mu^{H, N, K}_{d-1}$ for SETs
to ``$(d-2)$-simplices connecting to the additional vertex $h^*$'' paved onto a gapped boundary $\partial R$.
}
\label{fig:lattice-space-1}
\end{figure}

\begin{figure}[h!] 
\begin{center}
(1)\includegraphics[scale=0.335]{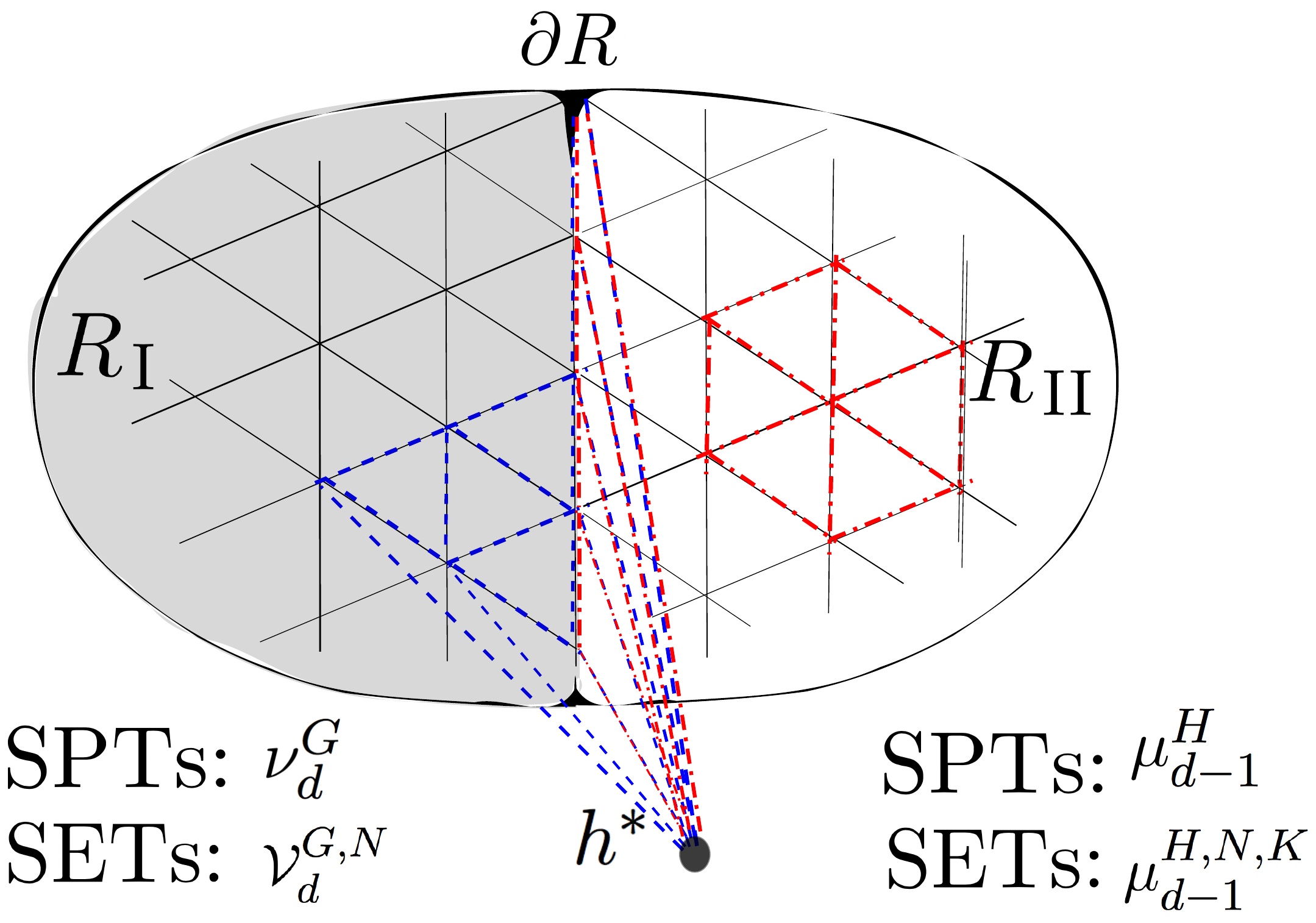}\;\;\;\;\;\;
(2)\includegraphics[scale=0.335]{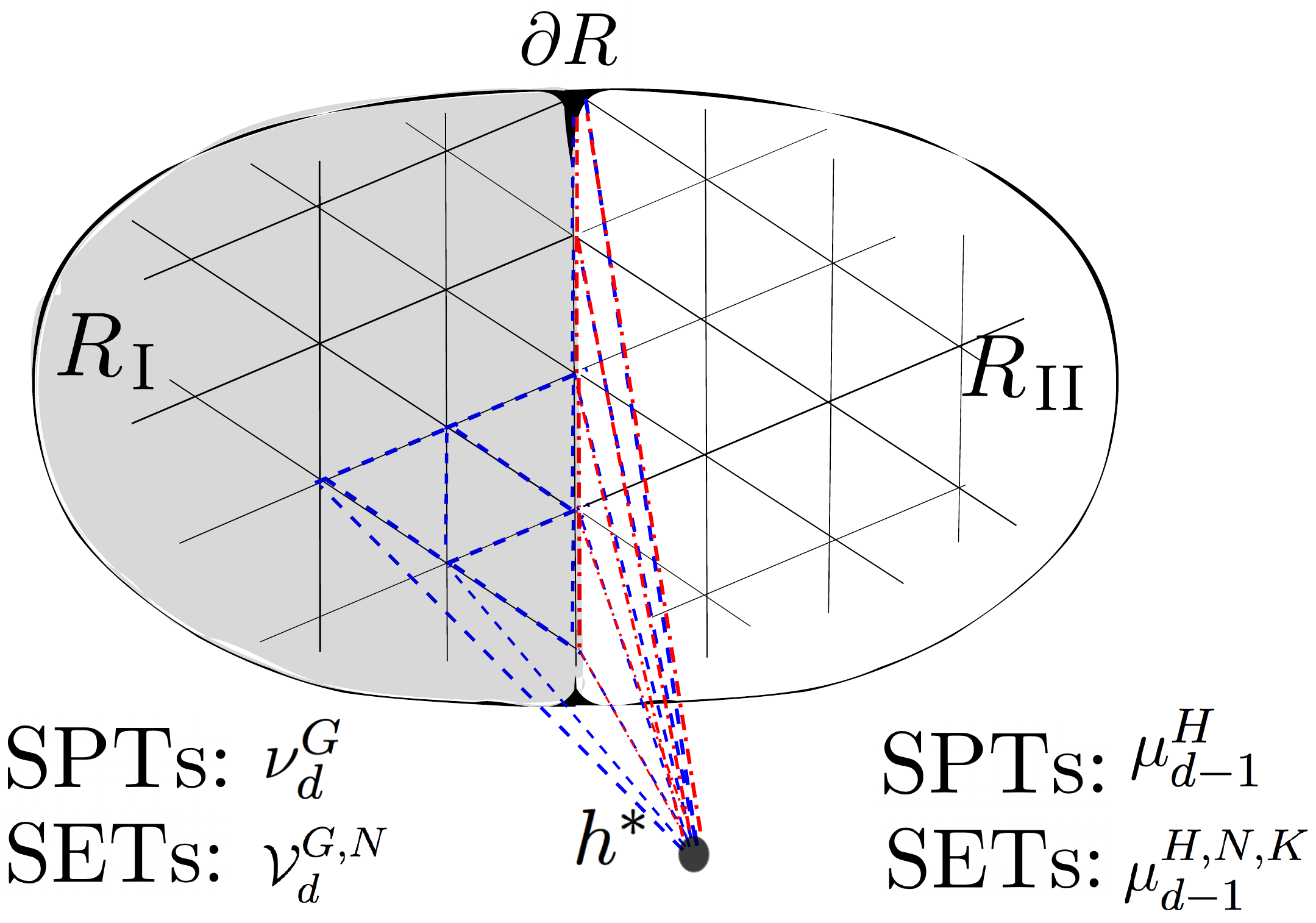}
\end{center}
\caption{
{Follow Fig.\ref{fig:lattice-space-1}, 
the fig.(1) shows that a wavefunction amplitude is the product of two contributions.
The first contribution is the filling of $d$-cocycles into the gapped bulk in $R_{\I}$ connecting to $h^*$.
The second contribution is  the filling of $(d-1)$-cochains onto a gapped boundary $\partial R$ connecting to $h^*$ and
 into the surface of the other complement bulk $R_{\II}$.
The combined result contributes to the fig.(2), where the $(d-1)$-cochains on the region $R_{\II}$ can be deformed to a trivial product state (as a trivial gapped vacuum)
under local unitary transformations without breaking the global symmetry.
We can \emph{remove} the wavefunction amplitude on $R_{\II}$ after a proper amplitude normalization.
Thus the wavefunction is well-defined simply in $R_{\I}$ and on $\partial R$. 
The explicit formula is derived in Sec.~\ref{sec:generalGBDWHamiltonian}.}
}
\label{fig:lattice-space-2}
\end{figure}

Schematically
Fig.\ref{fig:lattice-space-1} and Fig.\ref{fig:lattice-space-2} 
summarize how to define an exactly soluble lattice Hamiltonian and wavefunction on a spatial manifold.
Normally, a wavefunction of gapped topological phase is well-defined on a closed spatial manifold.
However, here in particular, some wavefunction of fully gapped topological phase
can also be well-defined in the gapped bulk on $R_{\I}$ with a gapped interface $\partial R$.

\subsubsection{Trivial product state and lattice Hamiltonian}
We can consider a total trivial product state wavefunction, where $\{g_i\}$ specifies the group element in a symmetry group $G$ and its assignment to a local site $i$ on a regularized 
$d$D spatial manifold $M$, the wavefunction has its coefficient:
$
\Phi_0(\{g_i\}_M)=1.
$
Its wave state-vector in the Hilbert space is:
\bea \label{eq:trivial_product_state}
|\Phi_0 \rangle \propto  \sum_{\{g_i\}_M} \Phi_0(\{g_i\}_M) |\{g_i\}_M \rangle=\sum_{\{g_i\}_M}  |\{g_i\}_M \rangle=
(\sum_{g_1} | g_1 \rangle) \otimes (\sum_{g_2} | g_2 \rangle) \dots \otimes (\sum_{g_i} | g_i \rangle) \otimes \dots,\eea
which we can properly normalize to have $\langle \Phi_0 |\Phi_0 \rangle =1$.
Note that $|\{g_i\}_M \rangle$ has a tensor product structure,
$
|\{g_i\}_M \rangle=\dots \otimes | g_i \rangle \otimes \dots,
$
here $i$ is the site index for some site $i$ distributed around the spatial manifold $M$.
To see that the state-vector is a trivial product state,
we notice that it is indeed a tensor product of $(\sum_{g_i} | g_i \rangle)$ on each site $i$, where $(\sum_{g_i} | g_i \rangle)$ sums over all group element bases.
The Hilbert space on each site $j$ is $\mathscr{H}_j$ with a Hilbert space dimension $|G|$ spanned by $| g_j \rangle$. The total Hilbert space is also a tensor product structure:
$
\mathscr{H}_{\text{total}}=\otimes_j \mathscr{H}_j.
$

Consider the site index $j$, we can write down the exactly soluble Hamiltonian whose ground state is $|\Phi_0 \rangle$:
\bea
\widehat{H}_j=- \ket{\phi_j}\bra{\phi_j}=-\sum_{g_j\in G}\ket{g_j} \sum_{g'_j\in G}\bra{g'_j}
=-\sum_{g_j, g'_j \in G} \ket{g_j} \bra{g'_j}.
\eea
Here $\widehat{H}_j=- \ket{\phi_j}\bra{\phi_j}$ is a local operator on each site $j$,
and $\ket{\phi_j}=\sum_{g_j\in G}\ket{g_j}$ is an equal-weight sum of all states of all group elements ${g_j}$ on each site.
Thus $\widehat{H}_j=- \ket{\phi_j}\bra{\phi_j}$ is proportional to a 
constant matrix 
$\Bigg(
\begin{smallmatrix}
1 & 1 &\dots\\
1 & 1 &\dots\\
\vdots &\vdots & \ddots
\end{smallmatrix}
\Bigg)$ in the group element basis $\ket{g_j}$
acting on each site.
Thus we construct a trivial product state and lattice Hamiltonian for a trivial insulator with a finite energy gap.

\subsubsection{Short-range/long-range entangled states and SPT/topologically ordered/SET lattice Hamiltonians}
\label{sec:generalSETHamiltonian}

Now we consider a gapped short-range or long-range entangled states for an anomaly-free Hamiltonian on a closed space
that is well-defined in $d-1$D spatial lattice. 
We can consider either (1) a $G$-SPTs for a cocycle $\nu^{G}_{d}$ in Sec.~\ref{sec:ZSPTclosed}, 
or (2) an $N$-gauge theory with intrinsic topological order for a cocycle $\omega^{N}_{d}$ in Sec.~\ref{sec:Zgaugeclosed},
or (3) a SETs prescribed by $1 \to N \to G \to Q \to 1$ for a cocycle $\cV^{G,N}_{d}$ in Sec.~\ref{sec:ZSETclosed}.

The SET state in Sec.~\ref{sec:ZSETclosed} is the most general containing all other cases  by eqn.\ref{eq:inhomonuSET}, 
thus we focus on the SETs below.
For a nontrivial non-product state wavefunction of SETs, we define a particular wavefunction coefficient on a closed space $M$ as:
 \bea   \label{eq:SETwavefunction}
 \Phi(\{g_i, n_{ij}\}_M) \equiv\prod_{\{...\}} \cV^{G,N \; s_{i_0 \dots i_*}}_{d}(g_{i_0},\cdots,g^*; n_{i_0i_1}, n_{i_1i_2},\cdots, n_{i_{d-1} i_*}) 
 \eea
where $\{g_i , n_{ij}\}_M$ are a set of site ($i$) and link ($ij$) variables on $M$, for $g_i \in G$ and $n_{ij} \in N$. 
{
Conventionally $\cV^{G,N}_{d}$ is a $U(1)$ phase, except that we set $\cV^{G,N}_{d}$ as zero if and only if any 
face of its simplex violates $n_{12}n_{23}n_{31} = 1$.}
The $g^*$ is fixed and assigned to an additional fixed point $i_*$ outside $M$. 
There are link variables $n_{i_j i_* }$ from any site $j$ on $M$ to $i_*$.
Given a wavefunction input parameter $\{g_i, n_{ij}\}_M$,
to determine the wavefunction $ \Phi(\{g_i, n_{ij}\}_M)$, the only input data we need are these two:
$$
g^*,\;\;\;\;  n_{i_0 i_*}.
$$
We only need to provide another input data $n_{ i_0  i_*}$, as a link variable connecting between a particular site  $i_0$ to $i_*$.
Any other variables $n_{i_j  i_*}$ are determined by a \emph{zero flux condition} through any closed loop $n_{i_j i_*} n_{ i_* i_0} n_{i_0 i_j} =1$, namely:
$n_{i_j i_* } =  n_{i_j i_0} n_{i_0  i_*}$. 
Here $\prod_{\{...\}}$ is a product over all simplices assigned with cocycles.
The zero flux condition through any closed loop constrains that the wavefunction has a trivial holonomy around any cycle of the closed manifold. 
Thus, we only generate a unique ground state so far. (We will comment how to generate other ground states with nontrivial holonomy for topological orders/SETs later.) 
This ground state as a vector in the Hilbert space is, up to a normalization: 
\bea \label{eq:SETstatevector}
|\Phi\rangle \propto  \sum_{\{g_i, n_{ij}\}_M} \Phi(\{g_i, n_{ij}\}_M) |\{g_i, n_{ij}\}_M \rangle.
\eea
The $|\{g_i, n_{ij}\}_M \rangle$ has a tensor product structure,
$
|\{g_i, n_{ij}\}_M \rangle=\dots \otimes | g_i \rangle  \otimes \dots \otimes | n_{ij}  \rangle \otimes \dots
=\otimes_i | g_i \rangle \otimes_{ij} | n_{ij}  \rangle
$

Now we construct an exactly soluble Hamiltonian for the above gapped ground state as
\bea\label{eq:SETHamiltonian}
  \widehat{H}=-\sum_v \widehat{A}_v-\sum_f \widehat{B}_f.
\eea
 The {first} term,  $\widehat{A}_v$ acts on the wavefunction of a constant-time slice through each vertex $v$ in the space by lifting 
 the {initial} state through an ``imaginary time'' evolution to a new state with a vertex $v'$ via
\bea
&& \widehat{A}_v=\frac{1}{|G|}\frac{1}{|N|}\sum_{\overset{[vv']=n\in N,}{g \in G}}\widehat{A}_v^{g, n}.\\
&&  \label{eq:Avgn}
\widehat{A}_v^{g, n} |g_v , n_{i v}, n_{v j}, \dots \rangle  \\
&& \;\;\;\;\;\; =\prod_{\{...\}} \cV^{G,N\; s_{...}}_{d}(g, \; g_v, \cdots; n, \; n_{i v} \cdot n, \;  n^{-1} \cdot n_{ v j},\cdots)  | g ,\;  n_{i v} \cdot n, \;  n^{-1} \cdot n_{ v j},  \dots\rangle.\nonumber \;\;\; 
\eea
We define $\widehat{A}_v^{g, n}$ operator above by its operation on a state-vector $|g_v , n_{i v}, n_{v j}, \dots \rangle$.
Under the $\widehat{A}_v^{g, n}$ operation, the group element assigned to $v$ as $| g_v \rangle$ has evolved to $v'$ as $| g \rangle$,
the link element assigned ${iv}$ as $| n_{i v} \rangle$ has evolved to $|n_{i v'} \rangle=| n_{i v} \cdot n \rangle$, 
and $| n_{ v j} \rangle$ has evolved to $|n_{ v' j} \rangle=| n^{-1} \cdot n_{ v j} \rangle$.

\begin{figure}[tb]
\begin{center}
\includegraphics[scale=0.8]{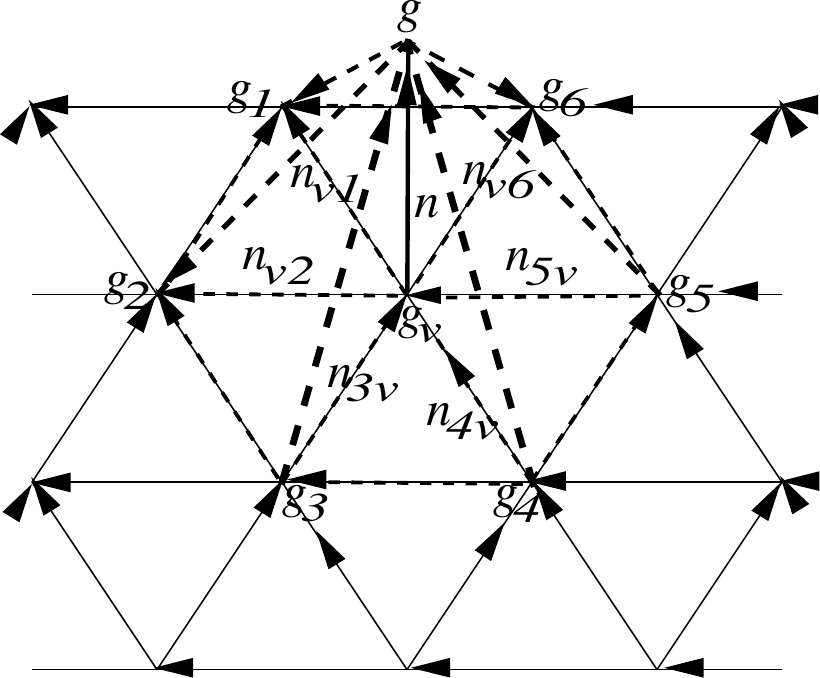}
\end{center}
\caption{The effective expression of that $\widehat{A}_v^{g, n}$ operation. Here we show $\widehat{A}_v^{g, n}$ acts on a 2D spatial lattice on a site $v$ and its neighbor links.
The explicit form is given in \Eqn{eq:Avgn2+1D}. 
The volume enclosed by dashed links contributes an amplitude filled by cocycles $\cV^{G,N}$.
A more general expression for any dimension is given in \Eqn{eq:Avgn}.
}
\label{fig:Avgn}
\end{figure}

In any dimension, we can construct $(d-1)$-simplices (that can be of irregular sizes) as a lattice to fill the space.
More explicitly, consider for example a 2+1D SETs, $\widehat{A}_v^{g, n}$ acts on a Hilbert space state-vector for a 2D spatial lattice system in Fig. \ref{fig:Avgn},
centered at the vertex $v$ and its six nearest-neighbored links:
\bea 
&&\widehat{A}_v^{g, n} |g_v,g_1,g_2,g_3,g_4,g_5,g_6; n_{v1},n_{v2}, n_{3v}, n_{4v},  n_{5v}, n_{v6} \rangle \nonumber \\
&& \;\;\;\; =\frac{\cV^{G,N}_{3}(g_4, g_5, g_v, g; n_{45}, n_{5 v}, n)  \cV^{G,N}_{3}(g_5, g_v, g, g_6; n_{5 v}, n, n^{-1}  n_{v6}) 
\cV^{G,N}_{3}(g_v, g, g_6, g_1; n, n^{-1} n_{v6}, n_{61}) }{
\cV^{G,N}_{3}(g_v, g, g_2, g_1; n, n^{-1}n_{v2}, n_{21}  ) \cV^{G,N}_{3}(g_3, g_v, g, g_2; n_{3v},n, n^{-1} n_{v2} ) 
\cV^{G,N}_{3}(g_4,  g_3, g_v, g; n_{43},n_{3v}, n )}  \nonumber \\
&& 
\;\;\;\;  | g ,g_1,g_2,g_3,g_4,g_5,g_6;  
n^{-1} \cdot n_{v1}, n^{-1} \cdot n_{v2}, n_{3v} \cdot n, n_{4v} \cdot n, n_{5 v} \cdot n,   n^{-1} \cdot n_{v6}\rangle. \label{eq:Avgn2+1D}
\;\;\;
\eea
We design the $\widehat{B}_f$ term as the zero flux constraint on each face / plaquette.
More explicitly, consider a face $f$ (in Fig. \ref{fig:Avgn}) with three vertices (assigned $g_1, g_2, g_v$) and three links (assigned $n_{v2}, n_{21}, n_{v1}$),
the $B_f$ acts on the corresponding state vector $| g_1, g_2, g_v; n_{v2}, n_{21}, n_{v1} \rangle$ as
\bea \label{eq:Bf}
\widehat{B}_f | g_1, g_2, g_v; n_{v2}, n_{21}, n_{v1} \rangle=(\delta_{n_{v2} n_{21} n_{1v}=1})\cdot  | g_1, g_2, g_v; n_{v2}, n_{21}, n_{v1} \rangle.
\eea
The $\delta_{n_{v2}, n_{21} n_{1v}=1}$ is a Kronecker delta which gives 1 if ${n_{v2}, n_{21} n_{1v}=1}$ is trivial in $N$; thus, the flux through the face $f$ is zero. 
The $\delta_{n_{v2}, n_{21} n_{1v}=1}$ gives 0 otherwise.
Even for SETs, the explicit zero flux condition is reduced to
$$(g_v^{-1}n_{v2} g_2)( g_2^{-1}n_{21} g_1 )(g_1^{-1} n_{1v} g_v)=n_{v2}n_{21}n_{21}=1,$$
the same as pure $N$-gauge theory of topological order.
For SPTs with a nontrivial $G$ but a trivial $N=1$, the zero flux always manifests as $(g_v^{-1} g_2)( g_2^{-1} g_1 )(g_1^{-1} g_v)=1$.
Some more remarks on the system are given as follows:
\begin{enumerate}
\item All $\widehat{A}_v^{g, n}$ and $\widehat{B}_f$ have mutually-commuting and self-commuting nice properties.
In principle, our model is an exactly soluble lattice model.

\item Since the SPTs always satisfies the zero flux on every face $f$, we can simplify the Hamiltonian \emph{without} the $\widehat{B}_f$ term:
$\widehat{H}_{\text{SPT}}=-\sum_v \widehat{A}_v.$
The additional $\widehat{B}_f$ term in \Eqn{eq:SETHamiltonian} for SETs and topological orders imposes the zero flux constraint at low energy.
However, at high energy, at the cost of an energy penalty, the zero flux condition does \emph{not hold at those faces $f$ with energetic anyon excitations.}
The anyon excitations are created at the end points of \emph{extended operators} (e.g. line operators in 2+1D). See also Remark {8}.

\item \emph{Hilbert space}:
The Hilbert space on each site $j$ is $\mathscr{H}_j$ with a Hilbert space dimension $|G|$ spanned by $| g_j \rangle$  for $g_j \in G$. 
The Hilbert space on each edge $ij$ is $\mathscr{H}_{ij}$ with a Hilbert space dimension $|N|$ spanned by $| n_{ij} \rangle$ for $n_{ij} \in N$. 
For our lattice Hamiltonian \Eqn{eq:SETHamiltonian}, the total Hilbert space is a tensor product structure:
\bea
\mathscr{H}_{\text{total}}=\otimes_j \mathscr{H}_j \otimes_{ij} \mathscr{H}_{ij}.
\eea
When we limit to a symmetric $G$-SPT, with $N=1$, we have a tensor product $\mathscr{H}_{\text{total}}=\otimes_j \mathscr{H}_j$ defined on sites.
When we limit to a gauge group $N$-topological order, with $G=1$, we have a tensor product $\mathscr{H}_{\text{total}}=\otimes_{ij} \mathscr{H}_{ij}$ defined on links.
Naively, one may ask that isn't that ``the discrete gauge theory description of topological order has no tensor product Hilbert space $\mathscr{H}_{\text{total}}\neq \otimes_{ij} \mathscr{H}_{ij}$?''
The answer is that the gauge theory description of topological order for our Hamiltonian \Eqn{eq:SETHamiltonian} only occurs at the lowest-energy ground states, when 
$\widehat{B}_f=1$ as zero flux on every face.
For those ground states of topological order, indeed, the Hilbert space is not a tensor product, $\mathscr{H}_{\text{total}}\neq \otimes_{ij} \mathscr{H}_{ij}$, due to
the requirement of projection constrained by $\widehat{B}_f=1$.
Thus, our Hamiltonian as a \emph{local bosonic lattice model}  at higher energy contains more than a discrete gauge theory. 
The same argument holds for SET states.
\item 
\emph{Gauge and global symmetries for Hamiltonians}: 
The Hamiltonian in \Eqn{eq:SETHamiltonian} is apparently invariant under the $N$-gauge 
\Eqn{eq:gauge_symmetry}
and $G$-global symmetry
\Eqn{eq:global_symmetry}
transformations.
For SETs and SPTs, each individual of $\widehat{A}_v^{g, n}$ and $\widehat{B}_f$ terms
is both $N$-gauge invariant and $G$-global invariant.
On the other hand, for a topological order of gauge group $N$ without any global symmetry (i.e. $G=1$), 
 the individual $\widehat{A}_v^{n}$ is not gauge invariant.
 For example, under a local gauge transformation $ \mathbf n_v$ applied on the vertex $v$, it transforms $\widehat{A}_v^{n} \to \widehat{A}_v^{(\mathbf n_v) \cdot n }$. 
If a local gauge transformation is applied on a neighbored vertex next to $v$, then $\widehat{A}_v^{n}$ is invariant. 
However the overall $\widehat{A}_v=\frac{1}{|N|}\sum_{{[vv']=n\in N}}\widehat{A}_v^{n}$ is gauge invariant. 
\item 
\emph{Gauge and global symmetries for wavefunctions}:
For the SET state vector $|\Phi\rangle$ of \Eqn{eq:SETstatevector}, we can apply 
symmetry transformations on either the wavefunction coefficient $\Phi(\{g_i, n_{ij}\}_M)$ or on the basis $|\{g_i, n_{ij}\}_M \rangle$;
 the two transformations are equivalent by an inverse transformation on another.
Thus, we focus on the transformations on the wavefunction $\Phi(\{g_i, n_{ij}\}_M)$.

$\bullet$ If $G$ is nontrivial, then we have either SPTs or SETs. It is easy to check that the cocycle $\cV^{G,N}$ is both gauge and global symmetry invariant
under 
$N$-gauge 
\Eqn{eq:gauge_symmetry}
and $G$-global symmetry
\Eqn{eq:global_symmetry} transformations.
Thus, apparently, the wavefunction
$$
\Phi(\{g_i, n_{ij}\}_M)=\Phi(\{ (\mathbf n_{i}) g_i, (\mathbf n_{i}) n_{ij} (\mathbf n_{j})^{-1}\}_M)
=\Phi(\{(\mathbf g) g_i, (\mathbf g) n_{ij} (\mathbf g)^{-1}\}_M )
$$ 
is gauge and global-symmetry invariant under 
transformations of \Eqn{eq:gauge_symmetry}
and 
\Eqn{eq:global_symmetry}.

$\bullet$ If $G=1$ is trivial and the gauge group $N$ is nontrivial, then we have a pure gauge theory with topological order.
The reduced inhomogeneous cocycle $\cV^{G,N}= \omega^{N}$ alone is \emph{not} gauge invariant,  
the wavefunction $\Phi(\{n_{ij}\}_M)$ is \emph{not} gauge invariant, either.
Even the ground state vector $|\Phi\rangle \propto  \sum_{\{ n_{ij}\}_M} \Phi(\{ n_{ij}\}_M) |\{ n_{ij}\}_M \rangle$
is  \emph{not}  gauge invariant,
and is  \emph{not}  gauge invariant up to a $U(1)$ phase.
Namely, each wavefunction obtains a different $U(1)$ phase $e^{\ii \theta(\{n_{ij}\}_M, \mathbf n_i)}$
that depends on the input $\{n_{ij}\}_M$ and gauge transformation $\mathbf n_i$, i.e.
$\Phi(\{n_{ij}\}_M) \to e^{\ii \theta(\{n_{ij}\}_M, \mathbf n_i)}\Phi(\{n_{ij}\}_M)$.
We define such a gauge transformed state vector as  $|\Phi\rangle \to | \Phi(\mathbf n_i)\rangle$.
However,  
as long as  any physical observable $\langle \widehat{O} \rangle =\langle \Phi | \widehat{O}| \Phi \rangle $ is strictly gauge invariant  as we show below,\footnote{
Recall that the gauge transformation can be implemented on the basis (a vector in the Hilbert space) or 
on the wavefunction (effectively a ``covector'').
The operator $\widehat{O}$ can be also implemented either on the basis as
$$
\widehat{O} | \{n_{ij}\}_M\rangle=  \sum_{ \{n_{ij}'\}} c_{\{n_{ij}'\}}^{\{n_{ij}\} } | \{n_{ij}'\}_M\rangle,
$$
or on the wavefunction
$$
\widehat{O} \Phi(\{n_{ij}\}_M)=  \sum_{\{\tilde{n}_{ij}\}} c_{\{{n}_{ij}\}}^{\{\tilde{n}_{ij}\} } \Phi(\{\tilde{n}_{ij}\}_M).
$$
In either case, we obtain the same consistent result for $\widehat{O}$ acting on the state vector $| \Phi \rangle$ as in \Eqn{eq:gaugeO}.
}
the theory is well-defined.
We find that $\langle \widehat{O} \rangle$ is indeed gauge invariant,
\bea
\langle \Phi | \widehat{O}| \Phi \rangle=
\sum_{\{{n}_{ij}\}} \sum_{\{\tilde{n}_{ij}\}} 
\Phi^{\dagger}(\{{n}_{ij}\}_M)  c_{\{{n}_{ij}\}}^{\{\tilde{n}_{ij}\} } \Phi(\{\tilde{n}_{ij}\}_M) 
= \langle \Phi(\mathbf n_i) | \widehat{O}|   \Phi(\mathbf n_i) \rangle,
\eea
where we have considered a generic operator $\widehat{O}$ defined by its operation on $| \Phi \rangle$:
\bea \label{eq:gaugeO}
\widehat{O}| \Phi \rangle =\widehat{O} \sum_{\{{n}_{ij}\}} \Phi(\{n_{ij}\}_M) | \{n_{ij}\}_M\rangle = 
\sum_{\{{n}_{ij}\}} \sum_{\{\tilde{n}_{ij}\}}
 c_{\{{n}_{ij}\}}^{\{\tilde{n}_{ij}\} } \Phi(\{\tilde{n}_{ij}\}_M) | \{n_{ij}\}_M\rangle
\eea
with generic $c_{\{{n}_{ij}\}}^{\{\tilde{n}_{ij}\} }$ coefficients.

\item 
\emph{Wavefunctions and their independence of input $g^*$ and $n_{i_0 i_*}$}:
Consider a wavefunction on a closed space $M$ defined in \Eqn{eq:SETwavefunction}. \\
$\bullet$ SPT wavefunction ${\Phi(\{g_i \}_M)}_{\text{SPT}}$ is \emph{independent} of the input choice $g^*$. Namely, changing $g^*$ to ${g'}^*\equiv (\mathbf g)^{-1} g^*$
\bea \label{eq:wavefunction_indept_g*}
{\Phi(\{g_i \}_M)}_{\text{SPT}} &=&\prod_{\{...\}} \nu^{G \; s_{i_0 \dots i_*}}_{d}(g_{i_0},\cdots, g_{i_{d-1}}, g^*)
  =\prod_{\{...\}} \nu^{G \; s_{\dots} }_{d}((\mathbf g) g_{i_0},\cdots, (\mathbf g)g_{i_{d-1}},  {g}^*)\nonumber \\
   &=&\prod_{\{...\}} \nu^{G \; s_{ \dots }}_{d}(g_{i_0},\cdots, g_{i_{d-1}}, (\mathbf g)^{-1} {g}^*)
   =\prod_{\{...\}} \nu^{G \; s_{\dots }}_{d}(g_{i_0},\cdots,g_{i_{d-1}}, {g'}^*).
\eea
Here we use the fact that ${\Phi(\{g_i \}_M)}_{\text{SPT}}$ is $G$-global symmetry invariant in the second equality. 
This proof 
$\frac{{\Phi(\{(\mathbf g)g_i \}_M)}_{\text{SPT}}}{{\Phi(\{g_i \}_M)}_{\text{SPT}}}=1$
requires the use of a $G$-cocycle condition, 
and we will show a complete proof in Sec.~\ref{sec:symmetry-preserving-wavefunction},
even in the presence of a gapped boundary/interface.
We also use that $\nu^{G}_{d}(\{ g_{i}\})=\nu^{G}_{d}(\{ (\mathbf g)^{-1} g_{i}\})$ due to the property of a homogeneous cocycle in the third equality.
One quick way to visualize this proof eqn.~(\ref{eq:wavefunction_indept_g*}) is that the ratio $\frac{{\Phi(\{(\mathbf g)g_i \}_M)}_{\text{SPT}}}{{\Phi(\{g_i \}_M)}_{\text{SPT}}}$
yields a term equivalent to a product of coboundary terms; fortunately the overall coboundary terms on a closed space $M$ must cancel out to be 1.\\
$\bullet$ Topological order and SET wavefunction $ \Phi(\{g_i, n_{ij}\}_M)_{\text{SET}}$ can be defined in such a way that it is 
\emph{independent} of the input $g^*$ and $n_{i_0 i_*}$. It is easier to prove that if we design and evaluate \Eqn{eq:SETwavefunction} in terms of homogeneous $G$ cocycles.
Below we show that replacing $g^*\to {g'}^*\equiv (\mathbf g)^{-1} g^*$ and $n_{i_* i_0} \to n'_{i_* i_0}\equiv (\mathbf n) n_{i_* i_0}$,
with a slight reordering of vertex indices and branch structure for our convenience, 
 the $\Phi(\{g_i, n_{ij}\}_M)_{\text{SET}}$ is still invariant:
\bea \label{eq:proofSETgindept}
&&\Phi(\{g_i, n_{ij}\}_M)_{\text{SET}} =\prod_{\{...\}} \nu^{G \; s_{i_* \dots i_{d-1}}}_{d}(g^*,  n_{i_*i_0} g_{i_0} , n_{i_*i_0 }n_{i_0 i_1} g_{i_1}, \dots,  n_{i_*i_0 }n_{i_0 i_1} \dots 
 n_{i_{d-2} i_{d-1}} g_{i_{d-1}}  ) \nonumber\\
 &&=\prod_{\{...\}} \nu^{G \; s_{ \dots }}_{d}(g^*,  (\mathbf g) n_{i_*i_0} g_{i_0} , (\mathbf g) n_{i_*i_0 }n_{i_0 i_1} g_{i_1}, \dots, (\mathbf g) n_{i_*i_0 }n_{i_0 i_1} \dots 
 n_{i_{d-2} i_{d-1}} g_{i_{d-1}}  ) \nonumber\\
 &&=\prod_{\{...\}} \nu^{G \; s_{ \dots}}_{d}( {g'}^*,  n_{i_*i_0} g_{i_0} , n_{i_*i_0 }n_{i_0 i_1} g_{i_1}, \dots,  n_{i_*i_0 }n_{i_0 i_1} \dots 
 n_{i_{d-2} i_{d-1}} g_{i_{d-1}}  ) \mid_{ {g'}^*\equiv(\mathbf g)^{-1}  g^*}.
\eea
\bea  \label{eq:proofSETnindept}
&&\Phi(\{g_i, n_{ij}\}_M)_{\text{SET}} =\prod_{\{...\}} \nu^{G \; s_{i_* \dots i_{d-1}}}_{d}(g^*,  n_{i_*i_0} g_{i_0} , n_{i_*i_0 }n_{i_0 i_1} g_{i_1}, \dots,  n_{i_*i_0 }n_{i_0 i_1} \dots 
 n_{i_{d-2} i_{d-1}} g_{i_{d-1}}  ) \nonumber\\
 &&=\prod_{\{...\}} \nu^{G \; s_{ \dots }}_{d}(g^*,  (\mathbf n) n_{i_*i_0} g_{i_0} , (\mathbf n) n_{i_*i_0 }n_{i_0 i_1} g_{i_1}, \dots, (\mathbf n) n_{i_*i_0 }n_{i_0 i_1} \dots 
 n_{i_{d-2} i_{d-1}} g_{i_{d-1}}  ) \nonumber\\
 &&=\prod_{\{...\}} \nu^{G \; s_{ \dots}}_{d}(g^*,  n'_{i_*i_0} g_{i_0} , n'_{i_*i_0 }n_{i_0 i_1} g_{i_1}, \dots,  n'_{i_*i_0 }n_{i_0 i_1} \dots 
 n_{i_{d-2} i_{d-1}} g_{i_{d-1}}  ) \bigg\rvert_{n'_{i_* i_0}\equiv (\mathbf n) n_{i_* i_0}}.
\eea
 The $\Phi(\{g_i, n_{ij}\}_M)_{\text{SET}}$ becomes that of topological order $\Phi(\{ n_{ij}\}_M)_{\text{TO}}$ if we set all $g =1$ for the trivial $G$.
 The proofs in \Eqn{eq:proofSETgindept} and \Eqn{eq:proofSETnindept} again require the use of a $G$-cocycle condition and the property of a homogeneous cocycle.

\item 
\emph{Local unitary transformation and the Hamiltonian}:
We can define a unitary transformation $\widehat U$ as
\bea \label{eq:Utransf}
\widehat U=
 \sum_{\{g_i, n_{ij}\}_M} \prod_{\{...\}} \cV^{G,N}_{d}(g_{i_0},\cdots,g^*; n_{i_0i_1}, n_{i_1i_2},\cdots, n_{i_{d-1} i_*})
\ket{ \{g_i, n_{ij}\}_M }\bra{ \{g_i, n_{ij}\}_M }.  
\eea
We can view that
the above $\cV^{G,N}_{d}$ is a $U(1)$ complex phase determined by local input data $\{g_{i_0}, \cdots; n_{i_0i_1}, \cdots\}$ 
that are given within a local $(d-1)$-simplex.
Since the $\widehat U$ sends the input state $\ket{ \{g_i, n_{ij}\}_M }$ to the same output state.
The overall $U(1)$ phase is determined by $\prod_{\{...\}} \cV^{G,N}_{d}$, which is a product of $U(1)$ phases assigned to each $(d-1)$-simplex.

$\bullet$ For SPTs, it is
\bea
&&\widehat U=
 \sum_{\{g_i\}_M} \prod_{\{...\}} \nu^{G}_{d}(g_i,...,g^*)\ket{\{g_i\}_M }\bra{\{g_i\}_M }.
\eea
For SPTs, actually this $\widehat U$ is a \emph{local unitary transformation} (LUT), because this $\widehat U$ is formed by a local circuit of many independent $\nu^{}_{d+1}$ on each local simplex. 
Overall $\widehat U$ is a unitary diagonal matrix acting on the full Hilbert space with diagonal elements
assigned with distinct $U(1)$ phases.
Under this LUT, 
the SPT's $| \Phi \rangle$ is deformed to $U^\dagger | \Phi \rangle=| \Phi_0 \rangle$ of \Eqn{eq:trivial_product_state} as a \emph{trivial product state}.
{However, such a LUT locally breaks the global $G$ symmetry of SPTs, because each $\nu^{G}_{d}(g \cdot g_i,...,g^*)$ is not $g$-invariant with a fixed $g^*$.}
The LUT can deform such a \emph{short-range entangled} state of SPTs to a trivial product state at the cost of breaking its global $G$ symmetry.

The SPT Hamiltonian (without the $\widehat{B}_f$ term) can be rewritten as 
\bea
\widehat{H}=\sum_{j} \widehat U \widehat{H}_j \widehat U^\dagger=-\sum_{j} \widehat U \ket{\phi_j}\bra{\phi_j} {\widehat U}^\dagger
=\sum_{j}  \widehat U(-\sum_{g_j, g'_j \in G}  \ket{g_j} \bra{g'_j}) {\widehat U}^\dagger. 
\eea 
The $\ket{\phi_j}=\sum_{g_j\in G}\ket{g_j}$ is an equal-weight sum of all states for all ${g_j}$ on each site.

$\bullet$ For topological orders/SETs, the $\widehat U$ defined in \Eqn{eq:Utransf} is not unitary for the total Hilbert space 
$\mathscr{H}_{\text{total}}=\otimes_j \mathscr{H}_j \otimes_{ij} \mathscr{H}_{ij}$, 
because $\cV^{G,N}_{d}(n_{12}, n_{23}, \dots)$ is defined to be $0$ when a closed loop $n_{12}n_{23}n_{31}\neq 1$.
We can artificially redefine $\widehat U'$ to design those zero $\cV^{G,N}_{d}$ terms to be $1$ by hand, and make $\widehat U'$ a new unitary matrix.
For example, one such unitary deformation sends to 
$${U'}^\dagger| \Phi \rangle = \mathrm{P}[  \sum_{\{g_i, n_{ij}\}_M} |\{g_i, n_{ij}\}_M \rangle]= \mathrm{P}[\otimes_i (\sum_{g_i} | g_i \rangle) \otimes_{ij} (\sum_{ n_{ij} } | n_{ij}  \rangle)],$$
where $\mathrm{P}$ is a projection operator imposing the zero flux condition through a closed loop as $n_{12}n_{23}n_{31}= 1$,
and $\mathrm{P}$ projects out any $n_{12}n_{23}n_{31}\neq 1$ state.
However, this final state is very different from a trivial product state, e.g.
$\otimes_i (\sum_{g_i} | g_i \rangle) \otimes_{ij} (\sum_{ n_{ij} } | n_{ij}  \rangle)$.
Regardless of how we design a unitary $\widehat U'$ matrix,
we \emph{cannot deform the ground state} $| \Phi \rangle$ \emph{of topological orders/SETs to a trivial product state through any local unitary transformation}.
This reason is due to a super-posed extended loop states as ground states of intrinsic topological orders are highly \emph{long-range entangled} ---
their information encoded in the projection $\mathrm{P}$ on the zero flux condition is incompatible with a trivial product state.
The LUT cannot deform a \emph{long-range entangled} state to a trivial product state.
Thus topological orders/SET Hamiltonian cannot be rewritten as $\widehat{H}=\sum_{j} \widehat U' \widehat{H}_j \widehat {U'}^\dagger$, for any unitary $\widehat {U'}$
and for some local Hamiltonian $\sum_{j} \widehat{H}_j$ whose ground state is a trivial product state.


\begin{figure}[tb]
\begin{center}
\includegraphics[scale=0.8]{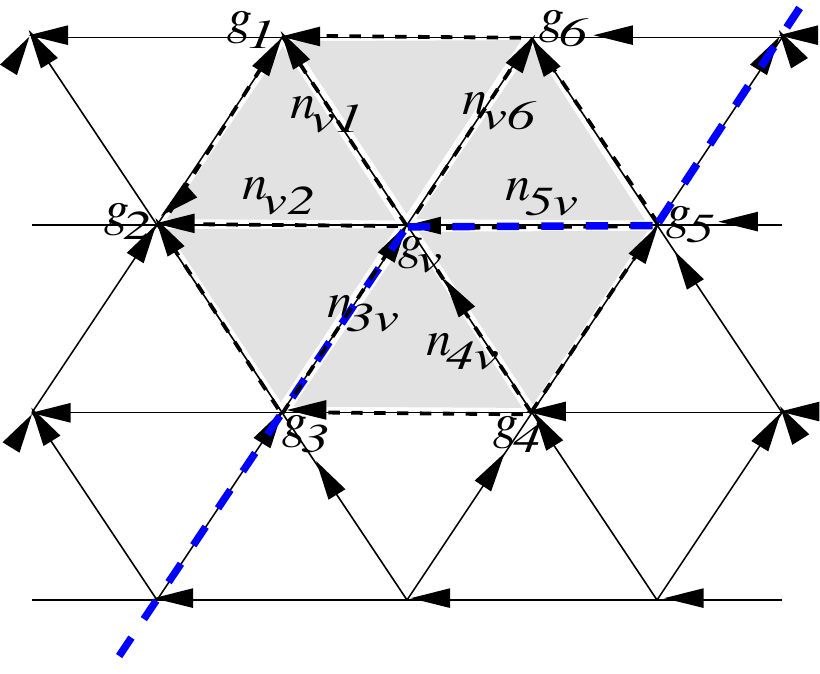}
\end{center}
\caption{An example of line operator
$\widehat{W}_{\CU}^{S^1_y} 
\equiv  \prod_{v} 
\widehat{\CU}_{\{g_v, g_{i}^{(v)}\}}^{
\{ n_{vi}^{(v)}, n_{ij}^{(v)}\} }$ acts along the blue dashed line.
The product of $v$ spans along all the vertices on the blue dashed line.
One of the most generic operators $\widehat{\CU}_{\{g_v, g_{i}^{(v)}\}}^{
\{ n_{vi}^{(v)}, n_{ij}^{(v)}\} }$ on this lattice centered at a vertex $v$ acts on a local Hilbert space of 7 $G$-vertices and 12 $N$-links on a shaded honeycomb region, thus
it acts on a Hilbert space of dimensions $|G|^7 |N|^{12}$.}
\label{fig:Woperator}
\end{figure}

\item 
\emph{Degenerate ground states with holonomies around non-contractible cycles}:
So far we focus only on a ground state $| \Phi \rangle$ that has \emph{no} holonomies around non-contractible cycles,
and that can be deformed to a trivial product state.
However, for gauge theories of topological orders and SETs, we have distinct degenerate ground states when the spatial topology is nontrivial (e.g. a  2D spatial torus $T^2_{xy}$).
Start from $| \Phi \rangle$, we can generate other degenerate ground states by inserting \emph{extended} operators 
as holonomies around non-contractible cycles. 
Without losing generality, let us consider a 2+1D system; we have generic line operators $\widehat{W}_{\CU}^{S^1}$ in a 2D spatial torus $T^2_{xy}$ with coordinates $x$ and $y$.
We can fully generate distinct ground states spanning the dimensions of Hilbert space on $T^2_{xy}$
by
\bea
\widehat{W}_{\CU}^{S^1_y} | \Phi \rangle \equiv  \prod_{v} 
\widehat{\CU}_{\{g_v, g_{i}^{(v)}\}}^{
\{ n_{vi}^{(v)}, n_{ij}^{(v)}\} }
| \Phi \rangle. 
\eea
Here ${S^1_y}$ in $\widehat{W}_{\CU}^{S^1_y}$ means that the line operator has a cycle around ${S^1_y}$, so the $\prod_{v}$ means a series of vertices $v$ spanning around
the ${S^1_y}$-cycle, for example along the blue dashed line in Fig.\ref{fig:Woperator}.
The $\widehat{\CU}_{\{g_v, g_{i}^{(v)}\}}^{
\{ n_{vi}^{(v)}, n_{ij}^{(v)}\} }$
is a shorthand expression
$\widehat{\CU}_{g_v, g_{1}^{(v)}, g_{2}^{(v)}, g_{3}^{(v)},g_{4}^{(v)}, g_{5}^{(v)}, g_{6}^{(v)}}^{
{
n_{v1}^{(v)},n_{v2}^{(v)}, n_{3v}^{(v)}, n_{4v}^{(v)},  n_{5v}^{(v)}, n_{v6}^{(v)}
},
{
n_{21}^{(v)},n_{32}^{(v)}, n_{43}^{(v)}, n_{45}^{(v)},  n_{56}^{(v)}, n_{61}^{(v)} } }$,
which acts on the honeycomb shaded region in Fig.\ref{fig:Woperator}.
Examples of $\widehat{\CU}_{\{g_v, g_{i}^{(v)}\}}^{
\{ n_{vi}^{(v)}, n_{ij}^{(v)}\} }$ include the $\widehat{A}_v^{g,n}$ and $\widehat{B}_f$ terms.
For example, for a $Z_2$ toric code \cite{KitaevToric} on a $T^2$ torus,
the expression for degenerate ground states $\widehat{W}_{\CU}^{S^1_y} | \Phi \rangle$ boils down to
$$
(\prod \sigma_{z})^{q} (\prod \sigma_{x})^{m} | \Phi \rangle,
$$
where $\sigma_{x}$ and $\sigma_{z}$ are the rank-2 Pauli matrices.
The product $\prod$ is along the ${S^1_y}$ line operator.
The $(q,m)$ are integer mod 2 values, with $(q,m)=(0,0),(1,0),(0,1),(1,1)$ are four distinct ground states.
Moreover, a generic $\widehat{\CU}_{\{g_v, g_{i}^{(v)}\}}^{
\{ n_{vi}^{(v)}, n_{ij}^{(v)}\} }$ does \emph{not} need to commute with $\widehat{A}_v^{g,n}$ and $\widehat{B}_f$,
and it can violate the zero flux condition of Remark 2.  Thus such a 
$\widehat{\CU}_{\{g_v, g_{i}^{(v)}\}}^{
\{ n_{vi}^{(v)}, n_{ij}^{(v)}\} }$ can
create \emph{anyon excitations} that cost higher energy.

\end{enumerate}

We can easily generalize the above discussion (2+1D)  to any spacetime dimension.

\subsubsection{Anomalous symmetry-preserving gapped boundary/interface of bulk SPTs and SETs}
\label{sec:generalGBDWHamiltonian}

Continued from Sec.~\ref{sec:generalSETHamiltonian}, we develop further to formulate a lattice wavefunction and Hamiltonian for topological phases with 
gapped boundaries/interfaces.
We first focus on a bulk $G$-SPTs on an open manifold while the gapped boundary has an anomalous $H$-SPTs that cannot exists without an extended bulk,
via a group extension $H/K=G$ in Sec.~\ref{sec:ZSPTSETG}. 
Along the way, we comment how to easily generalize to a bulk with SETs.

\begin{itemize}
\item
\emph{Wavefunction}: For wavefunction, we can simply adopt the $G$-SPT limit of \Eqn{eq:SETwavefunction} as
$\Phi(\{g_i\}_M) \equiv\prod_{\{...\}} \nu^{G \; s_{i_0 \dots i_*}}_{d}(g_{i_0},\cdots,g^*)$ defined first on a closed space $M \equiv M^{d-1}$ of $(d-1)$-spatial dimensions.
The $g^*$ is again some fixed value outside the $M^{d-1}$.    
We would like to keep the degrees of freedom on each site with Hilbert space dimensions $|G|$ on the gapped left region $R_{\I}$, and
 extend the site's Hilbert space dimensions to $|H|$ on the gapped right region $R_{\II}$
as well as on the interface $\partial R$ ($\equiv \partial R_{\I} \equiv \partial R_{\II}$ up to an orientation).
We denote the group element in $H$ assigned along $\partial R$ as $h^\partial \in H$.
We also extend the Hilbert space dimensions of $i_*$ from $|G|$ to $|H|$, and we choose $r(h^*)=g^*$.
The modified wavefunction defined on $M= R_{\I}  \cup R_{\II}$ is 
\bea
&&\Phi(\{g_{i}, h_j\}) \equiv \Phi(\{g_{i}\}_{R_{\I}}, \{h_j^{\partial}\}_{\partial R},  \{h_j\}_{R_{\II}})\nonumber\\
&&=\prod_{\{...\}} \nu^{G \; s_{i_a \dots i_*}}_{d}(\{g_{i_a}\}_{R_{\I}}, r(h^*))
\cdot\prod_{\{...\}} \nu^{G \; s_{i_a j_b \dots i_*}}_{d}(\{g_{i_a}\}_{R_{\I}}, \{ r(h_{j_b}^{\partial})\}_{\partial R}, r(h^*)) \nonumber\\
&&\;\;\;\;\; 
\cdot \prod_{\{...\}} \nu^{G \; s_{j_a j_b \dots i_*}}_{d}(\{ r(h_{j_a}^{\partial})\}_{\partial R}, \{ r(h_{j_b})\}_{R_{\II}},  r(h^*)) \label{eq:splitGSPTwavefunc1}\\
&&=\Big( \prod_{\{...\}} \nu^{G \; s_{i_a \dots i_*}}_{d}(\{g_{i_a}\}_{R_{\I}}, r(h^*))
\cdot \prod_{\{...\}} \nu^{G \; s_{i_a j_b \dots i_*}}_{d}(\{g_{i_a}\}_{R_{\I}}, \{ r(h_{j_b}^{\partial})\}_{\partial R}, r(h^*)) \Big) \nonumber\\
&&\;\;\;\;\; 
\cdot \Big( \prod_{\{...\}} \mu^{H \; s_{j_a \dots i_*}}_{d-1}(\{ h_{j_a}^{\partial}\}_{\partial R},   r(h^*)) \Big)
 \Big( \prod_{\{...\}} \mu^{H \; s_{j_a j_b \dots i_*}}_{d-1}(\{ h_{j_a}^{\partial}\}_{\partial R}, \{ h_{j_b}\}_{R_{\II}}) \Big)
 \;\;\;\;\;\;\;\;\;\;\; \label{eq:splitGSPTwavefunc2}\\
 && \equiv \Phi_{R_{\I}}(\{g_{i}\},\{h_j^{\partial}\}) \; \Phi_{\partial R}(\{h_j^{\partial}\})   \;
 \Phi_{R_{\II}}(\{h_j^{\partial}\}, \{h_j\}).
 \label{eq:splitGSPTwavefunc3}\\
 &&\xrightarrow{\;\;\text{LUT}\;\;}  \Big( \prod_{\{...\}} \nu^{G \; s_{i_a \dots i_*}}_{d}(\{g_{i_a}\}_{R_{\I}}, r(h^*))
\cdot\prod_{\{...\}} \nu^{G \; s_{i_a j_b \dots i_*}}_{d}(\{g_{i_a}\}_{R_{\I}}, \{ r(h_{j_b}^{\partial})\}_{\partial R}, r(h^*))  \Big) \nonumber\\
&&\;\;\;\;\; 
\cdot  \Big( \prod_{\{...\}} \mu^{H \; s_{j_a \dots i_*}}_{d-1}(\{ h_{j_a}^{\partial}\}_{\partial R},   r(h^*))  \Big)
 \;\;\;\;\;\;\;\;\;\;\; \label{eq:splitGSPTwavefunc4}\\
  && \equiv \Phi_{R_{\I}}(\{g_{i}\},\{h_j^{\partial}\}) \; \Phi_{\partial R}(\{h_j^{\partial}\}) \label{eq:splitGSPTwavefunc5}
\eea
where we have split the above $H$-coboundary $\nu^{G}_d(r(h))=\nu^{H}_d(h)$ in \Eqn{eq:splitGSPTwavefunc1} into $H$-cochains
$\mu^{H}_{d-1}$ in \Eqn{eq:splitGSPTwavefunc2}.
We define 
\bea
&&\Phi_{R_{\I}}(\{g_{i}\},\{h_j^{\partial}\}) \equiv
\Big( \prod_{\{...\}} \nu^{G \; s_{i_a \dots i_*}}_{d}(\{g_{i_a}\}_{R_{\I}}, r(h^*))
\cdot \prod_{\{...\}} \nu^{G \; s_{i_a j_b \dots i_*}}_{d}(\{g_{i_a}\}_{R_{\I}}, \{ r(h_{j_b}^{\partial})\}_{\partial R}, r(h^*)) \Big), \nonumber\\
&&\Phi_{\partial R}(\{h_j^{\partial}\}) \equiv
\Big( \prod_{\{...\}} \mu^{H \; s_{j_a \dots i_*}}_{d-1}(\{ h_{j_a}^{\partial}\}_{\partial R},   r(h^*)) \Big), \nonumber\\
&& \Phi_{R_{\II}}(\{h_j^{\partial}\}, \{h_j\})  
\equiv
 \Big( \prod_{\{...\}} \mu^{H \; s_{j_a j_b \dots i_*}}_{d-1}(\{ h_{j_a}^{\partial}\}_{\partial R}, \{ h_{j_b}\}_{R_{\II}}) \Big).
\eea
 Notice that $\Phi_{R_{\II}}(\{h_j^{\partial}\}, \{h_j\})$ is simplified to no dependence on $h^*$ because those $\mu^H_{d-1}$ 
 that depend on $h^*$ are pair cancelled out due to overlapping on the same $(d-1)$-simplex with opposite orientations $\pm 1$.
From \eqn{eq:splitGSPTwavefunc3} to \eqn{eq:splitGSPTwavefunc4}, the notation ``$\xrightarrow{\text{LUT}}$'' means that
we \emph{do a local unitary transformation} (LUT) {to deform} $\Phi_{R_{\II}}$ \emph{to a gapped trivial product state} $\Phi_{R_{\II}}=1$ \emph{without breaking any symmetry}.
 Thus, the simplified nontrivial wavefunction only resides on ${R_{\I}}$ and ${\partial R}$ as 
 $\Phi(\{g_{i}, h_j\}) \equiv \Phi_{R_{\I}}(\{g_{i}\},\{h_j^{\partial}\}) \; \Phi_{\partial R}(\{h_j^{\partial}\})$.
  
For example, more explicitly in 2+1D,
\begin{multline} \label{eq:splitGSPTwavefunc2+1D}
\Phi(\{g_{i}, h_j\}) \equiv \Phi_{R_{\I}}(\{g_{i}\},\{h_j^{\partial}\}) \; \Phi_{\partial R}(\{h_j^{\partial}\})\\
=\prod_{ {\{...\}} } \nu_{3}^{G \; s }( g_{i_1},  g_{i_2},  g_{i_3}, r(h^*)) 
\nu_{3}^{G  \; s}( r(h_{j_1}^{\partial}),   g_{i_2},  g_{i_3}, r(h^*))
\nu_{3}^{G \; s }( r(h_{j_1}^{\partial}),  r(h_{j_2}^{\partial}),  g_{i_3}, r(h^*)) \\
  \mu_{2}^{H \; s}( h_{j_1}^{\partial},   h_{j_2}^{\partial},  h^*)
 \mu_{2}^{H \; s}( h_{j_1}^{\partial},   h_{j_2}^{\partial},  h_{j_3})
  \mu_{2}^{H \; s}( h_{j_1}^{\partial},  h_{j_3},  h_{j_4})
  \mu_{2}^{H \; s}(  h_{j_3},  h_{j_4}, h_{j_5}) \\
  \xrightarrow{\;\;\text{LUT}\;\;} \prod_{ {\{...\}} } \nu_{3}^{G \; s }( g_{i_1},  g_{i_2},  g_{i_3}, r(h^*)) 
\nu_{3}^{G  \; s}( r(h_{j_1}^{\partial}),   g_{i_2},  g_{i_3}, r(h^*))
\nu_{3}^{G \; s }( r(h_{j_1}^{\partial}),  r(h_{j_2}^{\partial}),  g_{i_3}, r(h^*)) \\
\cdot \prod_{ {\{...\}} } \mu_{2}^{H \; s}( h_{j_1}^{\partial},   h_{j_2}^{\partial},  h^*). 
\end{multline}
Here the shorthand $s=\pm 1$ depends on the ordering of each assigned simplex.
We see that those $\mu_{2}^{H}$ that do not depend on $h^*$ can be 
\emph{deformed to a gapped trivial product state by local unitary transformation without breaking any symmetry} 
(again, we denote the procedure as ``$\xrightarrow{\text{LUT}}$'' ),
because the homogeneous cochain satisfies $\mu_{d-1}^{H }(  \{ (\mathbf h)\cdot h_{j} \})=\mu_{d-1}^{H }(  \{ h_{j} \})$.
Thus keeping only $ \mu_{2}^{H}( h_{j_1}^{\partial},   h_{j_2}^{\partial},  h^*)$ but removing other $\mu_{2}^{H}$,
we obtain the last simplified equality. In generic dimensions, we have \Eqn{eq:splitGSPTwavefunc5}.

\item
\emph{Lattice Hamiltonian}: The Hamiltonian for the above gapped ground state has the same form in the bulk region $R$ as $\widehat{H}=-\sum_v \widehat{A}_v-\sum_f \widehat{B}_f$ in
\Eqn{eq:SETHamiltonian}. However, we need to modify the boundary term on ${\partial R}$.
The {first} term $\widehat{A}_v$ on the boundary acts on the wavefunction of a constant-time slice through each vertex $v$ in the space by lifting 
 the {initial} state through an ``imaginary time'' evolution to a new state with a vertex $v'$ via
\bea
&& \widehat{A}_v=\frac{1}{|H|}\sum_{{h \in H}}\widehat{A}_v^{h}.\\
&&  
\widehat{A}_v^{h} |h_v, \; \{ h_j^\partial \},  \{g_i\} \rangle  \\
&& \;\;\;\;\;\; =\prod_{\{...\}} \nu^{G\; s_{...}}_{d}
(r(h), \; r(h_v), \{ r(h_j^\partial) \}, \{g_i\}) \prod_{\{...\}} 
\mu^{H\; s_{...}}_{d-1}( h, \; h_v, \{ h_j^\partial \} )  | h ,\; \{ h_j^\partial \},  \{g_i\}    \dots\rangle.\nonumber \;\;\; 
\eea

\begin{figure}[h!] 
\begin{center}
(1) \includegraphics[scale=0.8]{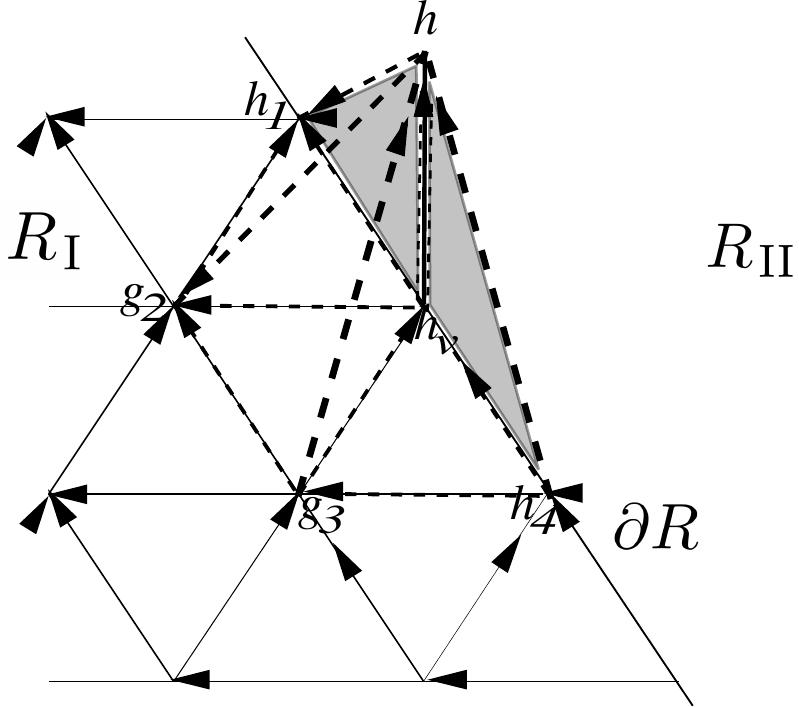} \;\;\;\; \;\;\;\; (2) \includegraphics[scale=0.8]{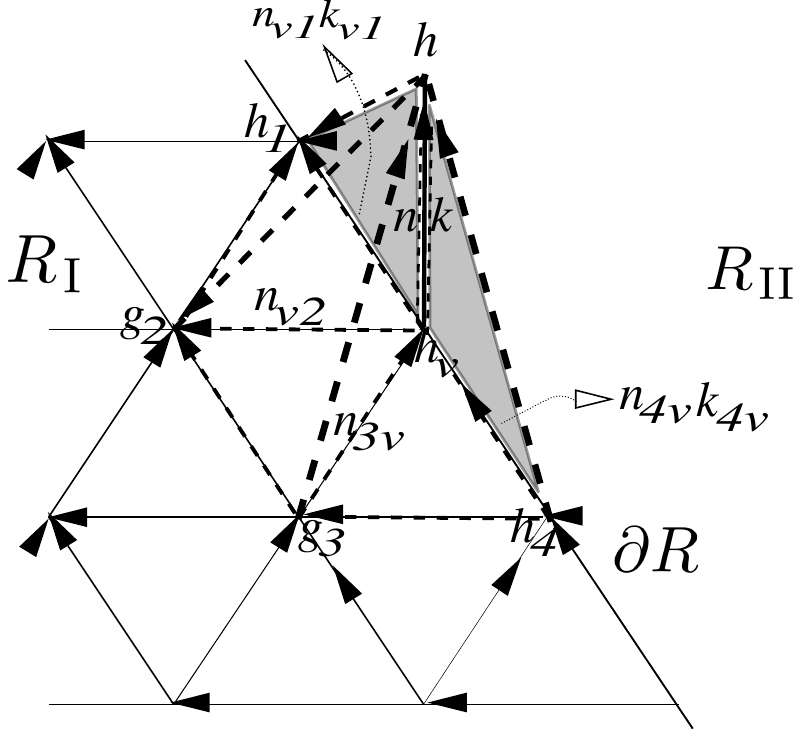}
\end{center}
\caption{(1) We consider a $G$-SPTs on the spatial region $R_{\I}$ with a lattice. We set a trivial vacuum on the spatial region $R_{\II}$, and 
 the gapped boundary of $H$-anomalous SPT on the boundary $\partial R$.
 The Hamiltonian $\widehat{A}_v^{h}$ acts on the state $|h_v,h_1,g_2,g_3,h_4\rangle$ and is given in \Eqn{eq:AvHGapBdry},
 which sends to a new state $| h, h_1,g_2,g_3,h_4\rangle$ with a $U(1)$ phase.
 (2) Now consider a $G$-SETs on the spatial region $R_{\I}$ lattice with a gapped boundary anomalous SETs, 
 the Hamiltonian $\widehat{A}_v^{h,n,k}$ is given in \Eqn{eq:AvGNHGapBdr}.
 }
 \label{fig:liftSPTgapbdryAv}
\end{figure}

More specifically, the effective 2+1D Hamiltonian term along the 1+1D gapped boundary $\partial R$, shown in Fig. \ref{fig:liftSPTgapbdryAv} (1), is written as:
\begin{multline}
\widehat{A}_v^{h} |h_v,h_1,g_2,g_3,h_4\rangle  \\
=\frac{{\mu^{H}_2(h_v,h,h_1)}\; {\mu^{H}_2(h_4,h_v,h) } } {
\nu^{G}_{3}(r(h_v), r(h), g_2, r(h_1) )\; \nu^{G}_{3}(g_3, r(h_v), r(h), g_2 ) \;
\nu^{G}_{3}(r(h_4),  g_3, r(h_v), r(h) )}  | h, h_1,g_2,g_3,h_4\rangle. 
 \label{eq:AvHGapBdry}
\end{multline}
The $\widehat{B}_f$ term imposes trivial $G$- and $H$- holonomies for the contractible loop.
But here $\widehat{B}_f$ does not play any role for SPTs, because SPTs always have trivial holonomy regardless the loop is contractible or not.

\item
\emph{More generic bulk/gapped boundary SET wavefunction and Hamiltonian}: 
We can consider more generic bulk SETs and boundary anomalous SETs as in Sec.~\ref{sec:ZSPTSETG} Remark 4 ---
a bulk SETs with global symmetry $Q$ and gauge symmetry ${N}$ via
$1 \to N \overset{\ta}{\to} G \to Q \to 1$, and a boundary anomalous SETs  with global symmetry $Q$ and gauge symmetry ${K \times N}$ via
$1 \to {K \times N}\to H \to Q \to 1$ where $\frac{H}{K \times N}=\frac{G}{N}\equiv Q$.
This also implies $1 \to {K}\to H \overset{r}{\to} G \to 1$.
The generic wavefunction is
 \bea  \label{eq:splitGSETwavefunc}
&&\Phi(\{g_{i}, n_{i_a i_b}, h_j, k_{j_a j_b}\}) \xrightarrow{\;\;\text{LUT}\;\;}  \Phi_{R_{\I}}(\{g_{i}\}, \{n_{i_a i_b}\}, \{h_j^{\partial}\}) \; 
\Phi_{\partial R}(\{h_j^{\partial}\}, \{n_{j_a j_b}\}, \{k_{j_a j_b}\}),\;\;\;\;\;\;\;\;\; \\
&& \text{where} \nonumber\\
&&
\Phi_{R_{\I}}(\{g_{i}\}, \{n_{i_a i_b}\}, \{h_j^{\partial}\}) \equiv
  \Big( \prod_{\{...\}} \cV^{G,N \; s_{ \dots i_*}}_{d} (\{g_{i}\}_{R_{\I}}, \{r(h_j^{\partial})\}_{\partial R}, r(h^*);  \{n_{i_a i_b}\}_{R_{\I}, {\partial R}} )  \Big),  \nonumber\\
&&
\Phi_{\partial R}(\{h_j^{\partial}\}, \{n_{j_a j_b}\}, \{k_{j_a j_b}\}) \equiv  \Big( \prod_{\{...\}} \mu^{H, N,K \; s_{ \dots i_*}}_{d-1}(\{ h_{j_a}^{\partial}\}_{\partial R},   r(h^*); \{n_{j_a j_b} k_{j_a j_b}\}_{\partial R})  \Big).
 \;\;\;\;\;\;\;\;\;\;\; \nonumber
\eea
Its Hamiltonian has the same form in the bulk region $R$ as $\widehat{H}=-\sum_v \widehat{A}_v-\sum_f \widehat{B}_f$ in
\Eqn{eq:SETHamiltonian}. But we need to modify the boundary term on ${\partial R}$ to
\bea
&& \widehat{A}_v=\frac{1}{|H||N| |K|}\sum_{{h \in H, n \in N, k \in K}}\widehat{A}_v^{h,n,k}.
\eea
\begin{multline}
\widehat{A}_v^{h,n,k} |h_v, \; \{ h_j^\partial \},  \{g_i\}; \{n_{i_a i_b}\}, \{k_{j_a j_b}\} \rangle  \\
 =\prod_{\{...\}} \cV^{G, N\; s_{...}}_{d}
(r(h), \; r(h_v), \{ r(h_j^\partial) \}, \{g_i\}; n, \{ n_{i_a i_b}\}) \\
 \prod_{\{...\}}
\mu^{H, N, K\; s_{...}}_{d-1}( h, \; h_v, \{ h_j^\partial \}; \{n_{j_a j_b}\}, \{k_{j_a j_b}\} )
 | h ,\; \{ h_j^\partial \},  \{g_i\} ; \{n'_{i_a i_b}\}, \{k'_{j_a j_b}\}\rangle.
\end{multline}
Here $n'_{i_a i_b}$ and $k'_{j_a j_b}$ are some modified link variables that may have $n$ and $k$ variables inserted.

{
The $\widehat{B}_f$ term imposes trivial holonomies for the contractible loops;
here, $\widehat{B}_f$ plays an important role to constrain ground states of SETs.
The bulk $\widehat{B}_f$ imposes trivial $G$- and $N$-holonomies for the contractible loops.
The boundary $\widehat{B}_f$ imposes trivial $H$-, $N$- and  $K$-holonomies for the contractible loops.
Similar to \Eqn{eq:Bf}, the bulk $\widehat{B}_f$ constrains that $(\delta_{n_{v2} n_{21} n_{1v}=1})$,
and the boundary $\widehat{B}_f$ constrains that $(\delta_{n_{v2} n_{21} n_{1v}=1})(\delta_{k_{v2} k_{21} k_{1v}=1})$ on each state vector associated to a 2-simplex triangle.
}

For example, more specifically, an effective 2+1D Hamiltonian term $\widehat{A}_v^{h,n,k}$ along the 1+1D anomalous SET gapped boundary $\partial R$, shown in Fig. \ref{fig:liftSPTgapbdryAv} (2), is written as:
\begin{multline}
\widehat{A}_v^{h,n,k} |h_v,h_1,g_2,g_3,h_4; n_{v1} k_{v1}, n_{v2}, n_{v3}, n_{4v} k_{4v} \rangle\\
=\frac{{\mu^{H}_2(h_v, n k h, n n_{v1} k k_{v1} h_1)}\; {\mu^{H}_2(h_4,  n_{4v} k_{4v} h_v, n_{4v} n k_{4v} k  h) } }{
\nu^{G}_{3}(r(h_v), \ta(n) r(h),\ta(n_{v2}) g_2, \ta(n_{v1}) r(h_1) )\; 
\nu^{G}_{3}(g_3,  \ta(n_{3v})  r(h_v), \ta(n_{3v} n) r(h), \ta(n_{3v} n_{v2} )  g_2 )}  \\
\frac{1}{\nu^{G}_{3}( r(h_4), \ta(n_{4v} n_{3v}^{-1})  g_3, \ta(n_{4v}) r(h_v), \ta(n_{4v} n) r(h) )} \\
 | h, h_1,g_2,g_3,h_4; n_{v1} n^{-1} k_{v1} k^{-1}, n^{-1} n_{v2}, n_{v3} n, n_{4v} n k_{4v} k \rangle. \label{eq:AvGNHGapBdr}
\end{multline}
Here $r(h) \in G$ and $r(h_{i_a}) \in G$  are aimed at emphasizing that they are obtained via the epimorphism $H \overset{r}{\to} G$.
The $\ta(n) \in G$ and $\ta(n_{i_a i_b}) \in G$ are aimed to emphasize that they are obtained via the monomorphism $N \overset{\ta}{\to} G$.
Since $N$ is a normal subgroup inside $G$, previously we have been abbreviating $\ta(n)=n \in G$ for $\forall n \in N$.
\end{itemize}

In the next section, we  analyze the symmetry-preserving property of such a gapped boundary system in Sec.~\ref{sec:symmetry-preserving-wavefunction}.

\subsubsection{Proof of the symmetry-preserving wavefunction with gapped boundary/interface} \label{sec:symmetry-preserving-wavefunction}

\begin{figure}[h!] 
\begin{center}
(1)\includegraphics[scale=0.7]{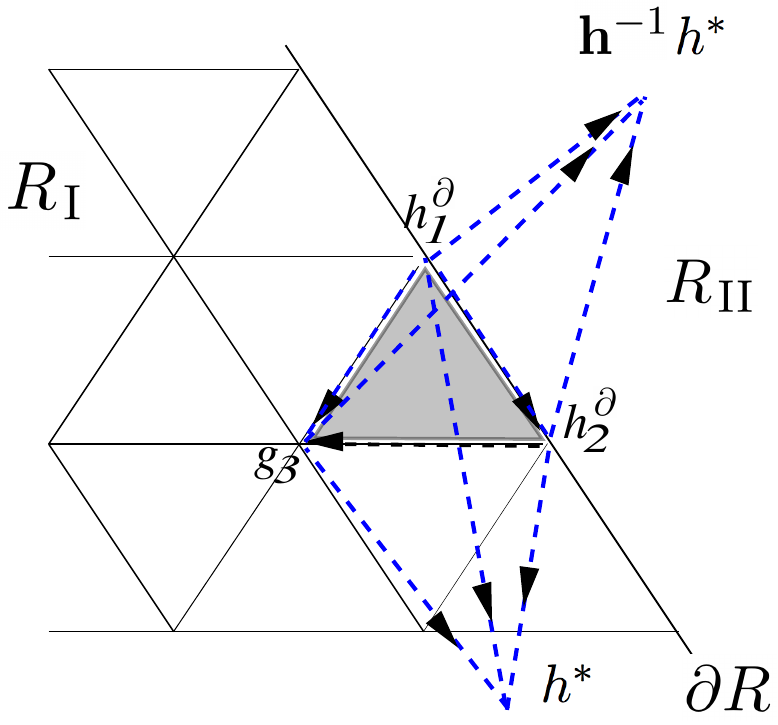} \;\;\;  (2)\includegraphics[scale=0.7]{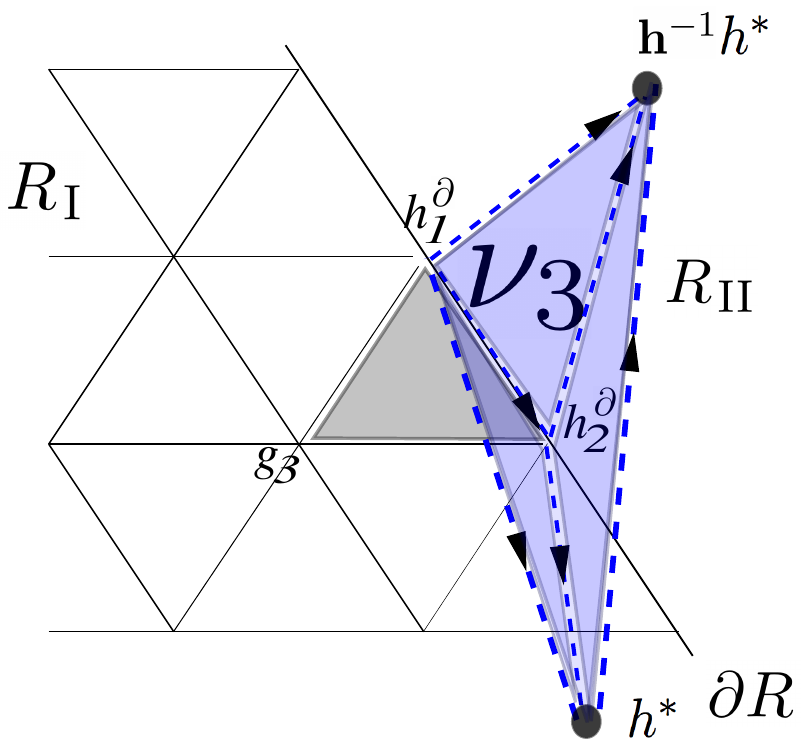}
\;\;\;  (3)\includegraphics[scale=0.7]{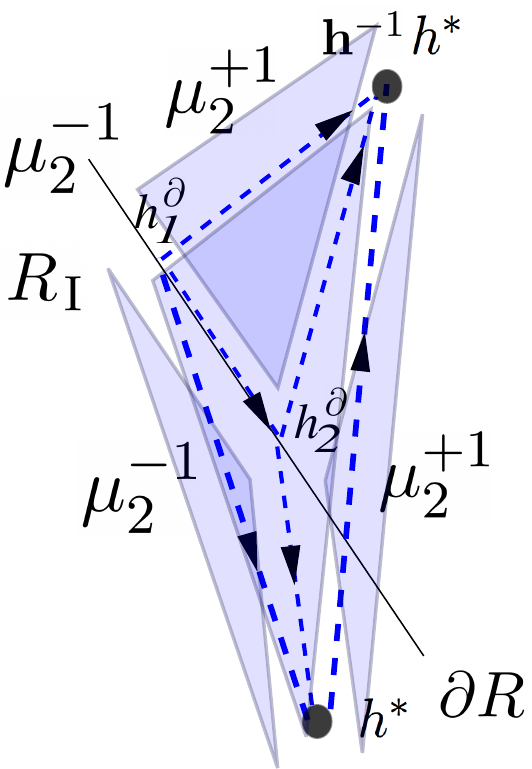}  
(4)\includegraphics[scale=0.7]{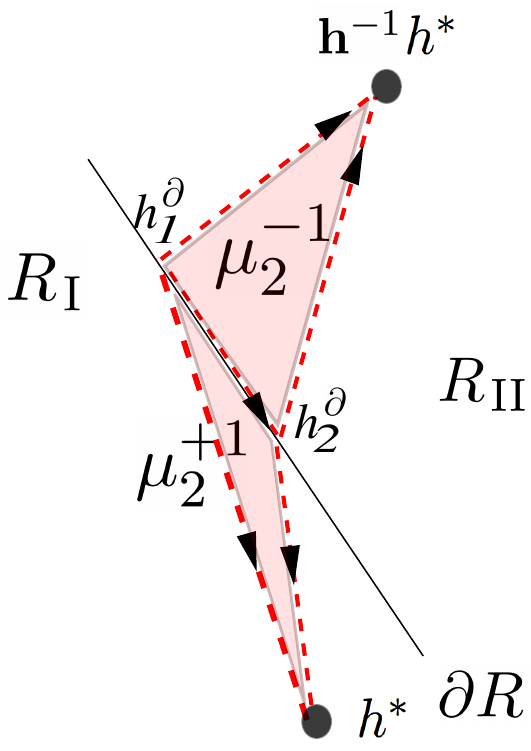} \;\;\;\;\;
(5)\includegraphics[scale=0.3]{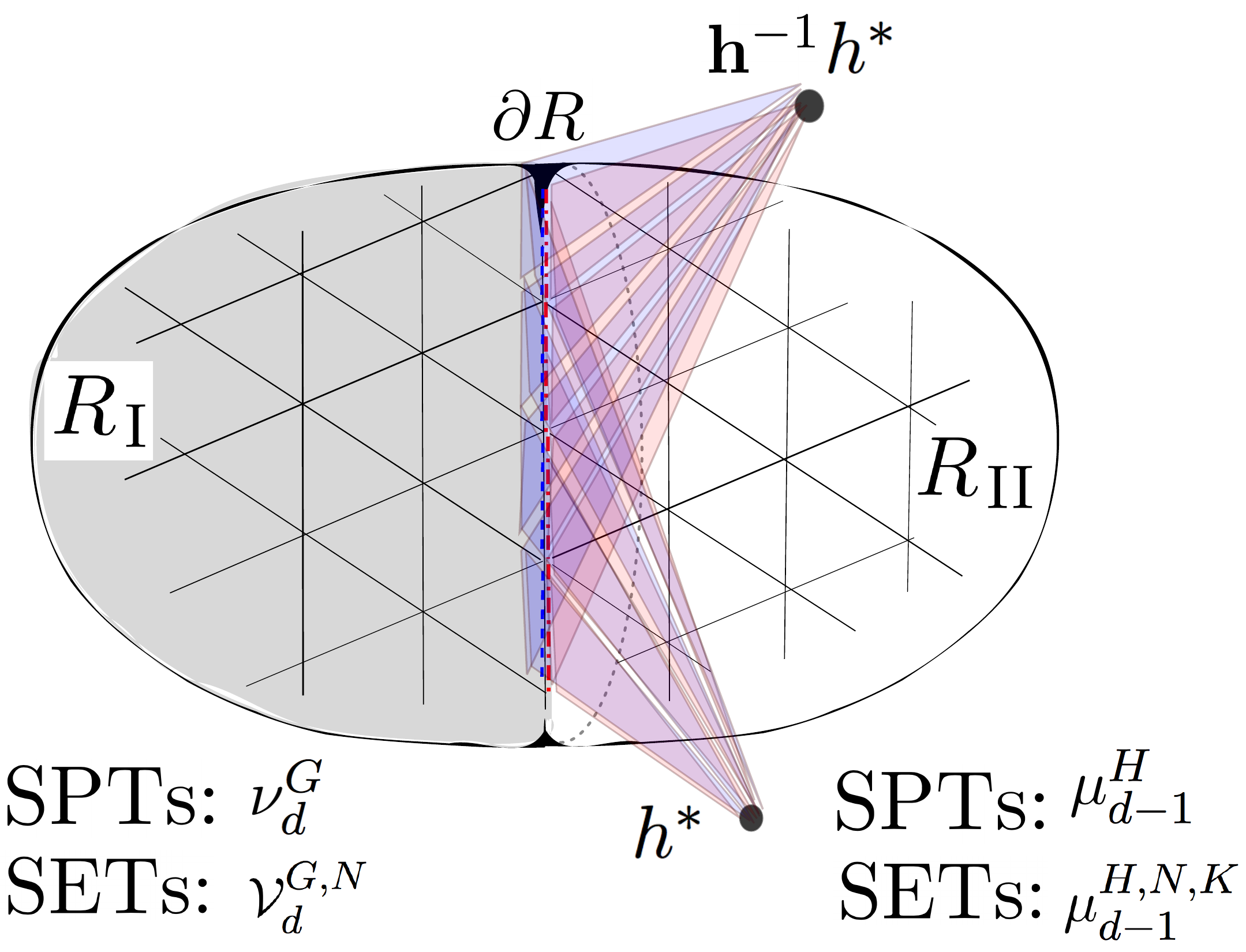} 
\end{center}
\caption{
We show geometry pictures how to understand the symmetry-transformation phase cancellation for the overall symmetry invariance in 2+1D/1+1D,
which can be easily generalized to any higher dimensional spacetime.
The fig.(1) shows how two pieces of $\nu_{3}$  in eqn.(\ref{eq:fig:sym-preserve1}) contribute to the left-region wavefunction $\Phi_{R_{\I}}$,
and then convert to the splitting of a $\nu_{3}$ into four pieces of 2-cochains in the fig.(2) and fig.(3) as in eqn.(\ref{eq:fig:sym-preserve2}).
The fig.(4) shows how two pieces of $\mu_{2}$  in eqn.(\ref{eq:sym_phase_mu2}) contribute to the interface-region wavefunction $\Phi_{\partial R}$.
The fig.(5) shows how, on a closed interface ${\partial R}$ (here an $S^1$), the
symmetry transformation on the combined wavefunction $\Phi_{R_{\I}} \cdot \Phi_{\partial R}$  canceling with each other to 1 as the symmetry invariance achieved in eqn.(\ref{eq:sym-cancel}).
}
\label{fig:proof:sym-preserve}
\end{figure}

Follow the setup in Sec.~\ref{sec:generalGBDWHamiltonian}, here we rigorously prove that the wavefunction \Eqn{eq:splitGSPTwavefunc5} of a bulk $G$-SPTs on an open manifold while the gapped boundary has an anomalous $H$-SPTs 
via a group extension $H/K=G$ (in Sec.~\ref{sec:ZSPTSETG}).
See Fig.\ref{fig:proof:sym-preserve} for a geometric illustration for the proof. 

We would like to interpret that the spatial bulk have two sectors $R_{\I} \equiv R_{\I}^d$ and $R_{\II} \equiv R_{\II}^d$, while the whole closed space is  $R_{\I}^d \cup R_{\II}^d= M^d$.
The SPTs of symmetry group $G$ is on the $R_{\I}$ side, a trivial vacuum is on the $R_{\II}$ side, 
while the gapped interface ($\equiv \partial R$) between the two phases is symmetry-enhanced to $H$.
This gapped $H$ interface can be viewed as a gapped boundary for the bulk $G$ SPTs.
Under the construction $1 \to K \to H \overset{r}{\longrightarrow} G \to 1$ 
of cocycle splitting, 
below we can have an exact
global $H$ symmetry transformation acting along the gapped interface, together with an exact
global $G$ symmetry transformation acting on the gapped left region $R_{\I}$, and no symmetry transformation on the trivial right region $R_{\II}$.
We consider the following setup:\\

\noindent
(1)  We assign a Hilbert space dimension $|H|$ on each site along the interface $\partial R$ 
between the $R_{\I}$ and the $R_{\II}$ regions, while the $R_{\I}$
region of the SPTs has a Hilbert space dimension $|G|$ on each site.\\

\noindent
(2) We require the dimension of Hilbert space on
the additional site $i_*$ assigned with $h^*$ outside $M^d$ has a Hilbert space dimension $|H|$. We also have an additional
virtual site $i_*'$ assigned with $\mathbf h^{-1} h^*$ for  $ \forall \mathbf h \in H$, such that 
$r(\mathbf h) \equiv \mathbf g$, $r( h^*) \equiv  g^*$
and 
$$r (\mathbf h^{-1} h^*)=r(\mathbf h^{-1}) r(h^*) = \mathbf g^{-1} g^*.$$ 
We also set that the site $i_*'$ has a
Hilbert space dimension $|H|$. The condition (2) is important in order
to split the cocycle on the $R_{\I}$ region that touches the interface.\\

\noindent
(3) We consider the algebraic structure preserving map 
from $H$ to $G$
with $r(h)=g$, the same map of $H  \overset{r}{\longrightarrow} G$. The symmetry transformation
sends $| g_j \rangle \to  |r( \mathbf h) g_j \rangle =| \mathbf g g_j \rangle$ when the dimension of Hilbert
space is $|G|$ on the site $j$. The symmetry transformation sends $| h_j \rangle
\to |\mathbf h h_j \rangle$ when the Hilbert space dimension is $|H|$ on the site $j$.\\

The exact global $G$ symmetry transformation on the left region $R_I$ 
and the exact global $H$
symmetry transformation along the interface yield global $U(1)$ phases to the
wavefunction, and the global $U(1)$ phases need to cancel out to 1. 
The cancellation of global $U(1)$ phases of $G$-symmetry and $H$-symmetry
transformations may be viewed as anomaly-free for the whole bulk and the interface. 
The wavefunction is only symmetry-invariant if we consider the whole system together.

Now consider the group manifold that has the left ($R_{\I}$) sector of group $G$ and the right sector of a trivial vacuum,
and all sectors can be lifted to the larger group $H$. 
Again we set that $g_{\I}^*=g_{\II}^*=g^*=r_V(h^*)=h^*=1$.
In general, we can easily generalize our result to any dimension.
Without losing generality,  let us take a specific example in 2+1D. 
And let us consider the 2-dimensional space lattice defined on a 2-sphere $S^2$. The $S^2$ can be regarded 
as two 2-disks $D^2$ glued together along the $S^1$ boundary. Let us call the two $D^2$ disks as $D^2_{R_{\I}}$ assigned with $G_{\I}$ on each site, and 
$D^2_{R_{\II}}$  assigned with $G_{\II}$ on each site. Along the $S^1$ boundary, we assign $H$ on each site.
The wavefunction on the whole $S^2$ surface is evolved from an additional point $i^*$ assigned $g^*=r(h^*)$. 
Thus the wavefunction can be determined by assigning the
3-cocycle into this spacetime volume of the $D^3$ ball (whose center is $i^*$ and whose spatial sector is $S^2$).

For SPTs, we use the  homogeneous cocycle denoted $\nu^{G_{s}}_{d}$ and cochain $\mu^{H}_{d-1}$, and we follow 
the wavefunction $\Phi(\{g_{i}, h_j^{\partial}\}) \equiv \Phi_{R_{\I}}(\{g_{i}\},\{h_j^{\partial}\}) \; \Phi_{\partial R}(\{h_j^{\partial}\})$ 
in \Eqn{eq:splitGSPTwavefunc2+1D}.
Here we arrange the wavefunction separated into a few parts:
\bea \label{eq:splitGSPTwavefunc2+1D2}
\Phi_{R_{\I}}(\{g_{i}\},\{h_j^{\partial}\}) &\equiv&  \prod_{\{...\}}  \nu_{3}^{G \; s }( g_{i_1},  g_{i_2},  g_{i_3}, r(h^*)) 
\nu_{3}^{G  \; s}( r(h_{j_1}^{\partial}),   g_{i_2},  g_{i_3}, r(h^*))
\nu_{3}^{G \; s }( r(h_{j_1}^{\partial}),  r(h_{j_2}^{\partial}),  g_{i_3}, r(h^*)), \nonumber \\
\Phi_{\partial R}(\{h_j^{\partial}\}) &\equiv& \prod_j \mu_{2}^{H \;s}( h_j^{\partial}, \;  h_{j+1}^{\partial},\;  h^*).
\eea 
Again there are orientations $s=\pm 1$ for each term.

Below we verify that the wavefunction $ \Phi(\{g_{i}, h_j^{\partial}\})$ is invariant under the global-symmetry transformation $\widehat{\textbf{S}}_{sym}$. 
It means that we can show the 
$\Phi(\{g_{i}, h_j^{\partial}\})$ is equal to
\bea
\widehat{\textbf{S}}_{sym} \Phi(\{g_{i},  h_j^{\partial}\})=
\Phi( \{ (r(\mathbf h) \cdot g_{i}),  \;(\mathbf h \cdot h_j^{\partial})\}).
\eea
We also denote the change $r(\mathbf h) \equiv \mathbf g$ in $G$.
The above shows the symmetry transformation acts on the wavefunction. Conversely, we can consider the equivalent dual picture that
the symmetry transformation acts on the state vector in the Hilbert space. Either way leads to the same conclusion.
Since
\bea
\widehat{\textbf{S}}_{sym} \Phi(\{g_{i}, h_j^{\partial}\})=
\big[\frac{\Phi( \{ (r(\mathbf h) \cdot g_{i}), \;(\mathbf h \cdot h_j^{\partial})\})}{\Phi(\{g_{i}, h_j^{\partial}\})} \big] {\Phi(\{g_{i}, h_j^{\partial}\})},
\eea
we need to show that the factor in the bracket $\big[ \dots \big]$ is 1 to prove the global symmetry preservation.
The $G$-symmetry on the region $R_{\I}$ must be able to be lifted
to some $H$-symmetry on the whole regions $R_{\I}$ including the interface $\partial R$,
based on the fact that $H \overset{r}{\to} G$ is surjective. 
We remind the readers that $\mathbf g \equiv r(\mathbf h)$, $g^* \equiv r( h^*)$.
Namely, it is effectively the $H$-symmetry transformation on the whole system.

In region $R_{\I}$, the wavefunction change 
$\frac{\nu_{3}( \mathbf g  \cdot g_{1}, \;  \mathbf g \cdot g_{2} ,\;  \mathbf g \cdot g_{3},\;  g^*)}{\nu_{3}( g_1, \; g_2,\;  g_3,\; g^*)}= \frac{\nu_{3}( g_{1}, \; g_{2} ,\;  g_{3},\; \mathbf g^{-1} \cdot g^*)}{\nu_{3}( g_1, \; g_2,\;  g_3,\; g^*)}$ can be simplified further based on a $d$-cocycle condition, 
\bea
&& {(\delta \nu_{3})(  g_i,\;  g_j,\; g^*,\; \mathbf g^{-1} \cdot g^*)}=1\\
&&\Rightarrow \frac{\nu_{3}( g_{1}, \; g_{2} ,\;  g_{3},\; \mathbf g^{-1} \cdot g^*)}{\nu_{3}( g_1, \; g_2,\;  g_3,\; g^*)}
=
 \frac{{\nu_{3}(  g_2,\;  g_3,\; g^*,\; \mathbf g^{-1} \cdot g^*)}\;{\nu_{3}(  g_1,\;  g_2,\; g^*,\; \mathbf g^{-1} \cdot g^*)}}{{\nu_{3}(  g_1,\;  g_3,\; g^*,\; \mathbf g^{-1} \cdot g^*)} } 
.
\eea
Here for convenience, let us denote $\overline{g_i g_j}$ as a link connecting two vertices $i$ and $j$, where two vertices are assigned with ${g_i}$ and ${g_j}$ respectively.
Notice the 3-cocycle ${\nu_{3}(  g_i,\;  g_j,\; g^*,\; \mathbf g^{-1} \cdot g^*)}$ which contains a link $\overline{g_i g_j}$
is cancelled out, because there exists a neighbor term which shares the same link $\overline{g_i g_j}$ and which contributes the same 
factor with opposite orientation thus opposite sign for $s=\pm 1$. The only subtle type of terms, 
 that survives and that requires further analysis, is  ${\nu_{3}(  r{(h_i^{\partial})},\;  r{(h_j^{\partial})},\; g^*,\; \mathbf g^{-1} \cdot g^*)}$
 which  contains a link with two vertices $\overline{h_i^{\partial} h_j^{\partial}}$ on the interface $\partial R$.
 If we approach from the region $R_{\I}$, we see that
\bea \label{eq:fig:sym-preserve1}
&&
\frac{\nu_{3}( r{(h_1^{\partial})}, \; r{(h_2^{\partial})},\;  g_3,\; \mathbf g^{-1} \cdot g^*)}{\nu_{3}( r{(h_1^{\partial})}, \; r{(h_2^{\partial})},\;  g_3,\; g^*)}
=
 \frac{{\nu_{3}(  r{(h_2^{\partial})},\;  g_3,\; g^*,\; \mathbf g^{-1} \cdot g^*)}\;{\nu_{3}(  r{(h_1^{\partial})},\;  r{(h_2^{\partial})},\; g^*,\; \mathbf g^{-1} \cdot g^*)}}{{\nu_{3}(  r{(h_1^{\partial})},\;  g_3,\; g^*,\; \mathbf g^{-1} \cdot g^*)} }
.\;\;\;\;\;\;\;\;
\eea
All the terms on the right-hand side cancel with some other terms in the product $\prod_{\{...\}}$ which share the same links connecting 
$\overline{h_1^{\partial} g_3}$ and $\overline{h_2^{\partial} g_3}$ on the same region $R_{I}$, except that
 ${\nu_{3}(  r{(h_1^{\partial})},\;  r{(h_2^{\partial})},\; g^*,\; \mathbf g^{-1} \cdot g^*)}$ term that touches the link $\overline{h_1^{\partial} h_2^{\partial}}$.
We would like to split 3-cocycle $\nu_{3}^{G}$ that touches the link $\overline{h_{i}^{\partial} h_{j}^{\partial}}$ into
2-cochains $\mu_{2}^H$:
\bea \label{eq:splitnu3G}
&&\nu_{3}^{G}( r{(h_1^{\partial})}, \; r{(h_2^{\partial})},\;   r(  h^*),\;   r(\mathbf h^{-1} \cdot h^*))
=\nu_{3}^{H}( h_1^{\partial}, \; h_2^{\partial},\;    h^*,\;   \mathbf h^{-1} \cdot h^*)\nonumber \\
&&=
(\delta \mu_{2}^H)( h_1^{\partial}, \; h_2^{\partial},\;    h^*,\;   \mathbf h^{-1} \cdot h^*)=
\frac{{\mu_{2}^H( h_2^{\partial}, h^*, \mathbf h^{-1} \cdot h^*)} {\mu_{2}^H(h_1^{\partial}, h_2^{\partial}, \mathbf h^{-1} \cdot h^*)} }{ {\mu_{2}^H(h_1^{\partial}, h^*, \mathbf h^{-1} \cdot h^*)} {\mu_{2}^H(h_1^{\partial}, h_2^{\partial},  h^*)}
 }.\;\;\;\;\;\;\;\;
\eea

We shall consider all such splitting terms along the interface. As an example, for the 1+1D interface on a spatial ring with a total number of $\mathrm{N}$ sites and
 $\mathrm{N}$ links ($\overline{h_j^{\partial} h_{j+1}^{\partial}}$) where 
$i = 1, \dots, \mathrm{N}$ (mod $\mathrm{N}$), we obtain:
\bea \label{eq:fig:sym-preserve2}
\prod_{j=1}^{\mathrm{N}} {\nu_{3}^{G}( r{(h_j^{\partial})}, \; r{(h_{j+1}^{\partial})},\;   r(  h^*),\;   r(\mathbf h^{-1} \cdot h^*))}
&=&
\prod_{j=1}^{\mathrm{N}}  
\frac{{\mu_{2}^H( h_{j+1}^{\partial}, h^*, \mathbf h^{-1} \cdot h^*)} }{ {\mu_{2}^H(h_j^{\partial}, h^*, \mathbf h^{-1} \cdot h^*)}}  
\prod_{j=1}^{\mathrm{N}}   \frac{{\mu_{2}^H(h_j^{\partial}, h_{j+1}^{\partial}, \mathbf h^{-1} \cdot h^*)}}{{\mu_{2}^H(h_j^{\partial}, h_{j+1}^{\partial},  h^*)} } \nonumber\\
  &=&  \prod_{j=1}^{\mathrm{N}} \frac{{\mu_{2}^H(h_j^{\partial}, h_{j+1}^{\partial}, \mathbf h^{-1} \cdot h^*)}}{{\mu_{2}^H(h_j^{\partial}, h_{j+1}^{\partial},  h^*)} }.
\eea
The first is based on \Eqn{eq:splitnu3G} on a ring. 
For the second equality, we use the fact that 
$\prod_{j=1}^{\mathrm{N}}  
\frac{{\mu_{2}^H( h_{j+1}^{\partial}, h^*, \mathbf h^{-1} \cdot h^*)} }{ {\mu_{2}^H(h_j^{\partial}, h^*, \mathbf h^{-1} \cdot h^*)}}$ $=1$ cancels out on a closed ring.
Combined with the fact that homogenous cochain does not change under symmetry transformation if inputs do not contain $h^*$, 
due to that the homogenous cocycle satisfies 
$\frac{\mu_{2}^H( \mathbf h \cdot  h_i, \; \mathbf h \cdot  h_j,\; \mathbf h \cdot  h_k )}{\mu_{2}^H( h_i, \;  h_j,\; h_k )}=1$,
so far we derive that
\bea   \label{eq:sym_phase_nu3}
\frac{\Phi_{R_{\I}}(\{ r_{}(\mathbf h) \cdot  g_{i}\},\{\mathbf h \cdot h_j^{\partial}\}) 
 }
{\Phi_{R_{\I}}(\{g_{i}\},\{h_j^{\partial}\}) }=
\prod_{j=1}^{\mathrm{N}} \frac{{\mu_{2}^H(h_j^{\partial}, h_{j+1}^{\partial}, \mathbf h^{-1} \cdot h^*)}}{{\mu_{2}^H(h_j^{\partial}, h_{j+1}^{\partial},  h^*)} }.
\eea
We can also see that the remaining part of wavefunction is
${\Phi_{\partial R}(\{h_j^{\partial}\})}=
\prod_{j=1}^{\mathrm{N}} {{\mu_{2}^H(h_j^{\partial}, h_{j+1}^{\partial},  h^*)} }^{-1}$, where the inverse with $s=-1$ is due to the opposite orientation
accounted from the other side ${R_{\II}}$.
Its symmetry transformation becomes:
\bea   \label{eq:sym_phase_mu2}
\frac{\Phi_{\partial R}(\{ \mathbf h \cdot h_j^{\partial}\}) 
}{\Phi_{\partial R}(\{h_j^{\partial}\})} \equiv 
\prod_{j=1}^{\mathrm{N}} (\frac{{\mu_{2}^H(h_j^{\partial}, h_{j+1}^{\partial}, \mathbf h^{-1} \cdot h^*)}}{{\mu_{2}^H(h_j^{\partial}, h_{j+1}^{\partial},  h^*)} })^{-1}.
\eea
Thus the phases in eqn.(\ref{eq:sym_phase_nu3}) and eqn.(\ref{eq:sym_phase_mu2}) cancel perfectly, 
and the whole wavefunction 
$\Phi(\{g_{i}, h_j^{\partial}\}) \equiv \Phi_{R_{\I}}(\{g_{i}\},\{h_j^{\partial}\}) \; \Phi_{\partial R}(\{h_j^{\partial}\})$
is invariant under the symmetry transformation:
\bea \label{eq:sym-cancel}
\widehat{\textbf{S}}_{sym} \Phi(\{g_{i}, h_j^{\partial}\})=
\frac{\Phi_{R_{\I}}(\{ r_{}(\mathbf h) \cdot  g_{i}\},\{\mathbf h \cdot h_j^{\partial}\}) 
 }
{\Phi_{R_{\I}}(\{g_{i}\},\{h_j^{\partial}\}) }
\frac{\Phi_{\partial R}(\{ \mathbf h \cdot h_j^{\partial}\}) 
}{\Phi_{\partial R}(\{h_j^{\partial}\})} 
\cdot\Phi(\{g_{i}, h_j^{\partial}\})
=
1 \cdot\Phi(\{g_{i}, h_j^{\partial}\}).
\eea
In Fig.\ref{fig:proof:sym-preserve}, we show a neat geometrical way to understand the symmetry-transformation phase cancellation for the symmetry invariance.
For any higher $d$-dimensional spacetime, we can give the same proof by replacing
${\mu_{2}^H}$ in  eqn.(\ref{eq:sym_phase_nu3}) and eqn.(\ref{eq:sym_phase_mu2}) with ${\mu_{d-1}^H}$.
It is easy to confirm that our proof on symmetry-preserving gapped interface holds for any higher-dimensional generalization (q.e.d).

We can apply a similar proof  for the global-symmetry-preserving property of the SET version of wavefunction \Eqn{eq:splitGSETwavefunc} to show
\bea
\widehat{\textbf{S}}_{sym} \Phi(\{g_{i}, n_{i_a i_b}, h_j, k_{j_a j_b}\})={\Phi( \{r_{}(\mathbf h) \cdot g_{i}, n_{i_a i_b}, \mathbf h \cdot  h_j, k_{j_a j_b}\})}= \Phi(\{g_{i}, n_{i_a i_b}, h_j, k_{j_a j_b}\}).  \;\;\;
\eea
To prove this, we may regard that $ \mathbf h \cdot  h_j \equiv   h_j  \cdot \mathbf h'$ where $ \mathbf h' = h_j^{-1} \mathbf h h_j$.
Similarly, $r_{}(\mathbf h) \cdot g_{i} \equiv \mathbf g \cdot g_{i} \equiv g_{i} \cdot \mathbf g'$, we find that $\mathbf g' =
g_j^{-1} \mathbf g  g_j = r(h_j^{-1}) r(\mathbf h) r(h_j) = r(h_j^{-1} \mathbf h h_j) =r(\mathbf h')$.
Regardless the branch structure for vertex ordering, we can convert the symmetry transformation, from acting on the left of the group elements to that 
acting on the right of group elements. This trick can facilitate the proof
that the SET wavefunction is invariant under global symmetry, even in the presence of gapped interfaces.

\subsubsection{More Remarks}   \label{sec:remarks}

Here are a summary and some more remarks:

\begin{enumerate}

\item \emph{Global enhanced $H$-symmetry invariant}: We have shown that the SPTs wavefunction on a whole system is invariant under $G$-symmetry transformation in the bulk ${R_{\I}}$ together under 
$H$-symmetry transformation on the interface ${\partial R}$.  The symmetry transformation is fixed by $H \overset{r}{\longrightarrow} G$, and
we may view that the symmetry is enhanced to $H$ for the whole system.

\item \emph{Global $K$-symmetry on the boundary/interface}: 
Under the construction $1 \to K \to H \overset{r}{\longrightarrow} G \to 1$ for $G$-bulk SPTs and an anomalous boundary $H$-SPTs,
the $K$ is trivial in the bulk as $r(k)=1 \in G$ for $k \in K$. How about $K$-symmetry transformation on the interface?
It is easy to check there is \emph{no local} $K$-\emph{symmetry} on the interface, since
${\Phi_{\partial R}(\{ \mathbf k_j \cdot h_j^{\partial}\}) } \neq {\Phi_{\partial R}(\{h_j^{\partial}\})}$ for arbitrary local $\mathbf k_j  \in K$ transformation on each site $j$.
However, below we can prove that there is a \emph{global} $K$-symmetry applying on the boundary/interface, namely
\bea
{\Phi_{\partial R}(\{ \mathbf k \cdot h_j^{\partial}\}) } = {\Phi_{\partial R}(\{h_j^{\partial}\})}.
\eea
Proof: Without losing generality, consider the 1+1D boundary of 2+1D SPTs. We see that
\bea
&&
{ \Phi_{\partial R}(\{ \mathbf k \cdot h_j^{\partial}\})=  \prod_{j=1}^{\mathrm{N}}
\frac{\mu_2^H (\mathbf kh_{j}^{\partial}, \mathbf k h_{j+1}^{\partial}, h^*)}{\mu_2^H (h_{j}^{\partial}, h_{j+1}^{\partial}, h^*)}  \cdot \Phi_{\partial R}(\{h_j^{\partial}\}) } \nonumber \\
&&
{=  \prod_{j=1}^{\mathrm{N}}
\frac{\mu_2^H (h_{j}^{\partial}, h_{j+1}^{\partial}, \mathbf k^{-1} h^*)}{\mu_2^H (h_{j}^{\partial}, h_{j+1}^{\partial}, h^*)} \cdot \Phi_{\partial R}(\{h_j^{\partial}\})=\Phi_{\partial R}}(\{h_j^{\partial}\}).
\eea
where in the last equality we use the fact of 3-cocycle splitting and $r(\mathbf k)=1 \in G$ so
\begin{multline}
1=\nu_{3}( r{(h_j^{\partial})}, \; r{(h_{j+1}^{\partial})},\;   r(  h^*)= g^*,\;   r(\mathbf k^{-1} \cdot h^*)= g^*)
=
\frac{ {\mu_2(h_{j+1}^{\partial}, h^*, \mathbf k^{-1} \cdot h^* )}}{{\mu_2( h_j^{\partial}, h^*, \mathbf k^{-1} \cdot h^*)}} 
\frac{\mu_2(h_j^{\partial}, h_{j+1}^{\partial}, \mathbf k^{-1} \cdot h^*)}{\mu_2(h_j^{\partial}, h_{j+1}^{\partial}, h^*)} \\
 \Rightarrow
1= \prod_{j=1}^{\mathrm{N}} 1 =
\prod_{j=1}^{\mathrm{N}} \frac{ {\mu_2(h_{j+1}^{\partial}, h^*, \mathbf k^{-1} \cdot h^* )}}{{\mu_2( h_j^{\partial}, h^*, \mathbf k^{-1} \cdot h^*)}} 
\prod_{j=1}^{\mathrm{N}} \frac{\mu_2(h_j^{\partial}, h_{j+1}^{\partial}, \mathbf k^{-1} \cdot h^*)}{\mu_2(h_j^{\partial}, h_{j+1}^{\partial}, h^*)}  \\
 \Rightarrow 1= 1 \cdot \prod_{j=1}^{\mathrm{N}} \frac{\mu_2(h_j^{\partial}, h_{j+1}^{\partial}, \mathbf k^{-1} \cdot h^*)}{\mu_2(h_j^{\partial}, h_{j+1}^{\partial}, h^*)}.
\end{multline}

\item \emph{Gauging SPTs to SETs}: Since there is a global $K$-symmetry on the boundary/interface, we can partially or fully gauge this $K$-symmetry.
We can also gauge a normal subgroup $N$ of the global $G$ symmetry of $G$-SPTs --- 
however, to gauge $N$ in the bulk we also need to gauge the $N$ for the anomalous $H$-SPTs on the boundary/interface.
By gauging the normal subgroups $N$ and $K$, this gives rise to SETs of Sec.~\ref{sec:ZSPTSETG} Remark 4. 

\item \emph{Degenerate ground states and holonomies for the boundary anomalous SETs}: 
If the gapped boundary is on a compact space with nontrivial cycles, there can be nontrivial holonomies for the gapped boundary anomalous SETs.
For example, for a 2+1D SPTs on a 2-disk $D^2$ and its 1+1D anomalous SETs on a 1-circle $S^1$, 
or, for a 3+1D SPTs on a solid torus $D^2 \times S^1$ and its 2+1D anomalous SETs on a 2-torus $T^2$,
their nontrivial boundary holonomies imply the ground state degeneracy (GSD). 
We will explicit compute such GSDs for some examples in Appendix \ref{sec:examples}, such as
$0 \to {Z}_2^K \to Z_4^H  \to Z_2^G \to 0$ in Sec.~\ref{sec:examplesHZ4GZ2GSD} and $1 \to Z_4^K \to Q_8^H  \to Z_2^G \to 1$ in Sec.~\ref{sec:examplesHQ8GZ2GSD}.

\item \emph{Gapped interfaces by folding trick}: Again based on the folding trick, 
we can construct a wavefunction and lattice Hamiltonian of gapped interfaces between two topological phases in Sec.~\ref{sec:ZSPTSETG1G2}, 
and we still can prove the symmetry-preserving wavefunction.
\end{enumerate}


\section{Conclusion} \label{sec:conclude}

Some concluding and additional remarks follow:

\begin{enumerate}

\item We provide a UV complete lattice regularization of the Hamiltonian and path integral definition of gapped interfaces based on the \emph{symmetry-extension} mechanism, partly rooted in Ref.\cite{Witten:2016cio}. 
Presumably, some of other phenomenon studied in Ref.\cite{Witten:2016cio} could also be examined based on our lattice regularized setting. 
 
 \item The \emph{anomalous non-onsite} $G$-symmetry at the boundary indicates that
if we couple the $G$-symmetric boundary to the weakly fluctuating background probed
gauge field of $G$, there is an anomaly in $G$ (in the same language as in
particle physics and high-energy theory) along the boundary.  The $G$-anomaly
can be a gauge anomaly (e.g. for an internal unitary $G$-symmetry), or a mixed
gauge-gravitational anomaly 
(e.g., for a $G$-symmetry that contains an anti-unitary time reversal symmetry $Z_2^T$).  The key ingredient
of our approach is based on the fact that certain non-perturbative global
anomalies in $G$ at the boundary become \emph{anomaly-free} in $H$, when $G$ is
pulled back to $H$ (see Sec.~\ref{sec:on-site-H}).

\item Given some bulk $G$-SPT states, 
our formulation finds their possible $H$-\emph{symmetry-extended} and $G$-\emph{symmetry-preserving} gapped boundaries, 
via a suitable group extension $1\to K \to H \to G \to 1$.\footnote{
To make a comparison, 
we remark that Refs.\cite{Kapustin:2014zva, Thorngren1511.02929, MCheng1606.08482, BeniniHsinSeiberg1702.07035} show a related physics by starting from a given anomalous boundary field theory, 
and finding the possible bulk TQFT. }
To construct an $H$-\emph{symmetry-extended} gapped boundary, we actually require a
\emph{weaker} condition on the group extension that $K$ may be a finite group or a continuous group, in any bulk dimension $\geq$ 1+1D.
To construct a $G$-\emph{symmetry-preserving} topologically ordered gapped boundary, 
we further require a \emph{stronger} condition on the group extension that $K$ is a finite group, 
in order to have a boundary deconfined $K$-gauge theory, for a 3+1D bulk and above.

\item
When $G$, $H$ and $K$ are finite groups, 
we can prove that there \emph{always exist} $H$-symmetry extended gapped boundaries (in any bulk dimension $\geq$ 1+1D) 
and there \emph{always exist} $G$-symmetry preserving gapped boundaries (for 3+1D bulk and above).
The gauge anomaly associated to a finite symmetry group $G$ must be a non-perturbative global anomaly.
The cohomology/cobordism group of a finite $G$ only contains the \emph{torsion} part, which indicates the non-perturbative anomalies.

We believe that the argument remains valid, even when $G$ and $H$
are infinite continuous compact groups, but $K$ remains a finite group.  In this
case, the boundary dynamics still yields a deconfined $K$-gauge theory, given that the bulk dimensions are larger or equal to 3+1D (see Sec.~\ref{sec:beyondGC}). 
(When the bulk is 2+1D, we comment in the next remark.)

When $G$ is a continuous group for the bulk $G$-SPTs, the boundary could have both \emph{perturbative} anomalies 
(e.g. 
captured by a 1-loop Feynman diagram), 
or \emph{non-perturbative global} anomalies, detected by coupling the boundary to $G$-gauge fields.\footnote{
\cblue{The \emph{free} part of the cohomology/(co)bordism group contributes the perturbative anomalies}.
\cblue{The \emph{torsion} part of the cohomology/(co)bordism group contributes the non-perturbative global anomalies}.
}
The \emph{perturbative} anomalies do not offer any symmetry-preserving surface topological orders.
In contrast, some of the \emph{non-perturbative global} anomalies can offer a symmetry-preserving surface topological order
as long as our construction trivializes the $G$-anomaly 
in $H$.\footnote{We demonstrate examples for each case 
in Appendices: First, a perturbative chiral anomaly of $U(1)$-SPTs in \ref{2+1DU1U1}
(with a bulk invariant $\exp(\ii  {2 \pi k}  \int (\frac{A}{2 \pi}) \; (c_1)^{d/2}$ of even dimensions $d$ and an integer $k$), 
do not offer a symmetry-preserving surface topological orders but only have a \emph{symmetry-enforced gapless boundaries}. 
Other \emph{symmetry-enforced gapless boundaries} also occurred in \cite{CWang1401.1142}.
Second, a global mixed gauge-gravitational anomaly on the boundary of 6+1D $U(1)$-SPTs in \ref{6+1DU1U1} (with a bulk invariant $\exp( \ii  2 \pi  \int \frac{1}{2} w_2 w_3 c_1)$), does allow a deconfined 5+1D $K=Z_2$ surface topological orders.}


\item We apply our symmetry-preserving gapped interface construction to the 2+1D bulk and 1+1D boundary.
For the 1+1D topologically ordered $K$-gauge theory on the boundary of a \emph{finite/continuous group symmetry} of 2+1D $G$-SPTs, 
we find an interesting phenomenon
that the 1+1D boundary deconfined $K$-gauge theory states
develop long-range orders that spontaneously break the $G$-symmetry (see Sec.~\ref{whichone}).
The 1+1D boundary deconfined and confined gauge theory states belong to the same phase; 
namely, they are both symmetry-breaking states connected without phase transitions.

Examples include those of a finite gauge group $K$, and 
a global symmetry $G$ containing discrete \emph{unitary or ant-unitary} global symmetry sectors that can be spontaneously broken.
For instance,
in Sec.~\ref{bdry3w}, and Appendix \ref{sec:deconfined-to-SSB} and \ref{sec:SSB_field-theory},
we show that the unitary $Z_2^G$-symmetry of a 1+1D $Z_2^K$ gauge theory is \emph{spontaneously broken}, on the boundary of 2+1D $Z_2^G$-SPTs.
In Appendix \ref{sec:SSB_field-theory},
we also show that the \emph{anti-unitary time reversal} $Z_2^T$-symmetry of a 1+1D $Z_2^K$ gauge theory is \emph{spontaneously broken}, 
on the boundary of 2+1D bosonic $U(1) \rtimes Z_2^T$-topological insulator and $Z_2 \rtimes Z_2^T$-topological superconductor.
This is, so far, consistent with the fact that there is \emph{no robust intrinsic topological order in 1+1D} robust 
against any local perturbations. \footnote{However, 
if we apply our construction for a \emph{continuous symmetry group} on a 1+1D boundary,
due to Coleman-Mermin-Wagner theorem, there shall be no spontaneous symmetry breaking for a continuous symmetry in 1+1D.
We may expect to find further interesting new physics. It will be illuminating to address this issue further in the future work. }

\item Our approach shall be applicable to obtain gapped interfaces of more
generic bosonic and fermionic topological states (other than the fermionic CZX
model in Appendix \ref{sec:fCZX}), including topological states from
the beyond-symmetry-group cohomology and cobordism approach
(Secs.~\ref{sec:beyondGC} and Sec.~\ref{sec:Cob}).  It will be interesting to
establish this result with more concrete examples.

\begin{table}[!h]
\begin{center}
\makebox[\textwidth][c] 
{
\begin{tabular}{|c||c|c|c|c|c}
\hline
Bulk/Interface Dim\; &\; $1\to K \to H \to G \to 1$  &  $\begin{matrix}\text{LHS}\\ \text{spectral}\\\text{sequence} \end{matrix}$ & \; $\begin{matrix}\text{Symmetric}\\ \text{Gapped}\\ \text{Boundary}\end{matrix}$ & \;  $\begin{matrix}\text{SPT Bulk inv.}\\ \text{($d$-cocycle $\omega_d$)} \end{matrix}$ \\ \hline
 \ref{sec:examplesHZ4GZ2}/\ref{sec:examples2+1DHZ4GZ2}: 2+1/1+1D & $0 \to {Z}_2 \to Z_4  \to Z_2 \to 0$& Yes  & No ($Z_2$-SSB) & $\omega_{3, \I}$, $\exp( \ii   \pi  \int (a_1)^3)$\\ \hline
 \ref{sec:examplesHZ4GZ2-any-dim}: $d+1$/$d$D & Even dim $d$: $0 \to {Z}_2 \to Z_4  \to Z_2 \to 0$& Yes  & Yes & $\omega_{d+1, \I}$, $\exp( \ii   \pi  \int (a_1)^{d+1})$\\ \hline
 \ref{sec:examplesHZ4TGZ2T}/\ref{sec:examples3+1DHZ4GZ2T}: 3+1/2+1D & $0 \to {Z}_2 \to Z_4^T  \to Z_2^T \to 0$ & -  & Yes & $\begin{matrix} \text{$Z_2^T$-cocycle,}\\ \exp( \ii   \pi  \int (w_1)^{4}) \end{matrix}$\\ \hline
 \ref{sec:examplesHZ4TGZ2T-any-dim}:  $d+1$/$d$D & Odd dim $d$: $0 \to {Z}_2 \to Z_4^T  \to Z_2^T \to 0$ & -  & Yes & $\begin{matrix} \text{$Z_2^T$-cocycle,}\\ \exp( \ii   \pi  \int (w_1)^{d+1}) \end{matrix}$ \\ \hline
  \ref{sec:examples3+1Dw2sq}:  3+1/2+1D & $1 \to {Z}_2 \to   \text{Pin}^+(\infty)  \to O(\infty) \to 1$ & -  & Yes & $\begin{matrix} \text{$Z_2^T$-cocycle,}\\ \exp( \ii   \pi  \int (w_2)^2) \end{matrix}$\\ \hline
   \ref{sec:examples3+1Dw2sq}:  3+1/2+1D & $1 \to {Z}_2 \to   \text{Pin}^-(\infty)   \to O(\infty) \to 1$ & -  & Yes & $\begin{matrix} \text{$Z_2^T$-cocycle,}\\ \exp( \ii   \pi  \int (w_1)^4+(w_2)^2) \end{matrix}$\\ \hline
 \ref{sec:examplesHZ4NGZ2}: 2+1/1+1D & $0 \to Z_{2N} {\to} Z_{4  N} {\to} Z_2 \to 0$ & Yes  & No (SSB) & $\omega_{3, \I}$, $\exp( \ii   \pi  \int (a_1)^3 )$\\ \hline
 \ref{sec:examplesHQ8GZ2}: 2+1/1+1D & $1 \to Z_4 \to Q_8  \to Z_2 \to 1$ & Yes  & No (SSB) & $\omega_{3, \I}$, $\exp( \ii   \pi  \int (a_1)^3  )$\\ \hline
 \ref{2+1DD4Z22}: 2+1/1+1D & $1 \to Z_2 \to D_4  \to (Z_2)^2 \to 1$   &Yes   & No (SSB) & $\omega_{3, \II}$, $\exp( \ii   \pi  \int a_1 \beta a_2  )$\\ \hline
 \ref{1+1DQ8Z22}: 1+1/0+1D & $1 \to Z_2 \to Q_8  \to (Z_2)^2 \to 1$ &No   & Yes & $\omega_{2, \II}$, $\exp( \ii   \pi  \int a_1 a_2)$ \\ \hline
 \ref{1+1DD4Z22}:  1+1/0+1D & $1 \to Z_2 \to D_4  \to (Z_2)^2 \to 1$  & Yes   & Yes & $\omega_{2, \II}$, $\exp( \ii   \pi  \int a_1 a_2)$\\ \hline
 \ref{2+1DD4Z2Z23}: 2+1/1+1D & $1 \to Z_2 \to D_4 \times Z_2  \to (Z_2)^3 \to 1$ & Yes  & No (SSB) & $\omega_{3, \III}$, $\exp( \ii   \pi  \int a_1 a_2 a_3)$\\ \hline
 \ref{3+1DD4Z22Z24}: 3+1/2+1D & $1 \to Z_2 \to D_4 \times (Z_2)^2  \to (Z_2)^4 \to 1$ & Yes  & Yes & $\omega_{4, \IV}$, $\exp( \ii   \pi  \int a_1 a_2 a_3 a_4)$\\ \hline
 \ref{3+1DD4Z22Z24}: d+1/dD & $1 \to Z_2 \to D_4 \times (Z_2)^{d-1} \to (Z_2)^{d+1} \to 1$  & Yes  & Yes & $\begin{matrix}\omega_{d+1,\text{Top}},\\ {\exp( \ii   \pi  \int \cup_{i=1}^{d+1} a_i)}\end{matrix}$\\ \hline
 \ref{2+1DD4Z2Z22}: 2+1/1+1D & $1 \to Z_2\times Z_2 \to D_4 \times Z_2  \to (Z_2)^2 \to 1$  & Yes   & No (SSB) & $\omega_{3, \II}$, $\exp( \ii   \pi  \int a_1 \beta a_2  )$\\ \hline
 \ref{3+1DD4Z22TypeII}: 3+1/2+1D & $1 \to (Z_2) \to D_4  \to (Z_2)^2 \to 1$   & Yes   & Yes & $\omega_{4, \II}$, $\exp( \ii   \pi  \int a_1 a_2 \beta  a_2  )$\\ \hline
 \ref{3+1DD4Z2Z23TypeIII}: 3+1/2+1D & $1 \to Z_2 \to D_4 \times (Z_2) \to (Z_2)^3 \to 1$ & Yes  & Yes & $\omega_{4, \III}$, $\exp( \ii   \pi  \int a_1 a_2 \beta  a_3  )$\\ \hline
 \ref{2+1DU1U1}: 2+1/1+1D & $1 \to Z_N \to U(1) \to U(1) \to  1$  & No   & No (Pert) & $\exp(  {\ii \; k} \int A {(\frac{dA}{2 \pi})}^{d/2})$ \\ \hline
 \ref{6+1DU1U1}: 6+1/5+1D & $\begin{matrix}1 \to Z_2 \to G  \to G \to  1,\\ G=U(1) \times SO(\infty) \end{matrix}$  & -   & Yes (Global) & $\exp( \ii   \pi  \int w_2 w_3 c_1)$ \\ \hline
 \ref{2+1DU1Z2T}: 2+1/1+1D & $\begin{matrix}1 \to Z_2 \to G  \to G \to  1,\\ G=U(1) \rtimes Z_2^T \end{matrix}$  & -   & $\begin{matrix}\text{No ($Z_2^T$-SSB)} \\ \text{(Global)}\end{matrix}$ & $\exp( \ii   \pi  \int w_1 c_1)$ \\ \hline
 \ref{2+1DU1Z2T}: 2+1/1+1D & $\begin{matrix}1 \to Z_2 \to G  \to G \to  1,\\ G=Z_2 \rtimes Z_2^T \end{matrix}$  & -   & $\begin{matrix} \text{No ($Z_2^T$-SSB)}\\ \text{(Global)}\end{matrix}$ & $\exp( \ii   \pi  \int w_1 (a_1)^2)$ \\ \hline
 \ref{1+1DSU2SO3}:  1+1/0+1D & $1 \to Z_2 \to SU(2) \to SO(3) \to 1$ & Yes  & Yes & \parbox{1in}{Odd-integer\\ AF spin chain}\\ \hline
 \end{tabular}
 }
  \hspace*{0mm}
\end{center}
\caption{We outline the gapped boundary/interface results obtained in Appendix \ref{sec:examples}.
In several cases, we check the validity of two techniques mentioned in Sec.~\ref{sec:trivializeGcocycle}, based on
Lyndon-Hochschild-Serre (LHS) spectral sequence technique in Sec.~\ref{sec:LHS}.
The ``{Symmetric} {Gapped} Boundary'' column means that the ``symmetry preserving gapped interface'' is available or not.
For a discrete finite $G$, this ``{Symmetric} {Gapped} Boundary'' means the cochain solution is found.
``Pert.'' means the perturbative anomaly.
``$\bar{G}$-SSB'' means the spontaneous symmetry breaking in $\bar{G}$ (e.g. $\bar{G}=Z_2$, $Z_2^T$, etc.).
``Global'' means the global gauge/gravitational anomaly. ``AF'' means anti-ferromagnet.
The $d$-cocycle for a finite Abelian group $G$ with its type indices (written in Roman numerals) follows the notation defined in Ref. \cite{Wang1405.7689}. 
The $\beta$ is the Bockstein homomorphism.
}
  \label{table:example-app}
\end{table}


\item 
In Appendix \ref{sec:examples}, we systematically construct various \emph{symmetry-extended} gapped boundaries for topological states in various dimensions 
(choosing homogeneous cocycles for SPTs and inhomogeneous cocycles for topological orders), summarized in the Table \ref{table:example-app}.
We can also combine results in different subsections in Appendix \ref{sec:examples} and use the folding trick to obtain the gapped interfaces between topological states.

The previously known gapped interfaces for the  $Z_2$ toric code and $Z_2$ double-semion model can be achieved by 
certain (gauge-) \emph{symmetry-breaking} sine-Gordon cosine interactions
at strong couplings.
The previously known gapped interfaces of
2+1D twisted quantum double models  $D^{\omega_3}(G)$ and Dijkgraaf-Witten gauge theories
can also be obtained through 
such a (gauge-) \emph{symmetry-breaking} mechanism or 
\emph{anyon condensation} \cite{KapustinSaulina1008.0654, KitaevKong1104.5047, Wang1212.4863, Levin:2013gaa,
1305.7203BarkeshliJianQi, HungWan:2014tba1408.0014, Lan:2014uaa}, see Appendix \ref{sec:examples-gauge-sym-break}.
It is known that there are 2 types of gapped boundaries for $Z_2$ toric code,
1 type of gapped boundary for $Z_2$ double-semion model,
and 2 types of gapped interfaces between $Z_2$ toric code and $Z_2$ double-semion model \cite{Lan:2014uaa}.
More generally, 
we systematically show gauge symmetry-breaking gapped interfaces in any dimension, 
 in Appendix \ref{sec:examples-gauge-sym-break},
including 2+1D 
(ours reproduce the results in the previous literature) and the less-studied 3+1D.

However, we can construct other new types of gapped interfaces between $Z_2$ toric code and $Z_2$ double-semion models via a \emph{symmetry-extension} 
mechanism,
such as examples given in Appendices \ref{sec:examplesHZ4GZ2}'s 2+1/1+1D  under $0 \to {Z}_2^K \to Z_4^H  \to Z_2^G \to 0$,
and
\ref{sec:examplesHQ8GZ2}'s 2+1/1+1D under $1 \to Z_4^K \to Q_8^H  \to Z_2^G \to 1$, and more, etc.
Our new gapped interface has an \emph{enhanced} Hilbert space and to certain degree an \emph{enhanced} gauge symmetry,
the first new type of gapped interface has $H=Z_4$ and
the second new type of gapped interface has $H=Q_8$. 
Through  a \emph{symmetry extension} 
mechanism,
we can construct new types of gapped boundaries/interfaces in 2+1D, 3+1D and any higher dimensions.\footnote{However, the fate of some of
gauge symmetry-extended interfaces turns out to be  the same phase as the 
gauge symmetry-breaking interface. This was later explored  in Sec.~7 of \cite{2018arXiv180105416W}, where  their  dual description and equivalence are found.
}

More generally, our framework encompasses the mixed \emph{symmetry breaking}, \emph{symmetry extension}, and \cred{\emph{dynamically gauging}}
mechanisms to generate gapped interfaces.

\item \emph{Fermionic symmetry extension and fermionic spin-TQFT gapped boundaries/interfaces}:
Previously fermionic gauge-symmetry-breaking 1+1D gapped boundary of 2+1D fermionic TO is examined in \cite{Wan:2016php1607.01388}.
More recently, the 2+1D symmetric anomalous gapped fermionic surface topological order of 3+1D fermionic SPT is examined in \cite{Fidkowski:2018lhf1804.08628}. 
A step toward a possible framework to incorporate these fermionic symmetry-breaking or fermionic symmetry-extension constructions of
gapped boundaries/interfaces is pursued in \cite{Guo:2018vij1812.11959}.

\item \emph{Higher-symmetry extension and higher-gauge theory/TQFT gapped boundaries/interfaces}:
Ordinary global symmetries can be regarded as the 0-form global symmetries which has a 0-dimensional charged object measured by codimension-1 (thus $(d-1)$-dimension)
charge operator.  
Generalized global symmetries \cite{Gaiotto:2014kfa1412.5148} include the $q$-form higher global symmetries which has a $q$-dimensional charged object measured by codimension-($q+1$) (thus $(d-q-1)$-dimension) charge operator.
An interesting research direction is to apply the 
symmetry breaking or symmetry extension method of constructing topological boundaries/interfaces
to higher global symmetries, including the higher SPT/SET states. Recent developments along this direction can be found in
\cite{Thorngren1511.02929}, \cite{Wan:2018djl1812.11955}, \cite{Wan:2018bns1812.11967} and References therein.

\item Future application: Gapped interfaces via \emph{gauge symmetry breaking} or anyon condensations have recently found their applications in 
topological quantum computation (see \cite{CongChengWang1609.02037} and Reference therein for 2+1D bulk systems).
We hope that our new types of gapped interfaces via global/gauge \emph{symmetry extensions} in any dimension have analogous potential applications, 
for science and technology, in the future. 
  

\end{enumerate}

\section{Acknowledgements}

J.W. thanks Irina Bobkova, Thomas Church, Tian Lan, Kantaro Ohmori and Pavel Putrov for conversations.
J.W. gratefully acknowledges the Corning Glass Works Foundation Fellowship and the NSF.  
X.-G.W's research is supported by NSF Grant No. DMR-1506475 and NSFC 11274192.
J.W.'s work was performed in part at the Aspen Center for Physics, which is supported by National Science Foundation grant PHY-1066293.
The research at IAS is supported by the NSF Grant PHY-1606531 and PHY-1314311.



\begin{center}
{\bf\LARGE{Appendix}}
\end{center}

\appendix

\section{Low energy effective theory for the boundaries of CZX model}

\label{sec:lowEbdryCZX}

\subsection{Low energy effective theory for 
the second boundary of the CZX model -- \\
A 1+1D model with
an {on-site} $Z_4^H$-symmetry}

\label{bdry2l}
In  Sec.~\ref{bdry2w}, we described a gapped boundary state of the CZX model in which the $Z_2^G$
bulk symmetry is extended to a $Z_4^H$ symmetry along the boundary.  The model as described there
is gapped in both bulk and boundary, and there is no hierarchy of energy scales:  The energy gaps
in bulk and along the boundary are comparable.

This is a physically sensible state of affairs in condensed matter physics, but
nonetheless one might ask what sort of model would have such a hierarchy of
scales.  In this section, we will describe several possibilities.  
{As a result, we obtain several pure 1+1D models as the effective boundary theories
for the CZX model.}

One approach is simply to reduce the coefficient of the boundary plaquette term
$H^\text{bdry}_{p}$ in the Hamiltonian.  In this limit (see Fig.
\ref{bdryB1}), the low-energy
degrees of freedom at the boundary are described by three spins per unit cell: $\v
\si_{i_-}$, $\v \si_{i_+}$, and a composite spin described by the two 
spins on the black dots next to $\v \si_{i_-}$ and $\v \si_{i_+}$, which are locked
due to the projector $P_p^r$ from the neighboring Hamiltonian.

\begin{figure}[h!]
\begin{center}
\includegraphics[scale=0.8]{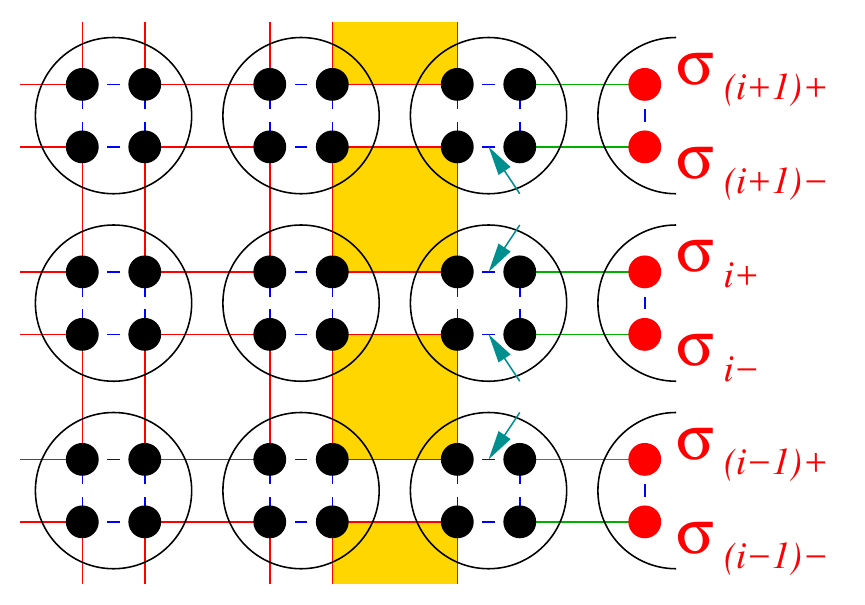}
\end{center}
\caption{
The filled dots are qubits (or spin-1/2's).  A circle (with dots inside)
represents a site. The bulk Hamiltonian \cblue{contains} terms that \cblue{force} the dots
connected by red and green lines to have the same $\si^z_i$ at low energies.
The dashed blue line connecting dots $i,j$ represents the phase factor $CZ_{ij}$
in the bulk $Z_2^G$ global symmetry transformation.  
}
\label{bdryEl}
\end{figure}

Here, we would like to reduce the boundary degrees of freedom further.
To do so, we will consider a slightly
different boundary, by omitting the $H^\text{bdry}_{p}$ terms in the Hamiltonian and
at the same time including some \cblue{projectors} at the boundary.  This gives
us another description of the second boundary of the CZX model (see Fig.
\ref{bdryEl}).  The bulk Hamiltonian of the model is still given by $H_{p}$
for each complete octagon in the bulk, with addition terms that force the
boundary spin $\v \si_{i_\pm}$'s to have the same  $\si^z$ value as the bulk
spins connected by the green lines.  
However, notice that the shaded squares are not  complete octagons
since the two spins to the right of the shaded squares
do not need to be parallel.
So the Hamiltonian for the  shaded squares needs to be modified:
\begin{align}
 H^\text{shaded}_{p} &=
- \cblue{H^{0}_{p}  P_p^{u} P_p^{d} P_p^{l} P_p^{r} 
+\t H^{0}_{p} P_p^{u} P_p^{d} P_p^{l} (1-P_p^{r})} 
\end{align}
where \cblue{$\t H^{0}_{p}$} is given by
\begin{align}
\label{tX4}
\cblue{
\t H^{0}_{p}} = \ii ( |\down\down\down\down\> \<\up\up\up\up| - |\up\up\up\up\> \<\down\down\down\down| ).
\end{align}

The above Hamiltonian has a $Z_4 \equiv Z_4^H$ global symmetry.  The $Z_4^H$ symmetry is
generated by
\begin{align}
\label{bUZ4}
\si^x_{i_-} \si^x_{i_+} \cblue{U_{CZ,{i_-,i_+}}}
\end{align}
when acts on a boundary site, and by
\begin{align}
\cblue{U_{X,s} U_{CZ,s}}=\si^x_{i_1} \si^x_{i_2} \si^x_{i_3} \si^x_{i_4} 
\cblue{U_{CZ,{i_1,i_2}} U_{CZ,{i_2,i_3}} U_{CZ,{i_3,i_4}} U_{CZ,{i_4,i_1}} }
\end{align}
when acts on a bulk site, where $i_1$, $i_2$, $i_3$, and $i_4$ label the four
spins on the bulk site.  Note that the $Z_4$ symmetry is actually a $Z_2$
symmetry in the bulk since 
\begin{align}
\cblue{(U_{X,s} U_{CZ,s})}^2=1.
\end{align}
So here we are actually considering a  model with on-site $Z_2^G$ symmetry in the
bulk, and the symmetry is promoted to $Z_4^H$ symmetry on the boundary, since
\begin{align}
(\si^x_{i_-} \si^x_{i_+} \cblue{U_{CZ,{i_-,i_+}}})^2 =  - \si^z_{i_-} \si^z_{i_+} \neq 1.
\end{align}
The total symmetry generator is given by
\begin{align}
 \widehat U_{Z_4}= \prod_i \si^x_{i_-} \si^x_{i_+}  \cblue{U_{CZ,{i_-,i_+}}}
\prod_\text{bulk sites $s$} \cblue{U_{X,s} U_{CZ,s}}.
\end{align}

To see that $ H^\text{shaded}_{p} $ is invariant under $\widehat U_{Z_4}$, we
first note that \cblue{$ H^{0}_{p} P_p^{u} P_p^{d} P_p^{l} P_p^{r}$} is invariant under
$\widehat U_{Z_4}$.  Rewriting \cblue{$\t H^{0}_{p} P_p^{u} P_p^{d}P_p^{l} (1-P_p^{r})$} as
\cblue{$\ii H^{0}_{p} P_p^{u} P_p^{d}P_p^{l} (1-P_p^{r})  \si^z_{i_1}$}, we see that
$\si^z_{i_1}$ anti-commutes with $\widehat U_{Z_4}$.  
\cblue{$H^{0}_{p} P_p^{u} P_p^{d}P_p^{l} (1-P_p^{r})$} also \cblue{anti-commutes} with $\widehat
U_{Z_4}$.
Thus  $ H^\text{shaded}_{p} $ is invariant
under $\widehat U_{Z_4}$.



The low energy boundary excitations have a basis labeled by $\si^z_{i_\pm}$
values of the boundary spins:
\begin{align}
 |\{ \si^z_{i_\pm}\}\>_\text{whole} =
 |\{ \si^z_{i_\pm}\}\>_\text{bdry}\times |\text{bulk}\> ,
\end{align}
Now, $|\text{bulk}\>$ is given by
\begin{align}
 |\text{bulk}\> = 
\otimes_\text{squares} |\text{square}\>
\otimes_\text{shaded-squares} |\text{shaded-square}\>
\end{align}
where $|\text{square}\> \equiv \frac{1}{\sqrt 2} (|\up\up\up\up\>+|\down\down\down\down\>)$ is the
spin state for the four spins connected by a red square in Fig. \ref{bdryEl} 
as determined by $H_{p}$, and
\begin{align}
\label{bulkw}
&|\text{shaded-square}\> \equiv \frac{|\up\up\up\up\>+|\down\down\down\down\>}{\sqrt 2}
 \text{ if } \si^z_{i_+}\si^z_{(i+1)_-}=1,
\\
&|\text{shaded-square}\> \equiv \frac{|\up\up\up\up\>-\ii |\down\down\down\down\>}{\sqrt 2}
 \text{ if } \si^z_{i_+}\si^z_{(i+1)_-}=-1,
\nonumber 
\end{align}
is the spin state for the four spins connected by a shaded red square in Fig.
\ref{bdryEl}, as determined by $H^\text{bdry}_{p}$.

Under the $\widehat U_{Z_4}$,
$ \frac{|\up\up\up\up\>+|\down\down\down\down\>}{\sqrt 2}
$ is unchanged for $\si^z_{i_+}\si^z_{(i+1)_-}=1$.
But for $\si^z_{i_+}\si^z_{(i+1)_-}=-1$,
$\widehat U_{Z_4}$ \cblue{changes}
$|\up\up\up\up\> \to |\down\down\down\down\>$ and
$|\down\down\down\down\> \to -|\up\up\up\up\>$.
The extra $-$ sign comes from the two uncanceled \cblue{$CZ$} factors
to the right of the plaquette (see Fig. \ref{bdryEl} where
the \cblue{$CZ$} factors are pointed out by arrows).
Therefore, under the $\widehat U_{Z_4}$,
$ \frac{|\up\up\up\up\>-\ii|\down\down\down\down\>}{\sqrt 2}
$ is changed to
\begin{align}
 \frac{|\down\down\down\down\>+\ii|\up\up\up\up\>}{\sqrt 2}
= \ii \frac{|\up\up\up\up\>-\ii|\down\down\down\down\>}{\sqrt 2}.
\end{align}


So, under the  $Z_4$ on-site transformation to the whole
system, the bulk state $|\text{bulk}\>$ changes into itself up to a phase
factor:
\begin{align}
 |\text{bulk}\> \to \ee^{\ii \th} |\text{bulk}\>.
\end{align}
The phase factor $\ee^{\ii \th}$ depends on the boundary spins $\si^z_i$ and is
given by
\begin{align}
\label{eth}
 \ee^{\ii \th} &= 
\prod_i \ii^{(1-\si^z_{i_+}\si^z_{(i+1)_-})/2} \cblue{U_{CZ,{i_-,i_+}}}
.
\end{align}
The $CZ_{i_-,i_+}$ factors in \eqn{eth} and \eqn{bUZ4} cancel each other.
Therefore, the effective $Z_4^H$ transformation on the boundary low-energy
subspace is given by
\begin{align}
\label{UZ4E}
 \widehat U_{Z_4} 
&= \prod_i \si^x_{i_-} \si^x_{i_+}\ii^{(1-\si^z_{i_+}\si^z_{(i+1)_-})/2}
\nonumber\\
&= \prod_i \si^x_{i_+} \cblue{\si^x_{(i+1)_-}}\ii^{(1-\si^z_{i_+}\si^z_{(i+1)_-})/2}
,
\end{align}
which is an on-site symmetry if we view $(i_+,(i+1)_-)$ as a site.  This means
that if we view the CZX model as a model with $Z_4$ symmetry, it is actually a
trivial $H=Z_4^H$-SPT state (since the effective $Z_4^H$ transformation on the
boundary is on-site \cblue{and} anomaly-free).  

\cblue{
To summarize, the original model in the Sec.~\ref{bdry2w} describes a \emph{gapped boundary} where 
the boundary plaquette term
$H^\text{bdry}_{p}$
has the same order as the bulk plaquette term. Now in this Sec.~\ref{bdry2l}, 
we reduce the boundary plaquette term
$H^\text{bdry}_{p}$ to only some newly-introduced projectors on the green links in Fig.\ref{bdryEl}.
For certain small or zero $H^\text{bdry}_{p}$, the boundary spins may have no 
constraint
in the whole wavefunction
$ |\{ \si^z_{i_\pm}\}\>_\text{whole} = |\{ \si^z_{i_\pm}\}\>_\text{bdry}\times |\text{bulk}\>$,
which can describe a \emph{gapless boundary}.
We have also obtained the effective $Z_4^H$ symmetry transformation on the boundary.
}

\subsection{The low energy effective theory for the fourth boundary of the CZX model -- A 1+1D exactly soluble
emergent $Z_2^K$-gauge theory} 

\label{bdry4l}

In the last subsection, we have constructed a boundary of the CZX model that \cblue{has} a
$Z_4^H$ symmetry. In this section, we \cblue{are going to modify} the above construction to
obtain a boundary that \cblue{has} the same $Z_2^G$ symmetry as the bulk.  We will
obtain a low energy effective theory for the fourth boundary of the CZX model
discussed in subsection \ref{bdry4w}.

\subsubsection{The boundary $Z_2^K$-gauge theory with an anomalous $Z_2^G$ global symmetry}

\begin{figure}[h!] 
\begin{center}
\includegraphics[scale=0.8]{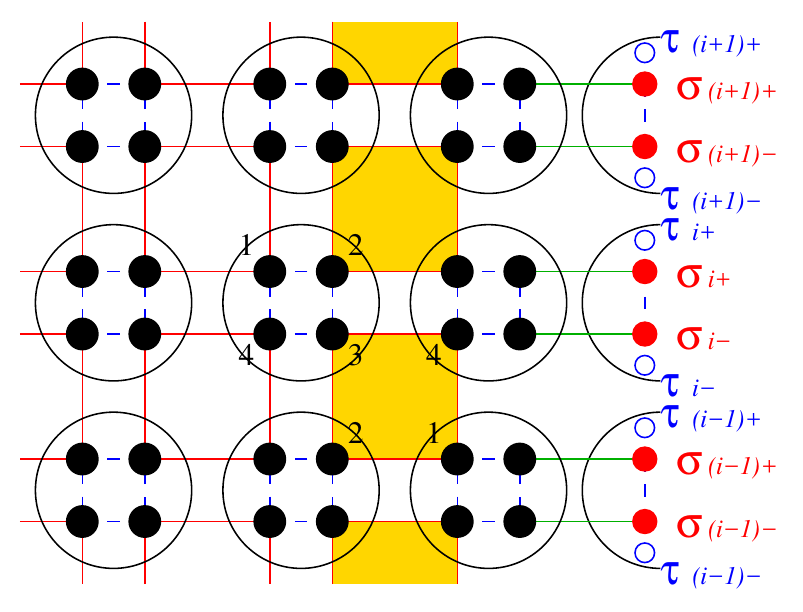}
\end{center}
\caption{
The filled dots are qubits $\up, \down$ (or spin-1/2's).  The open blue dots
are qubits $\pm 1$ representing $Z_2^K$-gauge degrees of freedom.  A circle (with
dots inside) represents a bulk site.  The bulk Hamiltonian \cblue{contains} terms that
\cblue{forces} the dots connected by red and green lines to have the same $\si^z_i$ at
low energies.  The dash blue line connecting dots $i,j$ represents the phase
factor \cblue{$U_{CZ,{ij}}$} in the $Z_2^G$ global symmetry transformation.  The open dots on
the boundary are the qubits $\v \tau_{i_\pm}$.
}
\label{bdryEge}
\end{figure}

We start with the boundary model obtained in last Sec.~\ref{bdry2l}, and add
qubits described by $\v \tau_{i_\pm}$ (see Fig. \ref{bdryEge}).  However, the
boundary physical Hilbert space is the subspace that satisfies a local gauge
constraint 
\begin{align}
\widehat U^\text{gauge}_i \equiv
\cblue{-}  \si^z_{i_+}\si^z_{i_-}  
  \tau^z_{i_+}\tau^z_{i_-}  
=1.
\end{align}

The symmetry generator is the same as before when acting on $\v \si_{i_\pm}$ 
spins. The symmetry generator acts on the $\v \tau_{i_\pm}$ 
spins as
\begin{align}
\prod_i  
\ee^{\ii \frac{\pi}{4} \tau^z_{i_-}}
\ee^{-\ii \frac{\pi}{4} \tau^z_{i_+}}
\end{align}
As we have discussed in Sec.~\ref{bdry4w}, such a symmetry generator
generates an on-site global $Z_2^G$ symmetry, in the $Z_2^K$-gauge-invariant
physical Hilbert space.

Using the effective boundary $Z_4^H$-symmetry calculated in the last subsection
\ref{bdry2l} (see \eqn{UZ4E}), plus an additional term $ \ee^{\ii \frac{\pi}{4}
\tau^z_{i_-}} \ee^{-\ii \frac{\pi}{4} \tau^z_{i_+}}$ acting on the new
$\tau_{i_\pm}$ spins, we find that the boundary effective symmetry generator is
given by 
\begin{align}
\label{UZ2Ege}
\widehat U_{Z_2} = \prod_i  \si^x_{i_+} \si^x_{(i+1)_-} \ii^{(1-\si^z_{i_+}\si^z_{(i+1)_-})/2} 
\ee^{\ii \frac{\pi}{4} \tau^z_{i_-}}
\ee^{-\ii \frac{\pi}{4} \tau^z_{i_+}}
.
\end{align}
$\widehat U_{Z_2}$ satisfies
\begin{align}
 \widehat U^2_{Z_2} &= 
\prod_i \si^z_{i_+} \si^z_{(i+1)_-}  \ii \tau^z_{i_-} (-\ii) \tau^z_{i_+} 
\nonumber\\
& =
\prod_i \Big(-\si^z_{i_+} \si^z_{(i+1)_-} \tau^z_{i_-} \tau^z_{i_+} \Big)
\prod_i (-1) 
=1.
\end{align}
in the constraint $Z_2^K$-gauge-invariant subspace.  
\cblue{Here we encounter the even-odd lattice site effect again, 
we assume that the total number of the boundary sites is always even, $\prod_i (-1)=1$, including
the example that the whole system is on a disk with only a single boundary.}
We have
turned the $Z_4^H$ symmetry in the last subsection into a $Z_2^G$ symmetry.

Next, let us include a boundary interaction term $-U_\tau\sum_i  \tau^z_{i_+}
\tau^z_{(i+1)_-}$.  In the following, we will take the $U_\tau\to +\infty$ limit.
In this case, the interaction locks $\tau^z_{i_+}= \tau^z_{(i+1)_-}$.  In the
low energy subspace, we introduce
\begin{align} \label{eq:E_and_V}
 E_{i+\frac 12}= \tau^z_{i_+}= \tau^z_{(i+1)_-}, \ \ \
 V_{i+\frac 12}= \tau^x_{i_+} \tau^x_{(i+1)_-} ,
\end{align}
that satisfies
\begin{align}
 E_{i+\frac12} V_{i+\frac12}=-V_{i+\frac12}E_{i+\frac12} .
\end{align}
Now the $Z_2^K$-gauge constraint becomes
\begin{align} \label{eq:EssEgauge}
\cblue{- E_{i-\frac12} \si^z_{i_-} \si^z_{i_+} E_{i+\frac12}} =1.
\end{align}
The effective $Z_2^G$ symmetry generator becomes
\begin{align}
\label{UZ2Eg}
\widehat U_{Z_2} &= \prod_i  \si^x_{i_+} \si^x_{(i+1)_-} \ii^{(1-\si^z_{i_+}\si^z_{(i+1)_-})/2} 
.
\end{align}

After \cblue{obtaining} the  effective $Z_2^G$ symmetry on the boundary, we can write down a
\cblue{global} $Z_2^G$ symmetric (under \eqn{UZ2Eg}) and local $Z_2^K$-gauge symmetric (under \eqn{eq:EssEgauge}) boundary effective Hamiltonian:
\begin{align}
\label{Z2H}
  H &=  
 - \sum_i  V_{i+\frac12} (|\up\up\>\<\down\down|\negthinspace+ |\down\down\>\<\up\up|)_{i_+,(i+1)_-}  
\nonumber\\
&\ \ \ \ \ \ \ -J\sum_i \si^z_{i_+} \si^z_{(i+1)_-} - U \sum_i E_{i+\frac12} 
\\
&=  
- \sum_i  V_{i+\frac12} 
 \big( \si^+_{i_+} \si^+_{(i+1)_-}  \cblue{+ } \; \si^-_{i_+} \si^-_{(i+1)_-}\big)  
\nonumber\\
&\ \ \ \ \ \ \ 
 -J\sum_i \si^z_{i_+} \si^z_{(i+1)_-} - U \sum_i E_{i+\frac12} 
.
\nonumber 
\end{align}
This is our fourth boundary of the CZX model discussed in Sec.~\ref{bdry4w}, \cblue{but now it becomes a} 1+1D lattice $Z_2^K$-gauge theory with an anomalous
(non-on-site) global $Z_2^G$-symmetry.

\subsubsection{Confined $Z_2^K$-gauge state -- A spontaneous symmetry breaking state}

In general, a large $U$ in the above Hamiltonian will give us a $Z_2^K$-gauge
confined phase (which will be discussed later in more detail).  In the
$Z_2^K$-gauge confined phase induced by  \cblue{a}  large $U$, we have $E_{i+\frac12}=1$.  
\cblue{In this case, 
because of \eqn{eq:E_and_V} and \eqn{eq:EssEgauge},
} 
\cred{$\si^z_{i_-}\si^z_{i_+}=-1$} on every site, which reduces two spin 
\cblue{$\v \si_{i_-}$ and $\v \si_{i_+}$} into one spin $\v \si_{i}$. This reduces the $Z_2^G$
symmetry transformation into
\begin{align} 
\widehat U_{Z_2} = \prod_i  \cblue{\tilde{\si}^x_{i}} \prod_i
\ii^{(1-\cblue{ (\tilde{\si}^z_{i}) (-\tilde{\si}^z_{i+1}) })/2} 
\end{align} 
which is a non-on-site (anomalous) $Z_2^G$-symmetry transformation.  
\cblue{Here $\tilde{\si}^x_{i}$ is a redefinition of $\si^x_{i_-} \si^x_{i_+}$  for the composite spin.
More precisely, due to the gauge constraint {$\si^z_{i_-}\si^z_{i_+}=-1$},
$\tilde{\si}^x_{i}$ flips the composite spin as
$\tilde{\si}^x_{i} | \up\>_{i_-} | \down\>_{i_+}=| \down\>_{i_-} | \up\>_{i_+}$ 
and  $\tilde{\si}^x_{i} |\down  \>_{i_-} | \up \>_{i_+}=|   \up\>_{i_-} | \down  \>_{i_+}$. 
Since the two spins are locked {$\si^z_{i_-}\si^z_{i_+}=-1$} in the same site, 
we can also simply define $\tilde{\si}^z_{i} \equiv \si^z_{i_+}$,
so that $\tilde{\si}^z_{i+1} \equiv \si^z_{(i+1)_+}=-\si^z_{(i+1)_-}$.
}
So in \cblue{the} large
$U$ limit, the lattice $Z_2^K$-gauge theory, at low energies, reduces to the
boundary of the CZX model constructed in Sec.~\ref{bdry1}.  When $J>0$, the
confined $Z_2^K$-gauge state is a \cblue{ferromagnetic} state, that spontaneously
\cblue{breaks} the global $Z_2^G$-symmetry.


\subsubsection{Deconfined $Z_2^K$-gauge state in 1+1D}

The model \eqn{Z2H} is exactly soluble. This is because in the big Hilbert
space before projecting into the $Z_2^K$-gauge-invariant subspace, the Hamiltonian $H$ in
\eqn{Z2H} is a sum of non overlapping local terms:
$H=\sum_i H_{i,i+1}$ with
\begin{align}
H_{i,i+1} &= - V_{i+\frac12}  
\big[ \si^+_{i_+} \si^+_{(i+1)_-}  +\si^-_{i_+} \si^-_{(i+1)_-} \big]
\nonumber\\
&\ \ \ \ \
 -J \si^z_{i_+} \si^z_{(i+1)_-}
- U E_{i+\frac12}
\end{align}
So the energy spectrum of $H$ can be obtained exactly from that of $H_{i,i+1}$.
The $Z_2^K$-gauge transformation
\begin{align}
\label{gtrans}
 \widehat U^\text{gauge}_i=-(E_{i-\frac12} \si^z_{i_-})(\si^z_{i_+} E_{i+\frac12})
\end{align}
\cblue{commutes} with $H$.  So the energy spectrum of $H$ in the $Z_2^K$-gauge-invariant
subspace is a subset of the spectrum in the big unconstrained Hilbert space.

In the deconfined state at $U=J=0$, $V_{i+\frac12}=\pm1$ and does not fluctuate
before we apply the $Z_2^K$-gauge constraint (\ie $V_{i+\frac12}$ does not
fluctuate in the big Hilbert space before projecting into the $Z_2^K$-gauge
invariant subspace, since $[V_{i+\frac12}, H]=0$).  The ground state wave
function on each link is $(|\up\up\> +
v_{i+\frac12}|\down\down\>)_{i_+,(i+1)_-}\otimes |v_{i+\frac12}\> $, where
$|v_{i+\frac12}=\pm 1\>$ are the eigenstates of $V_{i+\frac12}$.  The
gauge-invariant ground \cblue{states $ |\Psi_{\mathrm{gs}}(\pm)\>$ are two distinct holonomy sectors labeled by $\prod_i v_{i+\frac12} =\pm 1$, explicitly as:}
\begin{align}
\label{Psigs}
& |\Psi_{\mathrm{gs}}(\pm)\>
=\sum_{\{v_{i+\frac12}\}, \cblue{\prod_i v_{i+\frac12} =\pm 1}}  \cblue{c_{\{v_{i+\frac12}\}}}
 \bigotimes_i
(|\up\up\> + v_{i+\frac12}|\down\down\>)_{i_+,(i+1)_-}\otimes |v_{i+\frac12}\> \cblue{.}
\end{align}
\cblue{Here the coefficient $c_{\{v_{i+\frac12}\}}$ is determined in the same way as eqns.(\ref{eq:gs+}) and (\ref{eq:gs-})
with alternating $\pm 1$ signs set by the gauge-invariant constraint on the ground states $ |\Psi_{\mathrm{gs}}(\pm)\>$.}

Under the $\widehat U_{Z_2}$ \cblue{global symmetry operation \eqn{UZ2Eg},} 
\begin{align}
 |\up\up\> + v_{i+\frac12}|\down\down\>
\to v_{i+\frac12} (|\up\up\> + v_{i+\frac12}|\down\down\>).  
\end{align}
Thus
\begin{align}
  \widehat U_{Z_2} |\Psi_{\mathrm{gs}}(\pm)\>
  =  \prod_i (v_{i+\frac12})|\Psi_{\mathrm{gs}}(\pm)\> \cblue{.}
\end{align}

From the above results, we see that \emph{the global $Z_2^G$ charge and the
$Z_2^K$-gauge flux $\prod_i v_{i+\frac12}$ are locked}.  In other words, the
deconfined state has two degenerate ground states on the ring and a finite
energy gap. One ground state carries the global $Z_2^G$ charge $0$ and no
$Z_2^K$-gauge flux through the ring. The other carries the global $Z_2^G$ charge
$1$ and {the} $\pi$ $Z_2^K$-gauge flux through the ring.  Near the end of
the next section, we will show that the above deconfined states spontaneously break
the global $Z_2^G$-symmetry, which is another way to understand the two
degenerate ground states on the ring.

\subsubsection{Deconfined and confined $Z_2^K$-gauge states
belong to the same phase that spontaneously breaks the $Z_2^G$ global symmetry} 

\label{sec:deconfined-to-SSB}


We note that for the following four spin states $|\up \up\>+|\down \down\>$,
$|\up \up\>-|\down \down\>$, $|\up \down\>$, and $|\down\up\>$ are common
eigenstates of $ \si^+_{i_+} \si^+_{(i+1)_-}  +\si^-_{i_+}
\si^-_{(i+1)_-}$ and $\si^z_{i_+} \si^z_{(i+1)_-}$ with eigenvalues $(1,
1)$, $(-1, 1)$, $(0, -1)$, and $(0, -1)$.

For $U,J>0$, the ground states have a 2-fold degeneracy, which is given by
\cblue{
\begin{align}
&|\psi_1 \>= (|\up \up\>+|\down \down\>)_{{i_+},{(i+1)_-}}\otimes (\cos(\th)|1\> +\sin(\th)|-1\>)_{{i+\frac12}}, 
\nonumber\\
&|\psi_2 \>= (|\up \up\>-|\down \down\>)_{{i_+},{(i+1)_-}}\otimes (\sin(\th)|1\> +\cos(\th)|-1\>)_{{i+\frac12}},    
\end{align}
}
where $|\pm 1\>$ are eigenstates of $V_{i+\frac12}$ with eigenvalues $\pm 1$.
\cblue{In order to have the two states as ground states,
$\theta$ is constrained to be the function of $U$ as $\theta=\frac{1}{2}\tan^{-1}U$
.}

The energy of the two ground states is $E= -\sqrt{1+U^2} -J$. Also $\th=0$ for $U=0$
(the $Z_2^K$-gauge deconfined case) and $\th\to \pi/4$ for $U\to +\infty$ (the
$Z_2^K$-gauge confined case).  The first excited states also  have a 2-fold
degeneracy, which is given by
\cblue{
\begin{align}
& |\up \down\>_{{i_+},{(i+1)_-}}\otimes (|1\> +|-1\>)_{{i+\frac12}},
\nonumber\\
\text{and }  \ \ \ \ 
& |\down\up \>_{{i_+},{(i+1)_-}} \otimes (|1\> +|-1\>)_{{i+\frac12}}, 
\end{align}}
with energy $E= -|U|\negthinspace+J$, which is higher than the ground state energy
by at least $2J$ (note that we have assumed $J>0$).

We note that
\begin{align}
&\ \ \ (|\up \up\>+|\down \down\>)\otimes (\cos(\th)|1\> +\sin(\th)|-1\>)
\nonumber\\
&+ (|\up \up\>-|\down \down\>)\otimes(\sin(\th)|1\> +\cos(\th)|-1\>)
\nonumber\\
&\;\cblue{\equiv}\; |++\>
\end{align}
is a common eigenstate of $(\si^z_{i_+}E_{i+\frac12},
E_{i+\frac12}\si^z_{(i+1)_-} )$ with eigenvalues $(+1,+1)$, and we denote it as
$|++\>$ \cblue{or $|++\>_{{i_+},i+\frac{1}{2},{(i+1)_-}}$}. Similarly,
\begin{align}
&\ \ \ (|\up \up\>+|\down \down\>)\otimes (\cos(\th)|1\> +\sin(\th)|-1\>)
\nonumber\\
&- (|\up \up\>-|\down \down\>)\otimes(\sin(\th)|1\> +\cos(\th)|-1\>)
\nonumber\\
& \;\cblue{\equiv}\; |--\>
\end{align}
is a common eigenstate of $(\si^z_{i_+}E_{i+\frac12},
E_{i+\frac12}\si^z_{(i+1)_-} )$ with eigenvalues $(-1,-1)$, and we denote it as
$|--\>$ \cblue{or $|--\>_{{i_+},i+\frac{1}{2},{(i+1)_-}}$}. 

A $Z_2^K$-gauge-invariant ground state (\ie $\widehat U^\text{gauge}_i=1$ state) on a
ring is given by the tensor product of those $|++\>$ and $|--\>$ states on the
$(i,i+1)$ links.  First we note that the gauge transformation in \eqn{gtrans} is a
product of two operators $E_{i-\frac12} \si^z_{i_-}$ and $\si^z_{i_+}
E_{i+\frac12}$ with an additional $-$ sign. \cblue{The} $|++\>$ and $|--\>$ are eigenstates of those operators.
Therefore, we have two $Z_2^K$-gauge-invariant ground states:

\begin{align} \label{eq:gs_theta}
|\Psi_1(\theta)\>= \cdots \otimes 
|++\>_{{(i-1)_+},i-\frac{1}{2},{i_-}} \otimes |--\>_{{i_+},i+\frac{1}{2},{(i+1)_-}}\otimes |++\>_{{(i+1)_+},i+\frac{3}{2},{(i+2)_-}} \otimes 
 \cdots,
\nonumber\\
|\Psi_2(\theta)\>= \cdots  \otimes
|--\>_{{(i-1)_+},i-\frac{1}{2},{i_-}} \otimes  |++\>_{{i_+},i+\frac{1}{2},{(i+1)_-}}\otimes |--\>_{{(i+1)_+},i+\frac{3}{2},{(i+2)_-}}\otimes  
 \cdots,
\end{align}
{up to a proper normalization factor.
Note that to get a $Z_2^K$-gauge-invariant state \cblue{under \eqn{gtrans}} we need to match $+$ to $-$ and
$-$ to $+$ in the neighboring links, as done in the above. 
However, the two ground states expressed in \eqn{eq:gs_theta} 
are not symmetric under the global $Z_2^G$ symmetry transformation in \eqn{UZ2Eg}:
$$
\widehat U_{Z_2} = \prod_i  \si^x_{i_+} \si^x_{(i+1)_-} \ii^{(1-\si^z_{i_+}\si^z_{(i+1)_-})/2} \equiv \prod_i U_{Z_2,i_+,(i+1)_-}
$$
In fact, $U_{Z_2,i_+,(i+1)_-}$ exchanges $|++\>$ and $|--\>$,
\begin{eqnarray}
&U_{Z_2,i_+,(i+1)_-} |++\>_{{(i-1)_+},i-\frac{1}{2},{i_-}} = |--\>_{{(i-1)_+},i-\frac{1}{2},{i_-}}, \;\;\;\;\; \\
&U_{Z_2,i_+,(i+1)_-} |--\>_{{(i-1)_+},i-\frac{1}{2},{i_-}} = |++\>_{{(i-1)_+},i-\frac{1}{2},{i_-}}.\;\;\;
\end{eqnarray}
The ground states that respect the global $Z_2^G$ symmetry transformation \eqn{UZ2Eg} are the linear combination of
\eqn{eq:gs_theta}:
\begin{eqnarray} \label{obo}
|\Psi_{gs,even}(\theta)\> =& \frac{1}{\sqrt 2}(|\Psi_1(\theta)\> + |\Psi_2(\theta)\>) \nonumber \\
|\Psi_{gs,odd}(\theta)\> =& \frac{1}{\sqrt 2}(|\Psi_1(\theta)\> - |\Psi_2(\theta)\>),
\end{eqnarray}
}
where the $|\Psi_{gs,even}(\theta)\>$ is $Z_2^G$-symmetry even by $\widehat U_{Z_2} |\Psi_{gs,even}(\theta)\>=+ |\Psi_{gs,even}(\theta)\>$,
and 
the $|\Psi_{gs,odd}(\theta)\>$ is $Z_2^G$-symmetry odd by $\widehat U_{Z_2} |\Psi_{gs,odd}(\theta)\>=- |\Psi_{gs,odd}(\theta)\>$.

{
When $\theta = 0$, the even/odd $Z_2^G$ symmetric ground states are identical to the even/odd $Z_2^K$-gauge holonomy sectors of ground states
 in \eqn{Psigs}
due to the locking of $Z_2^G$-charge and $Z_2^K$-holonomy:
\begin{eqnarray}
&&|\Psi_{gs,even}(\theta = 0)\> = \frac{1}{\sqrt 2}(|\Psi_1(0)\> + |\Psi_2(0)\>)= |\Psi_{\mathrm{gs}}(+)\>, \nonumber\\
&&|\Psi_{gs,odd}(\theta = 0)\> = \frac{1}{\sqrt 2}(|\Psi_1(0)\> - |\Psi_2(0)\>) = |\Psi_{\mathrm{gs}}(-)\>.\;\;\;\;\;\;
\end{eqnarray}
}
%
%
%
{
When $\theta =\frac{\pi}{4} $, we have the confined states:
}
\begin{align}
|\Psi_1(\theta=\frac{\pi}{4})\>= (\cdots \otimes 
| \up \up \>_{{(i-1)_+},{i_-}} \otimes |\down \down\>_{{i_+},{(i+1)_-}}\otimes |\up \up\>_{{(i+1)_+},{(i+2)_-}} \otimes 
 \cdots) \bigotimes_i (|1\> +|-1\>)_{{i+\frac12}},
\nonumber\\
|\Psi_2(\theta=\frac{\pi}{4})\>= (\cdots  \otimes
|\down \down\>_{{(i-1)_+},{i_-}} \otimes  |\up \up \>_{{i_+},{(i+1)_-}}\otimes |\down \down\>_{{(i+1)_+},{(i+2)_-}}\otimes  
 \cdots) \bigotimes_i (|1\> +|-1\>)_{{i+\frac12}},
\end{align}
{up to a proper normalization factor.
Below we aim to show that at $\theta=0$,
namely $U =0$ and $J> 0$, 
we have the deconfined state with spontaneous
$Z_2^G$-symmetry breaking;
at $\theta=\frac{\pi}{4}$,
namely $U \to +\infty$ and $J> 0$, 
we have the confined state with spontaneous
$Z_2^G$-symmetry breaking.
We demonstrate a strange property for this system: \emph{the deconfined state with spontaneous
$Z_2^G$-symmetry breaking and the confined state with spontaneous $Z_2^G$-symmetry
breaking belong to the same phase}.
In the next few paragraphs, we explain the 
meanings of the deconfined and confined phases,
and also the meanings of the spontaneous symmetry breaking. 
}

{First, we elaborate further on the physical meanings of the deconfined and confined phases.
The \emph{deconfined} phase ($U=0$) 
here means that the distinct holonomies or loop excitations (namely Wilson lines) can span the large system without causing extra energy.
Consider the expectation value $\langle 0 | W | 0\rangle$ of Wilson line operator $W \equiv  \prod_i V_{i+\frac12}$ for some ground state $| 0\rangle$,
the $\langle 0 | W | 0\rangle$ goes to some constant (proportional to the net holonomy $\prod_i v_{i+\frac12} = \pm 1$) in the Euclidean spacetime,  
and, thus, obeys the perimeter law instead of the area law \cite{Wilson1974_PhysRevD}. 
The two ground states with distinct holonomies in our case
imply that we are in the deconfined phase, even if the energy spectrum is gapped between the ground states and the first excitations.
On the other hand, the \emph{confined} phase ($U\to \infty$, $J>0$) has the gauge field variable $|v_{i+\frac{1}{2}}\>$ quantum disorder and strong fluctuations
in the state $(|1\> +|-1\>)_{{i+\frac12}}$.
The long-distance lines/holonomies are energy-disfavored. 
Consider the expectation value $\langle 0 | W | 0\rangle$ of Wilson line operator
$W$ for any ground state $| 0\rangle$, 
the $\langle 0 | W | 0\rangle$ exponentially decays to zero
in the Euclidean spacetime, thus obeys the area law, thus the phase is confined.  
The $Z_2^K$-gauge confined
phase for $U \to +\infty$ and $J> 0$ is a ferromagnetic along the link $\overline{{i_+}{(i+1)_-}}$ but anti-ferromagnetic between the neighbored links
between spin up and down.
There is no phase transition as $U$ goes from $0$ to $+\infty$ for $J > 0$,
since the energy gap above the ground state is always bigger than $2J$.  Thus
\emph{the $Z_2^K$-gauge deconfined state for $U=0$ and the $Z_2^K$-gauge confined
state for $U=+\infty$ belong to the same phase}. }  

{Second, we elaborate further on the physical meanings of the spontaneous symmetry breaking (SSB) and possible long-range orders.
Based on Ref.\cite{1406.5090}, we know that the SSB in a quantum system does not necessarily mean that its ground states break the
symmetry. 
Traditionally, we identify the symmetry-breaking order parameter and we compute the long-range order correlation functions to detect the symmetry-breaking. 
The better definition for SSB is based on the Greenberger-Horne-Zeilinger (GHZ)  entanglement \cite{greenberger_going_1989-1}.
Use GHZ form, we can probe the symmetry without knowing the symmetry or the Ginzburg-Landau symmetry-breaking order parameters. 
Use GHZ form, we can detect the symmetry-breaking hidden in the symmetric ground-state wavefunction.} 

{
Indeed, $|\Psi_1(\theta)\>$ and $|\Psi_2(\theta)\>$ are GHZ states,
\begin{eqnarray} \label{obbo}
|\Psi_{gs,even}(\theta)\> &=& \frac{1}{\sqrt 2}(|\Psi_1(\theta)\> + |\Psi_2(\theta)\>) \equiv |\text{GHZ}_+(\theta)\>  \nonumber \\
|\Psi_{gs,odd}(\theta)\> &=& \frac{1}{\sqrt 2}(|\Psi_1(\theta)\> - |\Psi_2(\theta)\>) \equiv |\text{GHZ}_-(\theta)\>. \;\;\;\;\;\;
\end{eqnarray}
Because the $Z_2^G$-global symmetry operator $\widehat U_{Z_2}$ acting on two states
gives rise to the symmetric charge $\pm 1$,
the following conditions for SSB of symmetry group $G$ are satisfied:
\begin{enumerate}
\item  $\widehat U_{Z_2}  |\text{GHZ}_{\pm }(\theta)\> = {\pm }|\text{GHZ}_{\pm }(\theta)\>.$
\item  The symmetric GHZ states have the same GHZ entanglement 
$|\text{GHZ}_{}\>=\sum_j c_j |\Psi_j \>$,  with $j \in G/G'$,  $G' \subset G$, where $|\Psi_j \>$ are locally distinguishable.
In our case, we have $G=Z_2$ and $G'$ is trivial.
\end{enumerate}
}

{To summarize, 
the symmetric many-body state has spontaneous symmetry breaking, which implies that the state has a GHZ entanglement.
Indeed, we can also show that the SSB here also implies the long-range order, consistent with what we observed in \eqn{zerof} in Sec.~\ref{bdry3w}. 
Defining the gauge-invariant operator $X_{i+1/2}=\sigma^z_{i+}E_{i+1/2}$
which is odd breaking the $Z_2^G$-symmetry,
we find $X_{i+1/2} |\Psi_1(\theta)\> = -|\Psi_1(\theta)\>$ and 
$X_{i+1/2} |\Psi_2(\theta)\> = +|\Psi_2(\theta)\>$. Moreover,
\bea
\label{eq:long-range-order-GHZ}
\langle \text{GHZ}_{\pm}(\theta) |X_{i+1/2}X_{j+1/2} |\text{GHZ}_{\pm }(\theta)\>= 1.
\eea
Thus the $G$-symmetry odd operator detects the long-range correlator of GHZ states, and 
we demonstrate the SSB through the long-range order.
In summary,
we show that \emph{the deconfined state
 and the confined state belong to the same phase without the phase transition by tuning the Hamiltonian coupling
 $U$ with the ground state parameter $\theta=\frac{1}{2}\tan^{-1}U$. 
All values of $U$ have the spontaneous $Z_2^G$-symmetry
breaking.}
This is possible since the $Z_2^K$-gauge
deconfined phase with no spin order has two-fold degenerate ground states with
opposite global $Z_2^G$ charge, the same as the ferromagnetic state with spin
order which also has two-fold degenerate ground states with opposite global
$Z_2^G$ charges.
}

We remind the readers that the fermionic version of the CZX model is studied in Appendix \ref{sec:fCZX}.
The boundary of the fermionic CZX model with emergent $Z_2^K$-gauge theory with anomalous global symmetry
is detailed in Appendix.\ref{sec:fCZXbdryZ2-gauge}.

One can read Sec.~\ref{rev} on more general boundaries of SPTs in any dimension.

\section{Fermionic CZX model} \label{sec:fCZX}

Consider a square lattice model with each single site endowed with four
fermion orbitals, each with eigenstates $|0\>$ and  $|1\>$  of the fermion number
operator $n_f=c^\dagger c$.  Thus a single site has a $2^4$-dimensional Hilbert
space.  We may call the single site a ``vertex,'' and  the four individual fermion
orbitals in a site ``sub-vertices.''
In the fermionic model, we have the anti-commutation relation
$$
\{c_i, c_j^\dagger\}=\delta_{ij},
$$
where $i,j$ can be any local fermion degree of freedom, on the same site or on  different sites.
Fermion parity operator $P_f$ on each site (with 1,2,3,4 four sub-vertices):
\begin{align}
P_f=\prod_{i=1,2,3,4} (-1)^{n_{f,i}}=\prod_{j=1,2,3,4} \si^z_{j}.
\end{align}
Notice that
\begin{align}
(1-2 c^\dagger_i c_i)=\si^z_{j}, \;\;\;\; c^\dagger_i c_i =\frac{1-\si^z_{j}}{2}
\end{align}
Let us introduce a $Z_2$ generator $U_X$ as
a product of $c^\dagger_j+ c_j$ on the four sub-vertices:
\begin{align}
\label{UXf}
U_X&=
(c^\dagger_1+ c_1)(-1)^{n_1}
(c^\dagger_2+ c_2)(-1)^{n_1}(-1)^{n_2}
(c^\dagger_3+ c_3)(-1)^{n_1}(-1)^{n_2}(-1)^{n_3}
(c^\dagger_4+ c_4)
\nonumber\\
&= 
 \si^x_1 \si^x_2 \si^x_3 \si^x_4 ,\ \ \ \ U_X^2=1,
\end{align}
where we have used the Jordan-Wigner transformation to express
fermion operators in terms of spin operators, for example
\begin{align}
 c^\dagger_j+ c_j = (\prod_{i<j} \si^z_i) \si^x_j,
\end{align}
where $i<j$ refers to a particular ordering of the orbitals (see Fig
\ref{bdryCf}).  We have chosen an unusual definition of $U_X$
(instead of the more obvious $(c^\dagger_1+ c_1) (c^\dagger_2+ c_2) (c^\dagger_3+ c_3)
(c^\dagger_4+ c_4)$), because we want $U_X$ to have a simple form after
bosonization.

For any pair of qubits, we set  $CZ=|00\>\< 00|\negthinspace+|01\>\< 01|\negthinspace+|10\>\< 10|-  |11\>\< 11|=1-2cc^\dagger c' c'^\dagger$.
For each site, we define $U_{CZ}$ as the product of such operators over all successive pairs:
\begin{align}
U_{CZ}&=\prod_{j=1,2,3,4}  (1-2 c^\dagger_{j+1}  c_{j+1} c^\dagger_j  c_j)
\nonumber\\
&=\prod_{j=1,2,3,4} (1-\frac{(1-\si^z_{j+1})(1-\si^z_{j})}{2})
\nonumber\\
&=\prod_{j=1,2,3,4} (\frac{(1+\si^z_{j+1}+\si^z_{j}  -\si^z_{j+1}\si^z_{j})}{2}),
\end{align}
where $j=5 \mod 4=1  \mod 4$.
Now we introduce a $Z_2$ transformation in each site:
\begin{align}
U_{CZX}=U_XU_{CZ}, \ \ \ \
U_{CZX}^2=1.
\end{align}
The group super-cohomology predicts that there are four distinct fermionic SPTs with $G=Z_2 \times Z_2^f$ symmetry from
$H_{\text{super}}^3[Z_2 \times Z_2^f,U(1)]=\Z_4$.  
The model we will first focus is the one with the second class $\nu =2$ for $\nu \in \Z_4$.
The full classification for four distinct fermionic SPTs with $Z_2 \times Z_2^f$ symmetry 
is $\Z_8$ from the spin cobordism group  $\Omega^\text{Spin}_3(BZ_2)= \Z_8$; then, our model here is $\nu =4$ for $\nu \in \Z_8$.


The fermionic CZX Hamiltonian is essentially the same as the bosonic CZX Hamiltonian:
\begin{align}
H^f=\sum H_{p}.
\end{align}
\begin{align}
H_{p}=-X_4 P_2^{u} P_2^{d} P_2^{l} P_2^{r}
\end{align}
Here plaquettes are defined in the bosonic CZX model.
$X_4$ acts on the four sub-vertices in a plaquette, 
\begin{align}
X_4&=c_3c_4c_2c_1+c^{\dagger}_3c^{\dagger}_4c^{\dagger}_2c^{\dagger}_1
\nonumber\\
&=\si^-_4\si^-_3\si^-_2\si^-_1+\si^+_4\si^+_3\si^+_2\si^+_1
\\
&= ( |0000\>\< 1111|+  |1111\>\< 0000|\negthinspace)_{\text{plaquette}},
\nonumber 
\end{align}
and the projection operator $P_2$ acts on a pair of qubits adjacent to a plaquette
as \begin{align}
P_2&=c_ic^{\dagger}_ic_{i+1}c^{\dagger}_{i+1}+c^{\dagger}_ic_ic^{\dagger}_{i+1}c_{i+1}
\nonumber\\
&=( |00\>\< 00|+  |11\>\< 11|\negthinspace)_{\text{line}}
\end{align}

We see that, after  bosonization, both the Hamiltonian and the $Z_2$ symmetry for the fermionic CZX model map to those of the bosonic CZX model.
So the  ground state of the fermionic CZX model is the same as that of the bosonic
CZX model described in Sec.~\ref{bCZX}.

It is also obvious that $[\prod P_f, H_f]=0$ since $H_f$ conserves fermion
number mod 2 (in fact, $H_f$ conserves fermion number mod 4).  So the fermionic
CZX model $H_f$ has $Z_2\times Z_2^f$ symmetry generated by $\prod U_{CZX}$ and
$\prod P_f$.  The ground state is invariant under the symmetry.

\section{A boundary of the fermionic CZX model --
Emergent $Z_2^K$-gauge theory with an anomalous global symmetry, and Majorana fermions}

\label{sec:fCZXbdryZ2-gauge}

\begin{figure}[h!]
\begin{center}
\includegraphics[scale=0.8]{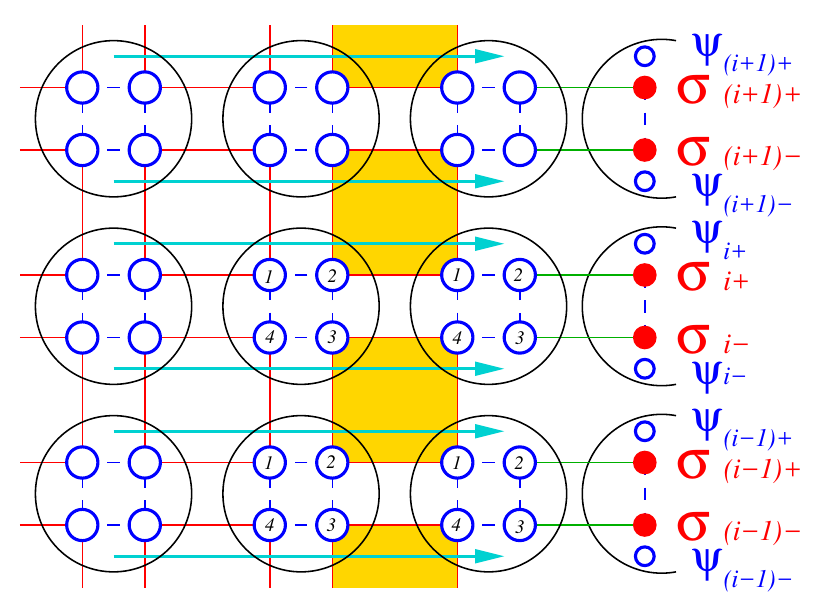}
\end{center}
\caption{
The filled dots are qubits (or spin-1/2's) described by $\v \si$.  The open
dots are fermion orbitals described by $c$ or $\psi$.  A circle (with dots
inside) represents a site.  The bulk Hamiltonian contains terms that force the
dots connected by red and green lines to have the same $(-1)^{n_i}$ or $\si^z_i$ 
at low energies.  The dashed blue line connecting dots $i,j$ represents the phase
factor $CZ_{ij}$ in the $Z_2^G$ global symmetry transformation.  The arrow
describes a particular ordering of all fermion orbitals.
}
\label{bdryCf}
\end{figure}

\begin{figure}[h!]
\begin{center}
\includegraphics[scale=0.8]{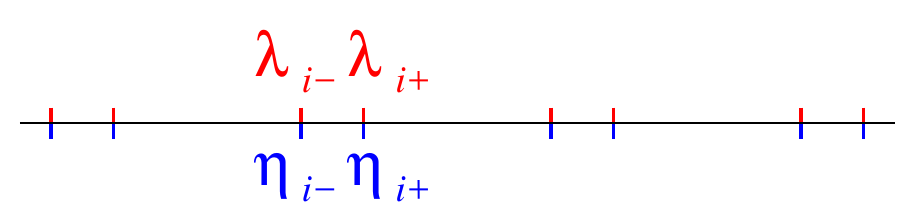}
\end{center}
\caption{
Emergent $Z_2^K$-gauge theory from Majorana fermions on the lattice.
}
\label{etala}
\end{figure}

To obtain a boundary of the fermionic CZX model, we start with the boundary model
described in Fig. \ref{bdryCf}.  On the boundary, we have qubits described by
$\v \si_{i_\pm}$ and fermions described by $\psi_{i_\pm}=\eta_{i_\pm}+\ii
\la_{i_\pm}$, where $\eta$ and $\la$ are Majorana fermion operators, see Fig.~\ref{etala}.

However, we assume that the boundary Hilbert space is not the one generated by $\v
\si_{i_\pm}$ and  $\psi_{i_\pm}$, but a subspace satisfying a local
$Z_2^K$-gauge constraint:
\begin{align}
 \widehat U^\text{gauge}_i = -\si^z_{i_+} \si^z_{i_-} (-1)^{n_{i_-}+n_{i_+}} =1, 
\end{align}
where
\begin{align}
n_{i_\pm} = \psi^\dag_{i_\pm} \psi_{i_\pm}
.
\end{align}
Thus, the boundary is a $Z_2^K$ lattice gauge theory.

The bulk Hamiltonian of the model is still given by $H^f_{p}$ for the
complete octagons in the bulk, with additional terms that force the boundary
qubits $\si_{i_\pm}^z$ to have the same value as the $(-1)^{n_i}$ for the bulk
fermions connected by the green lines.  However, notice that the shaded squares
are not  complete octagons since the two spins to the right of the shaded
squares do not need to be parallel.  So the Hamiltonians for the shaded squares
need to be modified:
\begin{align}
 H^{f,\text{shaded}}_{p} &=
- X_4 P_2^{u} P_2^{d} P_2^{l} P_2^{r} 
+\t X_4 P_2^{u} P_2^{d} P_2^{l} (1-P_2^{r}) 
\end{align}
where $\t X_4$ is given in \eqn{tX4}.
The $Z_2^G$-symmetry of the system is generated by
\begin{align}
\label{bUZ2}
\widehat U_{Z_2}= \prod_i \si^x_{i_-} \si^x_{i_+} CZ_{i_-,i_+} 
\ee^{\ii \frac{\pi}{4} (1-2n_{i_-})}
\ee^{-\ii \frac{\pi}{4} (1-2n_{i_+})}
\prod_\text{bulk} U_{CZX}.
\end{align}

After the bosonization via Jordan-Wigner transformation on Majorana fermion operators, 
\begin{align}
 \la_j = (\prod_{i<j} \tau^z_i) \tau^x_j, \ \ \ 
 \eta_j = (\prod_{i<j} \tau^z_i) \tau^y_j,
\end{align}
the above Hamiltonian
and the $Z_2^G$-symmetry map to those of the bosonic model discussed in subsection
\ref{bdry4l}.  So we can use the results there. First one can show that 
\begin{align}
\Big(\si^x_{i_-} \si^x_{i_+} CZ_{i_-,i_+} 
\ee^{\ii \frac{\pi}{4} (1-2n_{i_-})}
\ee^{-\ii \frac{\pi}{4} (1-2n_{i_+})} 
\Big)^2 =1
\end{align}
in the $Z_2^K$-gauge-invariant physical Hilbert space.  So $\widehat U_{Z_2}$
generates an on-site global $Z_2^G$-symmetry.  Second one can show that the
Hamiltonian is indeed $Z_2^G$ symmetric.  Third, one can find the low-energy
effective $Z_2^G$-symmetry on the boundary to be generated by
\begin{align}
\label{UZ4Ef}
 \widehat U_{Z_2} &= 
\prod_i \si^x_{i_+} \si^x_{(i+1)_-}
\ii^{(1-\si^z_{i_+}\si^z_{(i+1)_-})/2}
\ee^{\ii \frac{\pi}{4} (1-2n_{i_+})}
\ee^{-\ii \frac{\pi}{4} (1-2n_{(i+1)_+})} 
.
\end{align}

Next, let us include a boundary interaction term $-U_\tau\sum_i  (1-2n_{i_+})
(1-2n_{(i+1)_-})$ and take $U_\tau\to +\infty$ limit.
In this case, the interaction locks $n_{i_+}= n_{(i+1)_-}$.  In the
low energy subspace, we introduce
\begin{align}
 E_{i+\frac 12} &= 1-2n_{i_+}= 1-2n_{(i+1)_-}, 
\nonumber\\
 V_{i+\frac 12} &= \la_{i_+} (-1)^{n_{i_+}}\la_{(i+1)_-}.
\end{align}
After the bosonization on the boundary, the above becomes
\begin{align}
 E_{i+\frac 12}= \tau^z_{i_+}= \tau^z_{(i+1)_-}, \ \ \
 V_{i+\frac 12}= \tau^x_{i_+} \tau^x_{(i+1)_-}, 
\end{align}
which satisfies
\begin{align}
 E_{i+\frac12} V_{i+\frac12}=-V_{i+\frac12}E_{i+\frac12} .
\end{align}
Now the $Z_2^K$-gauge constraint becomes
\begin{align}
 -E_{i-\frac12} \si^z_{i_+}\si^z_{i_-} E_{i+\frac12} =1.
\end{align}
The effective $Z_2^G$ symmetry generator becomes
\begin{align}
\label{UZ2Egf}
\widehat U_{Z_2} &= \prod_i  \si^x_{i_+} \si^x_{(i+1)_-} \ii^{(1-\si^z_{i_+}\si^z_{(i+1)_-})/2} 
.
\end{align}
We can write down a $Z_2^G$ symmetric and local $Z_2^K$-gauge symmetric boundary
effective Hamiltonian:
\begin{align}
\label{Z2Hf}
  H &=  
 - \sum_i  V_{i+\frac12}  (|\up\up\>\<\down\down|\negthinspace + |\down\down\>\<\up\up|)_{i_+,(i+1)_-}  
\nonumber\\
&\ \ \ \ \ \ \ -J\sum_i \si^z_{i_+} \si^z_{(i+1)_-} - U \sum_i E_{i+\frac12} 
\\
&=  
- \sum_i  V_{i+\frac12} 
 \big( \si^+_{i_+} \si^+_{(i+1)_-}  +\si^-_{i_+} \si^-_{(i+1)_-}\big)  
\nonumber\\
&\ \ \ \ \ \ \ 
 -J\sum_i \si^z_{i_+} \si^z_{(i+1)_-} - U \sum_i E_{i+\frac12} 
.
\nonumber 
\end{align}
which is identical to the effective boundary Hamiltonian (\ref{Z2H}) in Appendix
\ref{bdry4l}.

Note that all the low-energy excitations at an energy scale much less than
$U_\tau$ are purely bosonic.  So the fermionic CZX model has a boundary that
can be identified as a boundary of bosonic CZX model, stacking with a fermionic
product state. This implies that the ground state of the fermionic CZX model
can also be viewed as a bosonic $Z_2^G$-SPT state, stacking with a fermionic
product states.

\section{Symmetry-extended gapped boundaries/interfaces: Comments, criteria and examples}   \label{sec:examples}

In this section, we aim to show many systematic examples of $G$-topological states, 
such that we can construct $H$-gapped boundary/interface through 
the \emph{symmetry extension} mechanism, based on a group homomorphism $r$ (a surjective epimorphism) by a short exact sequence
\bea\label{eq:gext-sec:10}
1 \rightarrow K \overset{}{\rightarrow}  H  \overset{{r}}{\rightarrow} G \rightarrow 1.
\eea 
In Sec.~\ref{pure}, we considered the mathematical set-up that $G$-cocycle is trivialized in $H$ based on homogeneous cocycles $\nu^G_d$, in order to consider SPT states.
In this Appendix \ref{sec:examples}, instead, we set-up the mathematics based on inhomogeneous cocycles $\omega^G_d$,
for the convenience of notations (which becomes more transparent later) 
and for more general topological phases (SET states and intrinsic topological orders).

The plan of this Appendix \ref{sec:examples} is the following. 
In Appendixes \ref{sym-extended-set-up}  and \ref{sec:sym-enhanced:DW}, we will give an overview of the set-up of problems
on the boundaries/interfaces.
In Appendix \ref{sec:LHS}, we show that the Lyndon-Hochschild-Serre (LHS) spectral sequence criteria, 
are helpful to  analytically derive some split $H$-cochains that can trivialize certain $G$-cochains (that can be $G$-cocycles) of one higher dimension. 
The advantage of this LHS approach, compared to Sec.~\ref{sec:extension-trivializeGcocycle}, is that we can obtain 
some analytic split $H$-cochains.\footnote{
Note that Sec.~\ref{sec:extension-trivializeGcocycle}'s 
approach  can only suggest the possible $K$ for a given $G$ and a given $G$-cocycle, but Sec.~\ref{sec:extension-trivializeGcocycle} cannot provide 
any analytic $H$-cochain easily.
}
However, the drawback of this LHS approach is that, in a few cases, the $G$-cochains may not always be the $G$-cocycles that we hoped for
(standing for nontrivial $G$-topological phases) 
but $G$-coboundaries (standing for a trivial vacuum).
Nevertheless we can still produce many valid successful examples through 
Appendix \ref{sec:LHS}'s LHS approach shown later in Appendix \ref{sec:examples}.
For all the examples given from \ref{sec:examplesHZ4GZ2} to \ref{1+1DSU2SO3}, 
all that we aim to provide are the data of the inhomogeneous {$G$-cocycle} $\omega_d^G(g)$ and its trivialization 
by finding the split {$H$-cochain} $\beta_{d-1}^H(h)$. 

\subsection{Symmetry Extension Setup: 
Trivialize a $G$-cocycle to an $H$-coboundary (split to lower-dimensional $H$-cochains) 
by lifting $G$ to a larger group $H$} \label{sym-extended-set-up}

\label{sec:sym-enhanced:bdry}

We switch to using the inhomogeneous version of $d$-cocycles $\omega_d$ and $d$-cochains $\beta_d$
for the convenience of notations.
The inhomogeneous version is more general and suitable even for gauge theories with nontrivial holonomies around non-contractible cycles.
Moreover, we can convert between $\nu^G_d$ and $\omega^G_d$ based on the well-known relation given in eqn.~({\ref{eq:inhomonuSET}}).
We can develop their path integrals, lattice Hamiltonians and wave functions suitable for many-body quantum systems as in Sec.~\ref{sec:general}.     

The setup of symmetry-extension eqn.~(\ref{eq:gext-sec:10}) for inhomogeneous cocycles goes as follows.
By pulling back a $G$-cocycle $ \omega^G_d$ back to $H$, it becomes an $H$-coboundary $\delta  \beta^H_{d-1}$.
Formally, we mean that a nontrivial $G$-cocycle
\bea
\omega_d^G(g) \in \cH^d(G,U(1))
\eea
becomes a trivial element $1$ (a coboundary) when it is pulled back (denoted as ${}^*$) to $H$
\bea
{r}^* \omega_d^{G} (g) = \omega_d^G(r(h)) = \omega_d^H(h)=\delta \beta_{d-1}^H(h)\in \cH^d(H,U(1)).
\eea
{This trivial element means a trivial group element $0$ in the cohomology group $\cH^d(H,U(1))$, or a coboundary $1$ for the $U(1)$ coefficient.}
The above variable $g$ (or $h$) in the bracket is a shorthand of many copies of group elements in a direct product group of $G$ (or $H$).
More precisely, we rewrite the above in terms of splitting a inhomogeneous $G$-cocycle:
\begin{align} \label{eq:trivializeHcoboundaryinhomo}
 \omega^G_d(g_{01},\cdots,g_{d-1d}) 
 &= \omega^G_d(r(h_{01}),\cdots,r(h_{d-1d})) = \omega^H_d(h_{01},\cdots,h_{d-1d}) \nonumber \\
 &=
(\beta^{H}_{d-1})^{s(h_{01})}(h_{12},\cdots,h_{i-1i},h_{ii+1},h_{i+1i+2},\cdots,\cdots,h_{d-1d}) \times\nonumber \\
 &
 {
\prod_{i={0}}^{d-2} \beta^{H(-1)^{i+1}}_{d-1}(h_{01},\cdots,h_{i-1i},h_{ii+1}h_{i+1i+2},h_{i+2i+3},\cdots,\cdots,h_{d-1d}) 
} \times  \nonumber \\
 & \beta^{H(-1)^{d}}_{d-1}(h_{01},\cdots,h_{i-1i},h_{ii+1},h_{i+1i+2},\cdots,\cdots,h_{d-2d-1}) \nonumber \\
&\equiv \delta  \beta^H_{d-1}.
\end{align}

Because of the property of the $G$-module for the cohomology group of $U(1)$ cocycles, 
we impose that $(\beta^{H}_{d-1})^{s(h)}=\beta^{H}_{d-1}$ for $h$ contains only a unitary group element,
and
$(\beta^{H}_{d-1})^{s(h)}=(\beta^{H}_{d-1})^{-1}$ for $h$ is an anti-unitary group element in $H$ such as an anti-unitary time-reversal symmetry group.

We call this approach ``\emph{symmetry extension}'' (or colloquially ``\emph{symmetry enhancement}''),
because $H$ is a larger group mapping surjectively to $G$.
For  quantum many-body systems, the dimension of Hilbert space is \emph{enhanced} from a $|G|$ per degree of freedom in the \emph{bulk} 
to a larger $|H|$ per degree of freedom on the \emph{boundary}.

Here we provide some useful information of the cohomology group $\cH^d( G,U(1))$ of $G$ that may be used later:
\begin{table}[!h]
\begin{center}
\begin{tabular}{c||c|c|c|c}
\hline
$G$\; &\; $\cH^1( G,U(1))$\; & \;$\cH^2( G,U(1))$ & \;$\cH^3( G,U(1))$ & \;$\cH^4( G,U(1))$ \\ \hline \hline
$D_{4}$\; &\; $(\Z_2)^2$\; & \;$\Z_2$ & \;$(\Z_2)^2 \times \Z_4$ & \;$(\Z_2)^2$ \\ \hline
$Q_{8}$\; &\; $(\Z_2)^2$\; & \;$0$ & \;$\Z_8$ & \; 0 \\ \hline
$Z_2$\; &\; $\Z_2$\; & \;$0$  & \;$\Z_2$ & \; $0$ \\ \hline
$Z_2^T$\; &\; $0$\; & \;$\Z_2$  & \;$0$ & \; $\Z_2$ \\ \hline
$(Z_2)^2$\; &\; $(\Z_2)^2$\; & \;$\Z_2$ & \;$(\Z_2)^3$ & \; $(\Z_2)^2$ \\ \hline
 \end{tabular}
 \end{center}
\caption{Some examples of cohomology group $\cH^d( G,U(1))$ for $G=D_{4}, Q_{8}, Z_2, Z_2^T$ and $(Z_2)^2$ that
can be used to construct $G$-topological phases.}
\label{table:cohomology-group}
\end{table}

\begin{table}[!h]
\begin{center}
\begin{tabular}{c||c|c}
\hline
subgroup $N$ \; &\; quotient group $Q$ \; & \;$G/N=Q$\\ \hline
$\{1\}$ & $D_4 /\{1\} = D_4$ & \; $D_4 /\{1\} = D_4$\\ \hline
$\{1,R^2\}$ (center) & $D_4 /\{1,R^2\} = (Z_2)^2$ & \; $D_4 /Z_2 = (Z_2)^2$\\ \hline
$\{1,\tx\}$  &  \text{No} & \text{No} \; \\
$\{1, \tx R^2\}$  & \text{No}   & \text{No} \; \\
$\{1, \tx R\}$  & \text{No}  & \text{No} \; \\
$\{1, \tx R^3\}$ & \text{No}  & \text{No} \; \\\hline
$\{1,\tx, R^2, \tx R^2\}$ & $D_4 /\{1,\tx, R^2, \tx R^2\}= Z_2$ & \; $D_4 /(Z_2)^2 = Z_2$\\ 
$\{1, \tx R, R^2,  \tx R^3\}$ & $D_4 /\{1, \tx R, R^2,  \tx R^3\}= Z_2$ & \; $D_4 /(Z_2)^2 = Z_2$\\ \hline
$\{1,R, R^2, R^3\}$ & $D_4 /\{1,R, R^2, R^3\}= Z_2$ & \; $D_4 /Z_4 = Z_2$\\ \hline
$D_4$ & $D_4 /D_4= 1$ & \; $D_4 /D_4 = 1$\\
\hline
 \end{tabular}
 \end{center}
\caption{Subgroup $N$ and quotient groups $Q$ of $G=D_4$.}
\label{table:D4}
\end{table}

\begin{table}[!h]
\begin{center}
\begin{tabular}{c||c|c}
\hline
subgroup $N$\; &\; quotient group $Q$\; & \;$G/N=Q$\\ \hline
$\{1\}$ & $Q_8 /\{1\} = Q_8$ & \; $Q_8 /\{1\} = Q_8$\\ \hline
$\{1,-1\}$ (center) & $Q_8 /\{1,-1\} = (Z_2)^2$ & \; $Q_8 /Z_2 = (Z_2)^2$\\ \hline
$\{1,i,-1,-i\}$ & $Q_8 /\{1,i,-1,-i\}= Z_2$ & \; $Q_8 /Z_4 = Z_2$\\ 
$\{1,j,-1,-j\}$ & $Q_8 /\{1,j,-1,-j\} = Z_2$ & \; $Q_8 /Z_4 = Z_2$\\ 
$\{1,k,-1,-k\}$ & $Q_8 /\{1,k,-1,-k\} = Z_2$ & \; $Q_8 /Z_4 = Z_2$\\ \hline
$Q_8$ & $Q_8 /Q_8= 1$ & \; $Q_8 /Q_8 = 1$\\
\hline
 \end{tabular} 
 \end{center}
\caption{Subgroup $N$ and quotient groups $Q$ of $G=Q_8$.}
\label{table:Q8}
\end{table}

We write the order-8 dihedral group as 
$$D_4=\langle \tx, R \mid R^4=\tx^2=1, \tx R \tx=R^{-1}\rangle$$ 
generated by $\tx$ and $R$.
We write the order-8 quaternion as 
$$Q_8=\langle x,y|x^2=y^2, xyx^{-1}=y^{-1},x^4=y^4=1\rangle$$
so that each element in $Q_8$ we can write uniquely as $x^q y^{n} $,
where $q\in\{0,1\}$ and $n \in\{0,1,2,3\}$.
For $(q,n)\in \{  (0,0),(0,1),(0,2),(0,3), (1,0),(1,1),(1,2),(1,3)\}$,
we can identify them as the well-known $Q_8$ notation as
$x^q y^{n}  \in \{ 1, i, -1, -i, j, -k, -j, k\}$.

For notation convention, we use the additive notation $0$ to denote the trivial group if all groups are finite Abelian groups such as in $0 \to {Z}_2^K \to Z_4^H  \to Z_2^G \to 0$.
We use the multiplicative notation $1$ to denote the trivial group if some group is non-Abelian such as in $1 \to Z_4^K \to Q_8^H  \to Z_2^G \to 1$.

For selective some examples below (from \ref{sec:examplesHZ4GZ2} to \ref{1+1DSU2SO3}), we will test the 
Lyndon-Hochschild-Serre (LHS) spectral sequence $d_2$ map technique in Sec.~\ref{sec:trivializeGcocycle}
and comment its validity to derive $H$-cochains for trivializing certain $G$-cocycles.

\subsection{Symmetry-extended gapped interfaces}
\label{sec:sym-enhanced:DW}

Consider the interface (i.e. domain wall) between two sides of phases labeled by groups $G_{\I}$ and $G_{\II}$ respectively. 
The two sides of phases could be
either both SPTs, both SETs or both topological orders. Below we present various systematic constructions for gapped interfaces.
The gapped boundary of $G$ can be regarded as a gapped interface between a $G$-topological state and a trivial vacuum.

\subsubsection{Symmetry-extension 
and the folding trick: Trivialize a $G_{\I} \times G_{\II}$-cocycle to an $H$-coboundary by splitting to lower-dimensional $H$-cochains}

\label{sec:sym-enhanced:DWG1G2}

Importantly the previous formulation of gapped boundary is also applicable to formulate the gapped interface, by using the \emph{folding trick}.
The strategy is that, by replacing the $G$ in Sec.~\ref{sym-extended-set-up} to $G_{\I} \times G_{\II}$, then we can determine
the gapped boundary between $G_{\I} \times G_{\II}$ and the vacuum, via trivializing a $G_{\I} \times G_{\II}$-cocycle to $H$-coboundary by splitting to lower-dimensional $H$-cochains. The surjective group homomorphism $r$ is given by 
$$1 \rightarrow K \overset{}{\rightarrow}  H  \overset{{r}}{\rightarrow} G_{\I} \times G_{\II}\rightarrow 1.$$
We can rewrite the above in terms of splitting an inhomogeneous $G=G_{\I} \times G_{\II}$-cocycle:
\begin{align} \label{eq:SEG1G2}
 \omega^{G_{\I} \times G_{\II}}_d(g)= \omega^{G_{\I} \times G_{\II}}_d(r(h)) = \delta  \beta^H_{d-1}(h).
\end{align}
Here $(g)$ is a shorthand of $(g_{01},\cdots,g_{d-1d})$ with each element in $G_{\I} \times G_{\II}$.
Generally $ \omega^{G_{\I} \times G_{\II}}$ is a cocycle in the cohomology group $\cH^{d}({G_{\I} \times G_{\II}},U(1))$.
K\"unneth theorem shows us that there exists a particular form of cocycle $\omega_{\I}^{G_{\I}}(g_{\I} ) \cdot \omega_{\II}^{G_{\II}}(g_{\II})^{-1}$,
obtained from $\omega_{\I}^{G_{\I}} \in \cH^{d}({G_{\I}},U(1))$ and $\omega_{\II}^{G_{\II}} \in \cH^{d}({G_{\II}},U(1))$.
Now, we see that the $G_{\I}$-symmetry action only acts on
$\omega_{\I}^{G_{\I}}(g_{\I} )$, 
while the $G_{\II}$-symmetry action only acts on
$\omega_{\II}^{G_{\II}}(g_{\II})$.
By folding $\omega_{\I}^{G_{\I}}(g_{\I} )$ and $\omega_{\II}^{G_{\II}}(g_{\II})$ to two different sides of the $H$-gapped boundary,
we obtain an $H$-gapped interface.

\subsubsection{Append a lower-dimensional topological state onto the boundary/interface}
\label{sec:append-lower}

For all the previous setups, we actually pick a trivialization of the pullback of the $G$-cocycle to $H$.   
The possible trivialization choices differed by a class in $\cH^{d-1}(H,U(1))$ 
physically imply that we can further append lower dimensional gapped topological states (that are well-defined in its own dimension) onto the boundary or the interface.
(See also Sec.~\ref{mixed} for a discussion.)
In general, it could be a SET of $(d-1)$-dimensions labeled by an $H$-cocycle with $H$-site and $K$-link variables:
\bea
\CV^{H,K}_{d-1}(\{h_{i}\}; \{k_{ij}\})
=\nu^{H}_{d-1}(h_{i_0}, k_{i_0i_1} h_{i_1}, \cdots, k_{i_0i_1} \cdots k_{i_{d-2} i_{d-1}} h_{i_{d-1}})
 \in \cH^{d-1}(H,U(1))
\eea 
and described by
$1 \to K \to H \to G \to 1$, with a total projective symmetry group $H$, a gauge group $K$, and a global symmetry group $G$. 
The $H$ cocycle obeys the cocycle condition:
$
\delta \CV^{H,K}_{d-1}
=\delta \nu^{H}_{d-1}=1$.
In different limit choices of $G$ and $K$, the topological phases of $\CV^{H,K}_{d-1}$ include SPTs, topological orders and SETs.

The proper choices of $G$ and $K$ on the boundary are also constrained by the choices of $G$ and $K$ in the bulk.
We will leave this issue as a case-by-case study.

In this Appendix \ref{sec:examples}, we use inhomogeneous cocycles as in Appendix \ref{sym-extended-set-up}, we replace $\CV^{H,K}_{d-1}$ by $\Omega^{H}_{d-1}$.
We see that
$$
\delta (\beta^H_{d-1}(h) \; \Omega^{H}_{d-1}(h)) =\delta (\beta^H_{d-1}(h))=  \omega^H_d(h)  = \omega^G_d(r(h))  =  \omega^G_d(g),
$$
where $\delta (\Omega^{H}_{d-1}(h))=1$.
It can also be appended on the interface, as in Appendix \ref{sec:sym-enhanced:DWG1G2}'s \Eqn{eq:SEG1G2}, 
$$ 
 \delta ( \beta^H_{d-1}(h) \Omega^{H}_{d-1}(h))= \delta  \beta^H_{d-1}(h)= \omega^{G_{\I} \times G_{\II}}_d(r(h)) =\omega^{G_{\I} \times G_{\II}}_d(g).
$$
Here the appended lower-dimensional  topological states (differed by $\Omega^{H}_{d-1}$, with $\delta (\Omega^{H}_{d-1}(h))=1$)  are all gapped.

\subsection{Criteria on trivializing the $G$-cocycle in a larger group $H$:\\ 
Lyndon-Hochschild-Serre spectral sequence} \label{sec:trivializeGcocycle}\label{sec:LHS}


We would like to provide a systematic way to determine the possible trivialization of the $d$-cocycle in $G$ by lifting to a larger group $H$, based on the setup of
the Lyndon-Hochschild-Serre (LHS) spectral sequence.
The question we would like to address here is that

``\emph{Given $1 \to K \to H  \overset{r}{\to} G \to 1$,
how can we analytically obtain the split $H$-cochain $\beta_{d-1}^H$ that satisfies that 
$\omega_d^G(r(h))=\omega_d^H(h)=\delta \beta_{d-1}^H(h)$
for some $G$-cocycle $\omega_d^G$?}'' 

An answer goes as follow.
For $1 \to K \to H  \overset{r}{\to} G \to 1$, with 
$G$ acting trivially on $\cH^*(K, U(1))$,\footnote{
If $K$ is contained in the center of $H$, it implies $G$ acts trivially on $\cH^*(K,U(1))$.} 
there is a spectral sequence $\{E^{p,q}_n, d_n\}$
with:\\
(a) $E^{p,q}_2 = \cH^p(G, \cH^q(K,U(1) ))$.\\
(b) The differential is defined as a map $d_n$: $E^{p,q}_n \to E^{p+n,q-n+1}_n$. We have $E^{p,q}_{n+1}=\frac{\text{Ker} (d_n)}{\Im (d_n)}$ at $E^{p,q}_{n}$. \\

We  focus on the $d_2$ differential of the $E_2$ page in the LHS spectral sequence 
\bea
&&d_2: E^{p,q}_2 \to E^{p+2,q-1}_2\\
&&\Rightarrow d_2:  \cH^p(G, \cH^q(K,U(1) )) \to  \cH^{p+2}(G, \cH^{q-1}(K,U(1) )),
\eea
in particular
\bea
 d_2:  \cH^{d-2}(G, \cH^{1}(K,U(1) )) \to  \cH^{d}(G, \cH^{0}(K,U(1) ))=\cH^{d}(G, U(1)).
\eea
If we want to trivialize the $d$-cocycle $\omega_d^G \in \cH^{d}(G, U(1))$, we
can look for a larger group $H$, where $H/K=G$ for some $K$.
The $d_2$ turns out to provide the following nice property. 
The image of the differential
$ d_2:  \cH^{d-2}(G, \cH^{1}(K,U(1) )) \to \cH^{d}(G, U(1))$
provides elements of $\omega_d^G \in \cH^{d}(G, U(1))$, such that
all such elements are guaranteed to vanish to be trivial as a coboundary in $\cH^{d}(H, U(1))$.
In other words, every element $\omega_d^G$ in the image of the $d_2$ map is guaranteed to be trivial in $\cH^{d}(H, U(1))$.\footnote{
{Namely, the image of the $d_2$ map is guaranteed to be contained in the kernel of the inflation map from $\cH^{d}(G, U(1))$ to $\cH^{d}(H, U(1))$.}
J.W. gratefully acknowledges Tom Church and Ehud Meir for illuminating the spectral sequence method \cite{brown2012cohomology, 249368}.
Given $d_2: H^{d-2}(G,H^1(K,U(1)))\to H^d(G,U(1))$, if $\omega_d^G$ is killed by $d_2$ (namely, $\omega_d^G$ is in the image of $d_2$), 
in other words $d_2(\alpha_{d-2})=\omega_d^G$, 
then $\omega_d^G$ becomes a coboundary in $H^d(H,U(1))$.
}
We have
\bea
\omega_d^G=\delta \beta_{d-1}^H,
\eea
or, more precisely,
\bea
\omega_d^G(r(h))=\omega_d^H(h)=\delta \beta_{d-1}^H(h),
\eea
where $\beta_{d-1}^H$ is determined by the $d_2$ differential and the map 
\bea
f: G^{d-2} \to \cH^{1}(K,U(1) ).
\eea
The $f$ is a function that relates to a cocycle 
\bea
\alpha_{d-2} \in \cH^{d-2}(G, \cH^1(K,U(1))).
\eea
If we know the data of $H$ are given by the pair $G$ and $K$, 
we can propose the $\beta_{d-1}^H$.\footnote{For example, we can write $\beta_{d-1}^H$ as a function F as  $\beta_{d-1}^H(h)=\text{F}(\alpha_{d-2}(g),k)$.
For many examples shown in this 
Appendix \ref{sec:examples}, we find that a candidate form of $\beta_{d-1}^H(h)$ is $\beta_{d-1}^H(h) \sim \alpha_{d-2}(g)^k$.} 
Notice that $d_2 (\alpha_{d-2})$ is in $\cH^{d}(G, U(1))$. 
The claim is that there exists the map $d_2: \cH^{d-2}(G, \cH^{1}(K,U(1) )) \to \cH^{d}(G, U(1))$
where every $G$-cocycle $\omega_d$ in the image of $d_2$ map is an $H$-coboundary that can be split to lower-dimensional $H$-cochains in $\cH^{d}(H, U(1))$.

By writing the group element $h \in H$ in terms of a pair $(k, g) \in (K, G)$ as $h=(k, g)$,
we can write down the further precise relation 
\bea
&&\omega_d^H(h)=\omega_d^H(h_1, h_2, \dots, h_d)=\omega_d^H( (k_1, g_1), (k_2, g_2), \dots, (k_d, g_d))
=\omega_d^G(  g_1, g_2, \dots,g_d)=\omega_d^G(g) \nonumber\\
&&=\delta ( \beta_{d-1}^H((k_1, g_1), (k_2, g_2), \dots, (k_{d-1}, g_{d-1})))=(\delta \beta_{d-1}^H)((k_1, g_1), (k_2, g_2), \dots, (k_d, g_d))\nonumber\\ 
&&=(\delta \beta_{d-1}^H) (h_1, h_2, \dots, h_d)=\delta \beta_{d-1}^H(h).
\eea
Such a construction of $\beta_{d-1}^H$ so that $\delta \beta_{d-1}^H(h) \sim d_2 (\alpha_{d-2})$ is a coboundary in $\cH^{d}(H, U(1))$, from the LHS spectral sequence.
This means that some $G$-cocycle $\omega_d^G(r(h))=\omega_d^H(h)=\delta \beta_{d-1}^H(h)$ can be split to lower dimensional $H$-cochains.
However, we  emphasize that some obtained $\omega_d^G(r(h))$ 
may be already a $G$-coboundary and may \emph{not be the specific non-trivial $G$-cocycle} that we originally aimed to trivialize.
We will show in Appendix \ref{sec:examples} (from \ref{sec:examplesHZ4GZ2} to \ref{1+1DSU2SO3}) 
how this LHS spectral sequence approach can help in constructing some examples, but not necessarily other examples.

\subsection{2+1/1+1D Bosonic $0 \to {Z}_2^K \to Z_4^H  \to Z_2^G \to 0$} \label{sec:examplesHZ4GZ2}

Consider the example where $G= Z_2$, $H=Z_4$ and $K=Z_2$, and denote them under $0 \to {Z}_2^K \to Z_4^H  \to Z_2^G \to 0$.
The twisted 3-cocycle is
\bea
&&\omega_3^{Z_2^G}(g_a , g_b, g_c)=\exp[\frac{\ii 2 \pi}{2^2} p \; [g_a]_2 ([g_b]_2+[g_c]_2-[[g_b]_2+[g_c]_2] ) ]
=(-1)^{g_a g_b g_c}
\eea
with $g \in {Z_2^G}$ and $p \in \cH^3({Z_2^G},U(1))=\Z_2$.
To have a nontrivial 3-cocycle, we set $p=1$.
This cocycle is equivalent to 
$\ee^{\ii 2 \pi \int \frac{1}{2} a_1 \cup a_1 \cup a_1}=(-1)^{\int  a_1 \cup a_1 \cup a_1}$ with a cup product form of $a_1 \cup a_1 \cup a_1$, in $\cH^3( Z_2,U(1))$.
The $a_1$ here is a $\Z_2$-valued 1-{cocycle} in $\cH^1(M^3, \Z_2)$ on the spacetime complex $M^3$.
For a discrete finite $G$, the principal $G$-bundle and the flat $G$ connection are effectively the same.
Here we consider $G=Z_2$, so in this context, we can view the nontrivial SPTs detectable by the principal $Z_2$-bundle and the flat $Z_2$-connection

We find that the analytic 2-cochain 
\bea
&&\beta_2(h_1,h_2) = \exp[  (\ii 2 \pi p / 4)  [h_1]_2  [h_2]_4 ].
\eea
splits $G$ 3-cocycle.
Alternatively, we can choose $\beta_2(h_1,h_2) = \exp[ ( \ii 2 \pi p/ 4) [h_1]_4  [h_2]_2 ]$ 
with $m,n \in {Z_4^H}$.

Furthermore, we find LHS technique in Appendix~\ref{sec:LHS} 
works successfully. 
For LHS technique of Appendix~\ref{sec:LHS},
we look for:
\bea
d_2:  \cH^{1}(G, \cH^{1}(K,U(1) )) &\to&   \cH^{3}(G, \cH^{0}(K,U(1) ))=\cH^{3}(G, U(1)). \nonumber\\
\Rightarrow d_2:  \cH^{1}(Z_2, Z_2)=\Z_2 &\to&  \cH^{3}(Z_2, U(1))=\Z_2. 
\eea
\bea
&&f: G \to \cH^{1}(K,U(1) ) \nonumber\\
&&\Rightarrow
 f: Z_2^G \to \cH^{1}(Z_2^K,U(1) )=\Z_2.
\eea
Because this $f$ maps to $\cH^{1}(Z_2^K,U(1) )=\Z_2$, the $\beta_2$ can be a base of $(-1)$.
We find that another 2-cochain that splits 3-cocycle is 
\bea
&&\tilde \beta_2(h_1, h_2)=f(g_2)^{k_1} =(-1)^{g_2 k_1}.
\eea
For $h=0, (g,k)=(0,0)$; $h=1, (g,k)=(1,0)$; $h=2, (g,k)=(0,1)$; $h=3, (g,k)=(1,1)$.
The group elements in $H$ satisfy 
$$
h_1 \cdot h_2 =(g_1, k_1)\cdot (g_2, k_2)=([g_1+g_2]_2,  [k_1+k_2 + g_1 g_2]_2).
$$
We would like to check that $(\delta \tilde \beta_2)(h_1, h_2, h_3) = (-1)^{g_1 g_2 g_3}$.
\bea
&&(\delta \tilde \beta_2)(h_1, h_2, h_3)=\frac{\tilde \beta_2(h_2, h_3) \tilde \beta_2(h_1, h_2 h_3)}{ \tilde \beta_2(h_1 h_2, h_3)\tilde \beta_2(h_1, h_2) }
=\frac{(-1)^{g_3 k_2} {(-1)^{[g_2+g_3]_2 k_1}}}{(-1)^{g_3  [k_1+k_2 + g_1 g_2]_2} (-1)^{g_2 k_1}}\\
&&
=\frac{(-1)^{g_3 k_2} {(-1)^{ (g_2+g_3) k_1}}}{(-1)^{g_3  (k_1+k_2 + g_1 g_2) } (-1)^{g_2 k_1}}
=(-1)^{g_1 g_2 g_3},
\eea
which is true. (Actually, both $\tilde \beta_2(h_1, h_2)=(-1)^{g_2 k_1}$ and $\tilde \beta_2(h_1, h_2)=(-1)^{g_1 k_2}$ work to trivialize $G$ 3-cocycle.)
We can rewrite
$\tilde\beta_2(h_1, h_2)=(-1)^{g_2 k_1}=(-1)^{g_2 \frac{h_1-[h_1] }{2}}=i^{g_2 ({h_1-[h_1]_2 })}=i^{[h_2]_2({[h_1]_4-[h_1]_2 })}$.
If we write $h\in H$ in terms of $h=(g,k)$, then
$\beta_2(h_1, h_2)  = \exp[ (2 \pi \ii/ 4)  ([h_1]_2)  ([h_2]_4) ] =i^{ [h_1]_2  [h_2]_4} = i^{ [g_1]_2  [g_2+2 k_2]_4}$.


If we consider the bulk to be fully gauged topologically ordered state, this becomes a gapped boundary for a bulk $2+1$D field theory of an action 
$\int  \frac{ 2}{2\pi}{B   d A} {{+}}  
\frac{1}{2\pi } A d A $, with $B$ and $A$ locally as 1-form gauge fields.

\subsubsection{Degeneracy on a disk and an annulus: Partition functions $Z(D^2 \times S^1)$ and $Z(I^1 \times S^1 \times S^1)$} \label{sec:examplesHZ4GZ2GSD}

Here we can put the 2+1/1+1D $0 \to {Z}_2^K \to Z_4^H  \to Z_2^G \to 0$ construction of topological states on a spatial $D^2$ disk or an annulus $I^1 \times S^1$ to count the degeneracy (GSD).
Whether we \emph{gauge the global symmetry $K$ and $H$ or not}, we have at least
three types of theories: \\
(i) Fully global symmetric SPTs (a bulk $G$-SPTs and a boundary anomalous $H$-SPTs), \\
(ii) Bulk SPTs/boundary SETs  (a bulk $G$-SPTs and a boundary anomalous $H$-SETs with a gauge group $K$), \\
(iii) Fully topological orders with dynamical gauge fields (a bulk $G$-topologically ordered gauge theory and a boundary anomalous $H$-gauge theory).
Since $K$ is a normal subgroup in $H$, we can label the $K$-holonomy in $H$. Thus below we write all holonomies $h$ in $H$.

Theory (i) is basically the second boundary discussed in Sec.~\ref{sec:bdryCZX} and \ref{rev}.
Theory (ii) is basically the third (hard-gauge) and fourth (soft-gauge) boundaries discussed in Sec.~\ref{sec:bdryCZX} and \ref{rev}.
Theory  (iii) is basically the fifth boundary mentioned in Table  \ref{table:KHGfinite} and  \ref{table:KfiniteHGcont}.
These three types of theories are also shown in the first three rows in Table \ref{table:summary1}.

We compute the partition function of Sec.~\ref{sec:ZSPTSETG1G2} on $Z(D^2 \times S^1)$ to evaluate GSD on a spatial $D^2$ disk in Table \ref{table:GSDD2HZ4GZ2}.
\begin{table}[!h]
 \begin{center}
\begin{tabular}{c||c|c|c}
\hline
Disk $D^2$&\; \begin{minipage}{1.25in}Theory (i)\\ (the second bdry) \end{minipage}  &\; 
\begin{minipage}{1.7in}Theory (ii)\\ (the third/fourth bdry) \end{minipage}  
&\; \begin{minipage}{1.25in}Theory (iii)\\ (the fifth bdry) \end{minipage}   \\ \hline
GSD &1  & 2  & 1\\ \hline
 \end{tabular}
 \caption{ For the theory (ii), GSD=2 from the holonomy $h=0$ and $h=2 \in H$.
{For the fully gauge theory (iii), GSD=1 from the holonomy $h=0 \in H$ 
.}
}
  \label{table:GSDD2HZ4GZ2}
 \end{center}
\end{table}

Note that the $h=0$ carries zero or an even $Z_2^G$ charge.
The $h=2$ carries an odd $Z_2^G$ charge.
For the theory (iii), when the $Z_2^G$ is gauged, the ground state for the whole system cannot carry an odd $Z_2^G$ charge, thus
$h=0 \in H$ implies GSD=1 on a disk. An important remark is that
we \emph{cannot} regard the 1+1D anomalous $Z_4^H$ gauge theory as a usual 1+1D discrete gauge theory --- because the
usual 1+1D $Z_4$ gauge theory has GSD$=|H|=4$ on a $S^1$ ring. In our case, the 2+1D bulk plays an important rule, which
causes the GSD reduces to GSD=1 for the theory (iii).

We compute the partition function of Sec.~\ref{sec:ZSPTSETG1G2} on $Z(I^1 \times S^1 \times S^1)$ to evaluate GSD on an annulus $I^1 \times S^1$ in Table \ref{table:GSDI1S1HZ4GZ2}:
\begin{table}[!h]
 \begin{center} 
\makebox[\textwidth][c] 
{
\begin{tabular}{c||c|c|c}
\hline
Annulus $S^1 \times I^1$ 
&\;  \begin{minipage}{1.25in}Theory (i)\\ (the second bdry) \end{minipage}  
&\; \begin{minipage}{1.7in} {Theory (ii)}\\ (the third/fourth bdry) \end{minipage}
&\; \begin{minipage}{1.25in} Theory (iii)\\ (the fifth bdry) \end{minipage} \\ \hline
GSD &\; 1  &\; 4 &\; 2  \\ \hline
 \end{tabular} 
}
\hspace*{0mm}
 \end{center}
 \caption{
{For the theory (ii), GSD=4 from the holonomies:
$(h_{\text{in}}, h_{\text{out}})$ with
$h_{\text{in}}, h_{\text{out}} \in \{0,2\}$
.}
{For the fully gauge theory (iii), GSD=2 from the holonomies of two sectors:
$(h_{\text{in}},h_{\text{out}})=(0,0),(2,2)$.
}
}
 \label{table:GSDI1S1HZ4GZ2}
\end{table}

Again the 2+1D bulk plays an important rule for the GSD reduction for the theory (iii) from 
GSD$=|H|^2=16$ to GSD=2 in Table \ref{table:GSDI1S1HZ4GZ2}.

\subsection{$d+1/d$D Bosonic $0 \to {Z}_2^K \to Z_4^H  \to Z_2^G \to 0$ for an even $d$} \label{sec:examplesHZ4GZ2-any-dim}

We can readily generalize Sec.~\ref{sec:examplesHZ4GZ2}
to consider a gapped boundary 
for the $d+1/d$D bosonic SPTs with a $G=Z_2$ symmetry for any even dimension $d$ under:
$0 \to {Z}_2^K \to Z_4^H  \to Z_2^G \to 0$. 
The twisted $(d+1)$-cocycle is
\bea
&&\omega_{d+1}^{Z_2^G}(g_1 , g_2, \dots, g_{d+1})
=(-1)^{g_1  g_2 \dots g_{d+1}}
\eea
with $g \in {Z_2^G}$ and $\cH^{d+1}({Z_2^G},U(1))=\Z_2$ for an even $d$.
This cocycle is equivalent to 
$\ee^{\ii 2 \pi \int \frac{1}{2} a_1 \cup a_1 \cup \dots \cup a_1}$ with a cup product form of $a_1 \cup a_1 \cup \dots \cup a_1$, in $\cH^{d+1}( Z_2,U(1))$.
The $a_1$ here is a $\Z_2$-valued 1-{cocycle} in $H^1(M^{d+1}, \Z_2)$ on the spacetime complex $M^{d+1}$.

As in Appendix~\ref{sec:examplesHZ4GZ2}, we write $h=(g,k) \in Z_4^H$ as a doublet where $g \in Z_2^G$ and $k \in Z_2^K$.
We find that the $d$-cochain that splits the $(d+1)$-cocycle in $H$ can be 
\bea
&&\tilde \beta_d(h_1, h_2, \dots, h_d) =(-1)^{g_2 \cdots g_d k_1}.
\eea
The group elements in $H$ satisfy 
$$
h_1 \cdot h_2 =(g_1, k_1)\cdot (g_2, k_2)=([g_1+g_2]_2,  [k_1+k_2 + g_1 g_2]_2).
$$
We would like to check that $(\delta \tilde \beta_d)(h_1, h_2, \dots, h_d, h_{d+1}) = (-1)^{g_1  g_2 \dots g_{d+1}}$ for an even $d$. Namely
\bea \label{eq:checkZ4Z2generalsplit}
&&(\delta \tilde \beta_d)(h_1, h_2, \dots, h_d, h_{d+1})=\frac{\tilde \beta_d(h_2, \dots, h_{d+1}) \dots \tilde \beta_d(h_1, h_2, \dots,  h_{d} h_{d+1})}{ 
\tilde \beta_d(h_1 h_2, \dots, h_{d+1}) \dots \tilde \beta_d(h_1, h_2, \dots, h_{d}) }
\nonumber \\
&&
=\frac{(-1)^{g_3 \cdots g_{d+1}  k_2} {(-1)^{ (g_2+g_3) g_4 \cdots g_{d+1} k_1}} \cdots {(-1)^{ g_2 \cdots (g_{d}+g_{d+1}) k_1}}}{(-1)^{g_3 \cdots g_{d+1}  (k_1+k_2 + g_1 g_2) } \cdots(-1)^{g_2  \cdots g_{d}  k_1}}= (-1)^{g_1  g_2 \dots g_{d+1}}
\eea
 is true. Moreover, since $\cH^{d} (Z_n, U(1)) = 0$ for any even dimension $d$,  
there is no any further lower-dimensional topological phase of the $H=Z_4$-cocycle that we can append on the gapped boundary
of an even spacetime dimension $d$. 
 
We find that the $d$+1D bosonic SPTs with $Z_2$ symmetry (the bosonic topological superconductor of $G=Z_2$)
have a $d$D symmetry-preserving surface deconfined $Z_2$ topologically ordered gauge theory, at least for $d \geq 4$.
When $d=2$, the boundary deconfined $Z_2$ gauge theory is a spontaneous symmetry breaking state crossing over to a confined state, 
thus we require fine tuning to have a deconfined gauge theory, shown in Sec.~\ref{sec:deconfined-to-SSB}.
 
If we consider the bulk to be fully gauged topologically ordered state, this becomes a gapped boundary for a bulk $d+1$D field theory of 
an action $\int  \frac{ 2}{2\pi}{B   d A} {{+}}  
\frac{1}{(2\pi)^{d/2}}   A (d A)^{d/2} $ 
with locally $A$ a 1-form gauge field and $B$ a $d$-form gauge field. 

\subsection{3+1/2+1D Bosonic $0 \to {Z}_2 \to Z_4^T  \to Z_2^T \to 0$  with $Z_2^T$ time-reversal symmetry} 
\label{sec:examplesHZ4TGZ2T}

We discussed this example in the main text of Sec.~\ref{sec:examples3+1DHZ4GZ2T} through a different method.
From Ref.\cite{XieSPT4} and Table.\ref{table:cohomology-group}, 
for an anti-unitary symmetry $Z_2^T$, we recall that
the cohomology groups for an odd dimension $d$ offer:
$\cH^{4}(Z_2^T, U_T(1)) = Z_2$.
The 4-cocycle  $\omega_{4}^{Z_2^T} \in \cH^{4}(Z_2^T, U_T(1))$ is of the similar form of the cocycle studied in the previous section.
The only new ingredient for the calculation involving $Z_2^T$ symmetry is the nontrivial anti-unitary action of $Z_2^T$ on the $Z_2^T$-module $ U_T(1)$.
This cocycle is equivalent to 
$\ee^{\ii 2 \pi \int \frac{1}{2} w_1^4}$  in $\cH^4( Z_2^T,U_T(1))$.
The $w_1$ here is a $\Z_2$-valued, the first Stiefel-Whitney class in $H^1(M^4, \Z_2)$ on the spacetime complex $M^4$.
The $w_1\neq 0$ holds on a non-orientable manifold.

We would like to check that $ \omega_{4}^{Z_2^T}(g_1, g_2, g_3, g_4)= (-1)^{g_1 g_2 g_3 g_4}=(\delta \tilde \beta_3)(h_1, h_2, h_3, h_4)$ for
some $\tilde \beta_3$.
Similar to Appendix~\ref{sec:examplesHZ4GZ2}, we write $h=(g,k) \in H=Z_4^T$ as a doublet where $g \in G= Z_2^T$ and $k \in K=Z_2$.
We propose $\tilde \beta_3(h_1, h_2, h_3)=(-1)^{g_2 g_3 k_1}$, which
splits the $G$-cocycle as an $H$-coboundary under $0 \to {Z}_2 \to Z_4^T  \to Z_2^T \to 0$.  Indeed we find
\bea
&&(\delta \tilde \beta_3)(h_1, h_2, h_3, h_4)=\frac{\tilde \beta_3(h_2, h_3, h_4) \tilde \beta_3(h_1, h_2 h_3, h_4) \tilde \beta_3(h_1, h_2, h_3)}{ \tilde \beta_3(h_1 h_2, h_3, h_4)\tilde \beta_3(h_1, h_2, h_3 h_4) }  \nonumber
\\
&&
=\frac{(-1)^{g_3  g_4 k_2} {(-1)^{ (g_2+g_3) g_4 k_1}} (-1)^{g_2  g_3 k_1} }{(-1)^{g_3  g_4 (k_1+k_2 + g_1 g_2) } (-1)^{g_2 (g_3+g_4) k_1}}
=(-1)^{g_1 g_2 g_3 g_4},
\eea
which is true.

We find that the 3+1D bosonic SPTs with $Z_2^T$ symmetry (the bosonic topological superconductor of $G=Z_2^T$)
have a 2+1D symmetry-preserving surface deconfined $Z_2$ topologically ordered gauge theory.

\subsection{$d+1/d$D Bosonic topological superconductor $0 \to {Z}_2 \to Z_4^T  \to Z_2^T \to 0$ for an odd $d$  with $Z_2^T$ time-reversal symmetry:
The $d$D $Z_2^K$-gauge theory boundary of $d+1$D bulk invariant $(-1)^{\int (w_1)^{d+1}}$} \label{sec:examplesHZ4TGZ2T-any-dim}

From Ref.\cite{XieSPT4} and Table.\ref{table:cohomology-group}, we recall that
the cohomology groups for an even dimension $d$ offer:
$$\cH^{d+1} (Z_2, U(1)) = Z_2,  \;\;\;\; \cH^{d+1} (Z_2^T, U_T(1)) = 0.$$
The cohomology groups for an odd dimension $d$ offer:
$$\cH^{d+1}(Z_2^T, U_T(1)) = Z_2, \;\;\;\; \cH^{d+1} (Z_2, U(1)) = 0.$$
We can readily generalize Appendix~\ref{sec:examplesHZ4TGZ2T}
to consider a gapped boundary 
for $d+1/d$D bosonic SPTs with a $G=Z_2^T$ symmetry for any odd dimension $d$ under:
$0 \to {Z}_2 \to Z_4^T  \to Z_2^T \to 0$. 
The twisted $(d+1)$-cocycle is
\bea
&&\omega_{d+1}^{Z_2^G}(g_1 , g_2, \dots, g_{d+1})
=(-1)^{g_1  g_2 \dots g_{d+1}}
\eea
with $g \in {Z_2^T}$ and $\cH^{d+1}({Z_2^T},U_T(1))=\Z_2$ for an even $d$.
This cocycle is equivalent to 
$\ee^{\ii 2 \pi \int \frac{1}{2} w_1^{d+1}}$  in $\cH^{d+1}( Z_2^T,U_T(1))$.
The $w_1$ here is a $\Z_2$-valued, the first Stiefel-Whitney (SW) class in $H^1(M^{d+1}, \Z_2)$ on the spacetime complex $M^{d+1}$.
Here we mean the SW class of the $O(d+1)$ bundle, where $O(d+1)$ is the structure group of the tangent bundle.
The $w_1\neq 0$ holds on a non-orientable manifold.

As in Appendix~\ref{sec:examplesHZ4GZ2}, we write $h=(g,k) \in H= Z_4^T$ as a doublet where $g \in G=Z_2^T$ and $k \in K=Z_2$.
We find that the $d$-cochain that splits the $(d+1)$-cocycle in $H$ can be 
\bea
&&\tilde \beta_d(h_1, h_2, \dots, h_d) =(-1)^{g_2 \cdots g_d k_1}.
\eea
The group elements in $H$ again satisfy 
$
h_1 \cdot h_2 =(g_1, k_1)\cdot (g_2, k_2)=([g_1+g_2]_2,  [k_1+k_2 + g_1 g_2]_2).
$
We can check that $(\delta \tilde \beta_d)(h_1, h_2, \dots, h_d, h_{d+1}) = (-1)^{g_1  g_2 \dots g_{d+1}}$ for an even $d$. Namely
\bea
&&(\delta \tilde \beta_d)(h_1, h_2, \dots, h_d, h_{d+1})=\frac{\tilde \beta_d(h_2, \dots, h_{d+1}) \dots 
\tilde \beta_d(h_1, h_2, \dots,  h_{d-1} h_{d}, h_{d+1}) \tilde \beta_d(h_1, h_2, \dots, h_{d})}{ 
\tilde \beta_d(h_1 h_2, \dots, h_{d+1}) \dots  \tilde \beta_d(h_1, h_2, \dots,  h_{d} h_{d+1})  } \nonumber
\\
&&
=\frac{(-1)^{g_3 \cdots g_{d+1}  k_2}  \cdots 
{(-1)^{ g_2   \cdots (g_{d-1}+g_d) g_{d+1} k_1}} (-1)^{g_2  \cdots g_{d}  k_1} }{(-1)^{g_3 \cdots g_{d+1}  (k_1+k_2 + g_1 g_2) } \cdots {(-1)^{ g_2 \cdots g_{d-1} (g_{d}+g_{d+1}) k_1}}}
=(-1)^{g_1  g_2 \dots g_{d+1}},
\eea
 is true.
Moreover, since $\cH^{d} (Z_n^T, U(1)) = 0$ for any odd dimension $d$,  
there is no any further lower dimensional topological phase of the $H=Z_4^T$-cocycle that we can append on the gapped boundary
of an odd spacetime dimension $d$. 
 
We find that the $d$+1D bosonic SPTs with $Z_2^T$ symmetry (the bosonic topological superconductor of $G=Z_2^T$)
have a $d$D symmetry-preserving surface deconfined $Z_2$ topologically ordered gauge theory, at least for $d \geq 3$.

\subsection{3+1/2+1D Bosonic topological superconductor $1 \to {Z}_2 \to \text{Pin}^{\pm}(\infty)  \to O(\infty) \to 1$ with $Z_2^T$ time-reversal symmetry:
The $2+1$D $Z_2^K$-gauge theory boundary of $3+1$D bulk invariant $(-1)^{\int (w_2)^2}$ and $(-1)^{\int (w_1)^4+(w_2)^2}$} \label{sec:examples3+1Dw2sq}

There is an additional 3+1D time-reversal symmetric Bosonic topological superconductor (BTSC) beyond the previous $\cH^{4}({Z_2^T},U_T(1))=\Z_2$ class.
It can be captured either 
within the group cohomology of
$G\times SO_\infty$ \cite{Wen1410.8477}
under $\cH^{4}(Z_2^T \times SO(\infty), U_T(1)) = (\Z_2)^2$,\footnote{
The $\cH^{4}(Z_2^T \times SO(\infty), U_T(1)) = (\Z_2)^2$ classification \cite{Wen1410.8477} suggests a
bulk topological invariant $\ee^{\ii 2 \pi \int \frac{1}{2} p_1}=(-1)^{\int p_1}$, where the Pontryagin class $p_1$ is related by the Stiefel-Whitney class $w_2$ through the relation
$w_2^2=p_1 \mod 2$ on any closed oriented 4-manifold.
Moreover, the class with $w_2$ is related to $\pi_1(SO(\infty))=\Z_2$ and $\pi_1(O(\infty))=\Z_2$.} 
or the cobordism classification $\Omega^4_O(pt,U(1))= (\Z_2)^2$ \cite{K1467}.
It gives rise to 3+1D bulk topological invariants 
$\ee^{\ii 2 \pi \int \frac{1}{2} w_2^{2}}=(-1)^{\int (w_2)^2}$ or
$(-1)^{\int (w_1)^4+(w_2)^2}$.
 $w_i \equiv w_i(TM)$ is the $i$-th Stiefel-Whitney class of a tangent bundle $TM$ over spacetime $M$.
We would like to find out the surface $K$-gauge topological order through a short exact sequence.

First, notice that the spin group $\text{Spin}(n)$ is the double cover of the special orthogonal group $SO(n)$. There exists a short exact sequence 
\bea 
1 \to {Z}_2 \to \text{Spin}(n)  \to SO(n) \to  1.
\eea
In our case, for the 3+1D bulk SPT invariant $(-1)^{\int (w_2)^2}$ obtained through $G=Z_2^T \times SO(\infty)$ in $\cH^{4}(Z_2^T \times SO(\infty), U_T(1))$,\footnote{
For the $d+1$D bulk, precisely we should consider the groups $SO(d+1)$,  \text{Spin}$(d+1)$ and \text{Pin}$^{\pm}(d+1)$ in this context. Here we replace $d+1$ to $\infty$ in order to
follow the convention in
\cite{Wen1410.8477}.} 
one may attempt to use
the short exact sequence 
$
1 \to {Z}_2^K \to Z_2^T \times \text{Spin}(\infty)  \to Z_2^T \times SO(\infty) \to 1
$
to construct the surface ${Z}_2^K$-gauge theory.
However, we suggest that the more proper way to consider a trivialization of 
the bulk BTSC, is not based on $G=Z_2^T \times SO(\infty)$, but
based on $G=O(\infty)$ via
\bea \label{eq:Pin+-pull}
1 \to {Z}_2^K \to \text{Pin}^{\pm}(\infty)  \to O(\infty) \to 1.
\eea
We can also rephrase Sec.~\ref{sec:examplesHZ4TGZ2T-any-dim} into this framework via the group extension 
\bea \label{eq:SOZ4pull}
1 \to {Z}_2^K \to SO(\infty) \times \Z_4^T  \to SO(\infty) \times \Z_2^T \to 1.
\eea
In summary,
\begin{enumerate}
\item By \eqn{eq:Pin+-pull}, we can trivialize 3+1D $(-1)^{\int (w_2)^2}$ on the 2+1D boundary by pulling $G=O(\infty)$ back to $H= \text{Pin}^{+}(\infty)$.
Because the $\text{Pin}^{+}$-structure constrains $w_2(TM)=0$, so it trivializes the $(-1)^{\int (w_2)^2}$. Moreover, 
the $\text{Pin}^{+}$-structure implies the quasi-particles are Kramers doublets ($T^2=(-1)^F$) and fermions ($f$). This means the 2+1D boundary
${Z}_2^K$-gauge theory has an emergent dynamical $\text{Pin}^{+}$-structure, with electric and magnetic quasi-particles as $e^f_T m^f_T$.
\item By \eqn{eq:Pin+-pull}, we can trivialize $(-1)^{\int (w_1)^4+(w_2)^2}$ on the 2+1D boundary by pulling $G=O(\infty)$ back to $H= \text{Pin}^{-}(\infty)$.
Because the $\text{Pin}^{-}$-structure constrains $w_2(TM)+w_1(TM)^2=0$, so it trivializes the $(-1)^{\int (w_2+w_1^2)^2}=(-1)^{\int (w_1)^4+(w_2)^2}$. Moreover, 
the $\text{Pin}^{-}$-structure implies the quasi-particles are Kramers singlets ($T^2=+1$) and fermions ($f$). This means the 2+1D boundary
${Z}_2^K$-gauge theory has an emergent dynamical $\text{Pin}^{-}$-structure, with electric and magnetic quasi-particles as $e^f m^f$.
\item By \eqn{eq:SOZ4pull}, we can trivialize $(-1)^{\int (w_1)^4}$ on the 2+1D boundary by pulling $G=SO(\infty) \times \Z_2^T$ back to $H= SO(\infty) \times \Z_4^T$.
Because the $SO(\infty) \times \Z_4^T$-structure constrains $w_1(TM)^2=0$, so it trivializes the $(-1)^{\int (w_1)^4}$. Moreover, 
the $SO(\infty) \times \Z_4^T$-structure implies the quasi-particles are Kramers doublets ($T^2=(-1)^F$)  and bosons ($b$). This means the 2+1D boundary
${Z}_2^K$-gauge theory has an emergent dynamical $SO(\infty) \times \Z_4^T$-structure, with electric and magnetic quasi-particles as $e^b_T m^b_T$.

\end{enumerate}

Other detailed physics aspects along this approach are discussed in \cite{Wang:2018qoy}.
By picking a spin/$\text{Pin}^{+}$/$\text{Pin}^{-}$ structure on the boundary, it means the
boundary can have fermionic quasiparticles. The choice of spin structure can be viewed as a twisted
version of $Z_2^K$ gauge theory.

We note that the $e^f m^f$ and $e^f_T m^f_T$ 
surface topological order first proposed in \cite{VS1306} on the surface of this 3+1D $Z_2^T$-bosonic TSC
is also a 2+1D deconfined $Z_2$-gauge theory with quasiparticles of $Z_2$-gauge charge and $Z_2$-gauge flux, both with fermionic statistics.
However, we remark that the past literatures were not careful enough and tended to mishandle the correspondences between 
3+1D bulk SPTs $(-1)^{\int (w_2)^2}$/$(-1)^{\int (w_1)^4+(w_2)^2}$ and their boundary ${Z}_2^K$-gauge theory $e^f_T m^f_T$/$e^f m^f$ \cite{K1459}.
Our approach here and Ref.~\cite{Wang:2018qoy} makes this relation transparent and precise.

\subsection{2+1/1+1D Bosonic $0 \to Z_{2N}^K {\to} Z_{4  N}^H {\to} Z_2^G \to 0$}
\label{sec:examplesHZ4NGZ2}

For 
$
0 \to Z_{2N}^K \overset{2}{\to} Z_{4  N}^H \overset{r}{\to} Z_2^G \to 0,
$
again we want to trivialize a cocycle
$
\omega_3^{Z_2^G}(g_a , g_b, g_c) 
=(-1)^{g_a g_b g_c}
$ to a cochain.
Generically, we can still reduce $(\text{mod } 4N)$ to $(\text{mod } 4)$ in the exponent so that
$\beta_2(h_1, h_2)  = \exp[ (2 \pi \ii/ 4)  ([h_1]_2)  ([h_2]_4) ]$, or 
$\beta_2(h_1, h_2)= \exp[ (2 \pi \ii/ 4) ([h_1]_4)  ([h_2]_2) ]$ can be the successful split cochains.
%

\subsection{ { 2+1/1+1D Bosonic $1 \to Z_4^K \to Q_8^H  \to Z_2^G \to 1$}} \label{sec:examplesHQ8GZ2}

Trivialize the 3-cocycle in $\cH^3(Z_2^G,U(1))$.
The example that the $H=Q_8$ is a non-Abelian group, while $G=Z_2$, we write 
\bea
1 \to Z_4^K \overset{}{\longrightarrow} Q_8^H \overset{r}{\longrightarrow} Z_2^G \to 1.
\eea
Again,
$
\omega_3^{Z_2^G}(g_a , g_b, g_c)=(-1)^{g_a g_b g_c}.
$

Write the quaternion 
$Q_8=\langle x,y|x^2=y^2, xyx^{-1}=y^{-1},x^4=y^4=1\rangle$
so that each element in the group we can write uniquely as $x^g y^{k}$ with $g\in\{0,1\}$ 
corresponding to $\{ \{1,i,-1,-i\},  j \{1,i,-1,-i\}\}$ in ${Z}_2^G$,
and $k\in\{0,1,2,3\}$ corresponding to $\{1,i,-1,-i\}$ in ${Z}_4^K$. 
Use $yx=x y^{-1}$ and $y^{-1}x=xy$, we can rewrite the group operation as 
$$
x^{g_1} y^{k_1} x^{g_2} y^{k_2} =x^{g_1}  x^{g_2} y^{(-1)^{g_2} k_1} y^{k_2}
=x^{[g_1 + {g_2}]_2} y^{[(-1)^{g_2} k_1+k_2+ 2 g_1 g_2 ]_4}. 
$$
We can write $h=(g,k)$ of $H$ as a doublet from $G$ and $K$, then
\bea
h_1 h_2=(g_1, k_1)\cdot (g_2, k_2)=(g_1+ g_2, (-1)^{g_2} k_1+k_2+ 2 g_1 g_2  )
\equiv (g_1+ g_2, F(k_1,k_2,g_1,g_2)).
\eea
We find that LHS technique in Appendix~\ref{sec:LHS} 
works successfully. 
For LHS technique of Appendix~\ref{sec:LHS},
we look for:
\bea
&& d_2:  \cH^{1}(G, \cH^{1}(K,U(1) )) =\Z_2 \to  \cH^{3}(G, \cH^{0}(K,U(1) ))=\cH^{3}(G, U(1))=\Z_2.\\
&&f: G \to \cH^{1}(K,U(1) )  \Rightarrow
 Z_2^G \to \Z_4.
\eea
In this case, it is found that
\bea
\beta_2(h_1,h_2)=\beta_2((g_1, k_1),(g_2, k_2)) =f(g_2)^{k_1}=\ii^{g_2 k_1}.
\eea
Here $f(g_2^{-1})$ corresponds to a $U(1)$ function labeled by $g_2$, and provides a $U(1)$ function via $f: G \to \cH^{1}(K,U(1) )$.
This $U(1)$ function depends on $k_1 \in K$ for $\cH^{1}(K,U(1) )$, thus we have $\beta(h_1,h_2)= f(g_2^{-1}) (k_1)$.
 We look for the base of $\ii$ because $\cH^{1}(K,U(1))=\Z_4$ is generated by $\ii$ with $\ii^4=1$. 
  
We would like to find a 2-cochain that satisfies the desired 3-cocycle splitting property: 
\bea
&&\omega_3^{Q_8^H}(h_a, h_b, h_c)=\omega_3^{Z_2^G}(r(h_a), r(h_b), r(h_c))=(-1)^{r(h_a) r(h_b) r(h_c)}=(-1)^{g_a g_b g_c}=(\delta\beta_2)(h_1,h_2,h_3).\;\;\;\;\;\;\;\;\;\;
\eea
We write
\bea
(\delta\beta_2)(h_1,h_2,h_3)=\frac{\beta_2(h_2,h_3) \beta_2(h_1, h_2 h_3)}{\beta_2(h_1 h_2, h_3)\beta_2(h_1, h_2)}
= \frac{ {f(g_3)^{(k_2)}} {f(g_2g_3)^{(k_1)}} }{{f( g_3)^{(F(k_1,k_2,g_1,g_2))} } {f(g_2)^{(k_1)}} }.
\eea
Recall that ${f( g_2 g_3)(k_1)}$ is the cocycle of $\cH^{1}(K,U(1) )$ with a power $k_1$. 
We should be able to rewrite $f( g_2 g_3)$ based on the 1-cocycle condition:
\bea
\frac{f( g_2) f( g_3) }{f( g_2 g_3)}=1 \Rightarrow f( g_2 g_3)={f( g_2) f( g_3) },
\eea
so
\bea
&&(\delta\beta_2)(h_1,h_2,h_3)=
 \frac{ {f(g_3)^{(k_2)}} {f(g_2)^{(k_1)}} {f(g_3)^{(k_1)}} }{{f( g_3)^{(F(k_1,k_2,g_1,g_2))} } {f(g_2)^{(k_1)} }}
=
\frac{ {f(g_3)^{(k_2)}} {f(g_3)^{(k_1)}} }{{f( g_3)^{(F(k_1,k_2,g_1,g_2))} }  } \nonumber \\
&&=
\frac{ {f(g_3)^{k_2}} {f(g_3)^{k_1}} }{  {f(g_3)^{[(-1)^{g_2} k_1+k_2+ 2 g_1 g_2]_4}} }.
\eea
Further computation shows, indeed,
\bea
&&(\delta\beta_2) (h_a, h_b, h_c)= \frac{\beta_2(h_b, h_c) \beta_2(h_a, h_b h_c)}{\beta_2(h_a h_b, h_c) \beta_2(h_a, h_b)}
= \frac{\ii^{(k_b g_c)} \ii^{k_a [g_b+g_c]_2} }{ \ii^{[k_a (-1)^{g_b}+ k_b + 2 g_a g_b]_4 g_c} \ii^{(k_a g_b)}}
= (-1)^{g_a g_b g_c}.\;\;\; \quad
\eea
Because $\cH^2( Q_{8},U(1))= 0$, we do not have another lower-dimensional 1+1D $Q_8$-topological state to stack on the boundary.


If we consider the bulk to be fully gauged topologically ordered state, this becomes a gapped boundary for a bulk $2+1$D field theory of 
$\int  \frac{ 2}{2\pi}{B   d A} {{+}}  
\frac{1}{2\pi } A d A $.

\subsubsection{Degeneracy on a disk and an annulus: Partition functions $Z(D^2 \times S^1)$ and $Z(I^1 \times S^1 \times S^1)$}
\label{sec:examplesHQ8GZ2GSD}

Follow the set up Appendix~\ref{sec:examplesHZ4GZ2GSD},
we put the 2+1/1+1D $1 \to Z_4^K \to Q_8^H  \to Z_2^G \to 1$ 
construction of topological states on a spatial $D^2$ disk or an annulus $I^1 \times S^1$ to count the degeneracy (GSD).
Depend on \emph{gauging the global symmetry $K$ and $H$ or not}, we have at least
three types of theories.
Since $K$ is a normal subgroup in $H$, we can label the $K$-holonomy in $H$. Thus, below, we write all holonomies $h$ in $H$.
We consider the group homomorphisms:
\bea
Z_4^K = 
\begin{pmatrix}
1\\ i \\ -1 \\ -i
\end{pmatrix} 
\overset{1}{\longrightarrow} 
\begin{pmatrix}
1\\ i \\ -1 \\ -i
\end{pmatrix}  \subset Q_8^H  
\eea
\bea
Q_8^H = 
\begin{pmatrix}
1, i,  -1 , -i \\
j, k, -j,  -k
\end{pmatrix} 
\overset{}{\longrightarrow} 
\begin{pmatrix}
1\\ -1
\end{pmatrix}  =Z_2^G.    
\eea

We compute the partition function of Sec.~\ref{sec:ZSPTSETG1G2} on $Z(D^2 \times S^1)$ to evaluate GSD on a spatial $D^2$ disk in Table \ref{table:GSDD2HQ8GZ2}.
\begin{table}[!h]
 \begin{center}
\begin{tabular}{c||c|c|c}
\hline
Disk $D^2$&\; 
\begin{minipage}{1.25in} Theory (i) \\ (the second bdry) \end{minipage}  
&\;  \begin{minipage}{1.7in} Theory (ii) \\ (the third/fourth bdry) \end{minipage}
&\; \begin{minipage}{1.25in}  Theory (iii) \\ (the fifth bdry) \end{minipage}  \\ \hline
GSD &1  & 4  & 2\\ \hline
 \end{tabular}
 \caption{ For the theory (ii), GSD=4 from the holonomy $h=1,i,-1,-i$  in $K$ and also in $H$.
{For the fully gauge theory (iii), GSD=2 from the holonomy $h=1$ and
$h=i/-i$. Here $h=i$ and $h=-i$ each contributes 1/2 state, and the i/-i  together act like a
2-dimensional irreducible representation as a non-Abelian ground state.
The setup and notations follow Appendix~\ref{sec:examplesHZ4GZ2GSD}}
}
  \label{table:GSDD2HQ8GZ2}
 \end{center}
\end{table}

The usual 1+1D topological gauge theory has its GSD on an $S^1$ ring and can be computed as $Z(S^1 \times S^1)$ by
\bea
&&\GSD =\frac{1}{|H|}\sum_{h, t} 1 |_{\text{if $ht=th$}}
=\frac{1}{|H|}\sum_{h} (\# \text{ of elements in the centralizer $C_H(h)$ of $h$}) \nonumber \\
&&
=(\# \text{conjugacy classes of $H$})  \nonumber\\
&&
= (\# \text{ of irrep  of $H$}) \le |H|, 
\eea
reduced to a smaller number than $|H|$. 
The $\#$ stands for the number.
For $H=Q_8$, we have
$(\# \text{conjugacy classes of $H$})=(\# \text{ of irre rep of $H$})=5  < |H| =8$.
The 5 conjugacy classes 1, -1, $\{i,-i\}$, $\{j,-j\}$ and $\{k,-k\}$ yield 5 distinct holonomies for GSD=5 on $S^1$. 

We find that the $h=1$ carries zero or an even $Z_2^G$ charge.
The $h=i$ and $h=-i$ combined are also zero or an even $Z_2^G$ charge.
Other sectors of $h$ carry an odd $Z_2^G$ charge.
For the theory (iii), when the $Z_2^G$ is gauged, the ground state for the whole system cannot carry an odd $Z_2^G$ charge, thus
$h=0$ or $h=i/-i \in H$ implies GSD=2 on a disk. An important remark is that
we \emph{cannot} regard the 1+1D anomalous $Q_8^H$ gauge theory as a usual 1+1D discrete gauge theory --- because the
usual 1+1D $Q_8$ gauge theory has GSD$=5$ on a $S^1$ ring. In our case, the 2+1D bulk plays an important rule, which
causes the GSD reduces from 5 conjugacy classes to 2 conjugacy classes (1 and $\{i,-i\}$) of GSD=2 for the theory (iii).

We compute the partition function of Sec.~\ref{sec:ZSPTSETG1G2} on $Z(I^1 \times S^1 \times S^1)$ to evaluate GSD on an annulus $I^1 \times S^1$ in Table \ref{table:GSDI1S1HQ8GZ2}:
\begin{table}[!h]
 \begin{center} 
\makebox[\textwidth][c] 
{
\begin{tabular}{c||c|c|c}
\hline
Annulus $S^1 \times I^1$ &\; 
\begin{minipage}{1.25in}  Theory (i) \\ (the second bdry) \end{minipage}  &\; 
\begin{minipage}{1.7in}  {Theory (ii)} \\ (the third/fourth bdry) \end{minipage} &\; 
\begin{minipage}{1.25in} Theory (iii) \\ (the fifth bdry) \end{minipage} \\ \hline
GSD &\; 1  &\; 16 &\; 8  \\ \hline
 \end{tabular} 
}
\hspace*{0mm}
 \end{center}
 \caption{For the theory (ii) without symmetry twist, GSD=16 from the holonomies of sectors $(h_{\text{in}},h_{\text{out}})$ with 
$h_{\text{in}}, h_{\text{out}} \in \{1, i,  -1 , -i\}$.
For the theory (iii) fully gauge theory, GSD=8 from the holonomies $(h_{\text{in}},h_{\text{out}})$
= $(1,1), (-1,-1), (1,i/-i), (-1,i/-i), (i/-i, 1), (i/-i, -1)$ and two more states from $(i/-i, i/-i)$.
The set-up and notations follow Appendix~\ref{sec:examplesHZ4GZ2GSD}}
 \label{table:GSDI1S1HQ8GZ2}
\end{table}

Again the 2+1D bulk plays an important role for the GSD reduction for the theory (iii) from 
GSD$=|(\# \text{ of irre rep of $H$})|^2=25$ to GSD=8 in Table \ref{table:GSDI1S1HQ8GZ2}.

\subsection{ { 2+1/1+1D Bosonic $1 \to Z_2 \to D_4  \to (Z_2)^2 \to 1$}} \label{2+1DD4Z22}

We consider the construction $1 \to K=Z_2 \to H=D_4  \to Q=(Z_2)^2 \to 1$.
Here $D_4$ is a dihedral group of order 8, namely $|D_4|=8$.
Write the dihedral group
$D_4=\langle \tx, R |\tx^2=R^4=1, \tx R \tx=R^{-1}\rangle$
so that each element in the group we can write uniquely as $\tx^a R^b$
with $a\in\{0,1\}$ and $b\in\{0,1,2,3\}$. 
The quotient group is
$$\frac{D_4}{Z_2}= \frac{D_4}{ \{1, R^2\}}= \{ 1 \{1, R^2\} , \tx \{1, R^2\} , R  \{1, R^2\}, \tx R \{1, R^2\}   \}=(Z_2)^2.$$

Here we would like to trivialize the particular twisted 3-cocycle of $G=(Z_2)^2$:
\bea
&&\omega_2(g_a , g_b, g_c)=\exp(\frac{\ii 2 \pi}{2} \; [g_{a_1}]_2  [g_{b_2}]_2 [g_{c_2}]_2 ) =(-1)^{[g_{a_1}]_2  [g_{b_2}]_2 [g_{c_2}]_2 } , \;\;\;\;
\eea
where $g_a=(g_{a_1},g_{a_2}) \in G=(Z_2)^2$, and similarly for $g_b, g_c$.
This cocycle is equivalent to 
$\ee^{\ii 2 \pi \int \frac{1}{2} a_1 \cup a_1 \cup a_2}$ with a cup product form of $a_1 \cup a_1 \cup a_2$, in $\cH^3( (Z_2)^2,U(1))$.
The $a_1$ and $a_2$ here are $\Z_2$-valued 1-{cocycles} in $\cH^1(M^3, \Z_2)$ on the spacetime complex $M^3$.

We can write $h=(g,k)\in H$ where $g \in G$ and $k \in K$.
Let us write $h=\tx^a R^b \in D_4$ in terms of a triplet, $h_u=(k_u, g_{u_1},g_{u_2}) \in D_4$, 
such that
$$(k_u, g_{u_1},g_{u_2}) \cdot (k_v, g_{v_1},g_{v_2}) =(k_u+k_v+g_{u_1} g_{v_2}, g_{u_1}+g_{v_1}, g_{u_2}+g_{v_2}).$$

Note that the $R^2 =(1,0,0) \in D_4$.
The $D_4 \to (Z_2)^2$  maps 
$h_u=(k_u, g_{u_1},g_{u_2}) \in D_4$ 
to 
$(g_{u_1},g_{u_2}) \in  (Z_2)^2$.
We can view the $k_u$ generates $R^2$ in $D_4$,
while $g_{u_1}$ and $g_{u_2}$ generates $\tx$ and $R$ respectively. We would like to split
\bea
&&\omega_3^H(h_u , h_v, h_w)=\omega_3^G(r(h_u) , r(h_v), r(h_w)) =(-1)^{[ g_{u_1}]_2  [g_{v_2}]_2 [g_{w_2}]_2 } = (\delta \beta_2)(h_u, h_v, h_w), 
\quad
\eea
into a 2-cochain $\beta_2$. The LHS technique in Appendix~\ref{sec:LHS}
suggests that we look for
\bea
&& d_2:  \cH^{1}(G, \cH^{1}(K,U(1) )) \to  \cH^{3}(G, \cH^{0}(K,U(1) )) \nonumber\\
&&\Rightarrow d_2:    \cH^{1}((Z_2)^2, Z_2)=(\Z_2)^{{2}} \to \cH^{3}(G, U(1))=(\Z_2)^3. 
\eea
\bea
f: G \to \cH^{1}(K,U(1) ) \Rightarrow
 (\Z_2)^2 \to \cH^{1}(Z_2^K,U(1) )=\Z_2.
\eea
In this case, it is found that
\bea
\beta_2(h_u,h_v)=\beta_2((k_u, g_{u_1},g_{u_2}),(k_v, g_{v_1},g_{v_2}))= f(g_v)^{k_u}=(-1)^{k_u g_{v_2} }.
\eea
We can see that
\begin{equation}
\delta (\beta_2) =\frac{ \beta_2(h_v, h_w) \beta_2(h_u, h_v h_w)}{\beta_2(h_u  h_v, h_w ) \beta_2(h_u,  h_v)}
= \frac{(-1)^{k_{v} g_{w_2}} (-1)^{k_u (g_{v_2}+ g_{w_2})}}{(-1)^{(k_{u}+k_{v}+g_{u_1} g_{v_2}) g_{w_2}} (-1)^{k_{u} g_{v_2}}} =(-1)^{g_{u_1} g_{v_2} g_{w_2}}
=\omega_3^H(h_u , h_v, h_w).\;\;
\end{equation}
Similarly, it turns out that we can find another 2-cochain
$
\beta_2(h_u, h_v) =(-1)^{k_u g_{v_1}}
$ 
that splits a different 3-cocycle
$
\delta (\beta_2) = \frac{(-1)^{k_{v} g_{w_1}} (-1)^{k_u (g_{v_1}+ g_{w_1})}}{(-1)^{(k_{u}+k_{v}+g_{u_1} g_{v_2}) g_{w_1}} (-1)^{k_{u} g_{v_1}}}=(-1)^{g_{u_1} g_{v_2} g_{w_1}}.
$

Since $\cH^2( D_{4},U(1)) = Z_2$, we can have two distinct classes of 2-cochain differed by a 2-cocycle $\omega_2 \in \cH^2( D_{4},U(1))$ corresponding to
a 1+1D $D_{4}$-topological state on the boundary.


If we consider the bulk to be fully gauged topologically ordered state, this becomes a gapped boundary for a bulk $2+1$D field theory of 
$\int \sum_{I=1}^{2} \frac{ 2}{2\pi}{B_I   d A_I} {{+}}  
\frac{1}{2\pi } A_1 d A_2 $.

\subsection{ { 1+1/0+1D Bosonic $1 \to Z_2 \to Q_8  \to (Z_2)^2 \to 1$}} \label{1+1DQ8Z22}

Here we like to trivialize a particular twisted 2-cocycle of $G=(Z_2)^2$:
\bea
\omega_2(g_a , g_b)=\exp(\frac{\ii 2 \pi}{2} \; [g_{a_1}]_2  [g_{b_2}]_2 )=(-1)^{[g_{a_1}]_2  [g_{b_2}]_2 }, \;\;\;\;
\eea
where $g_a=(g_{a_1},g_{a_2}) \in G=(Z_2)^2$, and similarly for $g_b$.
This cocycle is equivalent to 
$\ee^{\ii 2 \pi \int \frac{1}{2} a_1 \cup a_2}$ with a cup product form of $a_1\cup a_2$, in $\cH^2( (Z_2)^2,U(1))$.
The $a_1$ and $a_2$ here are $\Z_2$-valued 1-{cocycles} in $H^1(M^2, \Z_2)$ on the spacetime complex $M^2$.

We consider the construction
$
1 \to K=Z_2 \to H=Q_8  \to G=(Z_2)^2 \to 1
$.
The quotient group can be realized as $Q_8 /\{1,-1\} = (Z_2)^2$.
We write each element in the group $H=Q_8$ uniquely as $h=x^\rh y^{\rh'}$ with $\rh \in\{0,1\}$ 
corresponding to $\{ 1\{1,i,-1,-i\},  j\{1,i,-1,-i\} \}$ and $\rh'\in\{0,1,2,3\}$ corresponding to $\{1,i,-1,-i\}$. 
By writing $h=x^\rh y^{\rh'}$, the $\rh=1$ and the $\rh'=1$ correspond to two generators of the quotient group $G=(Z_2)^2$.  
Apply the relation $yx=xy^{-1}$ and $x^2=y^2$, we find
$
x^{\rh_1} y^{\rh_1'} x^{\rh_2} y^{\rh_2'} = x^{[\rh_1+\rh_2]_2} y^{[\rh_1' (-1)^{\rh_2}+ \rh_2'+2 \rh_1 \rh_2]_4}. 
$
We can rewrite
\bea
&&\omega_2^{Q_8^H}(h_a, h_b)=\omega_2^{Z_2^G}(r(h_a), r(h_b))=(-1)^{[\rh_a']_2 {\rh_b}}.
\eea
We claim that the above 3-cocycle can be split by 2-cochains:
\bea
\beta_1(h)=\beta_1(x^{\rh} y^{\rh'}) = \ee^{\frac{\ii \pi}{ 2} (\rh+\rh')} = \ii^{(\rh+\rh')}.
\eea 
Indeed we find it works:
\bea
&&(\delta \beta_1) (h_a, h_b)= \frac{\beta_1(h_a)\beta_1(h_b)}{\beta_1(h_a h_b)}
= \frac{\ii^{(\rh_a+\rh_a')} \ii^{(\rh_b+\rh_b')} }{\ii^{([\rh_a+\rh_b]_2
+[{\rh_a' (-1)^{\rh_b}+ \rh_b'+ 2 \rh_a \rh_b}]_4)}} \nonumber\\
&&= \frac{\ii^{([\rh_a]_2+[\rh_a']_4)} \ii^{([\rh_b]_2+[\rh_b']_4)} }{\ii^{([\rh_a+\rh_b]_2
+[{\rh_a' (-1)^{\rh_b}+ \rh_b'+ 2 \rh_a \rh_b}]_4)}} 
= \frac{\ii^{([\rh_a']_4)} \ii^{([\rh_b']_4)} }{\ii^{(
[{\rh_a' (-1)^{\rh_b}+ \rh_b'}]_4)}}=\ii^{\rh_a' (1- (-1)^{\rh_b}) }=(-1)^{\rh_a' {\rh_b}}=(-1)^{[\rh_a']_2 {\rh_b}}\nonumber\\
&&=\omega_2^{Q_8^H}(h_a, h_b).
\eea
There are various legal 1-cochains that trivialize the $G$ 2-cocycle as 2-coboundary in $H$, such as
$\beta_1(h)=\beta_1(x^{\rh} y^{\rh'}) = \ii^{(\rh+\rh')}, \ii^{(\rh-\rh')},  \ii^{(-\rh+\rh')},  \ii^{(-\rh-\rh')}$.
These 1-cochains can be differed by a 1-cocycle $\omega^H_1$ in $H=Q_8$, such that
$\omega^H_1(h) \in \cH^1(Q_8, U(1)) =  (\Z_2)^2$ thus they differ by a 0+1D topological state on the boundary. 
Indeed, the 1-cocycle $\omega^H_1$ can be:
$$
\omega_1(x^\rh y^{\rh'})=(-1)^{\rh}, (-1)^{\rh'}, (-1)^{\rh+\rh'}
$$
One can check the following is true:
\bea
&&(\delta \omega_1) (h_a, h_b)= \frac{\omega_1(h_a) \omega_1(h_b)}{\omega_1(h_a h_b)}=1.
\eea
All these 1-cochains $\beta_1(x^{\rh} y^{\rh'}) = \ii^{(\rh+\rh')}, \ii^{(\rh-\rh')},  \ii^{(-\rh+\rh')},  \ii^{(-\rh-\rh')}$
are differed by each other via stacking 0+1D-topological states labeled by 1-cocycle $\omega_1 = (-1)^{\rh}, (-1)^{\rh'}, (-1)^{\rh+\rh'} \in \cH^1( Q_{8},U(1)) = Z_2 \times Z_2$.\\
  
The LHS technique in Appendix~\ref{sec:LHS}
suggests that we look for
\bea
&& d_2:  \cH^{0}(G, \cH^{1}(K,U(1) )) \to  \cH^{2}(G, \cH^{0}(K,U(1) )) \nonumber\\
&&\Rightarrow d_2:    \cH^{0}((Z_2)^2, Z_2)=\Z_2 \to \cH^{2}((\Z_2)^2, U(1))=\Z_2. 
\eea
\bea
f: G \to \cH^{1}(K,U(1) ) \Rightarrow
 (\Z_2)^2 \to \cH^{1}(Z_2^K,U(1) )=\Z_2.
\eea
In this case, it suggested that $\beta_1(h)=\beta_1((g, k))$ can be written as a base of $(-1)$, but we found the solution for a base of $\ii$ instead.
So LHS technique is \emph{not} helpful here.


If we consider the bulk to be fully gauged topologically ordered state, this becomes a gapped boundary for a bulk $1+1$D field theory of 
$\int \sum_{I=1}^{2} \frac{ 2}{2\pi}{B_I   d A_I} {{+}}  
\frac{1}{\pi } A_1 A_2 $.
  
\subsection{ { 1+1/0+1D Bosonic $1 \to Z_2 \to D_4  \to (Z_2)^2 \to 1$}}  \label{1+1DD4Z22}

Here we like to trivialize 
a particular twisted 2-cocycle of $G=(Z_2)^2$ based on $1 \to Z_2^K \to D_4   \overset{r}{\to} (Z_2)^2 \to 1$,
\bea
\omega_2(g_a , g_b)=\exp(\frac{\ii 2 \pi}{2} \; [g_{a_1}]_2  [g_{b_2}]_2 )=(-1)^{[g_{a_1}]_2  [g_{b_2}]_2 }, \;\;\;\;
\eea
where $g_a=(g_{a_1},g_{a_2}) \in G=(Z_2)^2$, and similarly for $g_b$.
This cocycle is equivalent to 
$\ee^{\ii 2 \pi \int \frac{1}{2} a_1  \cup a_2}$ with a cup product form of $a_1 \cup a_2$, in $\cH^2( (Z_2)^2,U(1))$.
The $a_1$ and $a_2$ here are $\Z_2$-valued 1-cochains in $H^1(M^2, \Z_2)$ on the spacetime complex $M^2$.

See Sec.~\ref{2+1DD4Z22}, the explicit group elements inside a quotient group can be written as:
$$\frac{D_4}{Z_2}= \frac{D_4}{ \{1, R^2\}}= \{ 1 \{1, R^2\} , \tx \{1, R^2\} , R  \{1, R^2\}, \tx R \{1, R^2\}   \}=(Z_2)^2.$$

We find the split 1-cochain as
$
\beta_1(h) =(-1)^{\rf(h)}.
$
This 1-cochain satisfies the desired 2-cocycle splitting property. Here
we can define the function $\rf$:
\bea
&& \rf(1)=\rf(\tx)=\rf(R) =\rf(\tx R)=0 \in Z_2^K,\\
&& \rf(R^2)=\rf(\tx \cdot R^2)=\rf(R \cdot R^2) =\rf(\tx R \cdot R^2)=1  \in Z_2^K. \nonumber
\eea
Let us write $h=\tx^a R^b \in D_4$ in terms of a doublet $h=(k, g)$, or a more precise triplet, $h_u=(k_u, g_{u_1},g_{u_2}) \in D_4$, such that
$(k_u, g_{u_1},g_{u_2}) \cdot (k_v, g_{v_1},g_{v_2}) =(k_u+k_v+g_{u_1} g_{v_2}, g_{u_1}+g_{v_1}, g_{u_2}+g_{v_2}).$
Note that the $R^2 =(1,0,0) \in D_4$.
The $D_4 \to (Z_2)^2$  maps 
$h_u=(k_u, g_{u_1},g_{u_2}) \in D_4$ 
to 
$(g_{u_1},g_{u_2}) \in  (Z_2)^2$, so that
\bea
{\rf(h)}={\rf(x^a R^b)}=\frac{b-[b]_2}{2}=k_u=\left\{
 \begin{array}{l}
 1, \text{ if } b=2, 3. \\ 
 0, \text{ if } b=0, 1. 
\end{array}
\right.
\eea 
\bea
\beta_1(h_u ) =(-1)^{\rf(h_u)}=(-1)^{k_u}.
\eea 
We can see that, indeed,
\bea
\delta (\beta_1) =\frac{\beta_1(h_u) \beta_1(h_v)}{\beta_1(h_u  h_v )}=
\frac{(-1)^{k_u} (-1)^{k_v}}{(-1)^{k_u+k_v+g_{u_1} g_{v_2}}}=(-1)^{g_{u_1} g_{v_2}}
=\omega_2^G ( r(h_u), r(h_v) )
=\omega_2^H ( h_u,  h_v ).
\eea

The LHS technique in Appendix~\ref{sec:LHS}
suggests that we  look for
\bea
&& d_2:  \cH^{0}(G, \cH^{1}(K,U(1) )) \to  \cH^{2}(G, \cH^{0}(K,U(1) )) \nonumber\\
&&\Rightarrow d_2:    \cH^{0}((Z_2)^2, Z_2)=\Z_2 \to \cH^{2}(G, U(1))=\Z_2. 
\eea
\bea
f: G \to \cH^{1}(K,U(1) ) \Rightarrow
 (\Z_2)^2 \to \cH^{1}(Z_2^K,U(1) )=\Z_2,
\eea
with a base of $(-1)$.
In this case, it is true that
$
\beta_1(h_u)=\beta_1((k_u, g_{u_1},g_{u_2}))=(-1)^{k_u}.
$


If we consider the bulk to be fully gauged topologically ordered state, this becomes a gapped boundary for a bulk $1+1$D field theory of 
$\int \sum_{I=1}^{2} \frac{ 2}{2\pi}{B_I   d A_I} {{+}}  
\frac{1}{\pi } A_1 A_2 $.

\subsection{ { 2+1/1+1D Bosonic $1 \to Z_2 \to D_4 \times Z_2 \to (Z_2)^3 \to 1$}} \label{2+1DD4Z2Z23}

Here we would like to trivialize the 3-cocycle of a cup product form $\ee^{\ii 2 \pi \int \frac{1}{2} a_1  \cup a_2 \cup a_3}$ in $\cH^3( (Z_2)^3,U(1))$
with $a_i \in H^1(M^3,Z_2)$ of  an $M^3$-spacetime complex, 
via $1 \to Z_2^K \to D_4 \times Z_2  \overset{r}{\to} (Z_2)^3 \to 1$.
The particular twisted 3-cocycle of $G=(Z_2)^3$ that we focus on is
\bea
\omega_3(g_a , g_b, g_c)
=(-1)^{[g_{a_1}]_2  [g_{b_2}]_2 [g_{c_3}]_2}, \;\;\;\;
\eea
where $g_a=(g_{a_1},g_{a_2},g_{a_3}) \in G=(Z_2)^3$, and similarly for $g_b$ and $g_c$.
Here $D_4$ is a dihedral group of order 8, namely $|D_4|=8$.
We write the dihedral group
$D_4=\langle \tx, R |\tx^2=R^4=1, \tx R \tx=R^{-1}\rangle$
so that each element in the group we can write uniquely as $\tx^a R^b$
with $a\in\{0,1\}$ and $b\in\{0,1,2,3\}$. 
Indeed the group homomorphism $D_4 \times Z_2 \to (Z_2)^3$
can be understood from a reduced map: $D_4  \to (Z_2)^2$.
We only need to understand the short exact sequence
$
1 \to Z_2^K \to D_4   \overset{r}{\to} (Z_2)^2 \to 1
$
in Appendix~\ref{1+1DD4Z22}.
Namely, we can take the $Z_2$ in $D_4 \times Z_2$ mapping directly to the third $Z_2$ component in $(Z_2)^3$, while
we only have to specify $D_4   \overset{r}{\to} (Z_2)^2$ such that
$\{1 \{1, R^2  \}, \tx \{1, R^2  \}, R \{1, R^2  \}, \tx R \{1, R^2  \}\} \overset{r}{\to}
(Z_2)^2$.
Meanwhile,
the normal subgroup $Z_2^K$ can be viewed as
$\{1, R^2  \}$ in $D_4$.

We denote the group elements of $h_u \in D_4 \times Z_2$ as $(k_u, g_{u_1},g_{u_2}, g_{u_3})$, 
where $(k_u, g_{u_1},g_{u_2}) \in D_4$, and $g_{u_3} \in Z_2$, 
such that
$(k_u, g_{u_1},g_{u_2}) \cdot (k_v, g_{v_1},g_{v_2}) =(k_u+k_v+g_{u_1} g_{v_2}, g_{u_1}+g_{v_1}, g_{u_2}+g_{v_2}).$
Follow the construction in a previous Appendix~\ref{1+1DD4Z22},
note that the $R^2 =(1,0,0) \in D_4$.
The $D_4 \times Z_2\to (Z_2)^3$  maps 
$h_u=(k_u, g_{u_1},g_{u_2}, g_{u_3}) \in D_4 \times Z_2$ 
to 
$(g_{u_1},g_{u_2}, g_{u_3}) \in  (Z_2)^3$.
We propose this 2-cochain satisfies the desired 3-cocycle splitting property: 
\bea
\beta_2(h_u  ,h_v ) =(-1)^{\rf(h_u) g_{v_3}}=(-1)^{k_u g_{v_3}}.
\eea 
We can indeed show 
\begin{multline}
(\delta \beta_2) (h_u,h_v,h_w) =\frac{\beta_2 (h_v, h_w) \beta_2(h_u , h_v h_w )}{\beta_2(h_u  h_v, h_w )\beta_2 (h_u, h_v)}
=\frac{(-1)^{k_v g_{w_3}} (-1)^{k_u (g_{v_3}+g_{w_3})}}{(-1)^{(k_u+k_v+g_{u_1} g_{v_2})g_{w_3}} (-1)^{k_u g_{v_3}}} =(-1)^{g_{u_1} g_{v_2} g_{w_3}}\;\;\;\;\;\;\;\;\;\\
=\omega_3^G ( r(h_u), r(h_v), r(h_w) )=\omega_3^H ( h_u,  h_v, h_w ). 
\end{multline}
The LHS technique in Appendix~\ref{sec:LHS} also gives the correct hint.


If we consider the bulk to be fully gauged topologically ordered state, this becomes a gapped boundary for a bulk $2+1$D field theory of 
$\int \sum_{I=1}^{3} \frac{ 2}{2\pi}{B_I   d A_I} {{+}}  
\frac{1}{\pi^2 }  A_1 A_2  A_{3}$.

\subsection{ { 3+1/2+1D Bosonic $1 \to Z_2 \to D_4 \times (Z_2)^2 \to (Z_2)^4 \to 1$} and \\
{$d+1/d$D Bosonic $1 \to Z_2 \to D_4 \times (Z_2)^{d-1} \to (Z_2)^{d+1} \to 1$}}  \label{3+1DD4Z22Z24}

We can easily generalize from Appendix~\ref{1+1DD4Z22} and \ref{2+1DD4Z2Z23} to any dimension.
For example, based on a 3+1/2+1D bosonic $1 \to Z_2 \to D_4 \times (Z_2)^2 \to (Z_2)^4 \to 1$ construction,
 we can trivialize the 4-cocycle of a cup product form  $\ee^{\ii 2 \pi \int \frac{1}{2} a_1  \cup a_2 \cup a_3 \cup a_4}$ in $\cH^4( (Z_2)^4,U(1))$,
 here $a_i \in H^1(M^4,Z_2)$
 of an $M^4$-spacetime complex.
We denote the group elements of $h_u \in D_4 \times (Z_2)^2$ as $(k_u, g_{u_1},g_{u_2}, g_{u_3}, g_{u_4})$, 
where $(k_u, g_{u_1},g_{u_2}) \in D_4$, and $(g_{u_3},  g_{u_4}) \in (Z_2)^2$.
We can define a 3-cochain in $H$
\bea
\beta_3(h_u  ,h_v, h_w ) =(-1)^{\rf(h_u) g_{v_3} g_{w_4}}=(-1)^{{k_u} g_{v_3} g_{w_4}}
\eea 
that indeed splits a nontrivial 4-cocycle
\bea
&&(\delta \beta_3) (h_u,h_v,h_w, h_z) =\frac{\beta_3 (h_v, h_w, h_z) \beta_3(h_u , h_v h_w, h_z ) \beta_3(h_u,  h_v, h_w ) }{\beta_3(h_u  h_v, h_w, h_z )\beta_3(h_u,  h_v, h_w h_z )} \nonumber\\
&&=
\frac{(-1)^{k_{v} g_{w_3} g_{z_4}} (-1)^{k_u (g_{v_3}+g_{w_3}) g_{z_4}}   (-1)^{k_{u} g_{v_3} g_{w_4}}}{
(-1)^{(k_u+k_v+g_{u_1} g_{v_2})g_{w_3} g_{z_4} } (-1)^{k_{u} g_{v_3} (g_{w_4}+g_{z_4})}} =(-1)^{g_{u_1} g_{v_2} g_{w_3} g_{z_4}}\nonumber\\
&&=\omega_4^G ( r(h_u), r(h_v), r(h_w), r(h_z) ) = \omega_4^H ( h_u,  h_v, h_w,h_z ). 
\eea
In general,
 based on a $d+1/d$D bosonic construction via $1 \to Z_2 \to D_4 \times (Z_2)^{d-1} \to (Z_2)^{d+1} \to 1$, 
  we can trivialize the $d+1$-cocycle of a cup product form 
 $\ee^{\ii 2 \pi \int \frac{1}{2} a_1  \cup a_2 \cup  \dots \cup a_{d+1}}$
  in $\cH^{d+1}( (Z_2)^{d+1},U(1))$.
We denote the group elements of $h_u \in D_4 \times (Z_2)^{d-1}$ as $(k_u, g_{u_1},g_{u_2}, g_{u_3}, \dots, g_{u_{d+1}})$, 
where $(k_u, g_{u_1},g_{u_2}) \in D_4$, and $(g_{u_3},  g_{u_4}, \dots, g_{u_{d+1}}) \in (Z_2)^{d-1}$.
We can write down the $d$-cochain
\bea
\beta_d(h_u, h_v, h_w, h_z, \dots) =(-1)^{\rf(h_u)  g_{v_3} g_{w_4} g_{z_5} \dots g_{._{d+1}} }=(-1)^{k_u g_{v_3} g_{w_4} g_{z_5} \dots g_{._{d+1}} }
\eea 
that splits a nontrivial $d+1$-cocycle in $\cH^{d+1}((Z_2)^{d+1},U(1))$. 
\bea
\omega_{d+1}^G ( r(h_u), r(h_v), r(h_w), r(h_z), \dots ) = \omega_{d+1}^H ( h_u,  h_v, h_w,h_z,  \dots )
=(-1)^{g_{u_1} g_{v_2} g_{w_3} g_{z_4}\dots  g_{._{d+1}} }.
\eea
Again the LHS technique in Appendix \ref{sec:LHS} also gives the correct hint.


If we consider the bulk to be fully gauged topologically ordered state, this becomes a gapped boundary for a bulk $d+1$D field theory of 
$\int \sum_{I=1}^{d+1} \frac{ 2}{2\pi}{B_I   d A_I} {{+}}  
\frac{1}{(\pi)^d } A_1 A_2 \dots A_{d+1}$.

\subsection{ { 2+1/1+1D Bosonic $1 \to (Z_2)^2 \to D_4 \times Z_2 \to (Z_2)^2 \to 1$}}  \label{2+1DD4Z2Z22}

Here we would like to trivialize a particular twisted 2-cocycle of $G=(Z_2)^2$ in $\cH^3( (Z_2)^2,U(1))$, 
\bea
\omega_3(g_a , g_b, g_c)=\exp(\frac{\ii 2 \pi}{2} \; [g_{a_1}]_2  [g_{b_2}]_2 [g_{c_2}]_2 )=(-1)^{[g_{a_1}]_2  [g_{b_2}]_2 [g_{c_2}]_2 }, \;\;\;\;
\eea
where $g_a=(g_{a_1},g_{a_2}) \in G=(Z_2)^2$, and similarly for $g_b$ and $g_c$.
The idea is extending the 1+1D example of Appendix \ref{1+1DD4Z22}'s via
$
1 \to Z_2^K \to D_4   \overset{r}{\to} (Z_2)^2 \to 1
$
in the normal subgroup side by $Z_2$, and we seek for a realization in 2+1D:
\bea
1 \to (Z_2)^2 \to D_4  \times Z_2 \overset{r}{\to} (Z_2)^2 \to 1.
\eea
Since we have discussed that in Appendix \ref{2+1DD4Z22} the 2+1D example of 
\bea
1 \to Z_2^K \to D_4   \overset{r}{\to} (Z_2)^2 \to 1
\eea
already trivializes the 3-cocycle of a cup product form $\ee^{\ii 2 \pi \int \frac{1}{2} a_1  \cup a_2 \cup a_2}$ in $\cH^3( (Z_2)^2,U(1))$,
then we can simply take
$D_4  \times Z_2 \overset{r}{\to} (Z_2)^2$
as the combination of $D_4    \overset{r}{\to} (Z_2)^2$ and $Z_2  \overset{r}{\to} 1$.
We denote the group elements of $h_u \in D_4 \times Z_2$ as $(k_u, g_{u_1},g_{u_2}, g_{u_3})$, 
where $(k_u, g_{u_1},g_{u_2}) \in D_4$, and $g_{u_3} \in Z_2$, 
such that
$(k_u, g_{u_1},g_{u_2}) \cdot (k_v, g_{v_1},g_{v_2}) =(k_u+k_v+g_{u_1} g_{v_2}, g_{u_1}+g_{v_1}, g_{u_2}+g_{v_2}).$
We propose the split 2-cochain
\bea
\beta_2(h_u, h_v) =(-1)^{k_u g_{v_2}}.
\eea 
We can see that
\begin{multline}
(\delta \beta_2) =\frac{ \beta_2(h_v, h_w) \beta_2(h_u, h_v h_w)}{\beta_2(h_u  h_v, h_w ) \beta_2(h_u,  h_v)}
= \frac{(-1)^{k_v g_{w_2}} (-1)^{k_u (g_{v_2}+ g_{w_2})}}{(-1)^{(k_u+k_v+g_{u_1} g_{v_2})  g_{w_2}} (-1)^{k_u g_{v_2}}}=(-1)^{g_{u_1} g_{v_2}  g_{w_2}}\;\;\;\\
=\omega_3^G(r(h_u) , r(h_v), r(h_w))=\omega_3^H(h_u , h_v, h_w).
\end{multline}
The LHS technique in \ref{sec:LHS} gives the correct hint.
Basically this shows the same result as in Appendix \ref{2+1DD4Z22}.

\subsection{ { 3+1/2+1D Bosonic $1 \to (Z_2) \to D_4  \to (Z_2)^2 \to 1$}} \label{3+1DD4Z22TypeII}

Here we like to trivialize a particular twisted 4-cocycle of $G=(Z_2)^2$  in $\cH^4( (Z_2)^2,U(1))$,
\bea
\omega_4(g_a , g_b, g_c, , g_d)=\exp(\frac{\ii 2 \pi}{2} \; [g_{a_1}]_2  [g_{b_2}]_2 [g_{c_2}]_2  [g_{d_2}]_2 )=(-1)^{[g_{a_1}]_2  [g_{b_2}]_2 [g_{c_2}]_2  [g_{d_2}]_2}.
\eea
We consider the construction via $1 \to Z_2 \to D_4   \to (Z_2)^2 \to 1$. Follow the earlier definition of $D_4$ group elements, we propose the split 3-cochain
\bea
\beta_3(h_u  ,h_v, h_w ) =(-1)^{\rf(h_u) g_{v_2} g_{w_2}}=(-1)^{{k_u} g_{v_2} g_{w_2}}.
\eea
We can check explicitly that the 3-cochain splits the 4-cocycle in $H$:
\bea
&&(\delta \beta_3) (h_u,h_v,h_w, h_z) =\frac{\beta_3 (h_v, h_w, h_z) \beta_3(h_u , h_v h_w, h_z ) \beta_3(h_u,  h_v, h_w ) }{\beta_3(h_u  h_v, h_w, h_z )\beta_3(h_u,  h_v, h_w h_z )} \nonumber\\
&&=
\frac{(-1)^{k_{v} g_{w_2} g_{z_2}} (-1)^{k_u (g_{v_2}+g_{w_2}) g_{z_2}}   (-1)^{k_{u} g_{v_2} g_{w_2}}}{
(-1)^{(k_u+k_v+g_{u_1} g_{v_2})g_{w_2} g_{z_2} } (-1)^{k_{u} g_{v_2} (g_{w_2}+g_{z_2})}} =(-1)^{g_{u_1} g_{v_2} g_{w_2} g_{z_2}} \nonumber\\
&&=\omega_{4,{\text{II}}}^G=\omega_4^G ( r(h_u), r(h_v), r(h_w), r(h_z) ) = \omega_4^H ( h_u,  h_v, h_w,h_z ). 
\eea
If we consider the bulk to be fully gauged topologically ordered state, this becomes a gapped boundary for a field theory of 
$\int \sum_{I=1}^2 \frac{ 2}{2\pi}{B_I   d A_I} {{+}}  
\frac{ 1 }{2{(\pi)^2 } }   
A_{1}  A_{2}  d A_{2}$.

\subsection{ { 3+1/2+1D Bosonic $1 \to Z_2 \to D_4 \times Z_2 \to (Z_2)^3 \to 1$}} \label{3+1DD4Z2Z23TypeIII}

Here we aim to trivialize the 4-cocycle of a particular twisted 4-cocycle of $G=(Z_2)^3$ in $\cH^4( (Z_2)^3, U(1))$,
\bea
\omega_4(g_a , g_b, g_c, , g_d)=\exp(\frac{\ii 2 \pi}{2} \; [g_{a_1}]_2  [g_{b_2}]_2 [g_{c_3}]_2  [g_{d_3}]_2 )=(-1)^{[g_{a_1}]_2  [g_{b_2}]_2 [g_{c_3}]_2  [g_{d_3}]_2}.
\eea
We consider the construction via $1 \to Z_2 \to D_4 \times Z_2 \to (Z_2)^3 \to 1$. Follow the earlier definition of $D_4$ group elements, we propose the split 3-cochain
\bea
\beta_3(h_u  ,h_v, h_w ) =(-1)^{\rf(h_u) g_{v_3} g_{w_3}}=(-1)^{{k_u} g_{v_3} g_{w_3}}
\eea
We can check explicitly that the 3-cochain splits the 4-cocycle in $H$:
\bea
&&(\delta \beta_3) (h_u,h_v,h_w, h_z) =\frac{\beta_3 (h_v, h_w, h_z) \beta_3(h_u , h_v h_w, h_z ) \beta_3(h_u,  h_v, h_w ) }{\beta_3(h_u  h_v, h_w, h_z )\beta_3(h_u,  h_v, h_w h_z )} \nonumber\\
&&=
\frac{(-1)^{k_{v} g_{w_3} g_{z_3}} (-1)^{k_u (g_{v_3}+g_{w_3}) g_{z_3}}   (-1)^{k_{u} g_{v_3} g_{w_3}}}{
(-1)^{(k_u+k_v+g_{u_1} g_{v_2})g_{w_3} g_{z_3} } (-1)^{k_{u} g_{v_3} (g_{w_3}+g_{z_3})}}
=(-1)^{g_{u_1} g_{v_2} g_{w_3} g_{z_3}}\nonumber\\
&&=\omega_{4,{\text{III}}}^G=\omega_4^G ( r(h_u), r(h_v), r(h_w), r(h_z) ) = \omega_4^H ( h_u,  h_v, h_w,h_z ). 
\eea
If we consider the bulk to be fully gauged topologically ordered state, this becomes a gapped boundary for a field theory of 
$\int \sum_{I=1}^3 \frac{ 2}{2\pi}{B_I   d A_I} {{+}}  
\frac{ 1 }{2{(\pi)^2 } }   
A_{1}  A_{2}  d A_{3}$.

\subsection{ { 2+1/1+1D to $d+1/d$D Bosonic $1 \to Z_N \to U(1) \to U(1) \to 1$}:  {Symmetry-enforced gapless boundaries protected by perturbative anomalies}} 
\label{2+1DU1U1}

It is tempting to ask for the construction of a 2+1/1+1D topological state via
\bea
1 \to Z_N \to U(1) \to U(1) \to 1,
\eea
where the bulk has  2+1D $U(1)$ SPTs obtained from $\cH^3(U(1),U(1))=\Z$, while the boundary has 1+1D SETs with a $U(1)$ global symmetry and an emergent exact $Z_N$ gauge symmetry.

Of course, this kind of group extension along the boundary is possible, in general.   But then the boundary theory is a 1+1D theory with a $U(1)$ global symmetry
that has a {\it perturbative} 't Hooft anomaly \cite{H8035}.   As in 't Hooft's original work on such matters, this obstructs the possibility of 
symmetrically gapping the boundary theory. 
Similar remarks apply for any even $d$ dimensional spacetime of the boundary theory.
\footnote{If the counter-statement was true, 
then we may have a 1+1D SETs where the $U(1)$ global symmetry cannot be spontaneously broken ---
this is due to that Coleman-Mermin-Wagner theorem asserts that there is no spontaneous symmetry breaking for continuous symmetry in 1+1D.
In this case, the degenerate ground states of the 1+1D anomalous SETs with emergent $Z_N$ gauge fields, may \emph{not} directly cross over to
symmetry breaking states
and may have a \emph{distinct} phase transition.
This continuous symmetry group protection 
will be a new phenomenon very different from the result from a discrete finite symmetry group in \ref{sec:deconfined-to-SSB}.
In this case, the 1+1D anomalous SETs may be a robust 1+1D anomalous topological order protected by a global symmetry.
If this was true, we can ask whether this example may be generalizable to higher dimensions $d+1/d$D since $\cH^{d+1}(U(1),U(1))=\Z$ when $d$ is even.

However, the above construction is \emph{invalid}. By coupling a $U(1)$ probed background gauge field to 1+1D boundary of 2+1D SPTs, 
the boundary exhibits a perturbative 
chiral anomaly. It is indeed a 
$U(1)$ gauge anomaly probable by the weak-coupling $U(1)$ gauge field.
In 1+1D, one can do the fermionization (or bosonization), a 1-loop Feynman diagram of the fermionized 1+1D boundary captures the $U(1)$-anomaly.
This 't Hooft $U(1)$-anomaly matching factor is equivalent to the effective quantum Hall conductance probed 
by external charged $U(1)$ gauge fields from the bulk \cite{Wang1307.7480}.
Such a perturbative 
anomaly \emph{cannot} be pulled back to another larger continuous group $H$ (here we have an $H=U(1)$ with $N$ times larger periodicity than $G=U(1)$)
with the $G$-anomaly eliminated to be anomaly-free in $H$. In this case, the $U(1)$-anomaly still remains robust in $H=U(1)$.
More generally, for $\cH^{d+1}(U(1),U(1))=\Z$ with an even $d=2,4,\dots$, prescribing the $d+1/d$D SPTs where
the $d+1$D bulk topological invariants are written in terms of Abelian Chern-Simons forms,
as 
\bea
\exp(\ii  \frac{k}{(2 \pi)^{d/2}} \int A (dA)^{d/2})=\exp(\ii  {2 \pi k}  \int (\frac{A}{2 \pi}) \; (c_1)^{d/2}),
\eea
probed by $A$ as a $U(1)$ 1-form gauge field, and $c_1 \equiv dA/2 \pi$ as the first Chern class, with $k \in \Z$. 
The boundary theories are enforced to be gapless by a continuous $U(1)$ symmetry and by a perturbative $U(1)$ anomaly for any even $d$ dimensions.
These are \emph{symmetry-enforced gapless} boundaries due to a perturbative anomaly.
(There are also \emph{symmetry-enforced gapless} boundaries due to a non-perturbative anomaly studied in \cite{CWang1401.1142}.)

Instead  we can find another scenario such that the SPTs is protected by a continuous $G$-global symmetry with a $\Z_N$ sub-classification, instead of a $\Z$ classification.
To make a comparison, the $\Z$ class indicates a \emph{perturbative} 
anomaly.
The $\Z_N$ class is obtained, for example, from the torsion (Tor) part in the universal coefficient theorem of group cohomology.
The $\Z_N$ indicates the boundary $G$-anomaly shall have \emph{global} gauge anomalies. 
For \emph{global} gauge anomalies, it is possible to find a larger continuous $H$ such that the global gauge $G$-anomaly becomes anomaly-free in $H$.
Our observation agrees with \cite{YouXu1502.07752}.
Appendix \ref{6+1DU1U1} and \ref{1+1DSU2SO3} provide two of such examples.
}

\subsection{ { 6+1/5+1D Bosonic $1 \to Z_2 \to U(1)  \times SO(\infty) \to U(1) \times SO(\infty) \to 1$}: {Surface topological order and global mixed gauge-gravitational anomaly }} 
\label{6+1DU1U1}

The previous Appendix~\ref{2+1DU1U1} discusses the $U(1)$-anomaly on the boundary of SPTs obtained from the group cohomology $\cH^{d+1}(U(1),U(1))=\Z$ of symmetry group $G=U(1)$.
However, there are $U(1)$ anomalies beyond the $\cH^{d+1}(G,U(1))$ but within $\cH^{d+1}(G \times SO(\infty), U(1))$ \cite{Wen1410.8477}. 
One example is the 3+1D \emph{perturbative mixed gauge-gravity} anomaly \cite{Wang1405.7689, {Wen1410.8477}} on the surface of 4+1D $U(1)$-SPTs characterized by
\bea
\exp( \ii  2 \pi \int \frac{1}{3}\frac{A}{2 \pi}p_1)
\eea
where $A$ is a $U(1)$ 1-form gauge field and $p_1$ is the first Pontryagin class of the tangent bundle of spacetime manifold.
Unfortunately, such anomaly has $\Z$ class (within $\cH^{5}(U(1) \times SO(\infty), U(1))=(\Z)^2$), it is still a perturbative 
anomaly protected to be \emph{symmetry-enforced gapless} that excludes symmetry-preserving gapped boundary (e.g. surface topological order).   

Another SPT theory with  6+1D bulk/5+1D boundary dimension can have a  $\Z_2$ anomaly 
(within $\cH^{7}(U(1) \times SO(\infty), U(1))=(\Z)^2 \times \Z_2$), 
labeled by the bulk topological invariant \cite{Wen1410.8477} on a 7-manifold ${M^7}$:
\bea 
\exp( \ii  2 \pi \int_{M^7} \frac{1}{2} w_2 w_3 \frac{dA}{2 \pi})= \exp( \ii  2 \pi  \int_{M^7} \frac{1}{2} w_2 w_3 c_1),
\eea
where the $w_i$ as the $i$th SW class. 
Here $w_i$ is a cohomology class with mod 2 coefficients. 
We can write $w_i=w_i(TM^7)$ of the spacetime tangent bundle $TM^7$.
This $\Z_2$ class indicates a non-perturbative global mixed gauge-gravitational anomaly from a continuous group $U(1)$.
We suggest that the 5+1D $Z_2$ gauge theory can be a surface topological order, via the construction $1 \to Z_2 \to U(1)  \times SO(\infty) \to U(1)  \times SO(\infty) \to 1$,
as a symmetry-preserving gapped boundary.
The $U(1)$ in the total group $H$ is the \emph{double cover} of that $U(1)$ in the quotient group $G$.
The boundary field theory could be
\bea 
\sum_{{b \in C^4( (\partial M)^6, \Z_2),}\atop{a\in C^1( (\partial M)^6, \Z_2)}} \exp( \ii  2 \pi \int_{(\partial M)^6} \frac{1}{2} \big( (b \delta a)  + w_2  w_3 a  +  b c_1  \big)).
\eea
The $C^d( \CM, \Z_n)$ contains all $d$-cochains of $\Z_n$ values assigned to a $d$-simplex on a triangulated manifold $\CM$.
Here $a$ is a 1-cochain and $b$ is a 4-cochain, both are integers with $\Z_2$ values. 
It is basically a 5+1D $Z_2$ gauge theory.
The ``gauge transformations'' are:
\bea 
w_2 \to w_2+ \delta \alpha,  \;  w_3 \to w_3+ \delta \beta,  \;  
  \lambda \equiv \alpha  \delta \beta + w_2   \beta +  \alpha w_3, \;  b \to b+\lambda, \;   c_1 \to c_1 + \delta \gamma, \;  a \to a- \gamma. \;\;
\;\;
\eea
Here $\lambda$, $\alpha$, $\beta$ and $\gamma$ are 4-cochain, 1-cochain, 2-cochain and 1-cochain respectively, all in $\Z_2$ values.
Effectively we view the normalized $U(1)$ probed gauge field $A/(2\pi)$ as a $\mathbb{R}$-valued 1-cochain $\tilde{A}$, such that 
the first Chern class $c_1 = \delta \tilde{A}$ becomes an integral 2-cochain on a triangulated manifold, 
so $c_1 \to c_1 + \delta \gamma$. 
We have the gauge transformation $w_2 w_3 \to w_2 w_3 + \delta \lambda = w_2 w_3 + w_2  \delta \beta + \delta \alpha w_3 +  \delta \alpha \delta \beta$, 
because the SW classes satisfy $\delta w_2 =\delta w_3=0$.
The whole partition function with bulk and boundary theories together is gauge invariant.
Since both $a$ and $b$ are $\Z_2$-valued cochains, coupled to $w_2, w_3$ and $c_1$ of the background $U(1)$ probed fields,
we can regard the 5+1D surface theory as a $Z_2$ gauge theory.

\subsection{ { 2+1D/1+1D Bosonic topological insulator} $1 \to Z_2^K \to U(1) \rtimes Z_2^T \to U(1) \rtimes Z_2^T \to 1$
and 2+1D/1+1D Bosonic topological superconductor of $Z_2^K \rtimes Z_2^T$:
Spontaneous $G$-symmetry breaking of boundary deconfined $K$-gauge theory}

\label{2+1DU1Z2T}

The bosonic SPTs with symmetry group $G=U(1)  \rtimes Z_2^T$ is called a bosonic topological insulator (BTI).
In 2+1D, we can obtain these SPTs from the group cohomology $\cH^{3}(U(1) \rtimes Z_2^T,U(1))= \Z_2$.
Let us focus on the nontrivial $\Z_2$ class, the bulk field theory on a 3-manifold ${M^3}$ is described by \cite{Wen1410.8477, K1459}
\bea 
\exp( \ii  2 \pi \int_{M^3} \frac{1}{2} w_1 \frac{dA}{2 \pi})= \exp( \ii  2 \pi  \int_{M^3} \frac{1}{2} w_1 c_1).
\eea
The boundary field theory is described by
\bea 
\sum_{{\phi \in C^0( (\partial M)^2, \Z_2),}\atop{a\in C^1( (\partial M)^2, \Z_2)}} \exp( \ii  2 \pi  \int_{(\partial M)^2} \frac{1}{2} (\phi \delta a+ w_1 a+ \phi c_1)),
\eea
where $\phi$ is a 0-cochain and $a$ is a 1-cochain, both in $\Z_2$ values. 
The ``gauge transformations'' are:
\bea 
w_1 \to w_1 + \delta \alpha, \;  \phi \to \phi+ \alpha, \;   c_1 \to c_1 + \delta \gamma, \;  a \to a- \gamma.  
\eea
Here $\alpha$ and $\gamma$ are 0-cochain and 1-cochain in $\Z_2$ values.
The $c_1$ is an integral 2-cochain defined as the same as in the previous Appendix \ref{6+1DU1U1}.
The boundary theory shows a $K=Z_2$ gauge theory in 1+1D coupled to $w_1$ and $c_1$.
In terms of $U(1)$-field $A$, we have the gauge transformation
$A \to A + 2 \pi \gamma$. This establishes our construction:
$$1 \to Z_2^K \to U(1) \rtimes Z_2^T \to U(1) \rtimes Z_2^T \to 1.$$

For this $Z_2^K$ gauge theory, there are a few topologically distinct sectors and gauge-invariant operators, shown in Table \ref{table:w1c1}: 
(1) The trivial sector is 1, with 
trivial quantum number $U(1)$ charge 0 and $T=+1$.
(2) The $Z_2^K$ gauge charge as $e$-sector corresponds to the line operator $e^{i \pi \int (a+ \frac{A}{2\pi})}$. 
Each of two ends of such an open line  $e^{i \pi \int_{x_1}^{x_2} (a+ \frac{A}{2\pi})}$ has an $e$-particle ($Z_2^K$ gauge charge $e$). 
Each of two ends  must attach with a $1/2$ $U(1)$ charge, due to its attachment to $U(1)$-field $A$.
Thus, the $e$-particle has quantum number $U(1)$ charge $1/2$ and $T=+1$.
(3) The $Z_2^K$ gauge flux as $m$-sector corresponds to the line operator $e^{i \pi (\phi (x_1)-\phi (x_2)+ \int_{x_1}^{x_2} w_1)}$,
where the vortex $e^{i \pi \phi}$ is an $m$-instanton insertion operator.
Similarly, each of the two ends of the open line must attach with an $m$ instanton with an eigenvalue of $T=-1$, due to $w_1$.
The $m$ instanton has a trivial eigenvalue of $U(1)$, namely 0.

\begin{table}[h!]
\begin{center}
\begin{tabular}{| c | c | c | c | }
\hline
Operators & Sectors (fractional objects) & $U(1)$ charge & {$T$ eigenvalue} \\ 
\hline
 1  & Trivial (none) & 0 & 1 \\ 
 \hline
 $e^{i \pi \int (a+ \frac{A}{2\pi})}$ & $Z_2$ gauge charge ($e$ particle)  & 1/2 & 1 \\  
 \hline
$e^{i \pi (\phi (x_1)-\phi (x_2)+ \int_{x_1}^{x_2} w_1)}$ & $Z_2$ gauge flux ($m$ instanton) & 0 & -1   \\ 
 \hline
\end{tabular}
\end{center}
\caption{The quantum numbers ($U(1)$ charge and $T$) of the $U(1)$ symmetry and $Z_2^T$ time reversal symmetry here are meant to associated to 
$e$-particle local excitations and $m$-instanton (the second column), \emph{not} to the entire line operators (the first column).}
\label{table:w1c1}
\end{table}

If we put either 2+1D SPTs on a spatial disk with a circular boundary,
and if the boundary $Z_2$ gauge theory is deconfined,
there are two fold degenerate ground states, labeled by a trivial (no) holonomy and a nontrivial holonomy of $Z_2$ gauge charge ($e$ particle)
winding an odd number of times, along the circular boundary.

Note that the SPTs with a smaller symmetry group $Z_2 \rtimes Z_2^T$ also renders the same class, due to $\cH^{3}(Z_2 \rtimes Z_2^T,U(1))= (\Z_2)^2$ ---
one of $\Z_2$ class coincides with $\cH^{3}(U(1) \rtimes Z_2^T,U(1))=\Z_2$.
The SPT invariant for that $\Z_2$ class in $\cH^{3}(Z_2 \rtimes Z_2^T,U(1))= (\Z_2)^2$ is 
\bea
\exp( \ii  2 \pi  \int_{M^3} \frac{1}{2} w_1 (a_1)^2),
\eea 
with
a  $\Z_2$-valued 1-cochain $a_1$.
This implies that the boundary physics of 2+1D $U(1) \rtimes Z_2^T$ SPTs can be understood in terms of that of 2+1D $Z_2 \rtimes Z_2^T$ SPTs. 
Even if the Coleman-Mermin-Wager theorem protects the continuous $U(1)$-symmetry against spontaneous symmetry breaking,
we may break $U(1)$ explicitly down to $Z_2$. The same physics is valid for both $U(1) \rtimes Z_2^T$ BTI and $Z_2 \rtimes Z_2^T$ SPTs.

{
For the $K=Z_2$ deconfined gauge theory on the 1+1D boundary of the above $U(1) \rtimes Z_2^T$ and $Z_2 \rtimes Z_2^T$ SPTs, 
we should have no spontaneous symmetry breaking, \emph{neither} on the $U(1)$ 
(supposing that Coleman-Mermin-Wager theorem still holds)
\emph{nor} on the $Z_2$ (because $U(1) \rtimes Z_2^T$ SPTs and $Z_2 \rtimes Z_2^T$ SPTs have the same physics).
It is likely that the boundary has spontaneous symmetry breaking on the time-reversal symmetry $Z_2^T$.
Below we provide arguments to support that the time-reversal symmetry $Z_2^T$ is spontaneously broken at the boundary.
}

\subsection{ { Spontaneous global symmetry breaking of boundary $K$-gauge theory}: 
$Z_2^G$-symmetry breaking on 2+1D $Z_2$-SPT's boundary
v.s. $Z_2^T$-symmetry breaking on 2+1D $U(1) \rtimes Z_2^T$-SPT's and $Z_2 \rtimes Z_2^T$-SPT's boundaries for $K=Z_2^K$.}

\label{sec:SSB_field-theory}

Here we like to show that 1+1D deconfined $K$-gauge theories with symmetry $G$ on the boundary of 2+1D bulk $G$-SPTs
can actually be spontaneous global $G$-symmetry breaking states. 
Some examples are in order.

\begin{enumerate}
\item
Our first example is already mentioned in the main text, Sec.~\ref{bdry3w}, as well as Appendix~\ref{sec:deconfined-to-SSB} and \ref{sec:examplesHZ4GZ2}.
Consider the 1+1D boundary of 2+1D $Z_2$-SPTs under the construction $0 \to {Z}_2^K \to Z_4^H  \to Z_2^G \to 0$.
This $\Z_2$-valued 3-cocycle of bulk SPTs is equivalent to 
$\ee^{\ii 2 \pi \int \frac{1}{2} a_1 \cup a_1 \cup a_1}=(-1)^{\int  a_1 \cup a_1 \cup a_1}$ with a cup product form of $a_1 \cup a_1 \cup a_1$, in $\cH^3( Z_2,U(1))$.
The $a_1$ is a $\Z_2$ valued 1-cochain. 
Through a field theory analysis, we can find a gauge-invariant partition function for the bulk on $M^3$ and boundary on $(\partial M)^2$. 
The boundary $Z_2^K$ gauge theory has a minimal coupling to the bulk fields, and its partition function is
\bea
\sum_{{\phi \in C^0( (\partial M)^2, \Z_2),}\atop{a\in C^1( (\partial M)^2, \Z_2)}} \exp( \ii  2 \pi  \int_{(\partial M)^2} \frac{1}{2} (\phi \delta a+ \phi  (a_1)^2+  a a_1)).
\eea
Here $\phi$ and $a$ are $\Z_2$ valued 0-cochain and 1-cochain fields respectively. 
The boundary  has a spin-1 electric gauge charge excitation associated to the $a$,
and a spin-0 magnetic instanton associated to the $\phi$. The gauge-invariant vortex operator has a nonzero vacuum expectation value with respect to ground states:
\bea \label{eq:vortex-vev-Z2}
\langle e^{i \pi (\phi (x_1)-\phi (x_2)+ \int_{x_1}^{x_2} a_1)} \rangle = \langle \Psig | e^{i \pi (\phi (x_1)-\phi (x_2)+ \int_{x_1}^{x_2} a_1)} | \Psig\rangle = \text{const.}
\eea
The const.~stands some constant value.
This statement shows the same physics as eqn.(\ref{zerof})'s 
$\langle \Psig (\pm) |X_{i+1/2}X_{j+1/2}|\Psig (\pm) \rangle = 1$.
The spin-0 vortex operator that is odd under $Z_2^G$-symmetry has a real expectation value,
and its two-point function develops a long-range order.
This implies that $Z_2^G$-symmetry is violated.
Thus the ground states of $Z_2^K$-gauge theory have spontaneous $Z_2^G$-symmetry breaking.

\item The second example is the main example of Appendix \ref{2+1DU1Z2T},
 the 1+1D boundary of 2+1D $U(1) \rtimes Z_2^T$-SPTs under the construction
$1 \to Z_2^K \to U(1) \rtimes Z_2^T \to U(1) \rtimes Z_2^T \to 1$.
Again the gauge-invariant vortex operator (see Table \ref{table:w1c1}) has a nonzero vacuum expectation value with respect to ground states:
\bea \label{eq:vortex-vev-Z2T}
\langle  e^{i \pi (\phi (x_1)-\phi (x_2)+ \int_{x_1}^{x_2} w_1)} \rangle = \langle \Psig | e^{i \pi (\phi (x_1)-\phi (x_2)+ \int_{x_1}^{x_2} w_1)} | \Psig\rangle = \text{const.}
\eea
The  
vortex operator that is odd under $Z_2^T$-symmetry has a real expectation value,
and its two-point function develops a long-range order.
This implies that $Z_2^T$-symmetry is violated.
Thus the ground states have spontaneous $Z_2^T$-symmetry breaking.
For the third example, we can also show that the 1+1D boundary of 2+1D $Z_2 \rtimes Z_2^T$-SPTs under the construction
$1 \to Z_2^K \to Z_2 \rtimes Z_2^T \to Z_2 \rtimes Z_2^T \to 1$
has the same two-point function as eqn.(\ref{eq:vortex-vev-Z2T}) and develops a long-range order for
$Z_2^T$-symmetry-odd vortex operators. 
Thus the ground states of $Z_2^K$-gauge theory have spontaneous $Z_2^T$-symmetry breaking.

\end{enumerate}

To summarize, the above field theory analysis suggests that the ground states of 1+1D deconfined $K$-gauge theory of 2+1D $G$-SPTs
have spontaneous $G$-symmetry breaking.
We expect that both its deconfined gauge theory and confined gauge theory, both have spontaneous $G$-symmetry breaking,
with crossover to each other without phase transitions, similar to the physics in Appendix~\ref{sec:deconfined-to-SSB}.

\subsection{ { 1+1/0+1D Bosonic $1 \to Z_2 \to SU(2) \to SO(3) \to 1$}} \label{1+1DSU2SO3}

In 1+1D, we have a nontrivial bosonic SPT state predicted by $\cH^2( SO(3),U(1))=\mathbb{Z}_2$.
This nontrivial class is exactly a 1+1D Haldane spin chain protected by the global symmetry $SO(3)$. 
For example, it is well-known that the 1+1D Haldane SPT state
is the ground state of the AKLT spin chain Hamiltonian:
\bea
H = \sum_j \left(\frac{1}{2}\left(\vec{S}_j \cdot \vec{S}_{j+1} + \frac{1}{3} \left(\vec{S}_j \cdot \vec{S}_{j+1}\right)^2\right) + 1/3\right).
\eea
Each site $j$ has a Hilbert space of a spin-1 degree of freedom, and the spin-1 operator $\vec{S}_j$ acts on each site $j$. 
The particular choice of Hamiltonian 
prefers the lowest-energy ground state such that the spin-1 on each site splits to two spin-1/2 qubits,
and the neighbor spin-1/2 spins between two sites have a total spin-0 singlet pairing.
In a closed chain, we have a gapped state with a unique ground state.
In an infinite-size open chain, we have a gapped state with two dangling spin-1/2 qubits at the two ends,
where the two dangling spin-1/2 of a spin-0 singlet and three spin-1 triplet states become 4-fold degenerate.

However, we can lift the 4-fold degeneracy of  a 1+1D open chain by adding two spin-1/2 qubits at the two ends. 
Formally, this is achieved by trivializing the 2-cocycle of $\cH^2( SO(3),U(1))$ by lifting $SO(3)$ to $SU(2)$ via
\bea
1 \to Z_2 \to SU(2) \to SO(3) \to 1.
\eea
The bulk topological term 
$$(-1)^{\int w_2(V_{SO(3)})}
$$ 
of the second SW class of principal $G=SO(3)$-bundle (or the associated vector bundle $V_{SO(3)}$ of SO(3))
becomes trivial when we lift 
the $SO(3)$ to the $SU(2)$-bundle.
The unique gapped ground state state is achieved when 
we introduce the edge Hamiltonian term pairing each of the old dangling spin-1/2 qubits
to the two newly added spin-1/2 qubits, 
such that the low-energy ground state favors the singlet spin-0 pairing sectors 
at the two ends.\footnote{This procedure has been shown explicitly in Ref.\cite{2018arXiv180411236P} recently.}

The LHS technique in \ref{sec:LHS}
suggests that we look for
\bea
&& d_2:  \cH^{0}(G, \cH^{1}(K,U(1) )) \to  \cH^{2}(G, \cH^{0}(K,U(1) )) \nonumber\\
&&\Rightarrow d_2:    \cH^{0}(SO(3), Z_2)=\Z_2 \to \cH^{2}(SO(3), U(1))=\Z_2. \\
&&
f: G \to \cH^{1}(K,U(1) ) \Rightarrow
 SO(3) \to \cH^{1}(Z_2^K,U(1) )=\Z_2,
\eea
with a 1-cochain of a suggested base of $(-1)$.

%


\section{Symmetry-breaking gapped boundaries/interfaces: Comments and criteria}   
\label{sec:sym-breaking:bdry}

The main focus of article is a new approach to define gapped interface via ``symmetry-extension:''
On lifting $G$ to a larger group $H$, as described in Sec.~\ref{sec:sym-enhanced:bdryDW} and Appendix \ref{sec:examples}, 
that trivialize $G$-cocycle to define a lower dimensional gapped boundary prescribed by the split $H$-cochain. 
On the other hand, there is another more familiar approach for a gapped interface known in the literature, by ``symmetry-breaking.''
Namely, the global or gauge symmetries are spontaneously or explicitly broken, described in Sec.~\ref{relsym}. 
For a finite group $G$, 
when the symmetry-breaking does not produce gapless Goldstone bosons, the boundary can be gapped.
Phenomenologically, one can achieve symmetry-breaking through the Higgs effect or through interactions such as sine-Gordon cosine potentials.

{The global symmetry-breaking mechanism is well-known in the fields of topological insulators and SPTs. 
For example, we can add a ferromagnet on the boundary of topological insulators to break time-reversal global symmetry to obtain a gapped anomalous surface quantum Hall state.}
{The gauge symmetry-breaking mechanism is also known in the literature.
The gapped boundary/interface criteria studied by Haldane\cite{Haldane1995}, Kapustin-Saulina\cite{KapustinSaulina1008.0654}, Kitaev-Kong\cite{KitaevKong1104.5047}, Lan-Wang-Wen \cite{Wang1212.4863, Lan:2014uaa} 
and many others can be viewed as gauge symmetry-breaking
\cite{Wang1212.4863, Levin:2013gaa, 1305.7203BarkeshliJianQi, Lan:2014uaa} or the Anderson-Higgs effect.}

{In particular, let us look at the symmetry-breaking mechanism in 2+1D Abelian bulk topological phases for simplicity. 
The bulk phase can be described by
an Abelian Chern-Simons theory with an action
$S_{bulk}= \frac{K_{IJ}}{4\pi}\int  a_{I} \wedge d a_{J}$ under a symmetric integral bilinear matrix $K$ and locally some 1-form gauge fields $a$.
The usual gapless boundary action is a $K$-matrix Luttinger liquid or a doubled-version chiral boson theory
$S_{\partial}= \frac{1}{4\pi} \int dt \; dx \; (K_{IJ} \partial_t \Phi_{I} \partial_x \Phi_{J} -V_{IJ}\partial_x \Phi_{I}   \partial_x \Phi_{J})$
with a non-universal velocity matrix $V_{IJ}$ and some scalar modes $\Phi$.
The gapped boundary conditions 
can be achieved through a set of sine-Gordon cosine terms $\int dt \; dx\;  \sum_{a} g_{a}  \cos(\ell_{a,I}^{} \cdot\Phi_{I})$
as a strong coupling $g_{a} \gg 1$ limit. Notice that the gapping cosine term indeed breaks the symmetry of $\Phi_{I} \to \Phi_{I} + \eta$ for some constant $\eta$.
Here the broken symmetry can be global symmetry \cite{LuVishwanath1205} or gauge symmetry \cite{KapustinSaulina1008.0654, KitaevKong1104.5047, Wang1212.4863, Levin:2013gaa,
1305.7203BarkeshliJianQi}, depending on the context.} 

The simplest example is that  $G'=1$ is a trivial group containing only the identity element. And $G' \to G$ is a map that the identity in $G'$ maps to the identity in $G$.
This can be regarded as breaking $G$ to nothing in $G'$. There are $G$-cocycles assigned in the bulk, but
the boundary becomes a trivial cocycle/cochain $1$ in $G'$. In terms of the inhomogeneous cochain $\beta^{G'}_{d-1}=1$.
The $G$-cocycle $\omega^G_d(g_{01},\cdots,g_{d-1d})$ that touches any boundary link, say, $g'_{01}$, must have
$\omega^G_d( \iota(g'_{01})=1,\cdots, g_{d-1d} )=1$.
This type of boundary condition works for any bulk defined by any discrete group $G$ with any cocycle.
The usual way that one would describe it is that the $G$ is spontaneously broken to nothing along the boundary.   

More generally, the symmetry-breaking mechanism involves breaking a $G$-topological phases of group $G$ down to a subgroup $G'$:
\bea \label{inj2}
G'  \overset{\iota}{\rightarrow} G
\eea
viewed through the injective map $\iota$.
If $G'$ is a subgroup of $G$, then we can define the symmetry-breaking gapped boundary of $G$-topological phases, if
the $G'$-cocycle becomes a $G'$-coboundary (with a similar expression as in \Eqn{eq:trivializeHcoboundaryinhomo})
\begin{align}
 \omega^{G}_d(\iota(g'_{01}),\cdots,\iota(g'_{d-1d}))  & = \omega^G_d(g_{01},\cdots,g_{d-1d})  = \omega^{G'}_d(g'_{01},\cdots,g'_{d-1d}) \nonumber 
 = \delta  \beta^{G'}_{d-1},
\end{align}
thus split to lower ${(d-1)}$ dimensional $G'$ cochains. 
Formally, we mean that a nontrivial $G$-cocycle
\bea
\omega_d^G \in \cH^d(G,U(1))
\eea
becomes a trivial element $1$ (a coboundary) when it is pulled back (denoted as ${}^*$) to $G'$
\bea
1=\iota^* \omega_d^{G} \in \cH^d(G',U(1)).
\eea
The dimension of Hilbert space is \emph{restricted} from a $|G|$ per degree of freedom in the \emph{bulk} 
to a smaller $|G'|$ per degree of freedom on the \emph{boundary}.

As an application of Appendix \ref{sec:sym-breaking:bdry}, we will count and classify distinct gauge symmetry-breaking gapped interfaces in various dimensions (e.g. 2+1D bulk and 3+1D bulk), in Appendix \ref{sec:examples-gauge-sym-break}.

\section{Dynamically gauged gapped interfaces of topologically ordered gauge theories}

\label{sec:gauged-DW-interface}

Because gauge symmetry is \emph{not} a physical symmetry but only a gauge redundancy,
the physical meanings of \emph{gauge symmetry breaking} and \emph{gauge symmetry extension} are rather different from their 
global symmetry counterparts. 
We would like to re-interpret the dynamically gauged gapped interfaces for topologically ordered gauge theories (such that the whole systems are topologically ordered without any global symmetries)
more carefully in \emph{any} dimensions.

Let us propose the generic gauged gapped interfaces of topologically ordered gauge theories as follows. 
Let $L$ be the gauge group of gauged interface, let $G_{\I}$ and  $G_{\II}$ be the gauge groups of the left sector and right sector relative to the interface respectively.
Let $L$ be a group with a group homomorphism map to $G_{\I} \times G_{\II}$, 
\bea \label{eq:gauged-DW-interface}
L \to G_{\I} \times G_{\II}
\eea
such that the product of the two cocycles of the two twisted gauge theories on left and right pulls back to a trivial cocycle in $L$.
Here we assume \emph{neither} a surjective map (as the gauge symmetry extension) \emph{nor} an injective map (as the gauge symmetry breaking), but we only require the group homomorphism
for $L \to G_{\I} \times G_{\II}$. 
Therefore such a construction actually includes mixed mechanisms of gauge symmetry extension and gauge symmetry breaking, but we do not require any global symmetry at all.
In eqn.(\ref{eq:gauged-DW-interface}), we view $L$ and $G_{\I} \times G_{\II}$ all as gauge groups.

In Appendix~\ref{sec:gauge-sym-breaking-interface}, we explore applications of gauge symmetry-breaking gapped interfaces.
In Appendix~\ref{sec:gauging-sym-extend-interface}, we explore applications of gauge symmetry-extended gapped interfaces,
and we make a comparison to gapped interfaces obtained from, first constructing global symmetry extended SPTs, and then dynamically gauging the system 
with various gauging procedures. The two subsections aim to demonstrate the generality of this
eqn.(\ref{eq:gauged-DW-interface}) for generic gauged interfaces.

\subsection{Gauge symmetry-breaking gapped interface via Anderson-Higgs mechanism --- Examples:
2+1D twisted quantum double models $D^{\omega_3}(G)$ and 3+1D gauge theories and Dijkgraaf-Witten gauge theories}
\label{sec:gauge-sym-breaking-interface}
\label{sec:examples-gauge-sym-break}

The motivation for this subsection is to construct and count 
gauge-symmetry breaking gapped interfaces for gauge theories, and to compare to the known methods and known examples in the past literature (mostly studied in the 2+1D bulk).
Then we can check consistency and further produce new concrete examples
for gauge symmetry-breaking gapped interfaces in \emph{any dimension}. 
Many examples are shown in this Appendix.

We consider Dijkgraaf-Witten (DW) gauge theories \cite{DW9093},
namely topologically ordered discrete $G$-gauge theories that allow ``twists'' by the cohomology group cocycle.
For a more specialized case, 
a gauge symmetry-breaking gapped boundary, this repeats the same setup in eqn.(\ref{inj2}) that we used in Appendix \ref{sec:sym-breaking:bdry}.
We only rewrite
eqn.(\ref{eq:gauged-DW-interface}) as
$G' \to G \times 1$ 
with $L=G'$, $G_{\I}=G$, and $G_{\II}=1$.

More generally, our strategy to \emph{construct} and \emph{count} 
distinct topological gapped interfaces between two given twisted gauge theories of $G_{\I}$ and $G_{\II}$ in \emph{any dimension}, under Anderson-Higgs gauge-symmetry breaking, is:\footnote{JW thanks Tian Lan for collaborating on a different approach in 2+1D \cite{Lan:2014uaa}.}
\begin{itemize}

\item 1st step: For gauge-symmetry breaking gapped interfaces, we consider eqn.(\ref{eq:gauged-DW-interface}), with 
an additional constraint that $L \subseteq G_{\I} \times G_{\II}$
be an unbroken gauge subgroup. The criteria are (similar to Appendix \ref{sec:sym-breaking:bdry} except that every group is gauge group) that
${G_{\I} \times G_{\II}}$-cocycle $\omega^{G_{\I} \times G_{\II}}=\omega_{\I}^{G_{\I}}(g_{\I} ) \cdot \omega_{\II}^{G_{\II}}(g_{\II})^{-1}$ (allowed by K\"unneth formula) in 
$\cH^{d}({G_{\I} \times G_{\II}},U(1))$ 
becomes a coboundary $1 \in \cH^{d}(L,U(1))$ when we restricted $G_{\I}$ (on the left) and $G_{\II}$ (on the right) to $L$ on the interface.

\item 2nd step: To fully implement the first step, one has to actually pick a trivialization of the cocycle $\omega^{G_{\I} \times G_{\II}}$.  
The choice is not unique and we can modify it by appending any
cocycle in $\cH^{d-1}(L,U(1))$, corresponding to a topological $L$-gauge theory on the boundary/interface, 
 following Appendix \ref{sec:append-lower}. This yields distinct new gauged interfaces.

\item 3rd step: Some of the gauged interfaces, constructed by the above two steps, can be identified. For example,
two different gauge groups $L_1$ and $L_2$ on the interfaces (between the same pair of bulk gauge groups)
with cocycles $\omega^{L_1}_{d-1}$ and $\omega^{L_2}_{d-1}$ can be identified as the same gapped interface if and only if the two interfaces are conjugate 
through the adjoint action of $G_{\I} \times G_{\II}$ \cite{Ostrik0111139}.
Namely, some element $g \in G_{\I} \times G_{\II}$ identifies two interfaces by
$g L_1 g^{-1}= L_2$.

\item 4th step: To construct and count all gauge-symmetry breaking gapped interfaces, we consider all the possible subgroups $L \subseteq G_{\I} \times G_{\II}$,
and all possible lower-dimensional distinct gauge theories in $\cH^{d-1}(L,U(1))$, and we identify the equivalence classes of them as in the third step.

\end{itemize}

Many examples of gauge interfaces are provided below in Appendix \ref{sec:examples-gauge-sym-break},
including 2+1D $G=Z_2$ gauge theory (namely, the $Z_2$ toric code and $Z_2$ topological order),
2+1D $G=Z_2$ twisted gauge theory (namely, the $Z_2$ double semions, or $U(1)_{2} \times U(1)_{-2}$-fractional quantum Hall states),
and more generic 2+1D Dijkgraaf-Witten discrete gauge theories, also written as twisted quantum double models
$D^{\omega_3}(G)$ of a gauge group $G$ with a twisted 3-cocycle $\omega_3$ for $G=(Z_2)^3,D_4,Q_8$.
We also consider 3+1D Dijkgraaf-Witten gauge theories of a gauge group $G$ with a twisted 4-cocycle $\omega_4$.

We show that the \emph{gauge symmetry-breaking} mechanism reproduces the previous results
on gapped boundaries/interfaces of 2+1D topological orders, either through the anyon condensation method or through the tunneling matrices constructed through 
modular $\cS$ and $\cT$ data, especially showing consistency with \cite{Lan:2014uaa}.
Furthermore we can systematically obtain gapped interfaces in any dimension, such as in 3+1D.

\subsubsection{Gauge symmetry-breaking boundaries/interfaces of $Z_2$ toric code and $Z_2$ double-semion}

\begin{enumerate}

\item 
Consider a 2+1D $G_{\I}=G=Z_2$ gauge theory (namely, the $Z_2$ toric code and $Z_2$ topological order) on the left, and $G_{\II}=1$ as a trivial vacuum on the right.
The 3-cocycle on the left is a trivial coboundary $\omega^{G}_3(g)=1$ and the cocycle on the right is also 1, but the Hilbert spaces of the left and right sides are different.
We can consider either subgroups $L=G'=1$ or $L=G'=Z_2$ so that $G' \to G$ both provides a trivial cocycle when pulling back to $G'$.
The $G'=1$ and $G'=Z_2$  define the 
famous $e$-condensed or $m$-condensed gapped boundaries, achieved by Anderson-Higgs gauge-symmetry breaking.
The two $e$- and $m$- gapped boundaries have been constructed explicitly on the lattice Hamiltonian model \cite{KitaevKong1104.5047},
and have been realized field theoretically through strong coupling sine-Gordon interactions at boundaries 
\cite{Wang1212.4863}.
Follow Appendix \ref{sec:sym-breaking:bdry},
given a bulk Abelian Chern-Simons action with a $K= \left( \begin{smallmatrix} 0 & 2 \\ 2 & 0\end{smallmatrix} \right)$ matrix for $Z_2$ gauge theory,
the $e$- or $m$- gapped boundaries are achieved by strong coupling interactions 
$\int dt \; dx\;  g  \cos(2 \Phi_{1})$
and $\int dt \; dx\;  g  \cos(2 \Phi_{2})$, on a Luttinger liquid boundary, respectively \cite{Wang1212.4863}.  
See 
Table \ref{table:Z2-gauge-bdry} for the details of these 2 gapped boundaries.

\item
Consider a 2+1D $G=Z_2$ twisted gauge theory (namely, the $Z_2$ double semions, or $U(1)_{2} \times U(1)_{-2}$-fractional quantum Hall states) on the left, and $G'=1$ as a trivial vacuum on the right.
The 3-cocycle on the left is nontrivial $\omega^{G}_3(g) \neq 1$ and the cocycle on the right is 1; again, the Hilbert spaces of the left and right sides are different.
We can consider only the subgroups $G'=1$ so that $G' \to G$ both provides a trivial cocycle when pulling back to $G'$.
The $G'=1$ defines the  semion-anti-semion condensed gapped interface by Anderson-Higgs gauge symmetry-breaking. 
Follow Appendix \ref{sec:sym-breaking:bdry},
given a bulk Abelian Chern-Simons action with a $K= \left( \begin{smallmatrix} 2 & 0 \\ 0 & -2\end{smallmatrix} \right)$ matrix for $Z_2$ twisted gauge theory,
the gapped boundary is achieved by the strong coupling interaction 
$\int dt \; dx\;  g  \cos(2 (\Phi_{1}+\Phi_{2}) )$, on a Luttinger liquid boundary \cite{Wang1212.4863}.  
Again, this 
unique gapped interface is also realized and consistent with earlier work \cite{KitaevKong1104.5047, Wang1212.4863, Levin:2013gaa,
1305.7203BarkeshliJianQi}.
See 
Table \ref{table:Z2-gauge-bdry} for the data of a gapped boundary.

\begin{table}[!h]
\begin{center}
\begin{tabular}{c||c|c|c}
\hline
$Z_2$'s subgroup $G'$  &  $\cH^2(G',U(1))$ &\; $\begin{matrix}\text{$Z_2$ toric code}\\ \text{\# of gauge boundaries} \end{matrix}$ 
&\; $\begin{matrix}\text{$Z_2$ double-semion}\\ \text{\# of gauge boundaries} \end{matrix}$ \\ \hline
$\{1\}=1$ &  \;  0 & 1  & 1 \\ \hline
$Z_2$ &  \;  0 & 1 & 0\\ \hline
&  \;   &   2 (total number)  &   1 (total number)\\
\hline
 \end{tabular}
 \end{center}
\caption{Subgroup $G'$ of a $Z_2$, $\cH^2(G',U(1))$ and gauge-symmetry-breaking boundaries in 2+1D.
Our result reproduces and agrees with the classification in \cite{Wang1212.4863}'s Table III and in \cite{Lan:2014uaa}'s Appendix I and II.}
\label{table:Z2-gauge-bdry}
\end{table}

\item Consider a $Z_2$ toric code on the left and a  $Z_2$ double-semion model on the right, as an example for gauge symmetry-breaking gapped interface.
Eqn.(\ref{eq:gauged-DW-interface})
becomes $L \to Z_2 \times Z_2$
with a trivial coboundary $\omega^{G_{\I}}_3=1$ of $G_{\I}=Z_2$ on the left, and a nontrivial cocycle $\omega^{G_{\II}}_3$ of $G_{\II}=Z_2$ on the right,
and gauge symmetry-breaking results in Anderson-Higgs to $L=1$ or $L=Z_2$. This is consistent with two gapped interfaces between 
 the $Z_2$ toric code and $Z_2$ double semions found in \cite{Lan:2014uaa}.

\end{enumerate}

\subsubsection{Gauge symmetry-breaking boundaries of $D(D_4)=D^{\omega_{3, \III}}((Z_2)^3)$}

\begin{table}[!h]
\begin{center}
\begin{tabular}{c||c|c}
\hline
$D_4$'s  subgroup $G'$  &\;  $\cH^2(G',U(1))$ &\; 
$\begin{matrix} D(D_4)=D^{\omega_{3, \III}}((Z_2)^3)\\
\text{\# of distinct gauge boundaries} \end{matrix}$ \\\hline
$\{1\}=1$ &  \;  0 & 1 \\ \hline
$\{1,R^2\}=Z_2$ &  \;  0 & 1\\ \hline
$\{1,\tx\} = R\{1, \tx R^2\} R^{-1} =Z_2$  &  \;  0 & 1 \\ 
$\{1, \tx R\}=R \{1, \tx R^3\} R^{-1}=Z_2$  &  \;  0 & 1 \\\hline
$\{1,\tx, R^2, \tx R^2\}=(Z_2)^2$ &  \;  $\Z_2$ &   2 \\ 
$\{1, \tx R, R^2,  \tx R^3\}=(Z_2)^2 $ &  \;  $\Z_2$ &   2\\ \hline
$\{1,R, R^2, R^3\}=Z_4$ & \;  0 &   1\\ \hline
$D_4$ &  \; $\Z_2$  &  2 \\ \hline
&  \;   &   11 (total number)\\
\hline
 \end{tabular}
 \end{center}
\caption{Subgroup $G'$ of a dihedral $D_4$, $\cH^2(G',U(1))$ and gauge symmetry-breaking boundaries in 2+1D.
Our result reproduces and agrees with the classification in \cite{Lan:2014uaa}'s Appendix XI.}
\label{table:D4-gauge-bdry}
\end{table}

Here we consider a 2+1D twisted quantum double model $D^{\omega_{3, \III}}((Z_2)^3)=D(D_4)$.
It can be described by a twisted Abelian gauge theory under a Type III 3-cocycle ${\omega_{3, \III}}$ (see its definition in \cite{Wang1405.7689}), 
or a non-Abelian topological field theory action $\int ( (\sum_{I=1}^3 \frac{2}{2 \pi} B_I dA_I)+ \frac{1}{\pi^2} A_1 A_2 A_3)$.
Alternatively, we can regard it as a discrete $D_4$ gauge theory, with $D_4$ a dihedral group of order $8$.
Now we aim to count the distinct types of topological gapped boundaries based on gauge symmetry breaking.
Follow eqns.(\ref{inj2}) and (\ref{eq:gauged-DW-interface}), we choose $G_{\I}=G=D_4$ and $G_{\II}=1$.
What are the possible unbroken subgroup $L=G'$?
In Appendix \ref{sec:examples} Table \ref{table:D4}, we show the subgroup data for the $D_4$ group.
Since $D(D_4)$ is an untwisted gauge theory with a trivial 3-cocycle $1 \in \cH^3(D_4,U(1))$,
when we pull $1$ back from $D_4$ to any subgroup $G' \subseteq D_4$, it is still a 3-coboundary $1 \in \cH^3(G',U(1))$.
Among the 10 subgroups of $D_4$, four of $Z_2$ subgroups are identified to two sets of conjugate subgroups under the adjoint action \cite{Ostrik0111139}. 
For two $(Z_2)^2$ subgroups and one $D_4$, each of them offers two distinct gapped boundaries by appending lower-dimensional topological states due to $\cH^2(G',U(1))=\Z_2$.
Thus the total distinct gauge symmetry-breaking gapped interfaces are 11 types, which is consistent with topological gapped boundaries
obtained from a different approach via modular $\cS$ and $\cT$ data in 2+1D \cite{Lan:2014uaa}.
See 
Table \ref{table:D4-gauge-bdry} for the details of these 11 gapped boundaries.

\subsubsection{Gauge symmetry-breaking boundaries of $D(Q_8)=D^{\omega_{3, \III}{\omega_{3, \I}}}((Z_2)^3)$ in 2+1D
and $Q_8$ gauge theory in 3+1D}

Let us now consider gapped gauge interfaces of discrete quaternion $Q_8$ gauge theories in 2+1D and 3+1D.

\begin{table}[!h]
\begin{center}
\begin{tabular}{c||c|c|c}
\hline
$Q_8$'s  subgroup $G'$\; &\; $\cH^2(G',U(1))$\; &\; $\cH^3(G',U(1))$\; & 
$\begin{matrix}   \text{$Q_8$ gauge theories}\\ 
\text{\# of distinct gauge boundaries} \\
\text{2+1D $D(Q_8)$ v.s. 3+1D}
\end{matrix}$ \\ \hline
$\{1\}=1$ & 0 & 0 & \; 1 v.s. 1 \\ \hline
$\{1,-1\}=Z_2$  & 0 & $\Z_2$ & \; 1 v.s. 2\\ \hline
$\{1,i,-1,-i\}=Z_4$ & 0 & $\Z_4$ & \; 1 v.s. 4\\ 
$\{1,j,-1,-j\}=Z_4$ & 0 & $\Z_4$ & \; 1 v.s. 4\\ 
$\{1,k,-1,-k\}=Z_4$ & 0  & $\Z_4$ & \; 1 v.s. 4\\ \hline
$Q_8$ & 0  & $\Z_8$ & \; 1 v.s. 8\\\hline
&  \;   & &   6 v.s. 23 (total number)\\
\hline
 \end{tabular} 
 \end{center}
\caption{Subgroup $G'$ of a quaternion $Q_8$, $\cH^2(G',U(1))$, $\cH^3(G',U(1))$ and gauge symmetry-breaking boundaries in 2+1D and 3+1D.
Our 2+1D result reproduces and agrees with the classification in \cite{Lan:2014uaa}'s Appendix XII.
Our 3+1D result may be new to the literature.}
\label{table:Q8-gauge-bdry}
\end{table}

\begin{enumerate}

\item
First, we consider a 2+1D twisted quantum double model $D^{\omega_{3, \III}{\omega_{3, \I}}}((Z_2)^3) =D(Q_8)$.
It can be described by a twisted Abelian gauge theory under Type III and Type I 3-cocycles ${\omega_{3, \III}} \cdot {\omega_{3, \I}}$ 
(see its definition in \cite{Wang1405.7689}), 
or a non-Abelian topological field theory action $\int ( (\sum_{I=1}^3 \frac{2}{2 \pi} B_I dA_I)+ \frac{1}{\pi^2} A_1 A_2 A_3 + \frac{1}{2 \pi} A_1d A_1)$.
Alternatively, we can regard it as a discrete $Q_8$ gauge theory, with $Q_8$ a quaternion group of order $8$.
Now we count the distinct types of topological gapped boundaries based on gauge symmetry breaking.
Follow eqns.(\ref{inj2}) and (\ref{eq:gauged-DW-interface}), we choose $G_{\I}=G=Q_8$ and $G_{\II}=1$.
What are the possible unbroken subgroups $L=G'$?
In Appendix \ref{sec:examples} Table \ref{table:Q8}, we show the subgroup data for $Q_8$ group.
When we pull $1 \in \cH^3(Q_8,U(1))$ for untwisted $D(Q_8)$ back from $Q_8$ to any subgroup $G' \subseteq Q_8$, it is still a 3-coboundary $1 \in \cH^3(G',U(1))$.
Among the 6 subgroups of $Q_8$, none is identified under the adjoint actions. 
None of them can append lower-dimensional topological states due to $\cH^2(G',U(1))=0$.
Thus, the total distinct gauge symmetry-breaking gapped interfaces have 6 types, which is consistent with topological gapped boundaries
obtained from a different approach via modular $\cS$ and $\cT$ data in 2+1D \cite{Lan:2014uaa}.
See 
Table \ref{table:Q8-gauge-bdry}'s 4th column for the details of these 6 gapped boundaries.

\item Second, we consider a 3+1D $Q_8$ gauge theory.
For an untwisted gauge theory with a trivial 4-cocycle $1 \in \cH^4(Q_8,U(1))$,
when we pull $1$ back from $Q_8$ to any subgroup $G' \subseteq Q_8$, it is still a 4-coboundary $1 \in \cH^4(G',U(1))$.
After appending lower dimensional topological states, 
see 
Table \ref{table:Q8-gauge-bdry}'s 4th column, we find 23 gapped boundaries.

\end{enumerate}

\subsubsection{Gauge symmetry-breaking boundaries of $G=Z_2$ or $(Z_2)^2$ twisted gauge theories in 3+1D}

Consider 3+1D Dijkgraaf-Witten gauge theories of a gauge group $G=Z_2$ and $(Z_2)^2$  with twisted 4-cocycle $\omega_4$.
 
\begin{table}[!h]
\begin{center}
\begin{tabular}{c||c|c|c}
\hline
$\begin{matrix}\text{$G$'s}\\ 
\text{subgroup $G'$}
\end{matrix}$ &  $\cH^3(G',U(1))$ &\; $\begin{matrix}\text{3+1D $G=Z_2$ gauge theory}\\ \text{\# of gauge boundaries} \end{matrix}$ 
&\; $\begin{matrix}\text{3+1D $G=(Z_2)^2$ twisted DW theory}\\ \text{\# of gauge boundaries} \end{matrix}$ \\ \hline
$\{1\}=1$ &  \;  0 & 1  & 1 \\ \hline
$Z_2^{(a)}$ &  \;  $\Z_2$ & 2 & 2\\ \hline
$Z_2^{(b)}$ &  \;  $\Z_2$ &  & 2\\ \hline
$(Z_2)^2$ &  \;  $(\Z_2)^3$ &  & 0\\ \hline
&  \;   &   3 (total number)  &    5 (total number)\\
\hline
 \end{tabular}
 \end{center}
\caption{For $G=Z_2=Z_2^{(a)} $ or $G=(Z_2)^2=Z_2^{(a)} \times Z_2^{(b)}$, we list down the
subgroup $G'$, $\cH^2(G',U(1))$ and gauge symmetry-breaking boundaries in 3+1D.}
\label{table:Z2-gauge-bdry-3+1D}
\end{table}

\begin{enumerate}

\item First, we consider a 3+1D $Z_2$ gauge theory,
described by a low energy $BF$ action $\int \frac{2}{2 \pi} B dA$ with 2-form and 1-form fields $B$ and $A$.
Follow eqns.(\ref{inj2}) and (\ref{eq:gauged-DW-interface}), we choose $G_{\I}=G=Z_2$ and $G_{\II}=1$.
What are the possible unbroken subgroup $L=G'$?
Since it is a untwisted gauge theory with a trivial 4-cocycle $1 \in \cH^4(Z_2,U(1))$,
when we pull $1$ back from $Z_2$ to any subgroup $G' \subseteq Z_2$, it is still a 4-coboundary $1 \in \cH^4(G',U(1))$.
There are two types of boundaries realized by condensing the $Z_2$'s charge $e$-particle and condensing the $Z_2$'s flux $m$-string on boundaries.
These two boundaries are $e$- and $m$-gapped boundaries, analogs to that of the 2+1D $Z_2$ toric code.
However, we can append a lower-dimensional topological state due to $\cH^3(Z_2,U(1))=\Z_2$, thus we find 3 gapped boundaries, as  shown in 
Table \ref{table:Z2-gauge-bdry-3+1D}'s third column.

\item Second, we consider a 3+1D $(Z_2)^2$ twisted gauge theory,
described by a low energy $BF$ action $\int (\sum_{I=1}^2 \frac{2}{2 \pi} B_I dA_I)+ \frac{2}{(2\pi)^2} A_1 A_2 dA_2$ with 2-form and 1-form fields $B$ and $A$.
Follow eqns.(\ref{inj2}) and (\ref{eq:gauged-DW-interface}), we choose $G_{\I}=G=(Z_2)^2$ and $G_{\II}=1$.
What are the possible unbroken subgroup $L=G'$?
For a twisted gauge theory with a 4-cocycle $\cH^4((Z_2)^2,U(1))$,
only limited subgroups $G'$ trivialize the cocycle after pulling $G$ back to $G'$.
After appending lower dimensional topological states, we find 5 gapped boundaries, as shown in 
Table \ref{table:Z2-gauge-bdry-3+1D}'s fourth column.

\end{enumerate}

To summarize, in this section, we provide many gauge-symmetry breaking gapped interfaces, and detailed data.
We find consistency with results obtained in previous literature (in 2+1D), but we can systematically obtain gapped interfaces in any dimension, such as 3+1D.

\subsection{Comparison to gapped interfaces
obtained from dynamically gauging the symmetry extended SPTs}


\label{sec:gauging-sym-extend-interface}

In Appendix \ref{sec:examples}, we had summarized how to construct symmetry-preserving gapped boundary for SPTs via eqn.(\ref{eq:gext-sec:10})'s 
symmetry-extension $1 \to K \to H \overset{r}{\to} G \to 1$.
In this section, we would like to explore various ways to dynamically gauge this SPT system to obtain different topologically ordered gauge versions of the system, and make comparison
with the generic gauge interface construction in eqn.(\ref{eq:gauged-DW-interface})'s $L \to G_{\I} \times G_{\II}$. 
The goal is to demonstrate that the gauge interface construction from $L \to G_{\I} \times G_{\II}$
is general enough to contain different dynamical gauging procedure of SPT system.
To narrow down the possibilities of outcomes, here we like to fully gauge the left side SPTs of group $G$ to be a twisted gauge theory of group $G$, and to fully gauge the interface of group $H$. 
What remains are the different but consistent choices of gauging the right side of the interface.
This corresponds to eqn.(\ref{eq:gauged-DW-interface}), where we choose $G_{\I}=G$, $L=H$, and leave $G_{\II}$ free for different choices.
Below we provide several examples for the different choices of $G_{\II}$, and interpret the construction from both perspectives of
(a) gauging of the symmetry-extended SPTs, and (b) the gauge interface of topologically ordered gauge theory systems,
in a generic $d$-dimensional spacetime.

\begin{enumerate}

\item Consider $H \to G \times 1$, where we choose $L=H$, $G_{\I}=G$ and  $G_{\II}=1$ in eqn.(\ref{eq:gauged-DW-interface}). The group homomorphism $H \to G \times 1$ is surjective,
sending $h \in H$ to $(r(h),1)=(g,1) \in G \times 1$. 
From the gauging SPTs perspective of (a), the construction is obtained by first doing a local unitary transformation on the right sector to a trivial product state, 
which thus can be removed and regarded as a trivial vacuum. 
We only dynamically gauge the left sector $G$-SPTs and the $H$-interface to their gauge theory counterparts, namely the $G$-twisted gauge theory (of Dijkgraaf-Witten) in $d$-dimensions,
and $H$-gauge theory with a $G$-anomaly in a lower $(d-1)$-dimensions. 
But we do not gauge the right sector thus $G_{\II}=1$.
From the gauge theory perspective of (b), the $H \to G \times 1$ construction means that
we have a nontrivial inhomogeneous
${G \times 1}$-cocycle $\omega^{G \times 1}=\omega_{\I}^{G}(g ) \cdot \omega_{\II}^{1}(1)^{-1} =\omega_{\I}^{G}(g ) \cdot 1$ for the gauge theory, and that can be pulled back to
$H$ as lower dimensional $H$-cochains to construct the interface gauge theory.

\item Consider $H \to G \times G$, where we choose $L=H$, $G_{\I}=G$ and  $G_{\II}=G$ in eqn.(\ref{eq:gauged-DW-interface}). 
It is not surjective but only a group homomorphism from $h \in H$ to a diagonal group $(r(h),r(h))=(g,g)\in G \times G$. 
From the gauging SPTs perspective of (a), the construction is obtained by first doing a local unitary transformation on the right sector to a trivial product state.
The dynamically gauging procedure on the left sector and the interface is the same as in the previous case, but we also gauge the right sector to an untwisted usual $G_{\II}=G$-gauge theory.
From the gauge theory perspective of (b), the $H \to G \times G$ construction means that
we have a nontrivial inhomogeneous
${G \times G}$-cocycle $\omega^{G \times G}=\omega_{\I}^{G}(g ) \cdot 1$ for the gauge theory with $\omega_{\II}^{G}=1$, and that $\omega^{G \times G}$ can be pulled back to
$H$ as lower dimensional $H$-cochains to construct the interface gauge theory.

\item Consider $H \to G \times H$, where we choose $L=H$, $G_{\I}=G$ and  $G_{\II}=H$ in eqn.(\ref{eq:gauged-DW-interface}).  
It is not surjective to $G \times H$, but it has a group homomorphism from $h \in H$ to $(r(h),h)=(g,h)\in G \times H$. 
From the gauging SPTs perspective of (a), the construction is obtained by first doing a local unitary transformation on the right sector to a trivial product state.
The dynamically gauging procedure on the left sector and the interface is the same as the previous case, but we also gauge the right sector to a untwisted usual $G_{\II}=H$-gauge theory.
From the gauge theory perspective of (b), the $H \to G \times H$ construction means that
we have a nontrivial inhomogeneous
${G \times H}$-cocycle $\omega^{G \times H}=\omega_{\I}^{G}(g ) \cdot 1$ for the gauge theory with $\omega_{\II}^{H}=1$, and that $\omega^{G \times H}$ can be pulled back to
$H$ as lower dimensional $H$-cochains to construct the interface gauge theory.

\end{enumerate}

More concretely, for a specific example, we can choose $G=Z_2$ and $H=Z_4$; 
from the perspective of gauging 2+1D SPTs (a) from eqn.(\ref{eq:gext-sec:10}), we choose $1 \to Z_2^K \to Z_4^H \overset{r}{\to} Z_2^G \to 1$.
The above constructions have the following implications.
The first item above offers $Z_4^H \to Z_2^G \times 1$, which indicates that the left sector is a 2+1D $Z_2$ double-semion model (i.e. a twisted $Z_2$ gauge theory),
the interface is a 1+1D $Z_4$ gauge theory (with a $Z_2^G$ anomaly), while the right sector is a trivial vacuum (no gauge theory).
The second item above offers $Z_4^H \to Z_2^G \times Z_2^G$, which indicates the left sector is a 2+1D $Z_2$ double-semion model,
the interface is a 1+1D $Z_4$ gauge theory (with a $Z_2^G$ anomaly), while the right sector is a 2+1D $Z_2$ toric code (i.e., a $Z_2$ gauge theory).
The second item above offers $Z_4^H \to Z_2^G \times Z_4^H$, which indicates the left sector is a 2+1D $Z_2$ double-semion model,
the interface is a 1+1D $Z_4$ gauge theory (with a $Z_2^G$ anomaly), and the right sector is a 2+1D $Z_4$ gauge theory.

The above construction requires a group homomorphism map, and we additionally need to impose the zero gauge flux constraint (more precisely, zero gauge holonomy for a shrinkable loop) everywhere,
on the left sector, the interface and the right sector. The previous three examples in Appendix \ref{sec:gauging-sym-extend-interface} all satisfy these constraints.
However, other proposals may fail the constraints, for example, by considering $H \to G \times K$ for the gauge interface construction.
This $H \to G \times K$ requests for a construction of a $d$-dimensional $G$-twisted gauge theory on the left,
a $(d-1)$-dimensional $H$ gauge theory (with $G$-anomaly) on the interface,
and a $d$-dimensional untwisted usual $K$-gauge theory on the right --- Will this be a valid construction?
If we consider the $H \to G \times K$ map as
$h \to (r(h), k) =(g, k)$, then it is not a group homomorphism, and the zero gauge flux constraint on the closed loop sitting between the interface (in $H$) and the right sector (in $K$) 
is generally non-zero. Thus $H \to G \times K$ is \emph{illegal} for a gauge interface construction between a $G$-twisted gauge theory and a $K$-gauge theory,
\emph{at least} from the perspective (a) of dynamically gauging a global symmetry extended SPTs.

However, we can make $H \to G \times K$ work for a gapped interface, if we consider it as a group homomorphism $H \times 1 \to G \times K$,
so $(h, 1) \in H \times 1 \to (r(h), 1)  \in G \times K$. This implies that
we have a gauge symmetry-extended construction from the left sector $H \to G$, but a gauge symmetry-breaking construction from the right sector $1 \to K$.
In short, the \emph{mixed} symmetry-extension and symmetry-breaking construction can support an $H$-gauge interface between
a $G$-twisted gauge theory on the left and a untwisted usual $K$-gauge theory on the right.

Overall, we show that the perspective (a) of gauging global symmetries of SPTs 
is within the construction of the perspective (b) of gauge interfaces of gauge theories based on eqn.(\ref{eq:gauged-DW-interface}).
This supports the generality of eqn.(\ref{eq:gauged-DW-interface}).

\bibliographystyle{arXiv_new}
\bibliography{SPTgb_ref_new,all,mybib,publst} 

\end{document}